\documentclass[a4paper,12pt]{report}
\usepackage{Stylefile}

\onehalfspace

\title
 {
\vspace*{1.0cm}
\LARGE{\bf Atomic and Molecular Aspects of Astronomical Spectra} \vspace*{3.0cm} \\
\Large{\bf Taha Sochi} \vspace*{4.0cm} \\
\large{\bf  A dissertation submitted to the \\
            Department of Physics \& Astronomy \\
            University College London \\
            in fulfilment of the requirements for the degree of \\
            Doctor of Philosophy}
 }

\date{September 2012}

\ifpdf
\fi

\makeindex

\begin{document}

\maketitle %

\phantomsection \addcontentsline{toc}{section}{Declaration}
%
%
\noindent
{\LARGE \bf \vspace{1.0cm} \\ Declaration} \vspace{0.5cm}\\
\begin{spacing}{2}
\noindent
The work presented in this thesis is the result of my own effort, except where otherwise stated.
All sources are acknowledged by explicit reference.

\noindent Taha Sochi .........................

\end{spacing}

\phantomsection \addcontentsline{toc}{section}{Abstract}
\noindent
{\LARGE \bf \vspace{3.0cm} \\ Abstract} \vspace{0.5cm}\\ 
\begin{spacing}{1.5}
\noindent

In the first section of this thesis, we present the atomic part of our investigation. A C$^{2+}$
atomic target was prepared and used to generate theoretical data required in the investigation of
recombination lines that arise from collisions between electrons and ions in thin plasma found in
planetary nebulae and other astrophysical objects. The \Rm-matrix method of electron scattering
theory was used to describe the C$^{2+}$ plus electron system. Theoretical data concerning bound
and autoionizing states were generated in the intermediate-coupling approximation by \Rm-matrix and
Autostructure codes and compared to experimental data. The comparison revealed very good agreement.
These theoretical data were then used to generate dielectronic recombination data for C$^+$ which
include transition lines, oscillator strengths, radiative transition probabilities, as well as
emissivity and \dierec\ coefficients. The data were cast in the form of a line list, called \SSo,
containing 6187 optically-allowed transitions which include many C\II\ lines observed in
astronomical spectra. The data were validated by comparison to C$^+$ recombination data obtained
from a number of sources in the literature. This line list was used to analyze the spectra from a
number of astronomical objects, mainly planetary nebulae, and identify their electron temperature
where the observational data were obtained from the literature. The electron temperature
investigation was also extended to include free electron energy analysis which uses observational
and theoretical data of FF and FB transitions to investigate the long-standing problem of
discrepancy between the results of recombination and forbidden lines analysis and its possible
connection to the electron distribution (Maxwellian or non-Maxwellian). In the course of this
investigation two elaborate methods, one for finding and analyzing resonances (\Km-matrix method)
and the other for analyzing and identifying electron temperature from astronomical spectra (least
squares minimization), were employed. A computer program for atomic transition analysis was also
developed and used as the main tool for obtaining the line list and analyzing the observational
spectra.

In the second section of the thesis we present the results of our molecular investigation; the
generation of a comprehensive, calculated line list of frequencies and transition probabilities for
the singly deuterated isotopologue of \htp, \htdp. The line list, which is the most comprehensive
one of its kind in existence, contains over 22 million rotational-vibrational transitions occurring
between more than 33 thousand energy levels and covers frequencies up to 18500 \wn. All energy
levels with rotational quantum number, $J$, up to 20 are considered, making the line list useful
for temperatures up to at least 3000~K. About 15\% of these levels are fully assigned with
approximate rotational and vibrational quantum numbers. The list is calculated using a previously
proposed, highly accurate, \abin\ model implemented in a high-accuracy computer code based on a
two-stage discrete variable representation (DVR) approach. Various consistency checks were carried
out to test and validate the results. All these checks confirmed the accuracy of the list. A
temperature-dependent partition function, valid over a more extended temperature range than those
previously published, and cooling function are presented. Temperature-dependent synthetic spectra
for the temperatures $T=100, 500, 1000$ and 2000~K in the frequency range 0-10000 \wn\ were also
generated and presented graphically.

\end{spacing}

\phantomsection \addcontentsline{toc}{section}{Acknowledgements and Statements} %
\noindent
{\LARGE \bf \vspace{3.0cm} \\ Acknowledgements and Statements} \vspace{0.5cm}\\ 
\begin{spacing}{1.5}
\noindent %

I would like to thank my supervisors Professor Peter Storey and Professor Jonathan Tennyson for
their kindness, guidance and advice. It is a great privilege and gratifying experience to work with
them over several years and learn from their deep knowledge and vast expertise in atomic, molecular
and astro physics.

I also would like to thank the external examiner Professor Nigel Badnell from the University of
Strathclyde, and the internal examiner Dr Sergey Yurchenko from the University College London for
their constructive remarks and corrections and valuable recommendations for possible future work.

Apart from \RMAT\ and \AS\ atomic codes and DVR3D molecular code which were used as the main
computing tools, I also used \RESMESH\ and another piece of code from Pete to calculate
$\OS$-values from \pcs s. The rest of tools and techniques belong to me. Anyway, I assume full
responsibility for any error or failure.

The emphasis in this thesis is on the practical aspects of my work in an attempt to provide a full
and credible account of how this investigation was conducted and how the conclusions were reached.
Bare minimum of theory will be given when necessary for motivation and background as the theory is
readily available in the literature and mainly belongs to other contributors.

Finally, I should acknowledge the encouragement and motivation which I received from family and
friends while doing this PhD with special tribute to my wife.

\vspace{1cm}

The following articles are based on results and activities related directly to the research
presented in this thesis:

\begin{itemize}

\item
Taha Sochi. Emissivity: A program for atomic transition calculations. Communications in
Computational Physics, 7(5):1118-1130, 2010.

\item
Taha Sochi; Jonathan Tennyson. A computed line list for the \htdp\ molecular ion. Monthly Notices
of the Royal Astronomical Society, 405(4):2345-2350, 2010.

\item
Taha Sochi; Peter J. Storey. Dielectronic Recombination Lines of C$^+$. Atomic Data and Nuclear
Data Tables (Accepted), 2012.

\item
Peter J. Storey; Taha Sochi. Electron Temperature of Planetary Nebulae from C\II\ Recombination
Lines. Submitted, 2012.

\end{itemize}

\end{spacing}

\phantomsection \addcontentsline{toc}{section}{Contents} %
\tableofcontents

\newpage
\phantomsection \addcontentsline{toc}{section}{List of Figures} %
\listoffigures

\newpage
\phantomsection \addcontentsline{toc}{section}{List of Tables} %
\listoftables

\newpage
\phantomsection \addcontentsline{toc}{section}{Nomenclature}
\noindent \vspace{1.0cm} \\
{\LARGE \bf Nomenclature and Notation}
\vspace{0.7cm} 

\begin{supertabular}{ll}
%
%
%

$\alpha$        &       indicial parameter in spherical oscillator \\
$\FSC$          &       \finstr\ constant ($= \ec^{2}/(\Dirac \C \, 4\pi \PFS) \simeq 7.2973525376 \times 10^{-3}$) \\
$\aTP$          &       autoionization transition probability (s$^{-1}$) \\
$\rTP$          &       radiative transition probability (s$^{-1}$) \\
$\rTPul$        &       radiative transition probability from upper state $u$ to lower state $l$ (s$^{-1}$) \\
$\EPS$          &       eigenphase sum \\
$\EPB$          &       eigenphase background \\
$\EPGK$         &       eigenphase gradient of \Km-matrix with respect to energy (J$^{-1}$) \\
$\FWHMe$        &       full width at half maximum for energy distribution (J) \\
$\RW$           &       width of resonance (J) \\
$\ED$           &       energy difference (J) \\
$\epsilon$      &       energy of resonance (J) \\
$\ECE$          &       energy of colliding electron (J) \\
$\PFS$          &       permittivity of vacuum ($\simeq 8.854187817 \times 10^{-12}$F.m$^{-1}$) \\
$\EMISS$        &       emissivity of transition line (J.s$^{-1}$.m$^{-3}$) \\
$\varepsilon^{nt}$ &       normalized theoretical emissivity \\
$\ENLul$        &       emissivity of transition line from state $u$ to state $l$ (J.s$^{-1}$.m$^{-3}$) \\
$\EffChar$      &       effective charge (C) \\
$\degFree$      &       number of degrees of freedom \\
$\elefun$       &       function of colliding electron \\
$\CF$           &       Coulomb function \\
$\kappa$        &       characteristic parameter of $\kappa$ distribution \\
$\WL$           &       wavelength (m) \\
$\WL_{lab}$     &       laboratory wavelength (m) \\
$\WL_{v}$       &       vacuum wavelength (m) \\
$\EV$           &       eigenvalue \\
$\mu$           &       reduced mass (kg) \\
$\F$            &       frequency (s$^{-1}$) \\
$\STOxii$       &       exponent in \STO\ formulation \\
$\Par$          &       parity \\
$\RCc$          &       \reccoe\ for capture process (m$^{3}$.s$^{-1}$) \\
$\RCf$          &       effective \reccoe\ (m$^{3}$.s$^{-1}$) \\
$\PICS$         &       \pcs\ (m$^{2}$) \\
$\LTS$          &       lifetime of state (s) \\
$\ATU$          &       atomic time unit ($= \Dirac / \Eh \simeq 2.418884326505 \times 10^{-17}$s) \\
$\vecofu$       &       vector coupled function \\
$\BSTfun$       &       function of bound-state type in \clocou\ expansion \\
$\GoF$          &       goodness-of-fit index \\
$\spiang$       &       spin and angle functions of colliding electron \\
$\tarfun$       &       function of $N$-electron target \\
$\WF$           &       wavefunction of \Schrodinger\ equation \\
$\omega_e$      &       fundamental frequency in Morse and spherical oscillators (s$^{-1}$) \\
 \vspace{-0.3cm} \\
%
%
$\BR$           &       Bohr radius ($\simeq 5.2917720859 \times 10^{-11}$m) \\
$\ansy$         &       anti-symmetrization operator \\
$\IRR$          &       inner region radius in \Rm-matrix formulation (m) \\
$\EAC$          &       Einstein A coefficient (s$^{-1}$) \\
{\AA}           &       angstrom \\
$\DC$           &       \depcoe\ \\
$\DCu$          &       \depcoe\ of state $u$ \\
$\EBC$          &       Einstein B coefficient (m.kg$^{-1}$) \\
\Bm             &       matrix in \QB\ method \\
$\C$            &       speed of light in vacuum (299792458 m.s$^{-1}$) \\
$\CCcoef$       &       coefficient in \clocou\ expansion \\
$\STOCi$        &       coefficient in \STO\ formulation \\
$\Cp$           &       contribution to polarizability \\
$\DO$           &       \dipope \\
$D_e$           &       dissociation energy in Morse oscillator (J) \\
$\ec$           &       elementary charge ($\simeq 1.602176487 \times 10^{-19}$C) \\
$\ele$          &       electron \\
$\E$            &       energy (J) \\
$\Eh$           &       \Hartree\ energy ($\simeq 4.35974394 \times 10^{-18}$J) \\
$\Ei$           &       energy point in \Km-matrix method (J) \\
$\Eo$           &       position of \Km-matrix pole on energy mesh (J) \\
$\Ep$           &       photon energy (J) \\
$\Er$           &       energy position of resonance (J) \\
$E_{x}$         &       characteristic energy of Druyvesteyn distribution \\
$\E_{\kappa}$   &       characteristic energy of $\kappa$ distribution \\
$\OS$           &       \oscstr \\
$\OSul$         &       \oscstr\ of transition from state $u$ to state $l$ \\
$\radfun$       &       radial function \\
$F$             &       energy flux (J.s$^{-1}$.m$^{-2}$) \\
$\radfunV$      &       vector of radial functions \\
$\PKF$          &       parameter in \Km-matrix method (J) \\
$\SW$           &       statistical weight in coupling schemes ($=2J+1$ for IC) \\
$\nsdf$         &       nuclear spin degeneracy factor \\
$\h$            &       \Planck's constant ($\simeq \Sci {6.6260693}{-34}$ J.s) \\
$\Dirac$        &       reduced \Planck's constant ($= h/2\pi \simeq \Sci {1.0545717}{-34}$ J.s) \\
$\Ham$          &       \Hamiltonian\ of the whole system \\
$\Hbp$          &       \BrePau\ \Hamiltonian \\
$\Hnr$          &       non-relativistic \Hamiltonian \\
$\Hrc$          &       relativistic correction \Hamiltonian \\
$\iu$           &       imaginary unit ($\sqrt{-1}$) \\
$I$             &       de-reddened flux (J.s$^{-1}$.m$^{-2}$) \\
$\IM$           &       identity matrix \\
$I^{no}$        &       normalized observational flux \\
$J$             &       total angular momentum quantum number \\
$J$             &       rotational quantum number \\
\BOLTZ          &       \Boltzmann's constant ($\simeq \Sci {1.3806505}{-23}$ J.K$^{-1}$) \\
$k$             &       the projection of $J$ along the body-fixed $z$-axis \\
\Km             &       reactance matrix in \Rm-matrix theory \\
$\K$            &       single-element \Km-matrix \\
$\SVKi$         &       scalar value of single-element \Km-matrix at energy $\Ei$ \\
$\BGCK$         &       background contribution to \Km-matrix \\
$l$             &       orbital angular momentum quantum number \\
$L$             &       total orbital angular momentum quantum number \\
$\m$            &       mass (kg) \\
$\me$           &       mass of electron ($\simeq 9.10938215 \times 10^{-31}$~kg) \\
\Mm             &       lifetime matrix \\
$\M$            &       single-element \Mm-matrix \\
$\Mmax$         &       maximum value of single-element \Mm-matrix \\
$\QN$           &       \priquanum \\
$\EQN$          &       \effquanum \\
$\ND$           &       number density (m$^{-3}$) \\
$\NDe$          &       number density of electrons (m$^{-3}$) \\
$\NDi$          &       number density of ions (m$^{-3}$) \\
$\SP$           &       \Saha\ population (m$^{-3}$) \\
$\NO$           &       number of observations \\
$\NP$           &       number of fitting parameters \\
$\pf$           &       partition function \\
$P_{nl}(r)$     &       radial function in \STO\ formulation \\
$\STOPi$        &       power of $r$ in \STO\ formulation \\
\Qm             &       matrix in \QB\ method \\
\Rm             &       resonance matrix in \Rm-matrix theory \\
$r$             &       radius (m) \\
$\RV$           &       radius vector \\
$r_1$           &       diatomic distance in triatomic coordinate system (m) \\
$r_2$           &       distance of third atom from the diatom center of mass (m) \\
$r_e$           &       equilibrium distance in Morse oscillator (m) \\
\Sm             &       scattering matrix in \Rm-matrix theory \\
$\Ss$           &       single-element \Sm-matrix \\
$S$             &       total spin angular momentum quantum number \\
$\Ls$           &       \linstr\ (J$^{-1}$) \\
$\ti$           &       time (s) \\
$\T$            &       temperature (K) \\
$\ElecT$        &       electron temperature (K) \\
$v$             &       vibrational quantum number \\
$\MPP$          &       multipole potentials (J.C$^{-1}$) \\
$\wei$          &       statistical weight in least squares procedure \\
$x$             &       characteristic parameter of Druyvesteyn distribution \\
$\cf$           &       cooling function (J.s$^{-1}$) \\
$\IECN$         &       ion of effective positive charge $n$ \\
$\BSRI$         &       excited bound state of ion with effective positive charge $n-1$ \\
$\DEAIS$        &       doubly-excited state with effective positive charge $n-1$ \\
$\ycal$         &       calculated value \\
$\yobs$         &       observed value \\
$\Rc$           &       residual charge \\
$\AN$           &       atomic number \\
$\nabla$        &       del operator \\
$\nabla^{2}$    &       Laplacian operator \\
$\verb|      |$ & \\
\end{supertabular}

\VS

{\LARGE \bf \noindent Abbreviations} \vspace{0.3cm} \\
\begin{supertabular}{ll}

[.]$^{*}$ \verb|   |     &       complex conjugate of matrix \\
$'$             &       upper \\
$''$            &       lower \\
{\AU}           &       atomic unit \\
BB              &       Bound-Bound \\
BJ              &       Balmer Jump \\
C               &       \Coulomb \\
CCD             &       Charge-Coupled Device \\
CEL             &       Collisionally-Excited Lines \\
DR              &       \DieRec\ \\
DVR             &       Discrete Variable Representation \\
\eV             &       electron-Volt \\
F               &       Faraday \\
FB              &       Free-Bound \\
FF              &       Free-Free \\
\fwhm           &       Full Width at Half Maximum \\
(.)$_f$         &       final \\
GHz             &       Gigahertz \\
IC              &       \IntCou \\
IUE             &       International Ultraviolet Explorer \\
(.)$_i$         &       initial \\
J               &       Joule \\
K               &       Kelvin \\
kg              &       kilogram \\
kpc             &       kiloparsec ($\simeq 3.1 \times 10^{19}$ meters) \\
(.)$_{l}$       &       lower \\
(.)$^{L}$       &       length form \\
m               &       meter \\
(.)$_{max}$     &       maximum \\
(.)$_{min}$     &       minimum \\
\NIST\          &       National Institute of Standards and Technology \\
ORL             &       Optical Recombination Lines \\
PJ              &       Paschen Jump \\
\PN(e)          &       \PlaNeb(e) \\
RR              &       \RadRec\ \\
\Ryd            &       \Rydberg \\
s               &       second \\
STO             &       \STO \\
THz             &       Terahertz \\
u               &       atomic mass unit \\
(.)$_{u}$       &       upper \\
(.)$_{ul}$      &       upper to lower \\
(.)$^{V}$       &       velocity form \\
$\forall$       &       for all \\
$\verb|      |$ & \\

\end{supertabular}

\vspace{0.5cm}

\noindent %
{\bf Note}: units, when relevant, are given in the SI system. Vectors and matrices are marked with
boldface. Some symbols may rely on the context for unambiguous identification. The physical
constants are obtained from the National Institute of Standards and Technology (\NIST)
\cite{NIST2010}.

%


\doublespace

{\setlength{\parskip}{6pt plus 2pt minus 1pt}

\pagestyle{headings} %
\addtolength{\headheight}{+1.6pt}
\lhead[{Chapter \thechapter \thepage}]%
      {{\bfseries\rightmark}}
\rhead[{\bfseries\leftmark}]%
     {{\bfseries\thepage}} 
\headsep = 1.0cm               


\newpage
\thispagestyle{empty} \vspace*{5.0cm} \phantomsection \addcontentsline{toc}{chapter}{\protect
\numberline{} Part I: Atomic Physics}

{\center

\LARGE{\bf Part I

\vspace*{1.0cm}

Atomic Physics: \vspace{0.3cm} \\
Dielectronic Recombination Lines of C\II\ and Their Application in Temperature Diagnosis of
Astronomical Objects} \vspace*{2.0cm} \\
}

\chapter{Introduction} \label{Introduction} 

Many atomic processes can contribute to the production of carbon spectral lines. These processes
include radiative and collisional excitation, and radiative and dielectronic recombination. The
current study is concerned with some of the carbon lines produced by dielectronic recombination and
subsequent radiative cascades. In the following we outline the recombination processes of primary
interest to our investigation.

\section{Recombination Processes} \label{RecPro}

The low-density \plasma\ of nebulae heated by energetic photons from their \progenitor s provides
an ideal environment for many physical processes. These include \radrec, \dierec, \thrbodrec\ and
\chatra. Recombination is believed to play a vital role in the physical processes that occur in
nebulae, and hence a deep understanding of its role is required for analyzing nebular spectroscopy
and modeling nebular evolution. The principal electron-ion recombination processes which are
responsible for the emission of recombination lines are \RadRec\ (RR) and \DieRec\ (DR).
Recombination of an electron and ion may take place through a background continuum known as
\radrec, or through a resonant recombination process involving doubly-excited states known as
\dierec. The latter can lead either to autoionization, which is a radiationless transition to a
lower state with the ejection of a free electron, or to stabilization by radiative decay of the
core to a lower bound state, possibly the ground state, with a bound electron. The RR and DR
processes are closely linked and the difference between them may therefore be described as
artificial; quantum mechanically they are indistinguishable. Recombination is mainly due to
dielectronic processes which are especially important at high temperatures when the autoionizing
levels are energetically accessible to the free electrons \cite{PequignotPB1991, NaharPZ2000,
BadnellOSABe2003}. In the following sections we give a brief account of these two important
recombination processes.

\subsection{Radiative Recombination} \label{RadRec}

\Radrec\ (RR) is the time inverse of direct photoionization where a single electron is captured by
an ion to a true bound state with the subsequent emission of photons. Symbolically, the RR process
is given by
\begin{equation}\label{radiative}
     \IECN + \ele \verb|    | \rightarrow \verb|    | \BSRI + \h \F
\end{equation}
where $\IECN$ is an ion of effective positive charge $n$, $\ele$ is an electron, $\BSRI$ is an
excited bound state of the recombined ion with effective positive charge ($n$$-$$1$), $\h$ is the
\Planck's constant and $\F$ is the frequency of radiation. The rate for RR rises as the free
electron temperature falls and hence it tends to be the dominant recombination process at low
temperatures. It is particularly significant in the cold ionized \plasma s found, for instance, in
some supernova remnants. \Radrec\ is also the only route to recombination for single electron
systems, such as H and He$^+$, because \dierec\ does not work on these systems
\cite{AldrovandiP1973, AldrovandiP1973b, PequignotPB1991, NaharPZ2000, BadnellOSABe2003}.

\subsection{\DieRec} \label{DieRec}

\Dierec\ (DR) is a two step process that involves the capture of a colliding continuum electron by
an ion to form a doubly excited state. Symbolically, DR can be described by the relation
\begin{equation}\label{dielectronic}
        \IECN + \ele \verb|   | \leftrightarrow \verb|   | \DEAIS
        \verb|   | \rightarrow \verb|   | \BSRI  + \h \F
\end{equation}
where $\DEAIS$ is a doubly-excited autoionizing state with effective positive charge ($n$$-$$1$).
This can be followed either by autoionization to the continuum with probability $\aTP$, or by
stabilization through a radiative decay of the captured electron or an electron in the parent to a
true bound state of the excited core ion with probability $\rTP$. The latter, which is the
dielectronic recombination process, is responsible for the production of many recombination lines.
Depending on the autoionizing state, the probabilities of these competing processes can go in
either direction, that is $\aTP > \rTP$ or $\aTP < \rTP$. These excited autoionizing systems
underly resonances in the photoionization cross sections. The resonance states are normally
unstable and hence they decay primarily by autoionization rather than radiative decay. The
resonance phenomenon is a highly important feature in atomic and molecular collisions as it plays a
vital role in electron scattering with atoms, ions and molecules. The Auger effect is one of the
early examples of the resonance effects that have been discovered and extensively investigated
\cite{AldrovandiP1973, AldrovandiP1973b}.

\Dierec\ dominates at high temperatures as it requires sufficiently energetic electrons to excite
the target, though it is still highly significant at low temperatures by recombination through
low-lying resonances near the ionization threshold, where at these temperatures the free electrons
can enter only these states. As the temperature rises, higher autoionizing levels can be reached
and the process becomes more complicated by the involvement of many autoionizing states. As the
energy of the resulting system in dielectronic recombination is greater than the binding energy of
the whole system, the state normally autoionizes spontaneously and hence it is usually described as
an autoionizing state. Because the dielectronic recombination and autoionization occur rapidly, the
population of the autoionizing state can be held at its local thermodynamic equilibrium level
relative to the ion-free electron system \cite{Ferland2003}.

DR is a fundamental atomic process that has many manifestations in astrophysics and fusion plasmas;
these include planetary nebulae and laboratory laser-induced plasmas. More specifically, the
recombination lines produced by this process can be used to determine elemental abundances and
electron temperatures in planetary nebulae and to diagnose plasma conditions in laboratory. The
process is also important for evaluating the energy balance and degree of ionization in hot and
cold plasmas as it contributes to the determination of state population and ionization balance at
various ranges of temperature and electronic number density. DR can occur through numerous
transitional autoionizing states; possibly the whole Rydberg series \cite{BadnellOSABe2003,
PengHWL2005}.

We may distinguish three types of \dierec\ mechanism with relevance at different temperatures

\begin{enumerate}

\item
High-temperature \dierec\ which occurs through Rydberg series of autoionizing states, and in which
the radiative stabilization is via a decay within the ion core.

\item
Low-temperature \dierec\ which operates via a few near-threshold resonances with radiative
stabilization usually through decay of the outer, captured electron. These resonances are usually
within a few thousand wave numbers of threshold, so this process operates at thousands to tens of
thousands of degrees Kelvin.

\item
Fine structure \dierec\ which is due to Rydberg series resonances converging on the fine structure
levels of the ground term of the recombining ion, and is necessarily stabilized by outer electron
decays. This process can operate at very low temperatures, down to tens or hundreds of Kelvin.

\end{enumerate}
In the current work we are concerned only with low-temperature \dierec\ and the resulting spectral
lines.

\section{Literature Review} \label{Literature}

In this section we present a short literature review to highlight some important developments in
the theory of recombination lines, followed by a brief account on the previous work on the carbon
lines in the spectra of \planeb e and similar astronomical objects.

\subsection{Theory of Recombination Lines} \label{LitRecThe}

Recombination of electrons and ions during scattering processes, which commonly occurs in many
gaseous and plasmic environments, is a fundamental atomic process that has a strong impact on many
physical systems. Recombination lines produced by such a process can be used as a diagnostic tool
to determine elemental abundances and electron temperatures in \planeb e. In the laboratory, these
lines can be used to probe the conditions of laser induced \plasma s and provide information that
is essential for \plasma\ modeling and fusion conditions. Therefore, there is a huge amount of
literature on all aspects of this important phenomenon. In this section we present a
non-comprehensive overview on some developments in the theory of recombination lines. As \dierec\
is the principal recombination process in most systems of interest to us and is the only one that
we considered in our recombination calculations, this overview will be dedicated to this process.

\Dierec\ was investigated by Massey and Bates \cite{MasseyB1942, BatesM1943} in connection with the
Earth's atmosphere using intuitive non-rigorous arguments focusing on recombination in \plasma\ at
rather low electron temperatures with consideration of low-lying resonances just above the
ionization threshold. Burgess \cite{Burgess1964} investigated \dierec\ in the solar corona and
highlighted the importance of this process at high temperatures in low-density hot \plasma\ making
it a possible dominant recombination process at coronal temperature. Goldberg \cite{Goldberg1968}
pointed out to the significance of \dierec\ at low temperatures in the context of analyzing the
intensity of transition lines in the spectrum of \planeb e.

The general theory of electron-ion scattering was developed by Seaton \cite{Seaton1969} in a study
to analyze resonance structures on the basis of quantum defect theory using the \Rm-matrix
formulation. In another study, Davies and Seaton \cite{DaviesS1969} investigated optical radiation
in the continuum caused by recombination of an electron with a positive ion using a radiation
damping approach for bound states transitions, and hence generalized this theory to the case of an
optical continuum. Their techniques are capable of solving the problem of overlapping resonances
found in \dierec. The results obtained from this investigation contributed to the subsequent
development of quantum mechanical treatment of DR.

Presnyakov and Urnov \cite{PresnyakovU1975} examined the problem of positive ion excitation by
electron collision, which leads to resonance structures in the cross sections, using Green
functions to describe the \Coulomb\ field. Their investigation concluded that the \dierec\
processes can lead to significant alteration in the rate coefficients of collisionally-excited
ions. Beigman and Chichkov \cite{BeigmanC1980} examined \dierec\ through the forbidden levels and
found that this process essentially increases the rate of total recombination at low temperatures.
They also observed that \dierec\ is significant for low-temperature highly-ionized thin plasma. The
importance of \dierec\ in general at low temperatures, typical of planetary nebulae, for the
low-lying autoionizing states was also established by Storey \cite{Storey1981} and Nussbaumer and
Storey (e.g. \cite{NussbaumerS1983}).

Using \quadef\ theory of Seaton \cite{Seaton1969} and radiation damping treatment of Davies and
Seaton \cite{DaviesS1969}, Bell and Seaton \cite{BellS1985} developed a theory for \dierec\ process
to accommodate resonance overlapping and the interaction with the radiation field. In a subsequent
study, Harmin \cite{Harmin1986} extended this theory to incorporate the effect of electric fields
in the scattering matrix where the \Rydberg\ series of autoionizing states have been analyzed
within a unified \Stark\ quantum-defect treatment.

In a succession of papers, mainly focused on the second-row elements of the periodic table,
Nussbaumer and Storey (e.g. \cite{NussbaumerS1987}) developed the theory of \dierec\ rate
coefficient calculations which is widely used nowadays in the computational work and simulation of
recombination processes. Their approach is based on the use of experimental data for autoionizing
energy levels with the possibility of using theoretical data obtained from atomic scattering and
structure codes, such as \RMAT\ and \AS, to fill the gaps in the experimental data. The current
work follows this approach in computing \dierec\ rate coefficients of C$^+$.

Badnell \cite{Badnell1986} developed a fast accurate method for computing the total \dierec\ rate
coefficients for He-like ions using his previously-developed \cite{Badnell1984} frozen-core
approximation, and applied this method to the highly charged ion Fe$^{24+}$. He also made other
contributions to the theory and application of \dierec\ which include calculating \dierec\ rate
coefficients for several ions (e.g. \cite{Badnell1986b, Badnell1988, Badnell2006}), assembly of a
comprehensive dielectronic recombination database \cite{BadnellOSABe2003}, and the development and
maintenance of the \AS\ code and its theoretical foundations \cite{BadnellAS2008}. \AS\ was used by
many researchers to compute theoretical data related to \dierec\ as well as other types of atomic
structure and scattering processes. This includes the current study where \AS\ was used, alongside
the \Rm-matrix code, to generate theoretical \dierec\ data for the SS1 line list.

Using the isolated resonance approximation and the distorted wave theory, LaGattuta \etal\
\cite{LaGattutaNH1987} proposed a theoretical approach for calculating the effect of static
electric fields on the target cross section in \dierec\ processes. In another theoretical study
based on a multichannel quantum defect theory of the \Stark\ effect, \dierec\ under the effect of
electric fields was examined by Sakimoto \cite{Sakimoto1987} with the derivation of simple formulae
for computing the cross sections in DR to account for the effect of externally-imposed electric
fields. Nahar and Pradhan \cite{NaharP1994, NaharP1995} presented a unified recombination \abin\
treatment which incorporates both radiative and dielectronic processes within the \clocou\
approximation of electron-ion scattering to enable the calculation of a single, total recombination
rate coefficient that may be employed for a variety of astrophysical applications.

\subsection{Recombination Lines of Carbon} \label{CLines}

There is an extensive literature on the applications of recombination lines theory. However, in
this section we only highlight some studies on the recombination lines of carbon relevant to the
applications of \PN e and similar astronomical objects with special interest in the C\II\ lines.

Leibowitz \cite{Leibowitz1972} investigated the energy level population of ions in \planeb e with
particular emphasis on the contribution of radiative excitation. The results were then applied to
the intensity ratios of C\IV\ lines in several \planeb e under various excitation conditions. In
another study \cite{Leibowitz1972b}, he investigated the polarization of C\IV\ emission lines in
the spectra of \planeb e. Investigating the structure of the \OrionNeb\ and the surrounding neutral
medium, Balick \etal\ \cite{BalickGD1974} mapped the recombination line C 85$\alpha$ of carbon
using high spatial resolution all over the nebula to determine the ionized carbon distribution and
its relationship to the H\II\ region.

Nussbaumer and Storey \cite{NussbaumerS1975} examined a number of recombination lines of carbon in
the context of studying the effect of the solar radiation field on the ionization balance for the
purpose of analyzing the transition region in the solar chromosphere-corona. In a subsequent study
\cite{NussbaumerS1984} they computed the effective \dierec\ coefficients for selected lines of ions
of carbon and other elements. The effective recombination coefficients were then fitted, as a
function of temperature, to a suitable empirical formula in a temperature range found in \planeb e
and similar astrophysical objects. Boughton \cite{Boughton1978} investigated several carbon
recombination lines observed in the direction of \OrionNeb\ to construct two-cloud models which can
replicate the observed intensities of the carbon lines and the velocities of the emitting regions.

In a series of papers, Harrington \etal\ \cite{HarringtonLSS1980} discussed the recombination
processes in the context of investigating the carbon abundance in the low-excitation \planeb\ IC
418\index{IC 418} using ultraviolet spectral observations. In another study, Harrington \etal\
\cite{HarringtonLS1981} investigated some recombination lines of carbon using in their analysis
results obtained earlier by Storey. In another paper of the same series, Clavel \etal\
\cite{ClavelFS1981} analyzed the C\II\ $\lambda$1335~\AA\ \dierec\ multiplet observed in the
spectra of the \planeb\ IC 418\index{IC 418}.

Clegg \etal\ \cite{CleggSPP1983} investigated a bright nebular region in the low-excitation
\planeb\ NGC 40\index{NGC 40} using optical and IUE spectra for the purpose of probing the physical
conditions and determining elemental abundances. This investigation led to the identification and
determination of relative intensity of several C\II\ recombination lines in NGC 40\index{NGC 40}
and IC 418\index{IC 418}. Hayes and Nussbaumer \cite{HayesN1984} examined the effect of \dierec\
process on the emissivity of the C\II\ $\lambda$1335~\AA\ multiplet observed in various
astronomical objects including \planeb e, and compared this to the effect of collisional
excitation.

Investigating the abundance of carbon in \planeb e, Kholtygin \cite{Kholtygin1984} calculated the
intensities of recombination lines of C$^{2+}$ ion and used the observed intensities of several
lines belonging to a number of carbon ions to determine their abundances, as well as the total
abundance of carbon, in 46 \planeb e. In a later study \cite{Kholtygin1998} he used the calculated
and observed intensities of carbon recombination and intercombination lines to find the effect of
small temperature and density fluctuations on the intensities in the spectra of \planeb e.
Bogdanovich \etal\ \cite{BogdanovichNRK1985} calculated the relative intensities of a number of
C\II\ recombination lines, and used the observed intensities of these lines to determine the
abundances of C$^{2+}$ ion in 12 nebulae.

Badnell \cite{Badnell1988} used the \AS\ code within an \intcou\ scheme to compute \conmix\
effective \dierec\ rate coefficients for the recombined C$^+$ ion at temperatures $\T =
10^{3}-\Sci{6.3}{4}$~K applicable to \planeb e, and the results were tabulated for the
quartet-quartet lines and some doublet-doublet lines. He also compared the total rate coefficients
for \dierec\ and \radrec\ where the former have been obtained within an \LS\ as well as an \intcou\
schemes. P\'{e}quignot \etal\ \cite{PequignotPB1991} computed effective and total radiative \reccoe
s for selected optical and ultraviolet transitions of ions of carbon and other elements that
produce most recombination lines observed in nebulae. The results were fitted to a four-parameter
empirical expression which is valid over a wide range of electron temperature and at low electron
number density.

In their investigation of the carbon abundance in \planeb e and the long-standing problem of
discrepancy between recombination and forbidden lines results, Rola and Stasi\'{n}ska
\cite{RolaS1994} examined the C\II\ $\lambda$4267~\AA\ recombination multiplet and analyzed the
observational data obtained from various sources for many \planeb e. Baluteau \etal\
\cite{BaluteauZMP1995} examined several recombination lines of carbon in the course of studying the
far-red spectrum of the bright and spectrum-rich NGC 7027\index{NGC 7027} \planeb. Davey \etal\
\cite{DaveySK2000} presented effective \reccoe s for C\II\ lines originating from transitions
between doublet states in \LS-coupling scheme using \Rm-matrix code. Their results span a wide
temperature range between 500-20000~K with an electron number density of 10$^{4}$~cm$^{-3}$
relevant to \planeb e.

Investigating the chemical abundances of \planeb e, Liu \etal\ \cite{LiuLBL2004} examined several
optical recombination lines of carbon observed in a number of \planeb e. Peng \etal\
\cite{PengWHL2004} investigated theoretically the C\II\ recombination line $\lambda$8794~\AA, which
is important for determining carbon abundance in \planeb e, using the \Rm-matrix method. In a
similar study \cite{PengHWL2005} they investigated \dierec\ processes of C$^{2+}$ ion in \planeb e
where several C\II\ recombination lines were examined in the course of this study. Zhang \etal\
\cite{ZhangLLPB2005} thoroughly examined the rich spectra of the \planeb\ NGC 7027\index{NGC 7027}.
The extensive line list presented in this study comprises over 1100 transition lines from many
atoms and ions of various elements and includes many recombination lines of carbon and its ions.

Extensive observational studies, similar to that of Zhang \etal, are those of Liu \etal\
\cite{LiuLBDS2001} on the M 1-42\index{M 1-42} and M 2-36\index{M 2-36} \planeb e, Sharpee \etal\
\cite{SharpeeWBH2003, SharpeeBW2004} related to the \planeb\ IC 418\index{IC 418}, Peimbert \etal\
\cite{PeimbertPRE2004} on NGC 5315\index{NGC 5315}, and Fang and Liu \cite{FangL2011} on the bright
\planeb\ NGC 7009\index{NGC 7009}. There are many other studies which are not as extensive as the
previous ones but contain some observational spectral data from \planeb e and similar objects that
include carbon recombination lines in general and C\II\ in particular. Examples of theses studies
are: Flower \cite{Flower1982}, Barker \cite{Barker1987}, Aller and Keyes \cite{AllerK1988},
Petitjean \etal\ \cite{PetitjeanBP1990}, Liu \etal\ \cite{LiuSBC1995, LiuSBDCB2000, LiuLBL2004,
LiuBZBS2006}, Esteban \etal\ \cite{EstebanPPE1995, EstebanPPE1998, EstebanPPR2002,
EstebanPRRPR2004, EstebanBPRPD2009}, De Marco \etal\ \cite{DemarcoSB1997}, Garnett and Dinerstein
\cite{GarnettD2001b}, Bernard-Salas \etal\ \cite{SalasPBW2001, SalasPFW2002, SalasPWF2003,
SalasPBW2003}, Peimbert \cite{Peimbert2003}, Wesson \etal\ \cite{WessonLB2003, WessonL2004,
WessonLB2005, WessonBLSED2008}, Tsamis \etal\ \cite{TsamisBLDS2003, TsamisBLDS2003b,
TsamisBLSD2004, TsamisWPBDL2008}, Ruiz \etal\ \cite{RuizPPE2003}, Garc\'{\i}a-Rojas \etal\
\cite{RojasEPRRP2004, RojasEPPRR2005, RojasEPCRe2006, RojasEPRPR2007, RojasPP2009}, Ercolano \etal\
\cite{ErcolanoWZBMe2004}, Stasi\'{n}ska \etal\ \cite{StasinskaGPHKS2004}, Robertson-Tessi and
Garnett \cite{TessiG2005}, Wang and Liu \cite{WangL2007}, Sharpee \etal\ \cite{SharpeeZWPCe2007},
and Williams \etal\ \cite{WilliamsJBZSe2008}. Some of these studies are used as sources of
observational data in our investigation, as will be examined later on in this thesis.

This review highlights the diversity of the work that has been done on the C\II\ recombination
lines although this topic has not been comprehensively investigated as there are many gaps to be
filled. From the perspective of atomic physics, the most comprehensive of these studies and most
relevant to the current investigation are those of Badnell \cite{Badnell1988} who calculated
configuration mixing effective dielectronic recombination coefficients for the recombined C$^+$ ion
at temperatures $T = 10^{3}-6.3\times10^{4}$~K applicable to planetary nebulae, and Davey \etal\
\cite{DaveySK2000} who computed effective recombination coefficients for C~{\sc ii} transitions
between doublet states for the temperature range 500-20000~K with an electron density of 10$^{4}$
cm$^{-3}$ relevant to planetary nebulae. Badnell performed calculations in both \LS\ and
intermediate coupling schemes and over a wide temperature range, using the \AS\ code
\cite{EissnerJN1974, NussbaumerS1978, BadnellAS2008} which treats the autoionizing states as bound
and the interaction with the continuum states by a perturbative approach. On the other hand, Davey
\etal\ performed their calculations using the \Rm-matrix code \cite{BerringtonEN1995} which
utilizes a more comprehensive recombination theory based on a unified treatment of bound and
continuum states but worked in the \LS-coupling scheme and hence their results were limited to
doublet states.

The aim of the current study is to build on these investigations and elaborate on them by using the
\Rm-matrix code with an intermediate coupling scheme. However, our investigation will be limited to
the \dierec\ concentrating our attention on the low-lying autoionizing states. On the observational
perspective, most of the previous studies have focused on a few recombination lines employing the
traditional method of using intensity ratio to analyze the physical conditions, especially electron
temperature and number density, of the line-emitting regions. The current study tries to broaden
some of these aspects first by using any reliable data, and second by adopting a more comprehensive
method of analysis, namely the least squares optimization, which is a collective method that can
involve several transitions simultaneously to analyze the physical conditions. However, we do not
use first-hand observational data and therefore we rely on data reported in the refereed
literature. Moreover, we will focus our attention on electron temperature diagnosis of the
line-emitting regions.

\section{Recombination versus Forbidden Lines Analysis} \label{RFLines}

In this section we briefly discuss the long-standing dilemma in the physics of nebulae demonstrated
by the discrepancy between recombination and forbidden lines results and the implication of this on
our investigation. Resolving this puzzle is extremely important not only for understanding the
nebular conditions but also for identifying possible flaws in the underlying physical models. The
collisionally-excited lines (CEL), also generally known in this context as forbidden lines, are
produced by collision between energetic electrons and atoms and ions with a subsequent excitation
and decay, while the optical recombination lines (ORL) are produced when electrons combine with
ions by one of the aforementioned recombination mechanisms. These lines are detected either in
emission or absorption modes. Both CELs and ORLs can in principle provide information about the
physical and chemical conditions of the emitting and absorbing regions such as temperature, number
density, pressure, elemental abundance, chemical composition and so on. Until recently, the main
means used in the \planeb\ investigation is the CELs. One reason is the relative ease of observing
and analyzing these lines as they are more abundant and intense than the ORLs. However, the ORLs
are also used in the elemental abundance determination studies by using their relative intensities
to the hydrogen lines \cite{KisieliusSDN1998, Ferland2003, TsamisBLDS2003b, TsamisBLSD2004}.

Although the forbidden lines are much stronger than the recombination lines, they are highly
dependent on temperature and density structure and hence can lead to large systematic errors when
used for instance in ionic abundance estimation. On the other hand, the faint recombination lines
can be easily contaminated by radiation from other excitation processes such as fluorescence.
Despite all these differences, there is a common feature between the results obtained from these
lines; that is for all the atomic species investigated so far (mainly from the second row of the
periodic table such as C, N, O and Ne) the forbidden lines in planetary nebulae normally produce
lower ionic abundances than the corresponding values obtained from the recombination lines. The
ratio of the ORLs to the CELs abundances is case dependent and can vary by a factor of 30 or even
more. This has cast doubt on the validity of the previously accepted CELs analysis results. The
systematic nature of this problem rules out the possibility of recombination line contamination by
resonance fluorescence or the inaccuracies in recombination coefficients as possible reasons as may
have been suggested. This problem may be correlated to the dichotomy between the temperature
obtained from the Balmer jump of H~{\sc i} recombination lines and that from the
collisionally-excited forbidden lines where the latter is systematically higher than the former
\cite{Kholtygin1998, Liu2002, TsamisWPBLD2007}. In fact, obtaining higher electron temperatures
from forbidden lines than those deduced from recombination lines is a general trend in nebular
studies.

Several explanations have been proposed to justify these discrepancies individually or
collectively, though no one seems to be satisfactory or universally accepted. One explanation is
the sensitivity of the collisionally-excited lines to temperature and structure fluctuations (e.g.
exponential temperature dependency of the emissivity of these lines) which amplifies the errors
that are intrinsically associated with these parameters. However, this reason alone cannot explain
why the forbidden lines always produce lower abundance values. Another explanation is that the
large temperature and density fluctuations within the nebular structure result in systematic
underestimation of the heavy element abundances from the forbidden lines. The existence of knots
depleted of hydrogen with high heavy element contents within the nebular gas has been proposed as a
possible reason for these fluctuations and subsequent implications. The temperature inside these
knots of high metallicity, and hence high opacity to stellar ultraviolet emissions and large
cooling rates, is expected to be too low for efficient production of forbidden lines though it is
still sufficiently high for the emission of recombination lines. Consequently, the recombination
and collisional lines originate in different regions of the nebular gas with different elemental
abundances and different temperatures. However, the existence and composition of these knots and
their effect on the selectivity of emission lines is unsettled issue and a matter of debate.
Moreover, in some cases the discrepancy between the collisional and recombination abundances is too
large to explain by temperature fluctuation alone although it may be partially responsible
\cite{LiuSBC1995, GarnettD2001, TsamisBLDS2003, LiuLBL2004}.

In a recent paper by Nicholls \etal\ \cite{NichollsDS2012} it is claimed that this long-standing
problem in \planeb e and H\II\ regions could arise from the departure of the electron energy
distribution from the presumed Maxwell-Boltzmann equilibrium condition, and hence it can be solved
by assuming a $\kappa$-distribution for the electron energy following a method used in the solar
data analysis. One possible criticism to this approach is that the large discrepancy obtained in
some cases may not be explicable by this adjustment to the electron distribution.
Electron distribution will be the subject of a fairly thorough investigation in
section~\ref{ElecDist} where we used lines originating from resonance states (that is FF and FB
transitions) to obtain a direct sampling of the electron energy distribution based on the
observational de-reddened flux with theoretically-obtained parameters such as the departure
coefficients of the involved autoionizing states and the radiative probabilities of these
transitions.

Amid the uncertainties highlighted by the CELs versus ORLs puzzle, it is extremely important to do
more research especially on the recombination lines theory to have a breakthrough in this
long-standing problem. We hope that our research on carbon recombination lines will be a valuable
contribution in this direction.

\chapter{Physics of Scattering and Recombination} \label{Physics} 
In this chapter we give a brief theoretical background about the physics of scattering and
recombination which forms the basis for our computational model. Our scattering quantum system
comprises an $N$-electron atomic or ionic target plus a colliding electron where the particles are
assumed to interact through electromagnetic forces only. The quest then is to solve the following
time independent \Schrodinger\ equation with suitable boundary conditions
\begin{equation}\label{SchroEq}
    \Ham \WF = \E \WF
\end{equation}
where $\Ham$ is the \Hamiltonian\ of the whole system (nucleus plus $N+1$ electrons), $\WF$ is the
quantum wavefunction, and $\E$ is the total energy of the system. In the case of light atoms and
ions, the relativistic effects can be ignored and hence $\Ham$ in atomic units is given by
\begin{equation}\label{HamTotEq}
    \Ham = \sum_{n=1}^{N+1} \left(- \, \frac{1}{2} \nabla_{n}^{2} \, - \,
    \frac{\AN}{r_{n}} + \sum_{m>n}^{N+1} \frac{1}{r_{nm}} \right)
\end{equation}
In this equation, $\nabla^{2}$ is the Laplacian operator, $\AN$ is the nuclear atomic number,
$r_{n}$ is the distance between the nucleus and the $n^{th}$ electron, and $r_{nm} = |r_{n} -
r_{m}|$ is the distance between the $n^{th}$ and $m^{th}$ electrons. The assumption here is that
the nucleus is a massive motionless point charge compared to the surrounding electrons. The first
term of Equation~\ref{HamTotEq} represents the total kinetic energy of the electrons, the second
term is the total potential energy due to the attractive Coulomb forces between the nucleus and
each electron, and the last one is the total potential energy due to the Coulomb repulsive forces
between each pair of electrons. The first two are called the one-electron terms while the third is
called the two-electron term \cite{BerringtonEN1995}.

There are several ways for solving this system, one of which is to employ the \clocou\ technique
which provides a powerful framework for performing scattering computations involving collisions
between an electron and an $N$-electron target. The method can be used to analyze such quantum
systems and obtain valuable information such as wavefunctions, energy of states, \pcs s, \oscstr s,
and so on. In the following we outline this method and its realization based on the \Rm-matrix
theory.

\subsection{\CloCou\ Approximation} \label{CloseCoupling}
In the \clocou\ approach, the wavefunction is expanded in the following form
\begin{equation}\label{CCbasic}
    \WF = \ansy \sum_{i} \tarfun_{i} \elefun_{i}
\end{equation}
where $\ansy$ is an anti-symmetrization operator accounting for the exchange effects between the
colliding electron and target electrons, $\tarfun_{i}$ are functions representing the target, and
$\elefun_{i}$ are functions of the free electron given by
\begin{equation}\label{eleFuncs}
    \elefun_{i} = \spiang_{i }\frac{1}{r} \radfun_{i}
\end{equation}
In the last equation, $\spiang_{i}$ are the spin and angle functions and $\radfun_{i}$ are the
radial functions of the scattered electron. The target functions are usually computed by employing
the \conint\ theory with the use of identical set of radial functions. The radial functions can be
either spectroscopic physical orbitals of the kind employed in a central filed approach, or pseudo
orbitals used to represent electron correlation effects and improve accuracy. On introducing
functions $\vecofu_{i}$, which are vector-coupled products of $\tarfun_{i}$ and $\spiang_{i}$, and
imposing an orthogonality condition, the \clocou\ expansion of Equation~\ref{CCbasic} becomes
\begin{equation}\label{CCexpansion}
    \WF = \ansy \sum_{i=1}^{I} \vecofu_{i} \frac{1}{r} \radfun_{i} +
    \sum_{j=1}^{J} \CCcoef_{j} \Phi_{j}
\end{equation}
where $\CCcoef_{j}$ are coefficients in this expansion and $\BSTfun_{j}$ are functions of
bound-state type for the entire system \cite{Seaton1985, BerringtonESSS1987, BerringtonBBSSTY1987,
BerringtonEN1995}.

Because Equation~\ref{CCexpansion} is a truncated expansion, as it consists of a finite number of
terms, $\WF$ is not an exact solution of the \Schrodinger\ equation given by
Equation~\ref{SchroEq}. The following variational condition can be used to derive the \clocou\
equations
\begin{equation}\label{varCond}
    ( \delta\WF \mid \Ham - \E \mid \WF) = 0
\end{equation}
where $\delta\WF$ is a variation in the wavefunction $\WF$ due to variations in the radial
functions $\radfun_{i}$ and the \clocou\ expansion coefficients $\CCcoef_{j}$. This results in a
system of integro-differential equations that should be satisfied by $\radfun_{i}$ and
$\CCcoef_{j}$ \cite{Seaton1985, BerringtonESSS1987}.

The total energy of the whole system, $\E$, in a certain state is given by
\begin{equation}\label{TotEneEq}
    \E = \E_{i} + \E_{\ele}
\end{equation}
where $\E_{i}$ is the energy of the target in state $i$ and $\E_{\ele}$ is the energy of the
colliding electron. When the total energy results in a bound state for the whole system, the
electron energy $\E_{\ele}$ is negative and the radial functions $\radfun_{i}$ exponentially decay
to zero as $r$ tends to infinity. However, for some $i$'s the collision states result in
$\E_{\ele}>0$. Channel $i$ is described as closed when $\E_{\ele}<0$ and open when $\E_{\ele}>0$.

If $\E_{i}$'s are ordered increasingly: $\E_{1}<\E_{2}< ... <\E_{I}$, then for a given total energy
$\E$ there are $I_{o}$ open channels such that \cite{BerringtonBBSSTY1987}
\begin{eqnarray}
  \nonumber
  \E_{\ele}>0    \hspace{1cm}    {\rm for}    \hspace{1cm}    i &=& 1  \verb|      | {\rm to}  \verb| | I_{o} \\
  \E_{\ele}<0    \hspace{1cm}    {\rm for}    \hspace{1cm}    i &=& (I_{o}+1)  \verb| | {\rm to}  \verb| | I
\end{eqnarray}
where $I$ is the total number of channels. A second subscript on the radial functions is normally
used, when some channels are open, to denote boundary conditions. The numerical calculations start
by computing real functions $\radfun_{ii'}$ subject to certain boundary conditions on the reactance
matrix
\begin{eqnarray}
  \nonumber
  \radfun_{ii'}(r) \underset{r \rightarrow \infty}\sim s_{i}(r) \delta(i,i') + c_{i}(r) K(i,i') \hspace{0.8cm}    {\rm for}    \hspace{0.8cm}    i = 1  \verb|      | {\rm to}  \verb| | I_{o} \\
  \radfun_{ii'}(r) \underset{r \rightarrow 0}\sim \verb| | 0 \hspace{5.1cm}   {\rm for}    \hspace{0.8cm}    i = (I_{o}+1)  \verb| | {\rm to}  \verb| | I
\end{eqnarray}
In the last equation, $s_{i}(r)$ and $c_{i}(r)$ are the components of the Coulomb functions
$\CF^{\pm}$ having an asymptotic form, and $i'= 1 \dots I_{o}$. These functions are defined by
\begin{equation}\label{CouFunEq}
    \CF^{\pm}(r) = c \pm \iu s
\end{equation}
where $c$ and $s$ represent cosine and sine functions respectively, and $\iu$ is the imaginary
unit. The \Sm-matrix functions also have asymptotic forms given by
\begin{eqnarray}
  \nonumber
  \radfun^{-}_{ii'}(r) \sim \frac{1}{2} \{\CF^{-}_{i}(r) \delta(i,i') - \CF^{+}_{i}(r) S(i,i')\}    \hspace{0.8cm}    {\rm for}    \hspace{0.8cm}    i = 1  \verb|      | {\rm to}  \verb| | I_{o} \\
  \radfun^{-}_{ii'}(r) \sim \verb| | 0 \hspace{5.9cm}   {\rm for}    \hspace{0.8cm}    i = (I_{o}+1)  \verb| | {\rm to}  \verb| | I
\end{eqnarray}
It can be shown that the scattering matrix, \Sm, is related to the reactance matrix, \Km, through
the following relation
\begin{equation}\label{SmatrixEq1}
    \Sm = \frac{\IM + \iu \Km}{\IM - \iu \Km}
\end{equation}
where $\IM$ is the identity matrix and $\iu$ is the imaginary unit. Moreover, the \Sm-matrix
functions, $\textbf{\em F}^{-}$, are related to the reactance matrix by
\begin{equation}\label{FFunEq}
    \textbf{\em F}^{-} = \frac{- \iu \textbf{\em F}^{-}}{\IM - \iu \Km}
\end{equation}
It should be remarked that the complex conjugate of the \Sm-matrix functions, $\textbf{\em F}^{+}$,
are used in the computation of \pcs s \cite{Seaton1985, BerringtonESSS1987, BerringtonBBSSTY1987,
BerringtonEN1995}.

\subsection{R-matrix Method} \label{Rmatrix}
One way of solving the \Schrodinger\ equation of the electron-atom system in the \clocou\
approximation is the \Rm-matrix method which is a computational technique based on the theory
developed originally by Burke and Seaton to provide a suitable framework for analyzing scattering
processes. The \Rm-matrix method is a computationally efficient way for solving the resulting
\clocou\ equations. Its main feature is the partitioning of the space into two disjoint regions:
inner and outer. These two regions are separated by a sphere centered at the system's center of
mass. The computational connection between these regions is established through the \Rm-matrix
which links them at the boundary. The main task then is to find solutions for the inner region
\cite{BerringtonESSS1987}.

The \Rm-matrix method starts by selecting a value $\IRR$ for the sphere radius $r$ such that the
functions $\vecofu_{i}$ and $\Phi_{j}$ are small outside the inner region. The focus of the method
then is to find solutions in this region. A many-body \Schrodinger\ equation should be solved when
the incident electron is inside the inner region, while a solution of only a two-body equation is
required in the outer region. This reduces the effort by restricting the mathematical and
computational complexities to a small region of space. A complete discrete set of basis functions
is used to express the wavefunction of the system in the scattering process. Initially, solutions
that fulfill certain conditions on the boundary between the two regions are sought. Such solutions,
$\WF = \tarfun_{n}$, which contain radial functions $\radfun_{i}(r) = f_{in}(r)$, do exist for a
discrete set of energies $\E = e_{n}$. A number of options are available for the boundary
conditions. One choice is that $f'_{in}(\IRR) = 0$ where the prime indicates a derivative with
respect to $r$. In this process, the functions $\tarfun_{n}$ are normalized to
\begin{equation}\label{PsiNorEq}
    (\tarfun_{n} | \tarfun_{n'})_{I} = \delta(n,n')
\end{equation}
with the evaluation of the matrix elements (...)$_{I}$ in the inner region. For a given energy
$\E$, the wavefunction $\WF_{\E}$ with radial functions $\radfun_{i\E}(r)$ can be expressed as
\begin{equation}\label{PsiEEq}
    \WF_{\E} = \sum_{n} \tarfun_{n} A_{n\E}
\end{equation}
where $A_{n\E}$ is given by
\begin{equation}\label{AnEEq}
    A_{n\E} = (e_{n} - \E)^{-1} \sum_{i} f_{in}(\IRR) \radfun'_{i\E}(\IRR)
\end{equation}
In the last equation, $\radfun'_{i\E}(\IRR)$ is chosen such that $\WF_{\E}$ meets predefined
boundary and normalization conditions. On substituting \ref{AnEEq} into \ref{PsiEEq} with some
algebraic manipulation, the following relation can be obtained
\begin{equation}\label{FiEEq}
    \radfun_{i\E}(\IRR) = \sum_{i'} R_{ii'}(\E) \radfun'_{i'\E}(\IRR)
\end{equation}
where
\begin{equation}\label{RiiEq}
    R_{ii'}(\E) = \sum_{n} f_{in}(\IRR) (e_{n} - \E)^{-1} f_{i'n}(\IRR)
\end{equation}
is the \Rm-matrix.

The target functions, $\tarfun_{n}$, are computed using expansions of the radial functions
$f_{in}(r)$ with the inclusion of orthogonality conditions to obtain good convergence. The
expansion coefficients are acquired from diagonalizing the \Hamiltonian\ matrix with a truncation
process. The \Buttle\ correction is then employed to compensate for the truncation
\cite{Seaton1985, BerringtonESSS1987, BerringtonBBSSTY1987, BerringtonEN1995}.

\subsubsection{Inner and Outer Regions} \label{InnerOuter}
As indicated earlier, the inner region, which may also be called the \Rm-matrix box, is identified
as the region inside a sphere of radius $\IRR$ centered on the center of mass of the entire system
and essentially contains all the target electrons, while the outer region is the rest of the
configuration space. The essence of the \Rm-matrix method is to find solutions in the inner region
where the system is considered a many-body problem. In the outer region, where electron correlation
and exchange effects are less important and hence can be ignored, the system is treated as a
two-body problem between the atomic target and incident electron, and the solution can then be
acquired from an asymptotic expansion or by a perturbative treatment. In this region, the
integro-differential equations are relegated to the following ordinary differential equations
\begin{equation}\label{OuterRegDE}
    \left( \DD {}{r} - \frac{l_{i}(l_{i}+1)}{r^{2}} + \frac{2 \Rc}{r} + \ECE \right)
    \radfun_{i} + \sum_{i'} \MPP \radfun_{i'} = 0
\end{equation}
where $\Rc$ is the ion's residual charge, $\ECE$ is the energy of the free electron, and $\MPP$ are
multi-pole potentials \cite{Seaton1985, BerringtonESSS1987, BerringtonBBSSTY1987,
BerringtonEN1995}.

\subsubsection{Matching of Solutions} \label{Matching}
The solutions in the inner and outer regions of the \Rm-matrix should be matched at the boundary
between these regions $r = \IRR$. Depending on the nature of the available channels (some open or
all closed), several matrices are defined and used to obtain and match the solutions. In this stage
of the \Rm-matrix calculations, the reactance matrix \Km\ is evaluated. This matrix, which
symbolizes the asymptotic nature of the whole wavefunction, is real and symmetric and carries
information from both regions \cite{BerringtonBBSSTY1987, BerringtonEN1995}.

\subsubsection{Bound States} \label{Bound}
Once the calculations of the inner-region of \Rm-matrix are carried out and the essential
\Rm-matrix data $f_{in}(\IRR)$ and $e_{n}$ are obtained, as outlined already, speedy computations
can be performed with minimum computational cost to find the energy and wavefunction of the bound
states \cite{Seaton1985}.

\subsubsection{Resonances} \label{Resonances}
On obtaining the reactance matrix \Km\ and the scattering matrix \Sm\ which is given by
Equation~\ref{SmatrixEq1}, resonances can be obtained following one of several approaches; some of
which are outlined in \S\ \ref{Methods}. An advantage of using the \clocou\ method is that
resonance effects are naturally delineated, since the interaction between closed and open channels
is already incorporated in the scattering treatment.

\subsubsection{\OscStr s} \label{OsciStr}
The \dipope s in length and velocity form, $\DOL$ and $\DOV$ respectively, are given by
\begin{equation}\label{DipOpeLVEq}
    \DOL = \sum_{n} \RV_{n}   \hspace{2cm}   \DOV = - \sum_{n} \nabla_{n}
\end{equation}
where $\RV$ is the radius vector, $\nabla$ is the del operator, and $n$ is an index running over
all electrons. Once these \dipope s are obtained, the \linstr\ for a dipole transition between
state $a$ and state $b$ can be computed in the following length and velocity forms
\begin{eqnarray}
  \nonumber
  \LsL_{ba} &=& | (b \parallel \DOL \parallel a) |^{2} \\
  \LsV_{ba} &=& 4 (\E_{b} - \E_{a})^{-2} | (b \parallel \DOV \parallel a) |^{2} \label{LinStrEqs}
\end{eqnarray}
where $\E_{a}$ and $\E_{b}$ are the energies of these states in \Ryd. The \oscstr, $\OS_{ba}$,
which is a dimensionless quantity, can then be obtained \cite{BerringtonBBSSTY1987}
\begin{equation}\label{OscStrEq}
    \OS_{ba} = \frac{(\E_{b} - \E_{a}) \Ls_{ba}}{3 \SW_{a}}
\end{equation}
where $\SW_{a}$ is the statistical weight of the initial state given by
$\SW_{a}=(2S_{a}+1)(2L_{a}+1)$ for \LS-coupling and $\SW_{a}=(2J_{a}+1)$ for \intcou. The \oscstr\
$\OS_{ba}$ can be in length or velocity form depending on the form of the \linstr\ $\Ls_{ba}$. When
exact functions are utilized in the equations, the two forms are equal, i.e. $\LsL = \LsV$ and
$\OSL = \OSV$. The use of approximate functions would normally lead to differences between the two
forms; the size of which may give a hint of the accomplished accuracy. However, agreement between
the two forms does not affirm accuracy as they can agree coincidentally \cite{BerringtonBBSSTY1987,
BerringtonEN1995}.

\subsubsection{\PCS s} \label{Photoionization}
For a transition from an initial bound state to a free final state a generalized \linstr\ $\Ls$,
similar to that of Equation~\ref{LinStrEqs}, can be introduced. The \pcs\ is then given by
\begin{equation}\label{PCSEq}
    \PICS = \frac{4 \pi^{2} \BR^{2} \FSC \Ep \Ls}{3 \SW}
\end{equation}
where $\BR$ is Bohr radius, $\FSC$ is the \finstr\ constant, $\Ep$ is the photon energy in \Ryd,
and $\SW$ is the degeneracy of the bound state. Again, in this formulation either the length or the
velocity operator can be used \cite{BerringtonBBSSTY1987, BerringtonEN1995}.

\subsubsection{Relativistic Effects} \label{Relativistic}
To account for relativistic effects, the previously-outlined non-relativistic treatment of the
\Rm-matrix method can be adjusted by adding terms from the relativistic \BrePau\ \Hamiltonian,
$\Hbp$, given by
\begin{equation}\label{BreitPauli}
    \Hbp = \Hnr + \Hrc
\end{equation}
where $\Hnr$ is the non-relativistic \Hamiltonian\ and $\Hrc$ is the relativistic correction terms
which include the one body mass, \Darwin\ and \spiorb\ terms; the two body \finstr\ terms; and the
two body non-\finstr\ terms. This is normally achieved by first computing the \Hamiltonian\
matrices in \LS-coupling followed by the application of a unitary transformation to convert to
\intcou. In the \RMAT\ code only some of these relativistic terms (i.e. mass correction, one body
\Darwin, and spin-orbit terms) are overtly held \cite{BerringtonEN1995}.

\subsection{\RMAT\ Code} \label{RMATCode}

The \Rm-matrix theory for atomic scattering is implemented in a computer code called \RMAT, which
is a general-purpose program for analyzing atomic processes such as collision of free electrons
with atoms and ions, radiative processes like transitions between bound states, polarizabilities,
and \pcs s. The \Rm-matrix calculations can be performed in \LS-coupling as well as in \intcou,
where the latter can be achieved through the inclusion of the one body correction terms from the
\BrePau\ relativistic \Hamiltonian, as outlined earlier. The program consists of
sequentially-coupled stages and is used in this study as the main tool for generating theoretical
data for scattering, recombination and subsequent atomic processes. These stages fall into two
major categories: inner region and outer region. The inner region code is composed of the following
stages \cite{BerringtonBBSSTY1987, BerringtonEN1995, BadnellRmax2002}

\begin{itemize}

\item
\STGO\ to compute the bound and continuum orbital basis and the associated radial integrals. The
target's radial functions should be provided as an input to this stage where they have been
obtained from atomic structure codes such as \AS\ and \CIV, as will be discussed later.

\item
\STGT\ to calculate angular algebra. This stage reads radial integrals produced by \STGO\ and
evaluates the \Hamiltonian\ and \dipope\ matrices. It can also diagonalize the \Hamiltonian\ matrix
of the target if required.

\item
\STGJK\ which is an optional stage that runs between \STGT\ and \STGTH\ to recouple the
\Hamiltonian\ in the case of \intcou\ scheme by transforming the \Hamiltonian\ matrix from
\LS-coupling to pair-coupling through the use of a unitary transformation to include \BrePau\
relativistic effects.

\item
\STGTH\ to diagonalize the \Hamiltonian\ matrix in the continuum basis and produce vital \Rm-matrix
data required for the outer region stages.

\end{itemize}
The outer region code consists of the following stages

\begin{itemize}

\item
\STGB\ to calculate wavefunctions and energy levels for bound states and produce a data set
required for radiative calculations of bound-bound and bound-free transitions.

\item
\STGF\ to calculate wavefunctions of free states. It can also compute \colstr s for inelastic
collisions and produce data needed for radiative calculations of free-free and bound-free
transitions.

\item
\STGBB\ to calculate \oscstr s for bound-bound transitions.

\item
\STGBF\ to calculate \pcs s for bound-free transitions.

\end{itemize}
An extra stage called \STGQB\ for finding and analyzing resonances was added to the \Rm-matrix code
by Quigley and co-workers \cite{QuigleyBP1998}.

\begin{figure}[!h]
  \centering{}
  \includegraphics[scale=0.8]{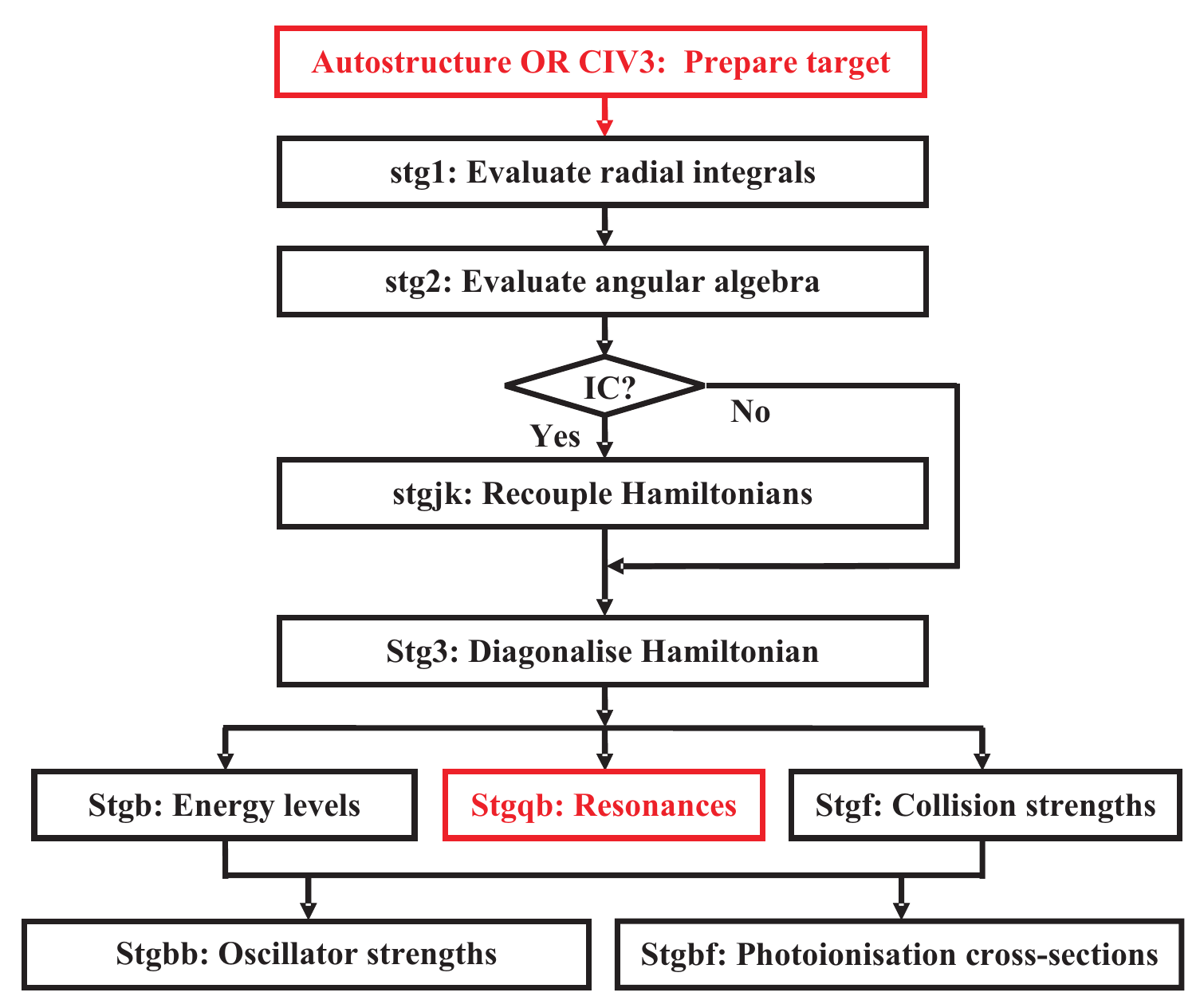}
  \caption{Flow chart of the \RMAT\ suite.}
  \label{Rmax}
\end{figure}

In Figure~\ref{Rmax} a flow chart of the \RMAT\ suite and its companions is presented. \RMAT\ needs
input data to define the atomic target. This can be provided either by \CIV\ code of Hibbert
\cite{Hibbert1975}, or by \SSt\ code of Eissner \etal\ \cite{EissnerJN1974}, or by \AS\ code. The
latter is a modified version of \SSt\ that incorporates extensions and modifications such as those
of Nussbaumer and Storey \cite{NussbaumerS1978} and Badnell \cite{BadnellAS2008}.

\CIV\ is a general multi-configuration atomic structure code that calculates \conint\ wavefunctions
in \LS\ and \intcou\ schemes. The wavefunctions can then be used for further computations related
to atomic processes such as finding \oscstr s. \CIV\ uses \STO\ (STO) formulation to specify the
\conint\ functions. The \STO s are defined by
\begin{equation}\label{STO}
    P_{nl}(r) = \sum_{i} \STOCi r^{\STOPi} e^{- \STOxii r}
\end{equation}
where $P_{nl}(r)$ is a radial function for the $nl$ orbital, $\STOCi$ is a coefficient and $\STOPi$
and $\STOxii$ are indicial parameters in this formulation, $r$ is the radius, and $i$ is a counting
index that runs over the orbitals of interest.

\SSt\ and \AS\ \cite{EissnerJN1974, NussbaumerS1978, BerringtonBBSSTY1987, BadnellAS2008} are
automatic general-purpose atomic structure codes based on the Thomas-Fermi-Dirac formulation. They
compute atomic radial functions using central potentials of the form
\begin{equation}\label{ASradialfunc}
    \CentPot(r) = - \frac{\EffChar(r)}{r}
\end{equation}
where $\EffChar(r)$ is a radius-dependent effective charge. The programs produce large quantities
of diverse atomic structure and transition data particularly useful for astronomical applications.
Minimum specification of the atomic system, the type of the requested data and the required degree
of approximation is needed as an input to \SSt\ and \AS. Among their many applications, they can be
used to calculate the target functions for the \Rm-matrix code and \conmix\ autoionization rates in
\LS\ or \intcou s, where \BrePau\ terms are used to account for relativistic corrections. \SSt\ and
\AS\ can also be used to calculate term energies, atomic and ionic energy levels, term-coupling
coefficients, cascade coefficients, atomic transition data such as forbidden and permitted
radiative transition probabilities, and so on. The programs employ multi-configuration type for
wavefunction expansions.

In our investigation of the C\II\ recombination lines, we used target data generated by \CIV\ for
the scattering calculations as an input to \STGO\ of the \Rm-matrix code, as presented in
Table~\ref{SevenorbitalsSTO} and \S\ \ref{RmaxDataSTGO}, while we used \AS\ to elaborate and
investigate related issues. \AS\ was also used to find the \oscstr s for the free-free transitions
between resonance states since the available \RMAT\ code does not contain a free-free stage. In
this approach the free-free transitions were treated as bound-bound transitions. \AS\ was also used
for generating $\OS$-values for free-bound and bound-bound transitions involving the 8 topmost
bound states, namely the \nlo1s2 2s2p(\SLP3Po)3d~\SLP4Fo and \SLP4Do levels, as these quartets with
their large \effquanum\ for the outer electron are out of range of the \Rm-matrix code validity.
The \AS\ input data files for these calculations are given in \S\ \ref{ASData} in Appendix
\ref{AppInData}.

\chapter{Dielectronic Recombination Lines of C\II} \label{CII} 
Carbon recombination lines are an important diagnostic tool that can provide valuable information
on the physical conditions in various astrophysical environments such as \planeb e and the
interstellar medium. Because the recombination rate depends on the electron temperature $\ElecT$
and number density $\NDe$, the variation of the carbon line intensities can be used in principle to
determine the temperature and density in the line-emitting regions \cite{SochiSCIIList2012}.

With regard to the recombination lines of C\II, there are many theoretical and observational
studies mainly within astronomical contexts. Examples are \cite{Leibowitz1972, Leibowitz1972b,
BalickGD1974, NussbaumerS1975, Boughton1978, ClavelFS1981, HarringtonLS1981, NussbaumerS19812,
CleggSPP1983, HayesN1984, NussbaumerS1984, BogdanovichNRK1985, YanS1987, Badnell1988,
PequignotPB1991, RolaS1994, BaluteauZMP1995, Kholtygin1998, DaveySK2000, LiuLBL2004, PengWHL2004,
PengHWL2005, ZhangLLPB2005}. The aim of the current study is to treat the near-threshold resonances
and subsequent radiative decays using the unified approach of the \Rm-matrix method in intermediate
coupling.

In this study we investigate the recombination lines of C\II, that is the emission lines produced
by the following recombination process
\begin{equation}\label{CIIRecLinesProcess}
    {\rm C}^{2+} + \ele \verb|  | \rightarrow \verb|  | {\rm C}^{+}
\end{equation}
and the subsequent radiative cascade decay which persists until the atom reaches the ground or a
metastable state. Our investigation includes all the autoionizing resonance states above the
threshold of C$^{2+}$ \nlo1s2 \nlo2s2 \SLP1Se with a \priquanum\ $n<5$ for the captured electron as
an upper limit. This condition was adopted mainly due to computational limitations. In total, 61
autoionizing states (27 doublets and 34 quartets) with this condition have been theoretically found
by the \Rm-matrix method. Of these 61 resonances, 55 are experimentally observed according to the
NIST database \cite{NIST2010}. More details will follow in the forthcoming sections.

The major work in this investigation can be split into three main parts

\begin{enumerate}

\item
Autoionizing and bound states calculations to generate the required theoretical data for the
resonances and bound states.

\item
Calculation of \oscstr, $\OS$-values, for various transitions.

\item
Recombination and emissivity calculations and analysis to find the emissivity of various transition
lines and compare the theoretical emissivity model to the observational data.

\end{enumerate}

In the next section we provide a general theoretical background which will be followed by three
sections in which we outline the theory and procedures that were followed in this investigation.

\section{Theoretical Background} \label{TheBac}
In a thermodynamic environment, an excited atomic state is populated by recombination and radiative
cascade from higher states, and depopulated by autoionization and radiative decay to lower states.
Many recombination lines originate from \dierec\ with subsequent cascade decay of autoionizing
resonance states just above the ionization threshold. These lines are dominated by low-temperature
\dierec, as outlined in \S\ \ref{DieRec}. The population of these resonances is established by the
balance between autoionization to the continuum and recombination with ensuing radiative decay to
lower levels. When autoionization dominates, the population is determined by the \Saha\ equation
\begin{equation}\label{SahaPop}
    \ND_{_{\BSRIt}} = \NDe \ND_{_{\IECN}} \frac{\SW_{_{\BSRIt}}}{2\SW_{_{\IECN}}}
    \left( \frac{\h^{2}}{2 \pi \me \BOLTZ \ElecT}  \right)^{3/2}
    e^{^{-\ED_{t} / \BOLTZ \ElecT}}
\end{equation}
where $\ND_{_{\IECN}}$ and $\ND_{_{\BSRIt}}$ are the population of ions with effective positive
charge $n$ and $(n-1)$ respectively, $\NDe$ is the number density of electrons, $\SW$ is the
statistical weight of the indicated state, $\BOLTZ$ and $\h$ are the constants of \Boltzmann\ and
\Planck\ respectively, $\me$ is the electron mass, $\ElecT$ is the electron temperature, and
$\ED_{t}$ is the energy of the recombined electron in the $\BSRIt$ state relative to the ionization
threshold \cite{BenoyMS1993, Ferland2003}.

The \Saha\ equation, which determines the degree of ionization of a gas, is only valid in local
thermodynamic equilibrium situations where the equilibrium is established through thermal collision
processes. In this case the velocity and energy distributions of the gas particles are described by
the \Maxwell\ and \Boltzmann\ distributions respectively and local temperature can be defined.
However, in many practical situations the assumption of local thermodynamic equilibrium is not
applicable in the energy range of interest due to significant involvement of processes other than
thermal collisions in establishing the particles distribution. Examples of these non-thermal
processes are recombination, photoionization, autoionization, and radiative decay. These processes
can shift the distribution significantly away from its thermodynamic equilibrium state if they have
substantial contribution. In these cases, particles distribution should be determined through more
complex balance equations in which the effect of these processes on the energy and velocity of the
particles is taken into account.

In the recombination calculations, to measure the departure of the state from its thermodynamic
equilibrium a \depcoe, $\DCu$, is defined as the ratio of autoionization probability to the sum of
radiative and autoionization probabilities, as given by Equation~\ref{departureCoef}. The value of
this coefficient is between 0 for radiative domination ($\aTP \ll \rTP$) and 1 for thermodynamic
equilibrium where autoionization dominates ($\aTP \gg \rTP$). A large value of $\DCu$ ($\simeq 1$)
is then required to justify the assumption of thermodynamic equilibrium and apply the relevant
physics.

In thermodynamic equilibrium situation the rate of radiationless capture is equal to the rate of
autoionization, yielding \cite{SeatonS1976}
\begin{equation}\label{depCoeBalTE}
    \NDe \NDi \RCc = \SPu \aTPu
\end{equation}
where $\NDe$ and $\NDi$ are the number density of electrons and ions respectively, $\RCc$ is the
\reccoe\ for the capture process, $\SPu$ is the \Saha\ population of the doubly-excited
autoionizing state and $\aTPu$ is the autoionization transition probability of that state. In
non-thermodynamic equilibrium situation, the balance is given by
\begin{equation}\label{depCoeBalnTE}
    \NDe \NDi \RCc = \NDu (\aTPu + \rTPu)
\end{equation}
where $\NDu$ is the non-thermodynamic equilibrium population of the indicated state and $\rTPu$ is
the total radiative transition probability of that state. The \depcoe\ $\DCu$ is a gauge of the
deviation of the population of an autoionizing state from its thermodynamic equilibrium value, and
hence is the ratio of the non-thermodynamic equilibrium population to the \Saha\ population.
Comparing Equations \ref{depCoeBalTE} and \ref{depCoeBalnTE} gives
\begin{equation}\label{depCoe}
    \DCu = \frac{\NDu}{\SPu} = \frac{\aTPu}{\aTPu + \rTPu}
\end{equation}

In the following we summarize the theoretical background for our recombination and emissivity
calculations as implemented in the \EmiCod\ code \cite{SochiEmis2010}

\begin{itemize}


\item
The calculations start by obtaining the radiative transition probability $\rTPul$ for all free-free
transitions as given by
\begin{equation}\label{rTPul}
    \rTPul = \frac{\FSC^{3} \Ep^{2} \SWl \OSul}{2 \SWu \ATU}
\end{equation}
where $\FSC$ is the \finstr\ constant, $\Ep$ is the photon energy in \Ryd, $\SWl$ and $\SWu$ are
the statistical weights of the lower and upper states respectively, $\OSul$ is the \oscstr\ of the
transition between the upper and lower states, and $\ATU$ is the atomic time unit. This is followed
by obtaining the radiative transition probability $\rTPul$ for all free-bound transitions, as in
the case of free-free transitions.


\item
The total radiative transition probability $\rTP_{u}$ for all resonances is then found. This is the
probability of radiative decay from an upper resonance state to all accessible lower resonances and
bound states. This probability is found by summing up the individual probabilities $\rTPul$ over
all lower free and bound states $l$ for which a transition is possible according to the \eledip\
rules, that is
\begin{equation}\label{rTP}
    \rTPu = \sum_{l} \rTPul
\end{equation}


\item
The \depcoe, $\DCu$, for all resonances is then obtained
\begin{equation}\label{departureCoef}
    \DCu = \frac{\aTPu}{\aTPu + \rTPu}
\end{equation}
where $\aTPu$ and $\rTPu$ are the autoionization and total radiative transition probabilities of
state $u$ respectively, and $\aTP$ is given by
\begin{equation}\label{aTP}
    \aTP = \frac{\RW}{\Dirac}
\end{equation}
where $\RW$ is the full width at half maximum of the autoionizing state and $\Dirac$ is the reduced
\Planck's constant.


\item
The next step is to calculate the population of resonances by summing up two components: the \Saha\
capture, and the radiative decay from all upper levels. In thermodynamic equilibrium the rate of
radiationless capture and the rate of autoionization are equal, that is
\begin{equation}\label{depCoeBalTERes}
    \NDe \NDi \RCc = \SPl \aTPl
\end{equation}
where $\NDe$ and $\NDi$ are the number density of electrons and ions respectively, $\RCc$ is the
\reccoe\ for the capture process, $\SPl$ is the \Saha\ population of the autoionizing state and
$\aTPl$ is the autoionization transition probability of that state. In non-thermodynamic
equilibrium situation, the population and depopulation of the autoionizing state due to radiative
decay from upper states and to lower states respectively should be included, and hence the balance
is given by
\begin{equation}\label{depCoeBalnTERes}
    \NDe \NDi \RCc + \sum_{u} \NDu \rTPul = \NDl (\aTPl + \rTPl)
\end{equation}
where $\NDl$ is the non-thermodynamic equilibrium population of the doubly-excited autoionizing
state, $\rTPl$ is the total radiative transition probability of that state, $\rTPul$ is the
radiative transition probability from an upper state $u$ to the autoionzing state $l$, and the sum
is over all upper states that can decay to the autoionzing state. Combining
Equations~\ref{depCoeBalTERes} and \ref{depCoeBalnTERes} yields
\begin{equation}\label{depCoeBalnTERes2}
    \SPl \aTPl + \sum_{u} \NDu \rTPul = \NDl (\aTPl + \rTPl)
\end{equation}
On manipulating Equation~\ref{depCoeBalnTERes2} algebraically, the following relation can be
obtained
\begin{eqnarray}\label{resonancePop2}
  \nonumber
  \NDl &=& \SPl \left( \frac{\aTPl}{\aTPl + \rTPl} \right) + \sum_{u} \frac{\NDu \rTPul}{\rTPl + \aTPl} \\
       &=& \SPl \DCl + \sum_{u} \frac{\NDu \rTPul}{\rTPl + \aTPl}
\end{eqnarray}
where $\DCl$ is the \depcoe\ of the autoionizing state.


\item
The next step is to calculate the radiative transition probabilities of the bound-bound
transitions, $\rTPul$. In these calculations the $\OS$-values can be in length form or velocity
form; however the length form values may be more reliable as they converge rather quickly
\cite{Hibbert1974}. The \EmiCod\ code therefore uses the length form of the \oscstr s where they
are read from the $\OS$-value file `FVALUE' produced by the \Rm-matrix code. The calculation of
$\rTPul$ for the bound-bound transitions is then followed by finding the total radiative transition
probability of the bound states, $\rTPu$, by summing up $\rTPul$ over all lower bound states $l$
accessible to an upper bound state $u$, as given earlier by Equation~\ref{rTP} for the case of
autoionizing states.


\item
The population of the bound states is then obtained
\begin{equation}\label{boundPop}
    \NDl = \sum_{u} \frac{\NDu \rTPul}{\rTPl}
\end{equation}
where the index $u$ includes all upper free and bound states that can decay to the bound state $l$.


\item
Finally, all possible free-free, free-bound and bound-bound transitions which are allowed by the
selection rules of \eledip\ approximation are found. The emissivity, $\EMISS$, of all the
recombination lines that arise from a transition from an upper state $u$ to a lower state $l$ is
then computed using the relation
\begin{equation}\label{emissivity1}
    \ENLul = \NDu \rTPul \h \F
\end{equation}
where $\NDu$ is the population of the upper state, $\rTPul$ is the radiative transition probability
between the upper and lower states, $\h$ is the \Planck's constant, and $\F$ is the frequency of
the transition line. The equivalent effective \reccoe\ $\RCf$, which is linked to the emissivity by
the following relation, can also be computed
\begin{equation}\label{EmissRecCoeff}
    \RCf = \frac{\EMISS} {\NDe \NDi \h \F}
\end{equation}
where $\NDe$ and $\NDi$ are the number density of the electrons and ions respectively.

\end{itemize}

\section{Autoionizing and Bound States Calculations}\label{AutBouCal}
The main computing tools that were used in this part to generate theoretical data is \RMAT\ code
and its companions (i.e. \Km-matrix implementation in stage \STGF\ and \STGQB\ code) though \AS\
was occasionally used to accomplish complementary tasks. In the first part of this section, three
methods for finding and analyzing resonances are presented. However, the main focus will be on the
two methods that were used in our investigation of the recombination lines of C\II. The remaining
parts will be dedicated to practical aspects.

\subsection{Methods for Investigating Resonances} \label{Methods}
The resonance phenomenon is a highly important feature in atomic collisions as it plays a vital
role in electron scattering with atoms, ions and molecules. The \Auger\ effect is one of the early
examples of the resonance effects that have been discovered and extensively investigated. There are
several methods for finding and analyzing resonances that arise during the recombination processes.
In the following sections we outline three of these methods; two of which are used in the current
study to find and analyze resonances of C\II. These two are the \Km-matrix method which was
developed and implemented by modifying stage \STGF\ of the \Rm-matrix code as part of this study,
and the \QB\ method of Quigley and Berrington \cite{QuigleyB1996}, which is implemented in the
\STGQB\ code of Quigley and coworkers \cite{QuigleyBP1998} as an extension to the \Rm-matrix code.
The essence of the \Km-matrix method is to identify the resonance position and width from locating
the poles of the \Km-matrix on the energy mesh, while the essence of the \QB\ method is to apply a
fitting procedure to the reactance matrix eigenphase near the resonance position using the analytic
properties of the \Rm-matrix theory. The third method is the \TimDel\ of Stibbe and Tennyson
\cite{StibbeT1998} which is implemented in a computer code called \TIMEDEL. This method is based on
the use of the lifetime eigenvalues to locate the resonance position and identify its width. We
tried to implement and use this method in an early stage of our investigation before developing the
\Km-matrix method, but this attempt was abandoned in favor of the \Km-matrix method.

\subsubsection{QB Method} \label{QBmethod}
A common approach for finding and analyzing resonances is to apply a fitting procedure to the
reactance matrix, \Km, or its eigenphase as a function of energy in the neighborhood of an
autoionizing state. However, fitting the \Km-matrix itself is complicated because the reactance
matrix has a pole at the energy position of the autoionizing state. An easier alternative is to fit
the arc-tangent of the reactance matrix. The latter approach was employed by Bartschat and Burke
\cite{BartschatB1986} in their fitting code RESFIT.

The eigenphase sum is defined by
\begin{equation}\label{EigSum}
    \EPS = \sum_{i=1}^{N} \arctan \EV_{i}
\end{equation}
where $\EV_{i}$ is an eigenvalue of the \Km-matrix and the sum runs over all open channels
interacting with the autoionizing state. The eigenphase sum is normally fitted to a Breit-Wigner
form
\begin{equation}\label{BreWig}
    \EPS = \EPB + \arctan \left(   \frac{\RW}{2\left(\Er - \E\right)}   \right)
\end{equation}
where $\EPB$ is the sum of the background eigenphase, $\RW$ is the resonance width, $\Er$ is the
energy position of the resonance, and $\E$ is the energy. This approach was used by Tennyson and
Noble \cite{TennysonN1984} in their fitting code RESON \cite{StibbeT1996}.

In theory, an autoionizing state exhibits itself as a sharp increase by $\pi$ radians in the
eigenphase sum as a function of energy superimposed on a slowly-varying background. However, due to
the finite width of resonances and the background variation over their profile, the increase may
not be by $\pi$ precisely in the actual calculations. A more practical approach then is to identify
the position of the resonance from the energy location where the increase in the eigenphase sum is
at its highest rate by having a maximum gradient with respect to the scattering energy, i.e.
$\left(d\EPS/d\E\right)_{max}$ \cite{TennysonN1984, QuigleyB1996}.

The \QB\ method of Quigley and Berrington \cite{QuigleyB1996} is a computational technique for
finding and analyzing autoionizing states that arise in atomic and molecular scattering processes
using eigenphase fitting. The merit of the \QB\ method over other eigenphase fitting procedures is
that it utilizes the analytical properties of the \Rm-matrix method to determine the variation of
the reactance matrix with respect to the scattering energy analytically. This analytical approach
can avoid possible weaknesses, linked to the calculations of \Km-matrix poles and arc-tangents,
when numerical procedures are employed instead. The derivative of the reactance matrix with respect
to the scattering energy in the neighborhood of a resonance can then be used in the fitting
procedure to identify the energy position and width of the resonance.

The \QB\ method begins by defining two matrices, \Qm\ and \Bm, in terms of asymptotic solutions,
the \Rm-matrix and energy derivatives, such that
\begin{equation}\label{QB}
    \D \Km \E = \Bm^{-1} \Qm
\end{equation}
The eigenphase gradients of the \Km-matrix with respect to energy can then be calculated. This is
followed by identifying the resonance position, $\Er$, from the point of maximum gradient at the
energy mesh, and the resonance width, $\RW$, which is linked to the eigenphase gradient at the
resonance position, $\EPGK(\Er)$, by the relation
\begin{equation}\label{QBwidth}
    \RW = \frac{2}{\EPGK(\Er)}
\end{equation}
This equation may be used to calculate the widths of a number of resonances in a first
approximation. A background correction due to overlapping profiles can then be introduced on these
widths individually to obtain a better estimate.

The \QB\ method was implemented in the \STGQB\ code of Quigley and coworkers \cite{QuigleyBP1998}
as an extension to the \Rm-matrix code. In the current study, \STGQB\ was used as a supplementary
tool for verifying the results of the \Km-matrix method.

\subsubsection{\TimDel\ Method} \label{TDmethod}
The \TimDel\ method of Stibbe and Tennyson \cite{StibbeT1998} is based on the \timdel\ theory of
Smith \cite{Smith1960}. According to this theory, the \timdel\ matrix \Mm\ is defined in terms of
the scattering matrix \Sm\ by
\begin{equation}\label{TDmatrixS2}
    \Mm = -\iu \, \Dirac \, \Sm^{*} \frac{d \Sm}{d \E}
\end{equation}
where \iu\ is the imaginary unit, $\Dirac$ ($=\h/2\pi$) is the reduced \Planck's constant,
$\Sm^{*}$ is the complex conjugate of $\Sm$, and $\E$ is the energy. It has been demonstrated by
Smith \cite{Smith1960} that the eigenvalues of the \Mm-matrix represent the collision lifetimes and
the largest of these eigenvalues corresponds to the longest \timdel\ of the scattered particle. For
a resonance, the \timdel\ has a Lorentzian profile with a maximum precisely at the resonance
position. By computing the energy-dependent \timdel\ from the reactance matrix, and fitting it to a
Lorentzian peak shape, the resonance position can be located and its width is identified.

This method, as implemented in the \TIMEDEL\ program of Stibbe and Tennyson \cite{StibbeT1998},
uses the reactance \Km-matrix as an input, either from a readily-available archived scattering
calculations or from dynamically-performed computations on an adjustable mesh. The \Sm-matrix is
then formed using the relation
\begin{equation}\label{SmatrixEq2}
    \Sm = \frac{\IM + \iu \Km}{\IM - \iu \Km}
\end{equation}
where $\IM$ is the identity matrix, and $\iu$ is the imaginary unit. The \timdel\ \Mm-matrix is
then calculated from Equation~\ref{TDmatrixS2}, with numerical evaluation of the \Sm-matrix
derivative, and diagonalized to find the eigenvalues and hence obtain the longest \timdel\ of the
incident particle. Approximate resonance positions are then identified from the energy locations of
the maxima in the \timdel\ profile, and the widths are estimated from the Lorentzian fit. On
testing the degree of overlapping of neighboring resonances, \TIMEDEL\ decides if the resonances
should be fitted jointly or separately.

\subsubsection{K-Matrix Method} \label{Kmethod}
Resonances arise from poles in the scattering matrix \Sm\ which varies slowly with energy. The
\Km-matrix method used in the current study is based on the fact that for the low-lying resonances
just above the C$^{2+}$ 1s$^2$ \nlo2s2 \SLP1Se\ ionization threshold, the scattering matrix \Sm\
has only one channel, and hence the reactance matrix, \Km, is a real scalar with a pole near the
resonance position at the energy mesh.

According to the collision theory of Smith \cite{Smith1960}, the lifetime matrix \Mm\ is related to
the \Sm-matrix by Equation~\ref{TDmatrixS2}. Now, a \Km-matrix with a pole at energy $\Eo$
superimposed on a background $\BGCK$ can be approximated by
\begin{equation}\label{Kmatrix}
    \SVKi = \BGCK + \frac{\PKF}{\Ei - \Eo}
\end{equation}
where $\SVKi$ is the value of the \Km-matrix at energy $\Ei$ and $\PKF$ is a physical parameter
with dimension of energy. In Appendix~\ref{AppKmatrix} we revealed that in the case of
single-channel scattering the \Mm-matrix is real with a value given by
\begin{equation}\label{mmatrix}
    \M = \frac{-2 \PKF}{(1+ \BGCK^{2})(\E-\Eo)^{2}+2 \BGCK \PKF (\E - \Eo) + \PKF^{2}}
\end{equation}
Using the fact demonstrated by Smith \cite{Smith1960} that the lifetime of the state is the
expectation value of $\M$, it can be shown from Equation~\ref{mmatrix} that the position of the
resonance peak $\Er$ is given by
\begin{equation}\label{Er}
    \Er = \Eo - \frac{\BGCK \PKF}{1 + \BGCK^{2}}
\end{equation}
while the full width at half maximum $\FWHMe$ is given by
\begin{equation}\label{FWHM}
    \FWHMe = \frac{|2 \PKF|}{1 + \BGCK^{2}}
\end{equation}

The two parameters of primary interest to our investigation are the resonance energy position
$\Er$, and the resonance width $\RW$ which equals the full width at half maximum $\FWHMe$. However,
for an energy point $\Ei$ with a \Km-matrix value $\SVKi$, Equation~\ref{Kmatrix} has three
unknowns, $\BGCK$, $\PKF$ and $\Eo$, which are needed to find $\Er$ and $\RW$. Hence, three energy
points at the immediate neighborhood of $\Eo$ are required to identify these unknowns. As the
\Km-matrix changes sign at the pole, the neighborhood of $\Eo$ is located by testing the \Km-matrix
value at each point of the energy mesh for sign change. Consequently, the three points are obtained
and used to find $\Er$ and $\RW$. Complete derivation of the \Km-matrix method is given in
Appendix~\ref{AppKmatrix}.

\subsection{Preparing Target} \label{Target}
We used the \Rm-matrix code \cite{BerringtonEN1995}, and \AS\ \cite{EissnerJN1974, NussbaumerS1978,
BadnellAS2008} to compute the properties of autoionizing and bound states of C$^+$. The first step
in the \Rm-matrix is to have a target for the scattering calculations. In our study, orbitals
describing the C$^{2+}$ target for the \Rm-matrix scattering calculations were taken from
Berrington \etal\ \cite{BerringtonBDK1977}, who used a  target comprising the six terms \nlo2s2
\SLP1Se, \nlo2s{}\nlo2p{} \SLP3Po, \SLP1Po and \nlo2p2 \SLP3Pe, \SLP1De, \SLP1Se, constructed from
seven orthogonal orbitals; three physical and four pseudo orbitals. These orbitals are: 1s, 2s, 2p,
$\po3s$, $\po3p$, $\po3d$ and $\po4f$, where the bar denotes a pseudo orbital. The purpose of
including pseudo orbitals is to represent electron correlation effects and to improve the target
wavefunctions. The radial parts, $P_{nl}(r)$, of these orbitals are Slater Type Orbital, defined by
Equation~\ref{STO}, generated by the \CIV\ program of Hibbert \cite{Hibbert1975}. The values of
these parameters are given in Table~\ref{SevenorbitalsSTO}.

In this work we construct a scattering target of 26 terms which include the 6 terms of Berrington
\etal\ \cite{BerringtonBDK1977} plus all terms of the configurations \nlo2s{}$\pon3l$ and
\nlo2p{}$\pon3l$, $l=0,1,2$.  This includes the terms outside the $n=2$ complex which make the
largest contribution to the dipole polarizability of the \nlo2s2 \SLP1Se and \nlo2s{}\nlo2p{}
\SLP3Po states of C$^{2+}$. \AS\ was used to decide which terms of the target have the largest
polarizability. This is based on the fact that the contribution to polarizability is proportional
to the ratio of the \oscstr\ of transition between levels $i$ and $j$ to the square of the
corresponding energy difference (i.e. $\Cp \propto \OS_{ij}/\Delta E^{2}_{ij}$). The \AS\ input
data file for this calculation is given in \S\ \ref{ASPloar} of Appendix~\ref{AppInData}.


\begin{table} [!b]
\caption{The seven orbitals used to construct the C$^{2+}$ target and their Slater Type Orbital
parameters. The bar marks the pseudo orbitals.} \label{SevenorbitalsSTO} \centering
\begin{tabular}{cccc}
\hline
   Orbital &   $\STOCi$ &   $\STOPi$ &  $\STOxii$ \\
\hline
        1s &   21.28251 &          1 &    5.13180 \\
           &    6.37632 &          1 &    8.51900 \\
           &    0.08158 &          2 &    2.01880 \\
           &   -2.61339 &          2 &    4.73790 \\
           & -0.00733  \VS &          2 &    1.57130 \\
        2s &   -5.39193 &          1 &    5.13180 \\
           &   -1.49036 &          1 &    8.51900 \\
           &    5.57151 &          2 &    2.01880 \\
           &    -5.25090 &          2 &    4.73790 \\
           & 0.94247  \VS &          2 &    1.57130 \\
   $\po3s$ &    5.69321 &          1 &    1.75917 \\
           &  -19.54864 &          2 &    1.75917 \\
           & 10.39428  \VS &          3 &    1.75917 \\
        2p &    1.01509 &          2 &    1.47510 \\
           &    3.80119 &          2 &    3.19410 \\
           &    2.75006 &          2 &    1.83070 \\
           & 0.89571  \VS &          2 &    9.48450 \\
  $\po3p$ &   14.41203 &          2 &    1.98138 \\
           & -10.88586  \VS &          3 &    1.96954 \\
  $\po3d$ & 5.84915  \VS &          3 &    2.11997 \\
  $\po4f$ &    9.69136 &          4 &    2.69086 \\
\hline
\hspace{3cm} & \hspace{3cm} & \hspace{3cm} & \hspace{3cm} \\

\end{tabular}
\end{table}

\subsection{Developing and Implementing \Km-matrix Method}\label{DevImpKM}
In the early stages of our theoretical investigation to C$^+$ autoionizing states, we experienced
numerical instabilities, convergence difficulties and failures from the \QB\ method as implemented
in the \STGQB\ code. It was necessary, therefore, to verify the results that we obtained from \QB\
by an independent method and fill the gaps left by the failure of \STGQB\ to converge in some cases
due to limitation on the width of resonances as it was established afterwards. This motivated the
development of \Km-matrix approach which we implemented in the \STGF\ stage by searching for
sign-change in the \Km-matrix at the poles. This came after an attempt to implement the \timdel\
method of Stibbe and Tennyson \cite{StibbeT1998} once within \STGF\ stage and another time as a
stand-alone program that processes the data produced by \STGF. However, this attempt was abandoned
in favor of the \Km-matrix approach which relies on the same physical principle as the \timdel\
method.

To improve the performance of the \Km-matrix method, which proved to be very successful and
computationally efficient, an interactive graphical technique was also developed to read the
\Km-matrix data directly and plot it against energy or arbitrary ordinal number while searching for
poles. With a reasonably fine mesh, a positive or negative pulse in the graph appears even when the
background is sufficiently large to prevent sign-change. As soon as this is detected, the search is
stopped and resumed at higher resolution by focusing on a very narrow energy band using a fine
mesh, and hence very small number of energy points are needed to find the resonance. The strategy
is to start the search with a coarse mesh over the suspected energy range. If the glitch failed to
appear, the search is repeated with finer mesh until the glitch is observed. This graphical
technique was essential for finding resonances in reasonably short time compared to the time
required by \STGQB. In a later stage, non-graphical tools for pole searching were developed and
used. The purpose of these tools is to search for any sudden increase or decrease in the background
of the \Km-matrix. As soon as this is detected, a search for poles with a finer mesh is resumed.
These non-graphic tools proved to be more efficient than the graphic tools, and hence they helped
substantially in finding most of the resonances with limited time and computational effort. It is
noteworthy that the \Km-matrix method as implemented in \STGF\ stage is completely automatic and
can find and identify resonances without user interaction. The purpose of these graphic and
non-graphic tools is to speed up the search and save the computing resources by reducing the
required CPU time.

In this paragraph we present a brief comparison between these two methods (i.e. \Km-matrix and \QB)
and assess their roles in this study and any possible future investigation. The main results of
these methods are presented in Table~\ref{RTableKQ}. As seen, \Km-matrix and \QB\ produce identical
results in most cases. However, \Km-matrix is computationally superior in terms of the required
computational resources, mainly CPU time, especially when coupled with the above-mentioned
interactive graphic and non-graphic techniques.
In fact we obtained most of the \QB\ results guided by the \Km-matrix results which were obtained
earlier. Without \Km-matrix it would be extremely hard and time consuming to obtain results for
some resonances, especially the very narrow ones, by \QB\ directly. In general, the purpose of
obtaining \QB\ results is to check the \Km-matrix results and compare the two methods. Another
advantage of the \Km-matrix method is that it has a wider range of validity with regard to the
resonance width, that is in principle it can be used to find resonances with any width. The \QB\
method fails to converge when the resonance width falls below a certain limit, whereas the only
observable constraint on the \Km-matrix method is machine precision. Nevertheless, the \QB\ method
is more general as it deals with multi-channel resonances, as well as single-channel resonances,
while the \Km-matrix method in its current formulation is restricted to single-channel resonances.

\subsection{Calculations} \label{Calculations}
Using the above-mentioned 26-term target, extensive calculations were carried out using the
\Rm-matrix code in \intcou\ (IC) scheme by including the one-body terms of the \BrePau\
\Hamiltonian\ as outlined in Chapter \ref{Physics}. The IC was achieved by utilizing stage \STGJK\
of the \Rm-matrix code with the levels and $J\Par$-symmetries identification. Sample calculations
in \LS-coupling were also performed in an early stage on typical case studies for validation and
test. The \intcou\ scheme is required for a comprehensive C\II\ recombination lines investigation
as many states and transitions do not exist within \LS-coupling. An important case is the quartets
as well as some doublet states which do not autoionize within \LS-coupling scheme where the
$LS\Par$ are not conserved. The requirement for the intermediate coupling arises from the fact that
in \LS-coupling the conserved quantities are $LS\Par$ and hence only the doublet states that
conserve these quantities, such as \SLP2Se and \SLP2Po, can autoionize. Therefore, in \LS-coupling
no autoionization is allowed for the quartet terms and some doublet states, such as \SLP2So and
\SLP2Pe. However, under \intcou\ scheme the conserved quantities are $J\Par$, therefore the
\LS-allowed and \LS-forbidden states with the same $J\Par$ mix up giving access to channels that
were formerly inaccessible, and hence these states can autoionize. This is schematically depicted
in Figure~\ref{quartets}.


\begin{figure}[!h]
\centering{}
\includegraphics[scale=0.6, trim = 0 0 0 0]{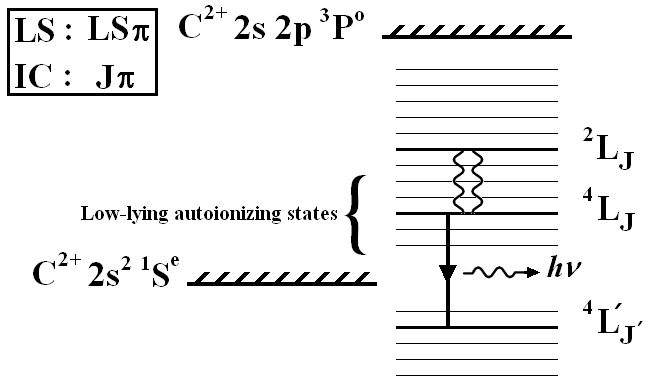}
\caption{Schematic diagram of the low-lying autoionizing quartet and doublet states and their
interaction in \intcou\ scheme.} \label{quartets}
\end{figure}

To investigate the convergence and numerical stability, the number of continuum basis orbitals used
to express the wavefunction in the inner region (MAXC in \STGO\ of the \Rm-matrix code) was varied
between 6-41 and the results were analyzed. It was noticed that increasing the number of basis
functions, with all ensuing computational costs, does not necessarily improve the results; moreover
a convergence instability may occur in some cases. It was decided therefore to use MAXC = 16 in all
calculations as a compromise between the computational resources requirement and accuracy. The
effect of varying the size of the inner region radius $\IRR$ in the \Rm-matrix formulation was also
investigated and a value of 10 atomic units was chosen on the basis of numerical stability and
convergence, that is the resulting values do not change on increasing the radius.

Concerning the investigation of resonances, initially all computational work was carried out using
the \STGQB\ code since this was the only available tool before developing and implementing the
\Km-matrix approach. Several parameters of the \QB\ method were systematically inspected. The
effect of varying the accuracy parameter, AC, in \STGQB\ was investigated and a value of $10^{-6}$
was chosen for all calculations as a deal between accuracy and speed. The effect of modifying the
step size for the \effquanum\ mesh was studied and beside missing some resonances when the step is
not appropriately small as expected, it is observed that in some cases the resonance position and
width may depend on the step size. Including the first order perturbation in the long-range
coupling potentials when calculating the reactance matrix was also examined by turning `IPERT' flag
on and off (i.e. IPERT = 1 and 0) but no tangible change was observed. We therefore concluded that
the perturbative treatment is not implemented in the \STGQB\ code.

The development of the \Km-matrix method provided a way for checking and filling the gaps of the
\QB\ results. In the initial formulation of the \Km-matrix the background was ignored, and although
there was good agreement between \QB\ and \Km-matrix in some cases where the background is small,
there was significant differences in the other cases. Consequently, the current formulation which
takes into account the effect of background was developed after some checking using \AS. The
results produced by the \Km-matrix method, which are presented in Table~\ref{RTableKQ}, are
perturbation-free, i.e. with `IPERT = NO' in the stage \STGF. Theoretical data with perturbative
treatment was also produced for some resonances at an early stage, but this work was abandoned. The
main reason is the highly time-consuming nature of the perturbative calculations. It is estimated
that such calculations are by a factor of 5-10 times slower than that with the perturbative
treatment off, i.e. when `IPERT=NO' \cite{BadnellRmax2002}. Our personal experience confirms this
estimation.

With regard to sampling the three points for the \Km-matrix calculations, it was observed that
sampling the points very close to the pole makes the energy position and width of resonances
susceptible to fluctuations and instabilities. Therefore, a sampling scheme was adopted in which
the points are selected from a broad range not too close to the pole. This approach was implemented
by generating two meshes, coarse and fine, around the pole as soon as the pole is found. To check
the results, several different three-point combinations for each resonance were used to find the
position and width of the resonance. In each case, the results from these different combinations
were compared. In all cases they were identical within acceptable numerical errors. This verified
that the sampling scheme is reliable. The results of \QB\ confirm this conclusion as they agree
with the \Km-matrix results as can be seen in Table~\ref{RTableKQ}.

The results for all bound and autoionizing states  are given in Tables~\ref{BTable} and
\ref{RTableKQ}. In total, 142 bound states belonging to 11 symmetries ($2J=1,3,5,7,9$ even and
$2J=1,3,5,7,9,11$ odd) and 61 resonances belonging to 11 symmetries ($2J=1,3,5,7,9,11$ even and
$2J= 1,3,5,7,9$ odd) were theoretically identified. As seen in Tables \ref{BTable} and
\ref{RTableKQ}, the theoretical results for both bound and resonance states agree very well with
the available experimental data both in energy levels and in fine structure splitting. Experimental
energies are not available for the very broad resonances as they are difficult to find
experimentally. The maximum discrepancy between experiment and theory in the worst case does not
exceed a few percent. Furthermore, the ordering of the energy levels is the same between the
theoretical and experimental in most cases. Order reversal in some cases is indicated by a minus
sign in the \finstr\ splitting. The input data for the \Rm-matrix which are used to perform these
calculations are given in \S\ \ref{RmaxData} in Appendix~\ref{AppInData}.

\section{$\OS$-Values Calculations} \label{fValues}
The \oscstr s for all types of transition (i.e. free-free, free-bound and bound-bound) are required
to find the radiative probabilities for these transitions. In the following three sections we
outline the procedures that we followed to produce the $\OS$-values for FF, FB and BB transitions.

\subsection{$\OS$-Values for Free-Free Transitions} \label{OfValFF}
As there is no free-free stage in the available \Rm-matrix code, the $\OS$-values for the FF
transitions were generated by \AS\ in the \intcou\ scheme where 60 electron configurations were
included in the atomic specification of the \AS\ input: 2s$^{2}$ $nl$ (2p$\leq nl \leq$7s), 2s2p
$nl$ (2p$\leq nl \leq$7s), 2p$^{3}$, and 2p$^{2}$ $nl$ (3s$\leq nl \leq$7s). An iterative procedure
was followed to find the orbital scaling parameters ($\lambda$'s) for \AS\ before producing the
required data. These scaling parameters are given in Table~\ref{lambdaTable}. The scaling
parameters, which are obtained by \AS\ in an automated optimization variational process, are
required to minimize the weighted energy sum of the included target states \cite{EissnerN1969,
EissnerJN1974, Storey1981}. The \AS\ input data for generating the $\OS$-values for the free-free
transitions is given in \S\ \ref{AsfVal} in Appendix~\ref{AppInData}. It should be remarked that
\AS, with the same input data, was also used to generate $\OS$-values for the free-bound
transitions and the bound-bound transitions that involve the 8 topmost bound states, namely the
1s$^2$ 2s2p(\SLP3Po)3d~\SLP4Fo and \SLP4Do levels, as these states have large \effquanum\ and hence
are not accessible to the \RMAT\ code. Because \AS\ produces data only to a certain level of
approximation (i.e. the $\OS$-value is ignored when it falls below a certain limit) some of the
entries were missing from the `olg' file of the \AS\ output and hence zero was inserted for these
missing entries in the \oscstr\ files required by the \EmiCod\ code.

\begin{table} [!h]
\caption{Orbital scaling parameters ($\lambda$'s) for \AS\ input. The rows stand for the principal
quantum number $n$, while the columns stand for the orbital angular momentum quantum number $l$.}
\label{lambdaTable} \vspace{0.2cm}

\centering
\begin{tabular}{|c|c|c|c|c|c|c|}
 \hline
 & s & p & d & f & g & h \\
 \hline
 1 & 1.43240 &  &  &  &  & \\
 \hline
 2 & 1.43380 & 1.39690 &  &  &  & \\
 \hline
 3 & 1.25760 & 1.20290 & 1.35930 &  &  & \\
 \hline
 4 & 1.25830 & 1.19950 & 1.35610 & 1.41460 &  & \\
 \hline
 5 & 1.26080 & 1.20020 & 1.35770 & 1.41420 & 1.32960 & \\
 \hline
 6 & 1.26370 & 1.20250 & 1.36210 & 1.41520 & 1.41420 & 2.34460 \\
 \hline
 7 & 1.26790 &  &  &  &  & \\
 \hline
\end{tabular}
\end{table}

\subsection{$\OS$-Values for Free-Bound Transitions}\label{OfValFB}
The $\OS$-values for more than 2500 free-bound transitions were computed by integrating the peaks
of the \pcs s (in mega barn) over the photon energy (in Rydberg). This was done for each bound
state and for all resonances in the corresponding cross-section. The area under the cross-section
curve comprises a background contribution, assumed linear with energy, and the contribution due to
the resonance, which is directly related to the bound-free oscillator strength. The background
contribution was therefore removed by subtracting the linear fit at the bottom of the profile from
the peak. The computation of oscillator strengths from integrating the peaks of the \pcs s is based
on the fact that the contribution of a resonance to the recombination rate correlates to the area
underneath, and hence it quantifies the \oscstr\ between the resonance and the interacting state,
while the width of a resonance depends on the strength of the interaction with the continuum
\cite{Storey1994}. Therefore, the first step in these computations is to obtain the \pcs\ of a
particular transition from a lower bound-state $l$ to the continuum, $\PICS_{(l \rightarrow u)}$.
Since the resonance contribution to a particular radiative process is proportional to the area
beneath $\PICS$ for that process, by integrating $\PICS$ over energy the $\OS$-value for a
transition from an upper resonance state $u$ to a lower bound state $l$ is found.

In the following points we outline the general procedure that was followed to compute the \oscstr s
for the FB transitions:

\begin{itemize}

\item
\RESMESH\ program of P.J. Storey [private communication] was used to create an energy mesh that
maps the resonances in the most precise way. The input data required by \RESMESH\ are the position
and width of resonances, the energy range of the mesh and an integral error index. In our case, the
positions and widths of resonances were obtained by the \Km-matrix method as implemented in stage
\STGF\ of the \Rm-matrix code. It should be remarked that a separate energy mesh was used to
generate \pcs\ data for each individual resonance with a refinement process to ensure correct
mapping and to avoid peak overlapping from different resonances.

\item
The mesh generated by \RESMESH\ was then used as an input to stage \STGF\ of the \Rm-matrix code to
create F-files which contain the required data for resonances.

\item
Stage \STGBF\ was then run to create `XSECTN' file which contains the data for \pcs s (in mega
barn) versus photon energies (in \Ryd). Plots of some representative cross-sections are shown in
Figure~\ref{GGs} which displays a number of examples of \pcs s of the indicated bound states close
to the designated autoionizing states versus photon energy on linear-linear graphs.

\item
The peaks of the \pcs s were extracted by graphic techniques. The numerical data were then used to
obtain the $\OS$-values by integrating the \pcs s over photon energy using a 3-point Simpson's
rule. A piece of code from P.J. Storey [private communication] was used to perform the integration
with the required scaling. Visual inspection to each peak during the extraction of data and
computing the \oscstr s was carried out to check the goodness of profile and verify that it
contains sufficient number of points. In general, the data points that define the peak profile and
which are used for the integrated cross sections were chosen to be as close as possible to the
background on both sides considering some sort of symmetry and avoiding mixing with other
resonances. The possibility of mixing between neighboring resonances was eliminated later on by
producing \pcs\ data (`XSECTN' files) for each resonance independently with the adoption of
one-resonance mesh approach, as will be discussed next.

\end{itemize}

\begin{figure}
\centering %
\subfigure[2s2p(\SLP3Po)3d \SLPJ4Po{1/2} resonance in 2s2p$^2$ \SLPJ4Pe{1/2} cross-section.]%
{\begin{minipage}[b]{1\textwidth} \centering \includegraphics[width=3.2in] {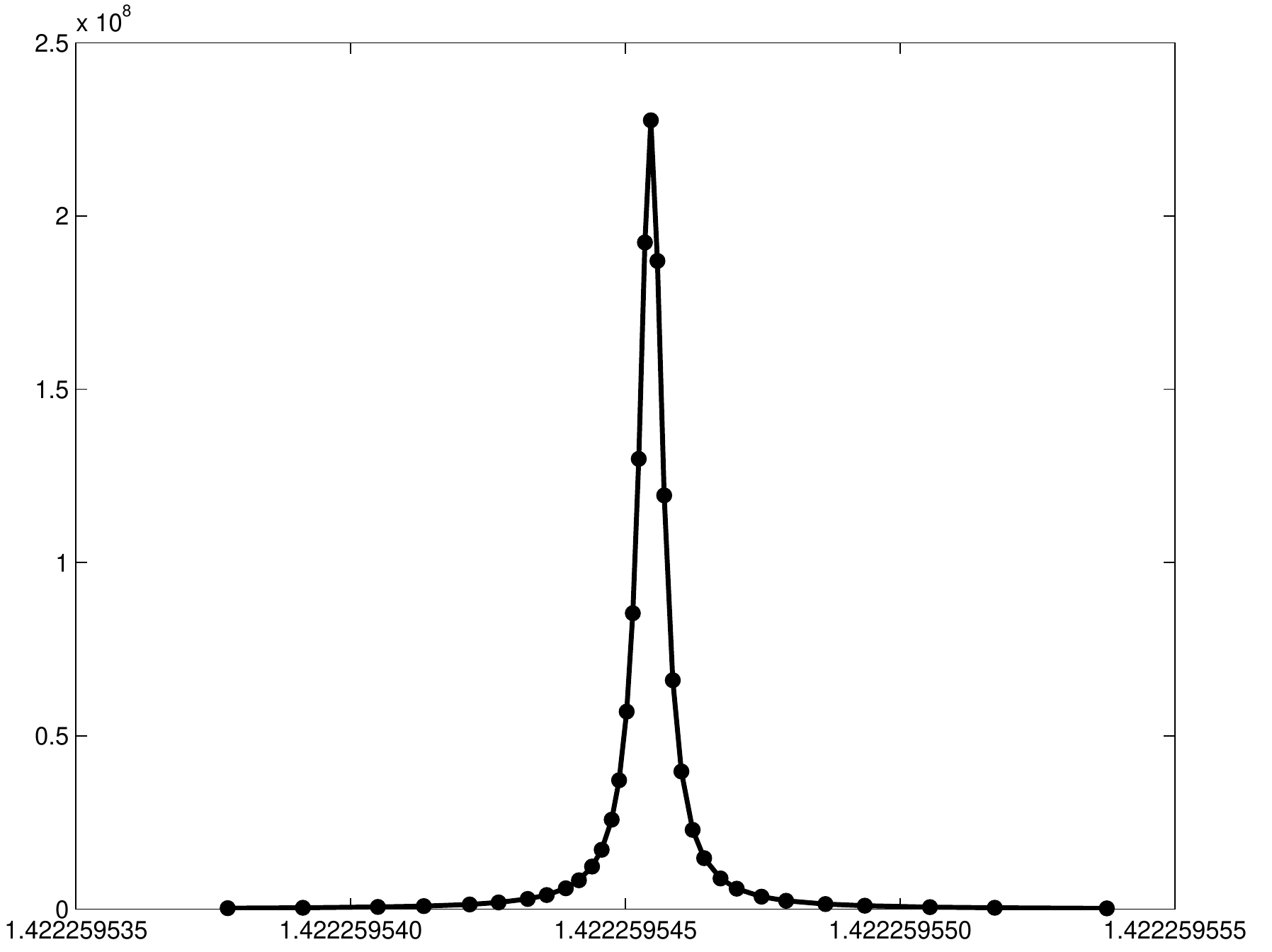}
\end{minipage}} \vspace{-0.1cm}
\centering %
\subfigure[2s2p(\SLP3Po)4d \SLPJ2Fo{5/2} resonance in 2s2p$^2$ \SLPJ4Pe{3/2} cross-section.]%
{\begin{minipage}[b]{1\textwidth} \centering \includegraphics[width=3.2in] {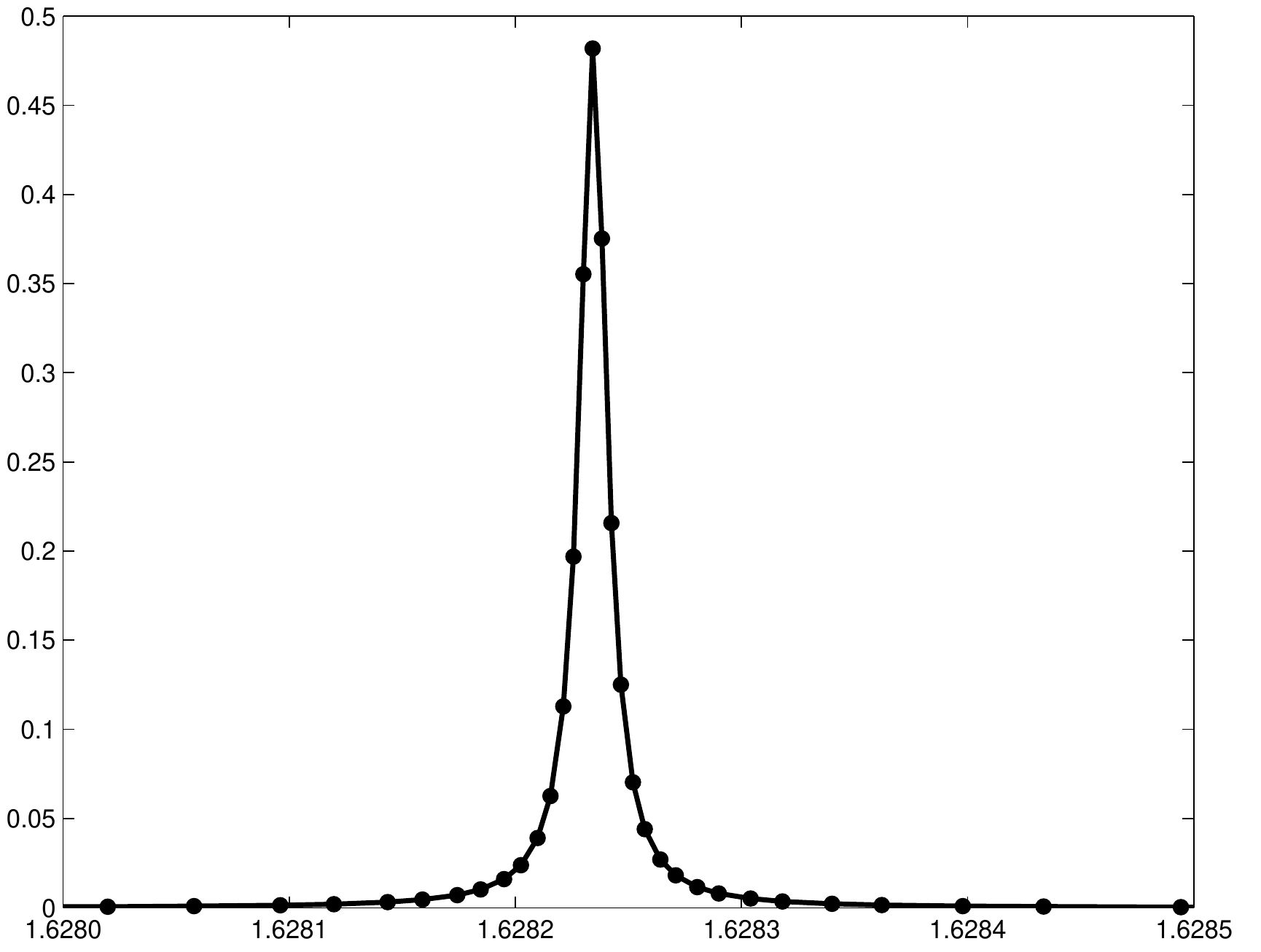}
\end{minipage}} \vspace{-0.1cm}
\centering %
\subfigure[2s2p(\SLP3Po)4p \SLPJ2Pe{3/2} resonance in  2p$^3$ \SLPJ2Do{5/2} cross-section.]%
{\begin{minipage}[b]{1\textwidth} \centering \includegraphics[width=3.2in] {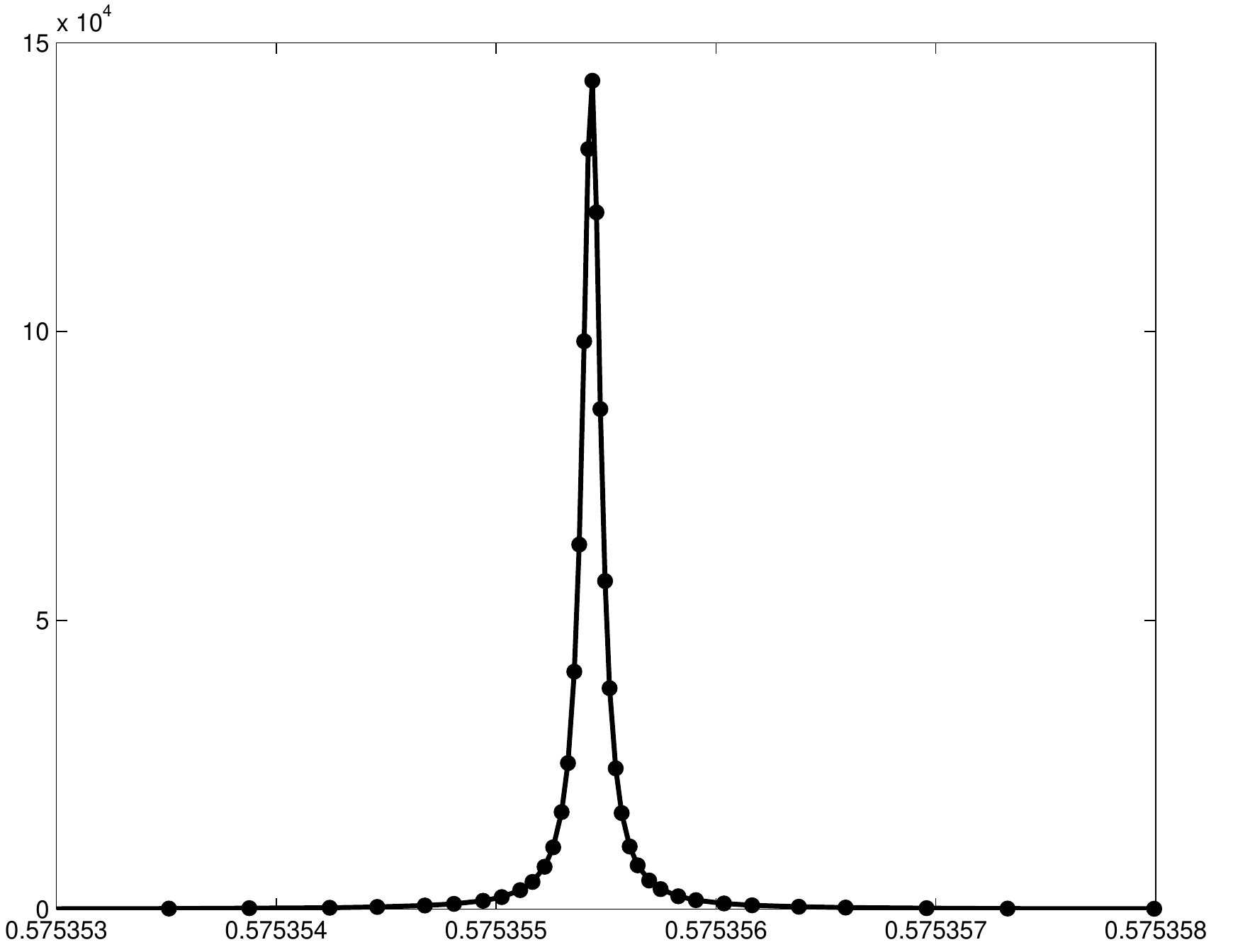}
\end{minipage}}
\caption{Examples of \pcs s in mega barn ($y$-axis) of the indicated bound states near to the
designated autoionizing states versus photon energy in Rydberg ($x$-axis). \label{GGs}}
\end{figure}

A refinement process was used to improve the positions of the resonances by inputting the results
of one run to the next during mesh generation and data creation. The purpose of refinement is to
optimize the position of resonances and obtain optimal resonance profiles. In the refinement
process the \RESMESH\ cycle, as outlined in the previous steps, was repeated several times (about
five) where the results of one cycle was fed to the next. To find the positions in each cycle, all
the peaks belonging to the resonance of interest are obtained from `XSECTN' file created in that
cycle. The energy position of the largest peak was then taken as an input to \RESMESH\ in the next
run. As the refinement progresses, the situation steadily improves in terms of peaks height and
shape until it stabilizes where no improvement is observed on performing more iterations. Most of
this process was carried out by automated procedures, and the results were thoroughly inspected and
checked.

Initially a single mesh was used to optimize the position of all resonances in a single run. This
is based on the assumption that the results will improve in general as the refinement progresses
from one cycle to the next. However, it was observed that although the results generally do improve
for some resonances they deteriorate for others. Moreover, there was a considerable overlap between
some resonances especially for those from multiplets belonging to different symmetries.

A new approach of refinement was then started by continuing the refinement for the ones that
continue to improve while holding the deteriorated ones at the position of their best \pcs\ value.
However, the results did not move in the anticipated direction and similar behavior to the previous
(i.e. deterioration for some resonances) was observed. The apparent conclusion is that the
resonances within a mesh interact, and therefore it is difficult to control the individual
resonances within the mesh. Another serious problem with the single mesh is the size of the data
files and the confusion created by the simultaneous presence of many resonances for each  data set
belonging to each bound state. Moreover, when resonances are too broad or too close to each other
they mix up creating more reasons for confusion and error. 
These factors made processing the data without serious errors unfeasible.

A different approach was therefore adopted by generating a mesh for each individual resonance. The
advantage of this approach is that it requires less effort in data extraction and produces more
reliable results. Moreover, the extraction of data can be highly automated. The biggest advantage
of the one-resonance mesh approach is that for the multiplets of the very broad resonances, such as
the doublet 2s2p(\SLP3Po)4s \SLPJ2Po{1/2} and \SLPJ2Po{3/2}, separate peak profiles are obtained
for each level, whereas only one peak profile which combines the profiles of these two levels was
obtained in the previous attempts using the single-mesh approach. However, the one-resonance mesh
requires more preparation work since preparing the data and running the code are repeated as many
times as the number of resonances. As indicated in \S\ \ref{OfValFF}, \AS\ was used to produce
$\OS$-values for the FB transitions that involve the 8 topmost bound states due to the \Rm-matrix
limitation related to the large \effquanum\ of these states.

\subsection{$\OS$-Values for Bound-Bound Transitions}\label{OfValBB}
No special effort was required to compute the $\OS$-values for the bound-bound transitions as these
calculations were performed using stage \STGBB\ of the \Rm-matrix code.
The input data for the \Rm-matrix stages which were used to generate these data are given in \S\
\ref{RmaxData} in Appendix \ref{AppInData}. As indicated in \S\ \ref{OfValFF}, \AS\ was used to
produce $\OS$-values for the BB transitions that involve the 8 topmost bound states due to the
\Rm-matrix limitation related to the large \effquanum\ of these states.

\section{Recombination and Emissivity Calculations} \label{EmiCal}

Emissivity and recombination calculations are the third major part of our work on the C\II\
recombination lines. We are concerned with spectral lines formed by dielectronic capture followed
by radiative decay. Including only these two processes and autoionization, the number density,
$\NDl$ of a state $l$ is given by
%
\begin{equation}\label{detbal}
    \NDe \NDi \RCc + \sum_{u} \NDu \rTPul = \NDl (\aTPl + \rTPl)
\end{equation}
where $\RCc$ is the rate coefficient for dielectronic capture to state $l$ and $\aTPl$ is the
autoionization probability of that state, related by
\begin{equation}
\RCc = \left( \frac{\NDl}{\NDe \NDi} \right)_S \aTPl
\end{equation}
where the subscript $S$ refers to the value of the ratio given by the Saha equation, and $\NDe$ and
$\NDi$ are the number densities of electrons and ions respectively. If state $l$ lies below the
ionization limit, $\RCc=\aTPl=0$. Equation~\ref{detbal} can be solved for the populations $l$ by a
stepwise downward iteration using the `Emissivity' code \cite{SochiEmis2010}. Full account of the
theoretical infrastructure of the `Emissivity' code, which include the computation of the
emissivities and their equivalent effective recombination coefficients which are the physical
parameters of primary interest to this investigation, is given in \S\ \ref{TheBac}.


As part of this investigation, the C~{\sc ii} lines from several observational line lists found in
the literature, such as that of Sharpee \etal\ \cite{SharpeeWBH2003} for the planetary nebula
IC~418, were analyzed using our theoretical line list and all correctly-identified C\II\
recombination lines in these observational lists were identified in our theoretical list apart from
very few exceptions which are outside our wavelength range. The analysis also produced an electron
temperature for the line-emitting regions of a number of astronomical objects in reasonably good
agreement with the values obtained by other researchers using different data and employing other
techniques. The details of this investigation are reported in Chapter~\ref{Temp}.

\subsection{Practical Aspects} \label{PracAsp}
%
Regarding the input data used in the transition lines calculations, all theoretical data for the
energy of resonances and bound states were replaced with experimental data from \NIST\ when such
experimental data were available. This is based on the approach of Nussbaumer and Storey which was
mentioned in Section \ref{LitRecThe}. The use of experimental positions removes the uncertainty and
errors which are usually introduced by \abin\ calculations.

It should be remarked that all transitions in our calculations are subject to the \eledip\ (E1)
rules. Also, the data of \SSo\ list are generated assuming electron and ion number density of
10$^{10}$~m$^{-3}$. Another remark is that the autoionization and radiative probabilities in the
emissivity calculations as outlined in \S\ \ref{TheBac} can also be obtained from \AS. In the early
stages of this research such an approach was adopted by producing some data by \RMAT\ and some by
\AS. According to the old approach, the widths of resonances are first obtained by the \Km-matrix
method; from which the autoionization transition probabilities, $\aTP$, are computed. The
corresponding radiative transition probabilities, $\rTP$, are then found by \AS. The configurations
that were used in the radiative calculations are the five most important ones for C$^+$: 1s$^{2}$
2s$^{2}$ 3d, 1s$^{2}$ 2s2p$^{2}$, 1s$^{2}$ 2s2p 3p, 1s$^{2}$ 2s2p 3d and 1s$^{2}$ 2s2p 4s. The \AS\
input data for these calculations are given in \S\ \ref{AsTraPro} in Appendix \ref{AppInData}. The
\depcoe s, $\DCu$, are then calculated and the states with $\DCu \ll 1$ are excluded from further
investigation because of their very low likelihood of existence. However, this approach was
abandoned afterwards because a better treatment within a more comprehensive theory can be obtained
from \Rm-matrix as we were able to produce all the required data by \RMAT\ and \EmiCod\ codes. The
main merit of the old approach is its simplicity and computational economy. Anyway, the method was
useful in checking the \Rm-matrix results by comparing the \depcoe s from \RMAT\ to those from \AS.
Good agreement were found in most cases between the two methods on a sample data set.

\section{Comparison to Previous Work} \label{Comparison}

In this section we make a brief comparison of some of our results against a sample of similar
results obtained by other researchers in previous studies. These include autoionization
probabilities, given in Table~\ref{TableDSBAI}, radiative transition probabilities, given in
Table~\ref{TableRTP}, and dielectronic recombination coefficients, given in
Table~\ref{TableBadnell}.

In Table~\ref{TableDSBAI} we compare our calculated autoionization probabilities with those of De
Marco \etal\ \cite{DemarcoSB1997}. They combined the \LS-coupled autoionization probabilities
calculated in the \clocou\ approximation by Davey \etal\ \cite{DaveySK2000} with one- and two-body
fine structure interactions computed with \SSt\ \cite{EissnerJN1974} to obtain autoionization
probabilities for four states that give rise to spectral lines seen in carbon-rich Wolf-Rayet
stars. For the three larger probabilities, there is agreement within 25\%, but for the
4f~\SLPJ2Ge{9/2} state there is a factor of about two difference. Because our autoionization
probabilities are computed directly from the resonance widths (Equation~\ref{aTP}) which are
obtained from two independent methods (i.e. \Km-matrix and \QB), we expect our values to be the
more reliable ones.

Radiative transition probabilities, as seen in Table~\ref{TableRTP}, generally show good agreement
between the various calculations for the strongest electric dipole transitions. There are some
significant differences for intercombination transitions, indicative of the increased uncertainty
for these cases. There are also significant differences between the present results and those of
Nussbaumer and Storey \cite{NussbaumerS19812} for some of the allowed but two-electron transitions
from the \nlo1s22s$^2$3p configuration, where we would expect our results to be superior, given the
larger scattering target.

In Table~\ref{TableBadnell} we compare our results with effective recombination coefficients from
Table~(1) of Badnell \cite{Badnell1988}. His results are tabulated between terms, having been
summed over the $J$ of the upper and lower terms of the transition, so results for individual lines
cannot be compared. Badnell only tabulates results for one transition in which the upper state is
allowed to autoionize in \LS-coupling (\nlo1s22s2p($^3$P$^{\rm o})$3d~$^2$F$^{\rm o}$) and here the
agreement is excellent as one might expect. On the other hand, the levels of
\nlo1s22s2p($^3$P$^{\rm o}$)3d~$^2$D$^{\rm o}$ have small autoionization widths and our
coefficients are generally smaller than Badnell's for transitions from this term. We note that the
fine structure splitting of this term is well represented in our calculation as is its separation
from neighboring states with large autoionization widths, giving us confidence in our result.

\begin{table} [!h]
\caption{Autoionization probabilities in s$^{-1}$ of four resonance states as obtained in the
current work (SS) compared to those obtained by De Marco \etal\ \cite{DemarcoSB1997} (DSB).
\label{TableDSBAI}} \vspace{-1cm}

\begin{center}
\begin{tabular}{c@{\hskip 4cm}c@{\hskip 4cm}c}
    \hline
    State & SS & DSB \\
    \hline
    \nlo1s22s2p(\SLP3Po)4f \SLPJ2De{5/2} & 9.860e10 & 8.263e10 \\

    \nlo1s22s2p(\SLP3Po)4f \SLPJ2Ge{9/2} & 8.601e08 & 1.623e09 \\

    \nlo1s22s2p(\SLP3Po)3d \SLPJ2Po{3/2} & 2.194e11 & 1.901e11 \\

    \nlo1s22s2p(\SLP3Po)3d \SLPJ2Fo{7/2} & 1.196e12 & 1.488e12 \\
    \hline
\end{tabular}
\end{center}
\end{table}

\clearpage

\begin{table} [!h]
{\tiny

\vspace{-1cm}

\caption{A sample of radiative transition probabilities in s$^{-1}$ as obtained from this work
compared to corresponding values found in the literature. The prime indicates an excited
\nlo1s22s2p(\SLP3Po) core. The 1s$^{2}$ core is suppressed from all configurations.
\label{TableRTP}}

\begin{center} \vspace{-0.3cm}

\begin{tabular}{ccccccccccc}
    \hline
    Upper & Lower & $\WL_{v}$ & SS & NS & LDHK & HST & GKMKW & FKWP & DT & DSB \\
    \hline
    2s2p$^2$ \SLPJ4Pe{1/2} & 2s$^2$2p \SLPJ2Po{1/2} & 2325.40 & 52.8 & 55.3 & 74.4 &  &  &  & 42.5 &  \\

    2s2p$^2$ \SLPJ4Pe{3/2} & 2s$^2$2p \SLPJ2Po{1/2} & 2324.21 & 1.79 & 1.71 & 1.70 &  &  &  & 1.01 &  \\

    2s2p$^2$ \SLPJ4Pe{1/2} & 2s$^2$2p \SLPJ2Po{3/2} & 2328.84 & 60.6 & 65.5 & 77.8 &  &  &  & 40.2 &  \\

    2s2p$^2$ \SLPJ4Pe{3/2} & 2s$^2$2p \SLPJ2Po{3/2} & 2327.64 & 9.34 & 5.24 & 12.4 &  &  &  & 8.11 &  \\

    2s2p$^2$ \SLPJ4Pe{5/2} & 2s$^2$2p \SLPJ2Po{3/2} & 2326.11 & 36.7 & 43.2 & 53.9 &  &  & 51.2 & 34.4 &  \\

    2s2p$^2$ \SLPJ2De{3/2} & 2s$^2$2p \SLPJ2Po{1/2} & 1334.53 & 2.40e8 & 2.42e8 & 2.38e8 &  &  &  & 2.41e8 &  \\

    2s2p$^2$ \SLPJ2De{5/2} & 2s$^2$2p \SLPJ2Po{3/2} & 1335.71 & 2.87e8 & 2.88e8 & 2.84e8 &  &  &  & 2.89e8 &  \\

    2s2p$^2$ \SLPJ2De{3/2} & 2s$^2$2p \SLPJ2Po{3/2} & 1335.66 & 4.75e7 & 4.78e7 & 4.70e7 &  &  &  & 4.79e7 &  \\

    2s2p$^2$ \SLPJ2Se{1/2} & 2s$^2$2p \SLPJ2Po{1/2} & 1036.34 & 7.75e8 & 7.74e8 &  &  &  &  & 7.99e8 &  \\

    2s2p$^2$ \SLPJ2Se{1/2} & 2s$^2$2p \SLPJ2Po{3/2} & 1037.02 & 1.53e9 & 1.53e9 &  &  &  &  & 1.59e9 &  \\

    2s2p$^2$ \SLPJ2Pe{1/2} & 2s$^2$2p \SLPJ2Po{1/2} & 903.96 & 2.76e9 & 2.74e9  &  &  &  &  & 2.63e9 &  \\

    2s2p$^2$ \SLPJ2Pe{3/2} & 2s$^2$2p \SLPJ2Po{1/2} & 903.62 & 6.92e8 & 6.86e8 &  &  &  &  & 6.58e8 &  \\

    2s2p$^2$ \SLPJ2Pe{1/2} & 2s$^2$2p \SLPJ2Po{3/2} & 904.48 & 1.39e9 & 1.38e9 &  &  &  &  & 1.33e9 &  \\

    2s2p$^2$ \SLPJ2Pe{3/2} & 2s$^2$2p \SLPJ2Po{3/2} & 904.14 & 3.46e9 & 3.43e9 &  &  &  &  & 3.30e9 &  \\

    2s$^2$3s \SLPJ2Se{1/2} & 2s$^2$2p \SLPJ2Po{1/2} & 858.09 & 1.26e8 & 1.31e8 &  &  &  &  & 3.69e7 &  \\

    2s$^2$3s \SLPJ2Se{1/2} & 2s$^2$2p \SLPJ2Po{3/2} & 858.56 & 2.46e8 & 2.58e8 &  &  &  &  & 1.11e8 &  \\

    2s$^2$3p \SLPJ2Po{1/2} & 2s2p$^2$ \SLPJ4Pe{1/2} & 1127.13 & 36.3 & 55.5 &  &  &  &  &  &  \\

    2s$^2$3p \SLPJ2Po{3/2} & 2s2p$^2$ \SLPJ4Pe{1/2} & 1126.99 & 58.28 & 97.8 &  &  &  &  &  &  \\

    2s$^2$3p \SLPJ2Po{1/2} & 2s2p$^2$ \SLPJ4Pe{3/2} & 1127.41 & 21.8 & 20.3 &  &  &  &  &  &  \\

    2s$^2$3p \SLPJ2Po{3/2} & 2s2p$^2$ \SLPJ4Pe{3/2} & 1127.27 & 14.0 & 22.3 &  &  &  &  &  &  \\

    2s$^2$3p \SLPJ2Po{3/2} & 2s2p$^2$ \SLPJ4Pe{5/2} & 1127.63 & 1.86e2 & 2.60e2 &  &  &  &  &  &  \\

    2p$^3$ \SLPJ4So{3/2} & 2s2p$^2$ \SLPJ4Pe{1/2} & 1009.86 & 5.77e8 & 5.82e8 &  &  &  &  & 5.76e8 &  \\

    2p$^3$ \SLPJ4So{3/2} & 2s2p$^2$ \SLPJ4Pe{3/2} & 1010.08 & 1.15e9 & 1.16e9 &  &  &  &  & 1.15e9 &  \\

    2p$^3$ \SLPJ4So{3/2} & 2s2p$^2$ \SLPJ4Pe{5/2} & 1010.37 & 1.73e9 & 1.74e9 &  &  &  &  & 1.73e9 &  \\

    2s$^2$3p \SLPJ2Po{3/2} & 2s2p$^2$ \SLPJ2De{5/2} & 1760.40 & 3.77e7 & 3.75e7 &  & 4.3e7 &  &  &  &  \\

    2s$^2$3p \SLPJ2Po{1/2} & 2s2p$^2$ \SLPJ2De{3/2} & 1760.82 & 4.19e7 & 4.16e7 &  & 5.2e7 &  &  &  &  \\

    2s$^2$3p \SLPJ2Po{3/2} & 2s2p$^2$ \SLPJ2De{3/2} & 1760.47 & 4.18e6 & 4.16e6 &  & 4e6 &  &  &  &  \\

    2p$^3$ \SLPJ4So{3/2} & 2s2p$^2$ \SLPJ2De{5/2} & 1490.38 & 1.03e2 & 45.0 &  &  &  &  &  &  \\

    2p$^3$ \SLPJ4So{3/2} & 2s2p$^2$ \SLPJ2De{3/2} & 1490.44 & 10.5 & 11.9 &  &  &  &  &  &  \\

    2s$^2$3p \SLPJ2Po{1/2} & 2s2p$^2$ \SLPJ2Se{1/2} & 2838.44 & 3.61e7 & 3.63e7 &  & 2.2e7 &  &  &  &  \\

    2s$^2$3p \SLPJ2Po{3/2} & 2s2p$^2$ \SLPJ2Se{1/2} & 2837.54 & 3.61e7 & 3.64e7 &  & 2.2e7 &  &  &  &  \\

    2p$^3$ \SLPJ4So{3/2} & 2s2p$^2$ \SLPJ2Se{1/2} & 2196.19 & 1.62 & 5.09 &  &  &  &  &  &  \\

    2s$^2$3p \SLPJ2Po{1/2} & 2s2p$^2$ \SLPJ2Pe{1/2} & 4739.29 & 5.85e4 & 7.70e4 &  & $<$3.0e5 &  &  &  &  \\

    2s$^2$3p \SLPJ2Po{3/2} & 2s2p$^2$ \SLPJ2Pe{1/2} & 4736.79 & 7.98e3 & 1.19e4 &  & $<$5e5 &  &  &  &  \\

    2s$^2$3p \SLPJ2Po{1/2} & 2s2p$^2$ \SLPJ2Pe{3/2} & 4748.61 & 2.35e4 & 3.23e4 &  & $<$2.3e5 &  &  &  &  \\

    2s$^2$3p \SLPJ2Po{3/2} & 2s2p$^2$ \SLPJ2Pe{3/2} & 4746.09 & 6.08e4 & 8.28e4 &  & $<$1.3e5 &  &  &  &  \\

    2p$^3$ \SLPJ4So{3/2} & 2s2p$^2$ \SLPJ2Pe{1/2} & 3184.42 & 16.4 & 14.9 &  &  &  &  &  &  \\

    2p$^3$ \SLPJ4So{3/2} & 2s2p$^2$ \SLPJ2Pe{3/2} & 3188.62 & 80.0 & 76.9 &  &  &  &  &  &  \\

    2s$^2$3p \SLPJ2Po{1/2} & 2s$^2$3s \SLPJ2Se{1/2} & 6584.70 & 3.53e7 & 3.68e7 &  & 2.9e7 & 3.29e7 &  &  &  \\

    2s$^2$3p \SLPJ2Po{3/2} & 2s$^2$3s \SLPJ2Se{1/2} & 6579.87 & 3.54e7 & 3.69e7 &  & 3.4e7 & 3.33e7 &  &  &  \\

    2p$^3$ \SLPJ4So{3/2} & 2s$^2$3s \SLPJ2Se{1/2} & 3923.19 & 6.35 & 10.2 &  &  &  &  &  &  \\

    2s$^2$3d \SLPJ2De{3/2} & 2s$^2$3p \SLPJ2Po{1/2} & 7233.33 & 3.46e7 &  &  &  & 4.22e7 &  &  &  \\

    2s$^2$3d \SLPJ2De{3/2} & 2s$^2$3p \SLPJ2Po{3/2} & 7239.16 & 6.91e6 &  &  &  & 8.49e6 &  &  &  \\

    2s$^2$3d \SLPJ2De{5/2} & 2s$^2$3p \SLPJ2Po{3/2} & 7238.41 & 4.15e7 &  &  &  & 5.10e7 &  &  &  \\

    2s$^2$3d \SLPJ2De{3/2} & 2s$^2$2p \SLPJ2Po{1/2} & 687.05 & 2.40e9 &  &  &  &  &  & 2.24e9 &  \\

    2s$^2$3d \SLPJ2De{3/2} & 2s$^2$2p \SLPJ2Po{3/2} & 687.35 & 4.81e8 &  &  &  &  &  & 4.50e8 &  \\

    2s$^2$3d \SLPJ2De{5/2} & 2s$^2$2p \SLPJ2Po{3/2} & 687.35 & 2.88e9 &  &  &  &  &  & 2.70e9 &  \\

    2p$^3$ \SLPJ2Do{3/2} & 2s2p$^2$ \SLPJ2De{3/2} & 1323.91 & 4.33e8 &  &  &  &  &  & 4.53e8 &  \\

    2p$^3$ \SLPJ2Do{5/2} & 2s2p$^2$ \SLPJ2De{3/2} & 1324.00 & 3.23e7 &  &  &  &  &  & 3.38e7 &  \\

    2p$^3$ \SLPJ2Do{3/2} & 2s2p$^2$ \SLPJ2De{5/2} & 1323.86 & 4.91e7 &  &  &  &  &  & 5.10e7 &  \\

    2p$^3$ \SLPJ2Do{5/2} & 2s2p$^2$ \SLPJ2De{5/2} & 1323.95 & 4.51e8 &  &  &  &  &  & 4.71e8 &  \\

    2p$^3$ \SLPJ2Do{3/2} & 2s2p$^2$ \SLPJ2Pe{1/2} & 2509.88 & 4.53e7 &  &  &  &  &  & 5.38e7 &  \\

    2p$^3$ \SLPJ2Do{3/2} & 2s2p$^2$ \SLPJ2Pe{3/2} & 2512.49 & 8.91e6 &  &  &  &  &  & 1.06e7 &  \\

    2p$^3$ \SLPJ2Do{5/2} & 2s2p$^2$ \SLPJ2Pe{3/2} & 2512.81 & 5.40e7 &  &  &  &  &  & 6.43e7 &  \\

    4f$'$ \SLPJ2De{5/2} & 3d$'$ \SLPJ2Po{3/2} & 5115.07 & 1.18e8 &  &  &  &  &  &  & 6.94e7 \\

    4f$'$ \SLPJ2Ge{9/2} & 3d$'$ \SLPJ2Fo{7/2} & 4620.54 & 2.29e8 &  &  &  &  &  &  & 1.84e8 \\

    3d$'$ \SLPJ2Po{3/2} & 3p$'$ \SLPJ2Pe{3/2} & 4966.12 & 3.14e7 &  &  &  &  &  &  & 2.89e7 \\

    3d$'$ \SLPJ2Fo{7/2} & 3p$'$ \SLPJ2De{5/2} & 8796.49 & 2.03e7 &  &  &  &  &  &  & 1.99e7 \\
    \hline
\end{tabular}

\end{center}
\vspace{-0.2cm} {SS = current work, NS = Nussbaumer and Storey \cite{NussbaumerS19812}; LDHK =
Lennon \etal\ \cite{LennonDHK1985}; HST = Huber \etal\ \cite{HuberST1984}; GKMKW = Glenzer \etal\
\cite{GlenzerKMKW1994}; FKWP = Fang \etal\ \cite{FangKWP1993}; DT = Dankwort and Trefftz
\cite{DankwortT1978}, DSB = De Marco \etal\ \cite{DemarcoSB1997}.} }
\end{table}

\clearpage

\begin{table} [!h]
{\tiny

\caption{Effective \dierec\ rate coefficients in cm$^3$.s$^{-1}$ of a number of transitions for the
given 10-based logarithmic temperatures. The first row for each transition corresponds to Badnell
\cite{Badnell1988}, while the second row is obtained from the current work. The superscripts
denotes powers of 10. The prime indicates an excited \nlo1s22s2p(\SLP3Po) core. The 1s$^{2}$ core
is suppressed from all configurations. \label{TableBadnell}}

\begin{center} \vspace{-0.3cm}

\begin{tabular}{l@{\hskip 0.25cm}l@{\hskip 0.25cm}l@{\hskip 0.25cm}l@{\hskip 0.25cm}l@{\hskip 0.25cm}l@{\hskip 0.25cm}l@{\hskip 0.25cm}l@{\hskip 0.25cm}l@{\hskip 0.25cm}l@{\hskip 0.25cm}l@{\hskip 0.25cm}l@{\hskip 0.25cm}l}
\hline
           &            &            &            &            &            & {\bf log(T)} &            &            &            &            &            \\
           \cline{3-12}
{\bf Upper} & {\bf Lower} &  {\bf 3.0} &  {\bf 3.2} &  {\bf 3.4} &  {\bf 3.6} &  {\bf 3.8} &  {\bf 4.0} &  {\bf 4.2} &  {\bf 4.4} &  {\bf 4.6} &  {\bf 4.8} \\
\hline
3d$'$ \SLP4Fo & 3p$'$ \SLP4De &   7.27\EE{-24} &   5.73\EE{-20} &   1.56\EE{-17} &   6.17\EE{-16} &   6.47\EE{-15} &   2.61\EE{-14} &   5.52\EE{-14} &   7.43\EE{-14} &   7.25\EE{-14} &   5.65\EE{-14} \\

           &            &   6.74\EE{-24} &   5.17\EE{-20} &   1.29\EE{-17} &   4.33\EE{-16} &   3.90\EE{-15} &   1.33\EE{-14} &   2.31\EE{-14} &   2.57\EE{-14} &   2.13\EE{-14} &   1.47\EE{-14} \\

3p$'$ \SLP4De & 3s$'$ \SLP4Po &   1.90\EE{-16} &   3.21\EE{-16} &   8.53\EE{-16} &   4.95\EE{-15} &   1.93\EE{-14} &   4.90\EE{-14} &   8.51\EE{-14} &   1.05\EE{-13} &   9.91\EE{-14} &   7.57\EE{-14} \\

           &            &   1.06\EE{-16} &   1.90\EE{-16} &   7.71\EE{-16} &   5.02\EE{-15} &   1.74\EE{-14} &   3.58\EE{-14} &   4.85\EE{-14} &   4.76\EE{-14} &   3.70\EE{-14} &   2.46\EE{-14} \\

3d$'$ \SLP2Do & 3p$'$ \SLP2Pe &   2.49\EE{-14} &   3.19\EE{-14} &   2.90\EE{-14} &   2.11\EE{-14} &   1.35\EE{-14} &   8.05\EE{-15} &   4.80\EE{-15} &   2.94\EE{-15} &   1.81\EE{-15} &   1.10\EE{-15} \\

           &            &   4.67\EE{-15} &   5.99\EE{-15} &   5.44\EE{-15} &   3.97\EE{-15} &   2.58\EE{-15} &   1.68\EE{-15} &   1.17\EE{-15} &   8.31\EE{-16} &   5.54\EE{-16} &   3.41\EE{-16} \\

3p$'$ \SLP4Pe & 3s$'$ \SLP4Po &   3.08\EE{-16} &   5.09\EE{-16} &   9.86\EE{-16} &   4.02\EE{-15} &   1.08\EE{-14} &   1.73\EE{-14} &   2.04\EE{-14} &   1.97\EE{-14} &   1.61\EE{-14} &   1.14\EE{-14} \\

           &            &   1.49\EE{-16} &   2.52\EE{-16} &   7.24\EE{-16} &   3.84\EE{-15} &   1.05\EE{-14} &   1.58\EE{-14} &   1.60\EE{-14} &   1.27\EE{-14} &   8.52\EE{-15} &   5.14\EE{-15} \\

4f$'$ \SLP2Fe & 3d$'$ \SLP2Do &   3.00\EE{-27} &   6.39\EE{-22} &   1.14\EE{-18} &   9.93\EE{-17} &   1.29\EE{-15} &   5.09\EE{-15} &   9.43\EE{-15} &   1.09\EE{-14} &   9.26\EE{-15} &   6.51\EE{-15} \\

           &            &   1.64\EE{-27} &   3.53\EE{-22} &   6.33\EE{-19} &   5.54\EE{-17} &   7.22\EE{-16} &   2.82\EE{-15} &   5.18\EE{-15} &   5.88\EE{-15} &   4.94\EE{-15} &   3.43\EE{-15} \\

4f$'$ \SLP4De & 3d$'$ \SLP4Po &   2.56\EE{-27} &   7.68\EE{-22} &   1.70\EE{-18} &   1.70\EE{-16} &   2.40\EE{-15} &   9.95\EE{-15} &   1.90\EE{-14} &   2.21\EE{-14} &   1.89\EE{-14} &   1.33\EE{-14} \\

           &            &   2.54\EE{-27} &   7.63\EE{-22} &   1.69\EE{-18} &   1.69\EE{-16} &   2.39\EE{-15} &   9.86\EE{-15} &   1.87\EE{-14} &   2.17\EE{-14} &   1.85\EE{-14} &   1.29\EE{-14} \\

4s$'$ \SLP4Po & 3p$'$ \SLP4Pe &   2.59\EE{-20} &   1.23\EE{-17} &   4.65\EE{-16} &   3.58\EE{-15} &   1.01\EE{-14} &   1.51\EE{-14} &   1.51\EE{-14} &   1.18\EE{-14} &   7.80\EE{-15} &   4.67\EE{-15} \\

           &            &   2.63\EE{-20} &   1.25\EE{-17} &   4.72\EE{-16} &   3.63\EE{-15} &   1.02\EE{-14} &   1.53\EE{-14} &   1.53\EE{-14} &   1.19\EE{-14} &   7.83\EE{-15} &   4.68\EE{-15} \\

4f$'$ \SLP4Fe & 3d$'$ \SLP4Do &   5.39\EE{-27} &   1.16\EE{-21} &   2.07\EE{-18} &   1.81\EE{-16} &   2.37\EE{-15} &   9.41\EE{-15} &   1.77\EE{-14} &   2.07\EE{-14} &   1.79\EE{-14} &   1.27\EE{-14} \\

           &            &   3.48\EE{-27} &   7.51\EE{-22} &   1.35\EE{-18} &   1.19\EE{-16} &   1.55\EE{-15} &   6.07\EE{-15} &   1.11\EE{-14} &   1.26\EE{-14} &   1.06\EE{-14} &   7.37\EE{-15} \\

4f$'$ \SLP4Ge & 3d$'$ \SLP4Fo &   5.42\EE{-27} &   1.50\EE{-21} &   3.15\EE{-18} &   3.05\EE{-16} &   4.24\EE{-15} &   1.74\EE{-14} &   3.30\EE{-14} &   3.86\EE{-14} &   3.32\EE{-14} &   2.34\EE{-14} \\

           &            &   2.98\EE{-27} &   8.25\EE{-22} &   1.73\EE{-18} &   1.68\EE{-16} &   2.33\EE{-15} &   9.49\EE{-15} &   1.78\EE{-14} &   2.06\EE{-14} &   1.75\EE{-14} &   1.22\EE{-14} \\

4s$'$ \SLP4Po & 3p$'$ \SLP4De &   2.86\EE{-20} &   1.36\EE{-17} &   5.14\EE{-16} &   3.95\EE{-15} &   1.11\EE{-14} &   1.67\EE{-14} &   1.67\EE{-14} &   1.30\EE{-14} &   8.61\EE{-15} &   5.15\EE{-15} \\

           &            &   3.15\EE{-20} &   1.49\EE{-17} &   5.65\EE{-16} &   4.35\EE{-15} &   1.22\EE{-14} &   1.83\EE{-14} &   1.83\EE{-14} &   1.42\EE{-14} &   9.37\EE{-15} &   5.59\EE{-15} \\

4d$'$ \SLP4Fo & 3p$'$ \SLP4De &   1.77\EE{-26} &   1.71\EE{-21} &   1.85\EE{-18} &   1.18\EE{-16} &   1.30\EE{-15} &   4.85\EE{-15} &   9.35\EE{-15} &   1.18\EE{-14} &   1.10\EE{-14} &   8.31\EE{-15} \\

           &            &   1.75\EE{-26} &   1.69\EE{-21} &   1.83\EE{-18} &   1.17\EE{-16} &   1.24\EE{-15} &   4.28\EE{-15} &   7.25\EE{-15} &   7.83\EE{-15} &   6.37\EE{-15} &   4.34\EE{-15} \\

2s2p$^2$ \SLP2De & 2s$^2$2p \SLP2Po &   2.45\EE{-12} &   5.89\EE{-12} &   8.34\EE{-12} &   8.17\EE{-12} &   6.33\EE{-12} &   4.36\EE{-12} &   3.01\EE{-12} &   2.21\EE{-12} &   1.66\EE{-12} &   1.19\EE{-12} \\

           &            &   1.85\EE{-12} &   5.12\EE{-12} &   7.66\EE{-12} &   7.67\EE{-12} &   6.00\EE{-12} &   4.09\EE{-12} &   2.61\EE{-12} &   1.60\EE{-12} &   9.38\EE{-13} &   5.28\EE{-13} \\

3d$'$ \SLP2Do & 2s2p$^2$ \SLP2Pe &   2.60\EE{-13} &   3.33\EE{-13} &   3.02\EE{-13} &   2.20\EE{-13} &   1.41\EE{-13} &   8.41\EE{-14} &   5.01\EE{-14} &   3.07\EE{-14} &   1.89\EE{-14} &   1.15\EE{-14} \\

           &            &   4.50\EE{-14} &   5.78\EE{-14} &   5.24\EE{-14} &   3.83\EE{-14} &   2.49\EE{-14} &   1.62\EE{-14} &   1.13\EE{-14} &   8.01\EE{-15} &   5.34\EE{-15} &   3.29\EE{-15} \\

4d$'$ \SLP2Do & 2s2p$^2$ \SLP2Pe &   6.56\EE{-27} &   1.10\EE{-21} &   1.70\EE{-18} &   1.35\EE{-16} &   1.66\EE{-15} &   6.33\EE{-15} &   1.16\EE{-14} &   1.33\EE{-14} &   1.13\EE{-14} &   8.02\EE{-15} \\

           &            &   1.98\EE{-27} &   3.29\EE{-22} &   5.03\EE{-19} &   3.97\EE{-17} &   4.85\EE{-16} &   1.82\EE{-15} &   3.26\EE{-15} &   3.64\EE{-15} &   3.03\EE{-15} &   2.09\EE{-15} \\

2s2p$^2$ \SLP2Pe & 2s$^2$2p \SLP2Po &   2.68\EE{-13} &   3.99\EE{-13} &   5.05\EE{-13} &   5.45\EE{-13} &   4.87\EE{-13} &   3.95\EE{-13} &   3.37\EE{-13} &   3.09\EE{-13} &   2.75\EE{-13} &   2.21\EE{-13} \\

           &            &   5.11\EE{-14} &   1.11\EE{-13} &   2.21\EE{-13} &   3.09\EE{-13} &   3.18\EE{-13} &   2.78\EE{-13} &   2.24\EE{-13} &   1.66\EE{-13} &   1.11\EE{-13} &   6.82\EE{-14} \\

3d$'$ \SLP2Do & 2s2p$^2$ \SLP2De &   7.38\EE{-13} &   9.47\EE{-13} &   8.59\EE{-13} &   6.26\EE{-13} &   3.99\EE{-13} &   2.39\EE{-13} &   1.42\EE{-13} &   8.71\EE{-14} &   5.38\EE{-14} &   3.25\EE{-14} \\

           &            &   1.36\EE{-13} &   1.75\EE{-13} &   1.59\EE{-13} &   1.16\EE{-13} &   7.53\EE{-14} &   4.90\EE{-14} &   3.43\EE{-14} &   2.42\EE{-14} &   1.61\EE{-14} &   9.95\EE{-15} \\

3s$'$ \SLP4Po & 2s2p$^2$ \SLP4Pe &   1.36\EE{-15} &   2.22\EE{-15} &   3.36\EE{-15} &   1.07\EE{-14} &   3.30\EE{-14} &   7.13\EE{-14} &   1.12\EE{-13} &   1.32\EE{-13} &   1.21\EE{-13} &   9.09\EE{-14} \\

           &            &   7.17\EE{-16} &   1.19\EE{-15} &   2.35\EE{-15} &   1.02\EE{-14} &   3.12\EE{-14} &   5.78\EE{-14} &   7.22\EE{-14} &   6.71\EE{-14} &   5.05\EE{-14} &   3.29\EE{-14} \\

3d$'$ \SLP2Fo & 2s2p$^2$ \SLP2De &   1.70\EE{-12} &   4.92\EE{-12} &   7.44\EE{-12} &   7.49\EE{-12} &   5.83\EE{-12} &   3.86\EE{-12} &   2.30\EE{-12} &   1.29\EE{-12} &   6.93\EE{-13} &   3.63\EE{-13} \\

           &            &   1.71\EE{-12} &   4.94\EE{-12} &   7.48\EE{-12} &   7.53\EE{-12} &   5.86\EE{-12} &   3.88\EE{-12} &   2.32\EE{-12} &   1.30\EE{-12} &   6.97\EE{-13} &   3.65\EE{-13} \\

4d$'$ \SLP2Do & 2s2p$^2$ \SLP2De &   2.30\EE{-26} &   3.86\EE{-21} &   5.94\EE{-18} &   4.72\EE{-16} &   5.81\EE{-15} &   2.22\EE{-14} &   4.05\EE{-14} &   4.65\EE{-14} &   3.97\EE{-14} &   2.81\EE{-14} \\

           &            &   9.59\EE{-27} &   1.59\EE{-21} &   2.43\EE{-18} &   1.92\EE{-16} &   2.35\EE{-15} &   8.82\EE{-15} &   1.58\EE{-14} &   1.76\EE{-14} &   1.47\EE{-14} &   1.01\EE{-14} \\

3d$'$ \SLP4Do & 2s2p$^2$ \SLP4Pe &   2.49\EE{-24} &   2.83\EE{-20} &   1.05\EE{-17} &   4.56\EE{-16} &   4.63\EE{-15} &   1.78\EE{-14} &   3.62\EE{-14} &   4.76\EE{-14} &   4.58\EE{-14} &   3.53\EE{-14} \\

           &            &   2.34\EE{-24} &   2.63\EE{-20} &   9.38\EE{-18} &   3.84\EE{-16} &   3.62\EE{-15} &   1.23\EE{-14} &   2.11\EE{-14} &   2.32\EE{-14} &   1.91\EE{-14} &   1.31\EE{-14} \\

3d$'$ \SLP4Po & 2s2p$^2$ \SLP4Pe &   6.71\EE{-14} &   1.08\EE{-13} &   1.13\EE{-13} &   9.07\EE{-14} &   6.36\EE{-14} &   4.86\EE{-14} &   4.61\EE{-14} &   4.51\EE{-14} &   3.87\EE{-14} &   2.84\EE{-14} \\

           &            &   3.41\EE{-14} &   5.51\EE{-14} &   5.78\EE{-14} &   4.63\EE{-14} &   3.38\EE{-14} &   2.97\EE{-14} &   3.10\EE{-14} &   2.93\EE{-14} &   2.30\EE{-14} &   1.55\EE{-14} \\

4s$'$ \SLP4Po & 2s2p$^2$ \SLP4Pe &   9.12\EE{-20} &   4.32\EE{-17} &   1.64\EE{-15} &   1.26\EE{-14} &   3.55\EE{-14} &   5.31\EE{-14} &   5.33\EE{-14} &   4.14\EE{-14} &   2.75\EE{-14} &   1.64\EE{-14} \\

           &            &   1.13\EE{-19} &   5.37\EE{-17} &   2.03\EE{-15} &   1.56\EE{-14} &   4.40\EE{-14} &   6.58\EE{-14} &   6.59\EE{-14} &   5.12\EE{-14} &   3.38\EE{-14} &   2.02\EE{-14} \\

4d$'$ \SLP4Do & 2s2p$^2$ \SLP4Pe &   4.87\EE{-26} &   6.33\EE{-21} &   8.27\EE{-18} &   5.94\EE{-16} &   6.90\EE{-15} &   2.57\EE{-14} &   4.72\EE{-14} &   5.54\EE{-14} &   4.85\EE{-14} &   3.49\EE{-14} \\

           &            &   1.73\EE{-26} &   2.26\EE{-21} &   2.95\EE{-18} &   2.12\EE{-16} &   2.43\EE{-15} &   8.79\EE{-15} &   1.53\EE{-14} &   1.69\EE{-14} &   1.39\EE{-14} &   9.52\EE{-15} \\

4d$'$ \SLP4Po & 2s2p$^2$ \SLP4Pe &   5.09\EE{-27} &   9.49\EE{-22} &   1.56\EE{-18} &   1.30\EE{-16} &   1.68\EE{-15} &   6.90\EE{-15} &   1.39\EE{-14} &   1.78\EE{-14} &   1.67\EE{-14} &   1.27\EE{-14} \\

           &            &   4.37\EE{-28} &   8.18\EE{-23} &   1.34\EE{-19} &   1.11\EE{-17} &   1.40\EE{-16} &   5.37\EE{-16} &   9.70\EE{-16} &   1.09\EE{-15} &   9.13\EE{-16} &   6.32\EE{-16} \\

3p$'$ \SLP2Pe & 2s$^2$2p \SLP2Po &   1.91\EE{-14} &   2.70\EE{-14} &   3.11\EE{-14} &   3.17\EE{-14} &   2.97\EE{-14} &   2.65\EE{-14} &   2.26\EE{-14} &   1.81\EE{-14} &   1.34\EE{-14} &   9.18\EE{-15} \\

           &            &   4.59\EE{-15} &   9.59\EE{-15} &   1.86\EE{-14} &   2.67\EE{-14} &   2.99\EE{-14} &   2.83\EE{-14} &   2.31\EE{-14} &   1.66\EE{-14} &   1.06\EE{-14} &   6.26\EE{-15} \\

4p$'$ \SLP2Pe & 2s$^2$2p \SLP2Po &   5.81\EE{-23} &   3.67\EE{-19} &   7.12\EE{-17} &   1.53\EE{-15} &   8.24\EE{-15} &   1.85\EE{-14} &   2.40\EE{-14} &   2.19\EE{-14} &   1.61\EE{-14} &   1.03\EE{-14} \\

           &            &   6.35\EE{-23} &   3.96\EE{-19} &   7.62\EE{-17} &   1.63\EE{-15} &   8.74\EE{-15} &   1.96\EE{-14} &   2.52\EE{-14} &   2.30\EE{-14} &   1.68\EE{-14} &   1.07\EE{-14} \\
\hline
\end{tabular}

\end{center}
}
\end{table}


%

\chapter{Electron Temperature of Astronomical Objects} \label{Temp} 
The study of transition lines of carbon and its ions in the spectra of astronomical objects, such
as planetary nebulae \cite{PottaschWD1978, NikitinSFK1981, BogdanovichLNRK1985}, has implications
for carbon abundance determination \cite{HarringtonLSS1980, HarringtonLS1981, NussbaumerS19812,
AdamsS1982, Kholtygin1984, BogdanovichNRK1985}, probing the physical conditions in the interstellar
medium \cite{Boughton1978, PengHWL2005}, and element enrichment in the CNO cycle \cite{AdamsS1982,
RolaS1994, CleggSWN1997}. The lines span large parts of the electromagnetic spectrum and originate
from various processes under different physical conditions. Concerning C$^+$, the subject of the
current investigation, spectral lines of this ion have been observed in many astronomical objects
such as planetary nebulae, Seyfert galaxies, stellar winds of Wolf-Rayet stars, symbiotic stars,
and in the interstellar medium \cite{StoreyS2012, NikitinY1963, HayesN1984, Kholtygin1984,
RolaS1994, DaveySK2000, PengWHL2004, ZhangLLPB2005, StanghelliniSG2005}.

There are many observational and theoretical studies on the recombination lines of carbon related
to the spectra of PNe \cite{Leibowitz1972, Leibowitz1972b, BalickGD1974, NussbaumerS1975,
NussbaumerS1984, Boughton1978, HarringtonLSS1980, HarringtonLS1981, Kholtygin1984, Kholtygin1998,
PequignotPB1991, BaluteauZMP1995, LiuLBL2004}; some of which are linked to the C$^+$ ion
\cite{ClavelFS1981, CleggSPP1983, HayesN1984, BogdanovichNRK1985, Badnell1988, RolaS1994,
DaveySK2000, PengWHL2004, PengHWL2005, ZhangLLPB2005}. The latter, however, are limited in number
and constrained in domain.

In this chapter we investigate the electron temperature in the emitting region of recombination
lines of C$^+$ ion in the spectra of a number of astronomical objects, mainly planetary nebulae,
using a least squares optimization method with theoretical data obtained from the recently-computed
theoretical line list, \SSo, of \cite{SochiSCIIList2012} and astronomical data gathered from the
literature. The theoretical list was generated using the \Rm-matrix \cite{BerringtonEN1995}, \AS\
\cite{EissnerJN1974, NussbaumerS1978, BadnellAS2008} and Emissivity \cite{SochiEmis2010} codes with
an intermediate coupling scheme where the transition lines are produced by dielectronic
recombination processes originating from the low-lying autoionizing states with subsequent cascade
decays.

The traditional method of identifying electron temperature in thin plasma, found for example in
planetary nebulae, is by taking the intensity ratio of two transition lines \cite{Ferland2003}. The
advantage of this method is simplicity and its reliance on limited data. The disadvantage is that
the lines should have strong intensity. They should also have considerable energy gap, typically of
several electron volts, if collisionally-excited transitions are employed. The method can be
subject to large errors if the observed intensity is contaminated with significant errors. On the
other hand, the least squares minimization is a collective method that normally involves a number
of lines and hence can be more reliable in identifying the electron temperature. The objective of
the least squares technique is to find the temperature at which the difference between the observed
data and the calculated model is minimal and hence identifying the temperature of the line-emitting
region. Least squares minimization is based on the fact that this difference tends to minimum when
the temperature-dependent theoretical data approaches the unknown temperature of the observation.
The main tool used in the emissivity analysis is the \EmiCod\ code \cite{SochiEmis2010}. The
analysis is based on minimizing the sum of weighted squared differences between the normalized
theoretical emissivity and the normalized observational flux of the C\II\ lines.

\section{Method}\label{TempMeth}

As indicated already, the theoretical data of the C\II\ \dierec\ transitions and subsequent cascade
decay are obtained from the \SSo\ line list of \cite{SochiSCIIList2012} which consists of 6187
optically-allowed transitions with their associated data such as emissivity and effective
recombination coefficients. The autoionizing states involved in the transitions of this list
consist of 64 resonances belonging to 11 symmetries ($J=1,3,5,7,9,11$ half even and $J= 1,3,5,7,9$
half odd) which are all the resonances above the threshold of C$^{2+}$ \nlo1s2 \nlo2s2 \SLP1Se with
a \priquanum\ $n<5$ for the combined electron. These include 61 theoretically-found resonances by
the \Km-matrix method plus 3 experimental ones which could not be found due to their very narrow
width. The bound states involved in these transitions comprise 150 energy levels belonging to 11
symmetries ($J=1,3,5,7,9$ half even and $J=1,3,5,7,9,11$ half odd). These include 142 theoretically
found by \Rm-matrix, which are all the bound states with \effquanum\ between 0.1-13 for the outer
electron and $0\le l \le 5$, plus 8 experimental top states which are the levels of the
1s$^2$2s2p(\SLP3Po)3d~\SLP4Fo and \SLP4Do terms.

The theoretical and computational backgrounds for the atomic transition calculations including the
emissivity thermodynamic model are given in Chapters \ref{Physics} and \ref{CII}.
The calculations were performed using an elaborate C$^{+2}$ ionic target under an intermediate
coupling scheme. The list has also been validated by various tests including comparison to
literature data related to autoionization and radiative transition probabilities and effective
dielectronic recombination coefficients. The theoretical data of the bound and resonance states
were also compared to the available experimental data \cite{NIST2010} and found to agree very well
both in energy levels and in \finstr\ splitting.

It should be remarked that processes other than \dierec\ have not been considered in the atomic
scattering and transition model of SS1 list, so the results of SS1 are incomplete for states likely
to be populated by radiative recombination or collisional excitation and de-excitation.
Consequently, the results for free-free and free-bound transitions can be used directly to predict
line intensities from low-density astrophysical plasmas such as gaseous nebulae but those between
bound states underestimate the line intensities in general and hence can only be used as part of a
larger ion population model including all relevant processes. Yes, there is an important exception
for the bound-bound transitions that is when the upper level has a doubly-excited core and hence it
occurs at the top of the cascade process where radiative contribution is minor. For these top-level
transitions the contribution of processes other than dielectronic recombination is negligible and
hence the theoretical data can be reliably used.

Therefore, with very few exceptions, only FF, FB, and BB with a doubly-excited upper state were
used in the current study. The main exception about the bound-bound transitions with doubly-excited
core is the $\WL$4267~\AA\ line which, being one of the strongest observed C\II\ recombination
lines and hence is a major line in almost all observational data sets, cannot be avoided
altogether. Therefore, recombination coefficients in the form of case B from Davey \etal\
\cite{DaveySK2000}, which are obtained within a more comprehensive theory that includes radiative
contribution as well as dielectronic recombination, were used.

It should be remarked that the use in our analysis of the lines originating at the top of the
cascade, which are excited in a very different and much simpler way than lines like $\WL$4267~\AA\
at the bottom of a complex cascade process, makes our analysis more reliable as it relies on a much
simpler and better-understood decay process. Our analysis may also provide a way, in some cases at
least, for testing and validating the results obtained from the low-lying transitions such as those
obtained from line $\WL$4267. Hence agreement between electron temperatures obtained without
$\WL$4267 with those using $\WL$4267, may provide a validation of recombination theory for line
$\WL$4267 and other low-lying transitions, as obtained previously by other researchers. For
example, the possibility of the existence of some unknown mechanism that overpopulates the levels
of the upper state of $\WL$4267 transition causing the enhancement of the ORL abundance may be
ruled out if the results with $\WL$4267 are consistent with those obtained without $\WL$4267. This
is one reason why in section~\ref{AstObj} we computed electron temperature from least squares
minimization once with and once without line $\WL$4267 when the selected observational data include
this line.

With regard to the observational data, we carried out a fairly thorough research for C\II\
recombination line data in the literature in which over 140 data sets related mainly to planetary
nebulae were collected and archived. However, most of these data sets were eliminated for various
reasons such as the failure in the reliability tests, the poor quality of the data, and the
consistence of only a single line and hence the data cannot be used in a least squares minimization
procedure which requires two lines at least. More important is that almost all data sets that
comprise only bound-bound transitions with no doubly-excited core upper state have been removed.

On eliminating most of the data sets on the previous grounds, the remaining data sets were
subjected to a refinement process in which the flux of all the observational lines in each data set
were normalized to the flux of a reference observational line in the set, which is usually chosen
as the brightest and most reliable, while the emissivity of all the theoretical lines in the set
were normalized to the emissivity of the corresponding theoretical line. The observational flux and
theoretical emissivity of the reference line were therefore unity. The ratio of the normalized
observed flux to the normalized theoretical emissivity of each line were then plotted on common
graphs as a function of temperature on log-linear scales. A sample of these graphs is presented in
Figure~\ref{NerNGC53151} for the planetary nebula NGC~5315\index{NGC 5315}. All lines that did not
approach the ratio of the reference line within an arbitrarily-chosen factor of 3 were eliminated.
Because the reference line was typically chosen as the brightest, and hence more reliable, any line
whose normalized ratio deviates largely from the normalized ratio of this line should be deemed
unreliable. The arbitrary factor of 3 was chosen as an appropriate limit considering practical
factors that contribute to error in the collection of observational data. The refinement process
also involved the utilization of graphs in which the ratio of theoretical emissivity to
observational flux of all lines in a certain data set was plotted on a single graph as a function
of electron temperature. A sample of these graphs are shown in Figures~\ref{EtoNGC7009} and
\ref{EtoIC418}.

Some other lines were also eliminated for various reasons related mainly to an established or
suspected misidentification of the line or its intensity. For example, the wavelength of the
alleged C\II\ line may not match with any known theoretical transition. Also, the absence of a
strong line in the observational data associated with the presence of a much weaker line with no
obvious reason, such as possible blending with a much stronger neighbor, casts doubt on the
identification of the present line. The line may also be eliminated because its intensity ratio
relative to another well-established line does not comply with the ratio obtained from theory. Very
few lines were also out of the wavelength range of our line list and hence were eliminated due to
lack of theoretical emissivity data. Other reasons for elimination include lack of intensity data,
blending or close proximity to other non-C\II\ lines.

The selected refined data sets were then subjected to a least squares optimization procedure which
is outlined in the following section. It should be remarked that the observed flux used in the
least squares procedure is the de-reddened flux obtained by correcting for extinction and other
sources of error as stated by the data source and not the raw flux data. Therefore, there should be
no ambiguity when we use `observed' flux in the following sections.

\subsection{Least Squares Minimization}

The main method for generalizing and summarizing a set of data is by fitting it to a model that
depends on adjustable parameters. A figure-of-merit function that measures the agreement between
the data and the model for a particular choice of parameters is then chosen or designed so that
small or large values indicate good agreement. The model parameters are then adjusted to achieve a
minimum or maximum in the merit function, yielding best fit parameters. Least squares minimization
is probably the most popular method in the physical sciences. It is widely used to determine the
best set of parameters in a model to fit a set of observational data. The objective of the least
squares technique is to minimize the difference between the observed data and the calculated model.
The goodness-of-fit index $\GoF$ is one of the most suggestive and commonly accepted figures of
merit. It is widely used as an indicator of the overall fit between a theoretical model and a set
of observational data. In general terms, it is given by \cite{PressTVF2002}
\begin{equation}\label{}
    \GoF =  \frac{\sum_{i} \wei_{i}(\yobs_{i} -  \ycal_{i})^{2}} {\degFree}
\end{equation}
where $\yobs_{i}$ and $\ycal_{i}$ are the observed and calculated values at step $i$, $\wei_{i}$ is
the corresponding weight, and $\degFree$ is the number of degrees of freedom given by $\degFree =
\NO - \NP$ where $\NO$ and $\NP$ are the number of observations and fitted parameters in the
calculated model, respectively. The summation index $i$ runs over all observed data points.

In our least squares calculations we have a single fitting parameter, which is the electron
temperature of the line-emitting regions in the astronomical objects (e.g. NGC 5315\index{NGC 5315}
nebula), while the observations are the flux data of the C\II\ recombination lines that we obtained
from the literature (e.g. observational data of Peimbert \etal\ \cite{PeimbertPRE2004}). The
purpose of the least squares procedure is to find the optimal temperature that gives the best fit
of the theoretical model to the observational data, and hence identifying the temperature of the
line-emitting region in the astronomical object. However, since the collected spectrum is usually
integrated over some spatial part of the nebula and along the line of sight through a
three-dimensional object, the obtained temperature normally is an average value. In the following
we outline the least squares procedure

\begin{itemize}

\item
The theoretical lines are mapped to their observational counterparts where the resolution of the
observational data is considered for this mapping when such data are available, i.e. all
theoretical lines within that resolution on both sides of the observational line are added up and
mapped to the observational line. This mapping scheme is justified by the fact that the lines
within the resolution limits are blended and hence cannot be identified separately by observational
means. The resolution of the observational data can be estimated from the graphs of spectra when
these graphs are available. Information reported in the data source, as well as other contextual
evidence, may also be used to determine the observational resolution.

\item
All blended C\II\ lines in the observational list are combined by considering them as a single line
with a single flux, while C\II\ lines blended with non-C\II\ lines are eliminated.

\item
To compare the theoretical emissivity to the observational flux, the theoretical emissivity of each
line is normalized to the total theoretical emissivity of all the lines involved in the least
squares procedure, while the observational flux of that line is normalized to the total
observational flux of these lines.

\item
When using line $\WL$4267~\AA, the theoretical emissivity is obtained from Davey \etal\
\cite{DaveySK2000} in the form of case B effective recombination coefficients, and option `4' of
the `NormalizationChoice' of the \EmiCod\ code is used, as explained in Appendix \ref{AppEmissMan}.
The reason for this special treatment is that this important line, which is the most intense line
of this ion in the visible part of the spectrum, comes from a BB transition, with no doubly-excited
upper state, for which our emissivity model is not very accurate as it considers dielectronic
contribution only.

\item
The normalized theoretical emissivities corresponding to a particular observational flux are added
up when multiple mapping occurs, that is when the observational flux is given for a multiplet
transition as a whole.

\item
The $\chi^{2}$ defined by the following equation
\begin{equation}
 \chi^{2}=\sum_{i=1}^{N}\frac{\left(I_i^{no}-\varepsilon_i^{nt}\right)^{2}}{\degFree \sigma_{I_i^{no}}^{2}}
\end{equation}
is then computed, where $i$ is an index running over all the $N$ lines involved in the least
squares procedure, $I_i^{no}$ and $\varepsilon_i^{nt}$ are the normalized observational flux and
normalized theoretical emissivity of line $i$ respectively, $\degFree$ is the number of degrees of
freedom, and $\sigma_{I_i^{no}}^{2}$ is the variance of $I_i^{no}$. This variance is computed from
the formulae derived in Appendix~\ref{AppErrAna} where for the data sets with given observational
errors the reported errors were used in conjunction with formula~\ref{EAf1} while for the data sets
with no reported error a Poisson distribution was assumed and hence Equation~\ref{EAPoisEq} was
used to estimate this variance. In some data sets, the observational error was given for some lines
only, and hence the average of the given errors was assigned to the missing ones. In some cases
when the reported error was unrealistically small resulting in large $\chi^2$, the $\chi^2$ curve
was scaled to unity at the minimum to obtain a more realistic error estimate. This scaling does not
affect the temperature at minimum $\chi^2$ but usually broadens the confidence interval.

\item
The temperature of the object is then identified from its value at the minimum $\chi^2$, while the
confidence interval is identified from the values of the temperature corresponding to the values of
$\chi^2_{min}\pm1$ on the lower and upper sides using linear interpolation. In some cases, the
$\chi^2$ curve was too shallow on one side and hence it resulted in a broad confidence interval on
that side.

\end{itemize}

In the following section we present some astronomical objects that have been the subject of
investigation in this study using the least squares minimization method.

\section{Astronomical Objects}\label{AstObj}

In this section we present a sample of the astronomical objects that have been investigated as part
of the current study. The objects are mainly \planeb e and the physical parameter of interest is
the electron temperature of the line emitting regions of the C\II\ \dierec\ and subsequent cascade
decay.

A Planetary Nebula (PN) is an expanding cloud of gas and dust ejected from a dying star due to
instabilities at the end of its life. The cloud is ionized by the highly energetic radiation from
the stellar core (progenitor). The spectra of planetary nebulae, which originate from various
physical processes such as collisional excitation and recombination, are very rich and span large
parts of the electromagnetic spectrum with manifestation in both emission and absorption. These
spectra involve many atomic and ionic species across the periodic table especially the light ones
such as helium, carbon, and nitrogen. The nebular spectra also contain transition lines from some
compound molecular species such as CO and CN. The normal temperature range in PNe is $10^3$-$2
\times 10^4$~K with a typical value of $10^4$ K, while the normal electron density range is
$10^9$-$10^{11}$ m$^{-3}$ with a typical value of 10$^{10}$ m$^{-3}$. Figure~\ref{PNFig} shows two
prominent examples of planetary nebulae. The PNe are essential for astronomical and astrophysical
studies because they are relatively abundant in the Milky Way galaxy and hence provide imperative
case studies. In other galaxies, they may be used to acquire useful information about the chemical
composition and the thermodynamic conditions. For cosmic processes, the PNe are very important
objects because they participate in the process of heavy element enrichment of the astronomical
objects and interstellar medium \cite{AllerK1988, IllingworthC2000, Matzner2001, ZhangLLPB2005,
TennysonBook2005}.

\begin{figure}[!h]
  \centering{}
  \includegraphics[scale=0.34]{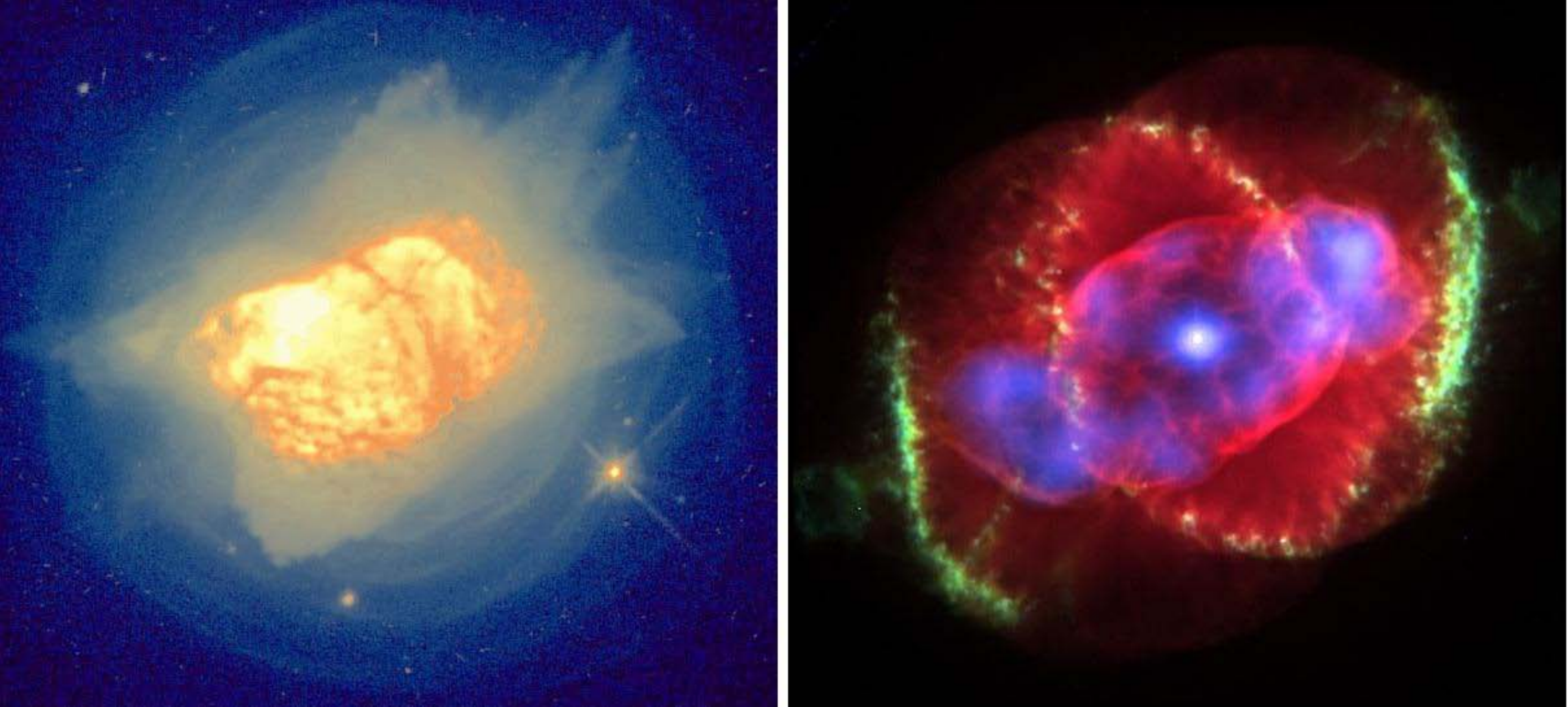}
  \caption[Two prominent planetary nebulae, NGC 7027 and Cat's eye]
          {Two prominent planetary nebulae, NGC 7027 (left) and Cat's eye (right) [from NASA].}
  \label{PNFig}
\end{figure}

\subsection{NGC 7009}

NGC 7009, also called Saturn nebula, is a bright \planeb\ in the Aquarius constellation. This
nebula has a double-ringed complex spatial structure with a rich recombination spectrum related to
some elements like oxygen; therefore it is expected to be an ideal environment for recombination
processes in general \cite{LiuSBC1995, MathisPP1998, LiuLBDS2001, FangL2011}. The observational
data of this object were obtained from Fang and Liu \cite{FangL2011} who collected very deep CCD
spectrum of this object in a wavelength range of 3040-11000\AA. Their observational line list
comprises about 1200 emission lines where approximately 80\% of these lines are identified as
originating from permitted transitions. On surveying this list for C\II\ recombination lines we
obtained about 64 lines attributed to C\II\ transitions. Following the refinement process, outlined
in \S\ \ref{TempMeth}, this list was reduced to just 9 lines. These lines are presented in
Table~\ref{DataFangL2011}. In this table, and other similar tables, $\WL_{lab}$ is the laboratory
wavelength, and $I$ is the flux (absolute or normalized to H$_{\beta}$), while the other columns
stand for the lower and upper terms, the lower and upper statistical weights, and transition type
(free-free, free-bound and bound-bound) respectively. For the multiplet transitions, the wavelength
is given for the first line while the statistical weights represent the sum of the individual
weights. The prime in the term designation indicates an excited core, i.e.
\nlo1s22s2p(\SLP3Po). In Figure~\ref{EtoNGC7009} the ratio of theoretical emissivity to
observational flux is plotted against electron temperature on a linear-linear graph for the
selected 9 lines. This graph, and other similar graphs for the other objects, were utilized to
select the suitable lines.

\begin{figure}[!h]
\centering{}
\includegraphics[scale=0.65]{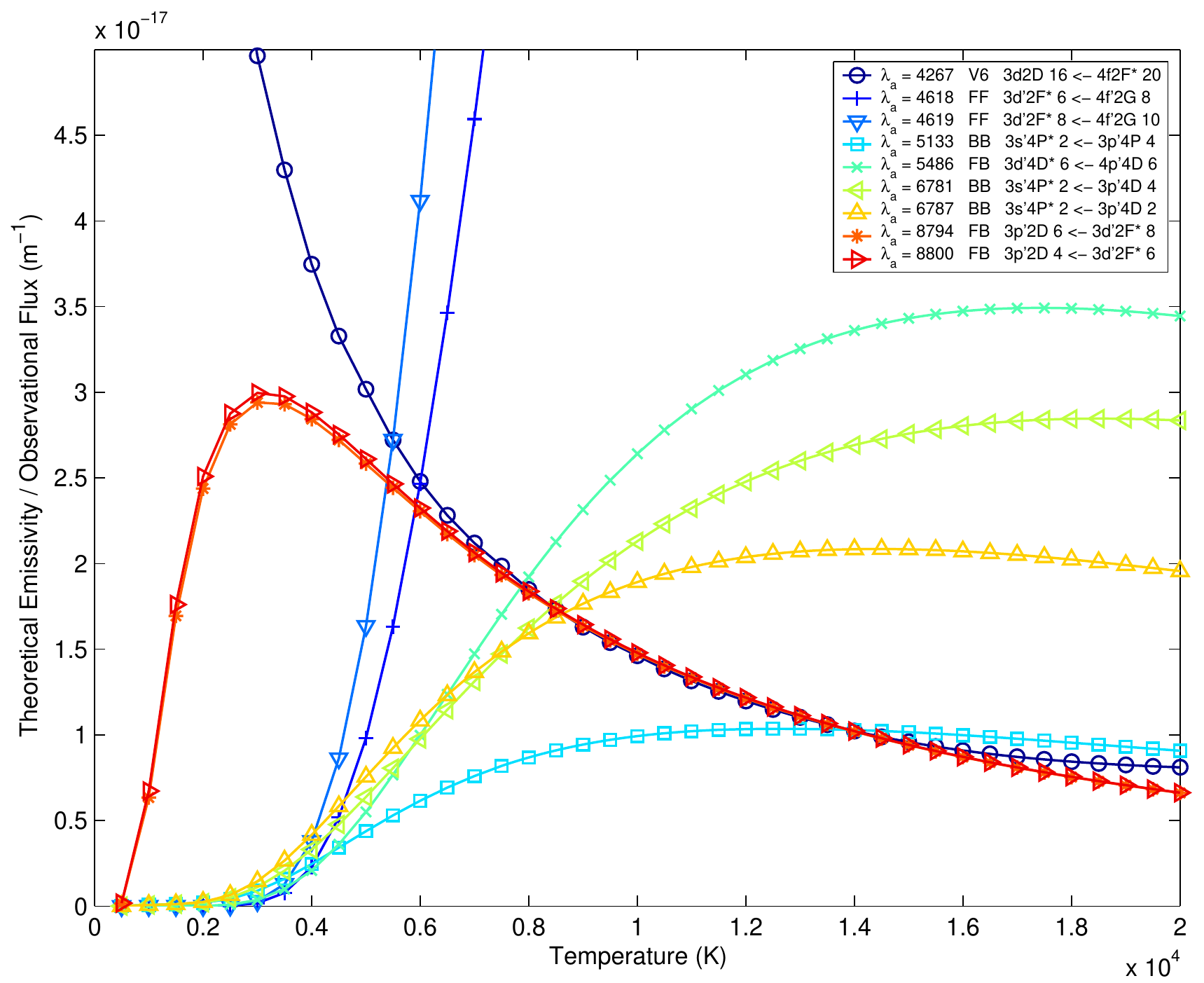}
\caption{The ratio of theoretical emissivity to observational flux as a function of temperature for
the selected C\II\ lines of NGC 7009 spectra of Fang and Liu \cite{FangL2011}.} \label{EtoNGC7009}
\end{figure}

\begin{table} [!h]
\caption{The observational C\II\ transition lines of Fang and Liu \cite{FangL2011} that were used
in the least squares minimization procedure to find the electron temperature of the line emitting
region in the planetary nebula NGC 7009. The given flux is the normalized to the H$_{\beta}$ flux. \label{DataFangL2011}}
\begin{center} 
\begin{tabular*}{\textwidth}{@{\extracolsep{\fill}}|c|c|c|c|c|c|c|}
\hline
$\WL_{lab}$ &       $I$ &      Lower &      Upper &     $\SW_l$ &     $\SW_u$ &       Type \\
\hline
   4267.00 &     0.8795 &       3d \SLP2De &      4f \SLP2Fo &         16 &         20 &         BB \\
\hline
   4618.40 &     0.0021 &     3d$'$ \SLP2Fo &      4f$'$ \SLP2Ge &          6 &          8 &         FF \\
\hline
   4619.23 &     0.0021 &     3d$'$ \SLP2Fo &      4f$'$ \SLP2Ge &          8 &         10 &         FF \\
\hline
   5132.94 &     0.0088 &     3s$'$ \SLP4Po &      3p$'$ \SLP4Pe &          2 &          4 &         BB \\
\hline
   5485.90 &     0.0004 &     3d$'$ \SLP4Do &      4p$'$ \SLP4De &          6 &          6 &         FB \\
\hline
   6780.61 &     0.0070 &     3s$'$ \SLP4Po &      3p$'$ \SLP4De &          2 &          4 &         BB \\
\hline
   6787.22 &     0.0052 &     3s$'$ \SLP4Po &      3p$'$ \SLP4De &          2 &          2 &         BB \\
\hline
   8793.80 &     0.0320 &      3p$'$ \SLP2De &     3d$'$ \SLP2Fo &          6 &          8 &         FB \\
\hline
   8799.90 &     0.0224 &      3p$'$ \SLP2De &     3d$'$ \SLP2Fo &          4 &          6 &         FB \\
\hline
\end{tabular*}
\end{center}
\end{table}

The observational data were normalized to the total observational flux while the theoretical to the
total theoretical emissivity, and $\chi^2$ was obtained and plotted against temperature over the
range $T=500-10000$~K in steps of 100~K. This process was repeated: once with inclusion of line
$\WL$4267 and another with exclusion of this line. The reason for this was discussed in \S\
\ref{TempMeth}. The $\chi^2$ graphs for these two cases are given in
Figures~\ref{Chi2NGC7009-With4267} and \ref{Chi2NGC7009-Without4267}. As seen, the first indicates
a temperature of about 5800~K while the second a temperature of about 5500~K, which are in good
agreement.

\begin{figure}[!h]
\centering{}
\includegraphics[scale=0.65]{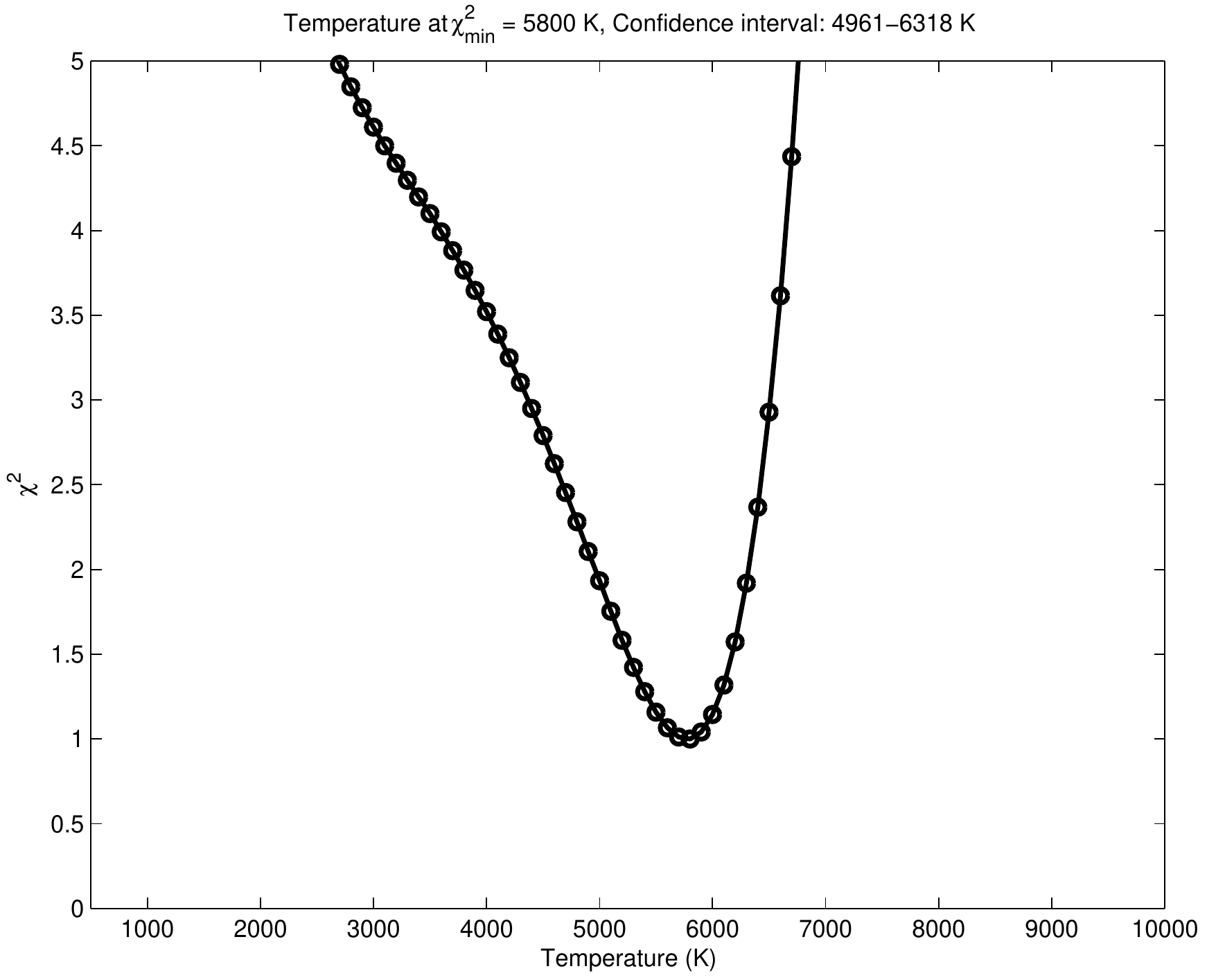}
\caption{Temperature dependence of $\chi^2$ for NGC 7009 with the inclusion of line $\WL$4267. The
temperature at $\chi^2_{min}$ and the confidence interval are shown.} \label{Chi2NGC7009-With4267}
\end{figure}

\begin{figure}[!h]
\centering{}
\includegraphics[scale=0.65]{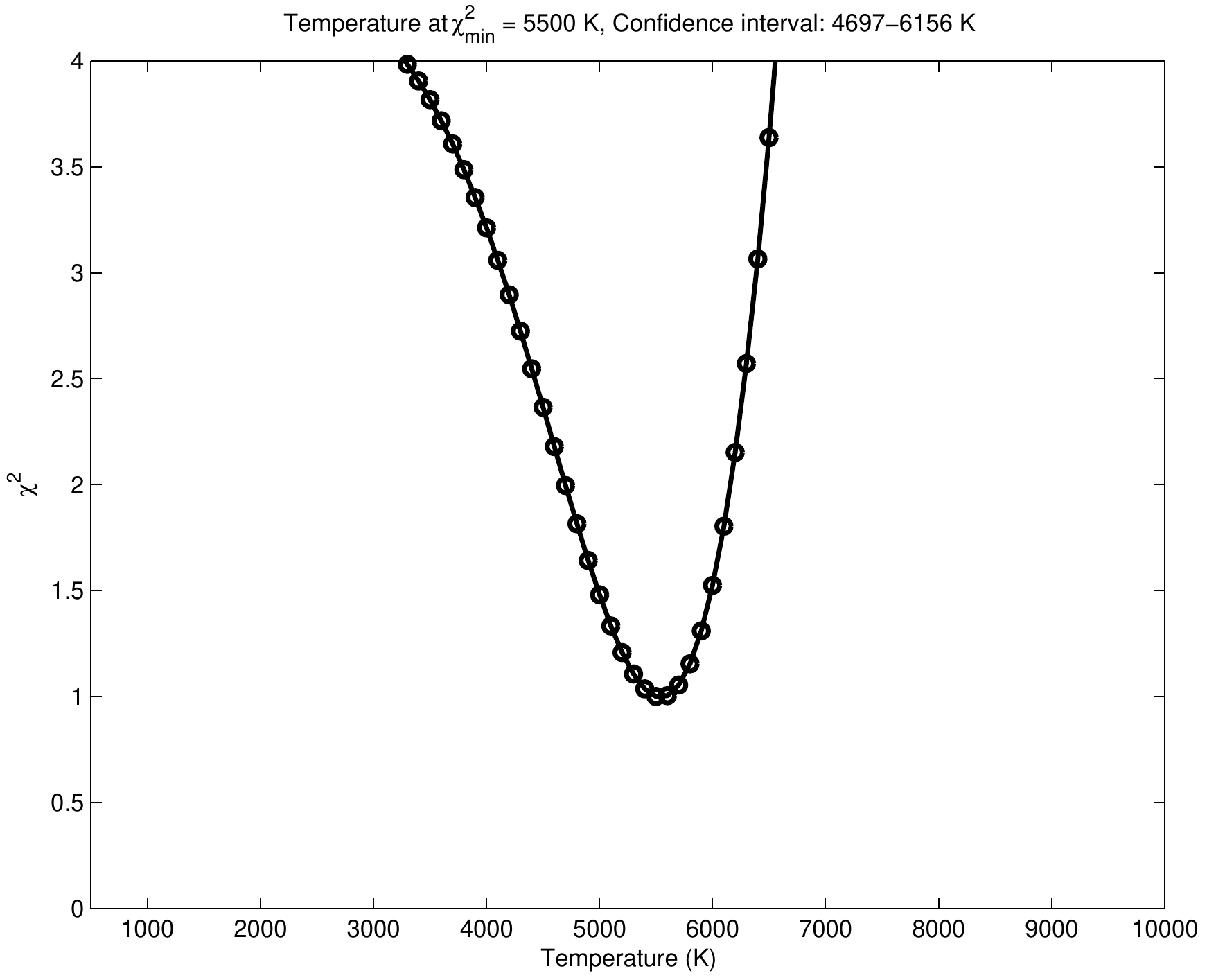}
\caption{Temperature dependence of $\chi^2$ for NGC 7009 with the exclusion of line $\WL$4267. The
temperature at $\chi^2_{min}$ and the confidence interval are shown.}
\label{Chi2NGC7009-Without4267}
\end{figure}

In Table~\ref{TempTableNGC7009} we listed the electron temperature from different types of
transitions related to various ions as obtained from the cited literature following a non-thorough
research. As can be seen, the values that we obtained from our least squares method is in a broad
agreement with the temperature obtained from several recombination lines. The lower electron
temperature from optical recombination lines (ORL) compared to the values obtained from
collisionally-excited lines (CEL) is consistent with the trend of the previous findings related to
the long-standing problem of discrepancy between the abundance and temperature results of ORL and
CEL (higher ORL abundance and lower electron temperature as compared to the CEL abundance and
temperature), as outlined in \S\ \ref{RFLines}. In this table, and other similar ones, some types
of transitions (e.g. C\II) have been assigned more than one temperature value in the same cited
reference; the reason normally is that more than one value has been obtained for that transition
type using different spectral lines. These tables may also contain non-first hand data, as the
purpose is to give a general idea about the values in the literature not to meticulously document
the results obtained from previous works. There may also be some repetitive values as they are
obtained from the same first hand source but reported in different literatures. It should be
remarked that no error estimate is given in these tables due to the limited column space; the cited
data sources should be consulted to obtain this information. Another remark is that we followed the
apparent identification and labeling of the authors without meticulous verification of the lines
and the type of transition they belong to. Since some authors use relaxed notation for ORL and CEL
lines there may be some cases of inaccurate labeling of these lines. For more reliable verification
our data sources, as cited in these tables, should be consulted to establish the origin of these
lines.

\begin{table} [!h]
\caption{The electron temperature in Kelvin of NGC 7009 from different atoms, ions and transitions
as obtained from the cited literature, where BJ stands for Balmer Jump and PJ for Paschen Jump. CW
refers to the current work. For reference \cite{HarringtonLS1981} we followed the apparent labeling
of the authors although the temperatures may have been derived from CEL lines.
\label{TempTableNGC7009}}
\begin{center} 
{\small
\begin{tabular*}{\textwidth}{@{\extracolsep{\fill}}|l|c|c|c|c|c|c|c|c|c|c|}
\hline
    Source & \cite{HarringtonLS1981} & \cite{MckennaKKWBA1996} & \cite{Kholtygin1998} & \cite{MathisPP1998} & \cite{LiuLBL2004} & \cite{ZhangLWSLD2004} & \cite{Liu2006} & \cite{PeimbertP2006} & \cite{TsamisWPBDL2008} &         CW \\
\hline
  H\Ii(BJ) &            &            &            &            &       8150 &       7200 &       7200 &       7200 &            &            \\
\hline
  H\Ii(PJ) &            &            &            &            &            &       5800 &            &            &            &            \\
\hline
     He\Ii &            &            &            &            &       5380 &            &       5040 &       8000 &            &            \\

           &            &            &            &            &            &            &            &       6800 &            &            \\

           &            &            &            &            &            &            &            &       5040 &            &            \\
\hline
      C\II &            &            &            &            &            &            &            &            &            &       5800 \\

           &            &            &            &            &            &            &            &            &            &       5500 \\
\hline
    C\III] &            &            &            &       8160 &            &            &            &            &            &            \\

           &            &            &            &       8350 &            &            &            &            &            &            \\
\hline
      N\II &      10800 &            &            &            &            &            &            &            &            &            \\
\hline
    [N\II] &            &      11040 &            &            &            &            &            &            &            &            \\
\hline
      O\II &            &            &            &            &       1600 &            &            &            &            &            \\
\hline
     O\III &       9600 &            &            &            &            &            &            &            &            &            \\

           &       8900 &            &            &            &            &            &            &            &            &            \\
\hline
    O\III] &            &            &            &       9570 &            &            &            &            &            &            \\

           &            &            &            &       8730 &            &            &            &            &            &            \\

           &            &            &            &      10340 &            &            &            &            &            &            \\

           &            &            &            &       9870 &            &            &            &            &            &            \\
\hline
   [O\III] &            &       9910 &            &      10380 &       9980 &       9980 &       9980 &      10000 &            &            \\

           &            &            &            &       9420 &            &       8350 &            &            &            &            \\

           &            &            &            &       9740 &            &            &            &            &            &            \\

           &            &            &            &       9930 &            &            &            &            &            &            \\
\hline
      Mean &            &            &       7419 &            &            &            &            &            &      10340 &            \\
\hline
\end{tabular*}
}
\end{center}
\end{table}

\clearpage
\subsection{NGC 5315}

NGC 5315 is a young dense planetary nebula in the southern constellation Circinus located at a
distance of about 2.6~kpc with an interesting complex flower shape appearance \cite{LiuBCDLe2001,
PeimbertPRE2004}. The observational data of this object were obtained from Peimbert \etal\
\cite{PeimbertPRE2004}. On inspecting this data source for C\II\ recombination lines we obtained
about 25 lines attributed to C\II\ transitions. Following the refinement process, outlined in \S\
\ref{TempMeth}, this list was reduced to just 4 lines. These lines are presented in
Table~\ref{DataPeimbertPRE2004}. Figure~\ref{NerNGC53151} displays the ratio of the normalized
observed flux to the normalized theoretical emissivity versus electron temperature for the selected
4 lines of NGC 5315 data. This graph, with similar graphs for the other objects were used in the
refinement process to include and exclude lines from the original list into the final one, as
explained in \S\ \ref{TempMeth}.

\begin{table} [!t]
\caption{The observational C\II\ transition lines of Peimbert \etal\ \cite{PeimbertPRE2004} that
were used in the least squares minimization procedure to find the electron temperature of the line
emitting region in the planetary nebula NGC 5315. The given flux is the normalized to the H$_{\beta}$ flux. \label{DataPeimbertPRE2004}}
\begin{center} 
\begin{tabular*}{\textwidth}{@{\extracolsep{\fill}}|c|c|c|c|c|c|c|}
\hline
$\WL_{lab}$ &       $I$ &      Lower &      Upper &     $\SW_l$ &     $\SW_u$ &       Type \\
\hline
   4267.00 &     0.6559 &       3d \SLP2De &       4f \SLP2Fo &         16 &         20 &         BB \\
\hline
   5145.16 &     0.0039 &     3s$'$ \SLP4Po &      3p$'$ \SLP4Pe &          6 &          6 &         BB \\
\hline
   6791.47 &     0.0068 &     3s$'$ \SLP4Po &      3p$'$ \SLP4De &          4 &          4 &         BB \\
\hline
   6800.68 &     0.0028 &     3s$'$ \SLP4Po &      3p$'$ \SLP4De &          6 &          6 &         BB \\
\hline
\end{tabular*}
\end{center}
\end{table}

\begin{figure}[!h]
\centering{}
\includegraphics[scale=0.7]{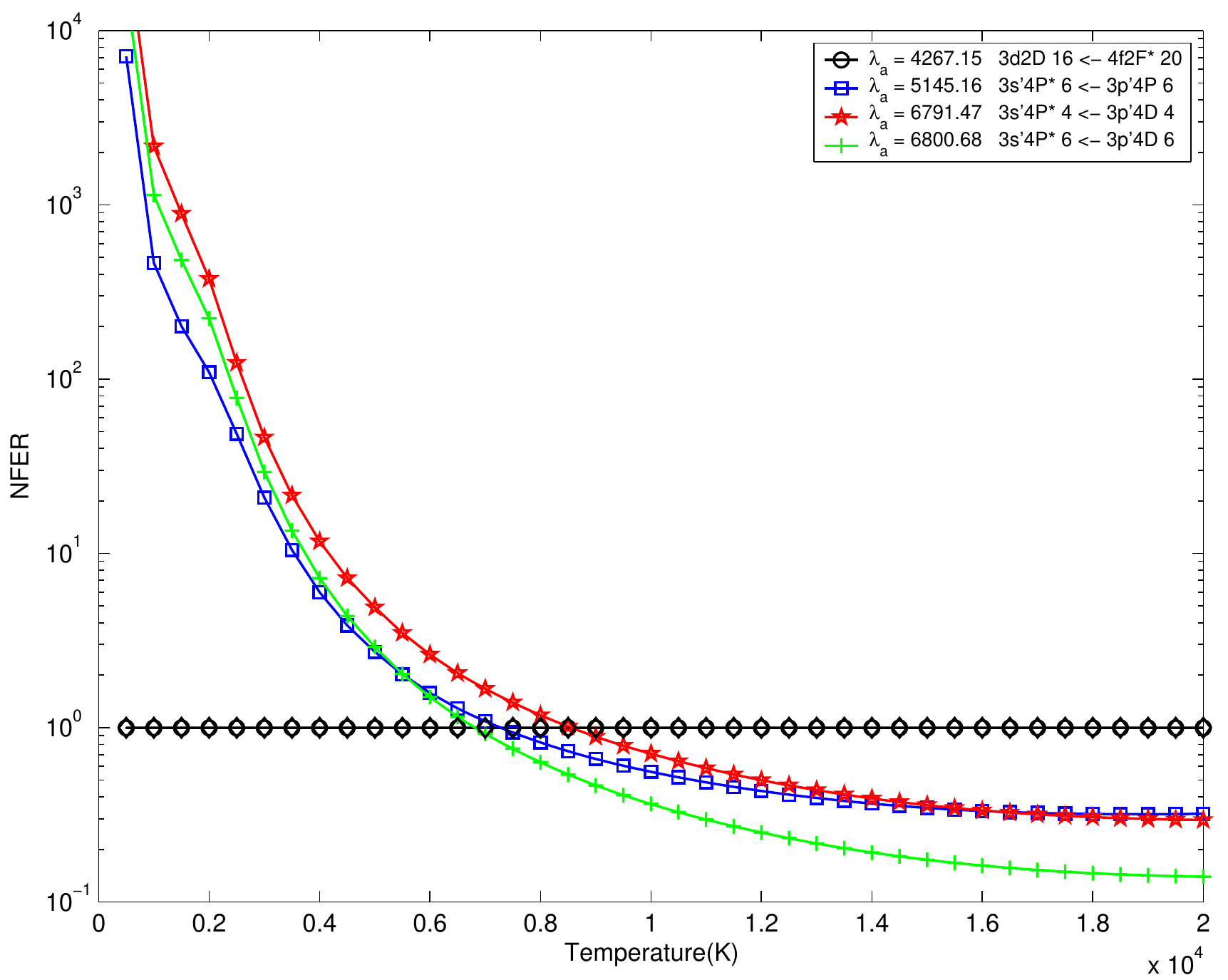}
\caption{Ratio of normalized observed flux to normalized theoretical emissivity (NFER) versus
temperature on log-linear scales for the \planeb\ NGC 5315.} \label{NerNGC53151}
\end{figure}

The data were normalized to the total flux and emissivity, as outlined earlier, and $\chi^2$ was
computed and plotted against temperature over the range $T=500-20000$~K in steps of 100~K. This
process was performed with and without line $\WL$4267, as for NGC 7009. The $\chi^2$ graphs for
these two cases are given in Figures~\ref{Chi2NGC5315-With4267} and \ref{Chi2NGC5315-Without4267}.
As seen, the first minimizes at $T\simeq7400$~K while the second at $T\simeq6500$~K. The shape of
these curves may indicate that the first value is more reliable. Table~\ref{TempTableNGC5315}
presents some values of the electron temperature of NGC 5315 as reported in the cited literature.
Our value of 7400~K compares very well with some of these values. Again, the temperature is lower
than that derived from the collisionally-excited lines.

\begin{figure}[!h]
\centering{}
\includegraphics[scale=0.65]{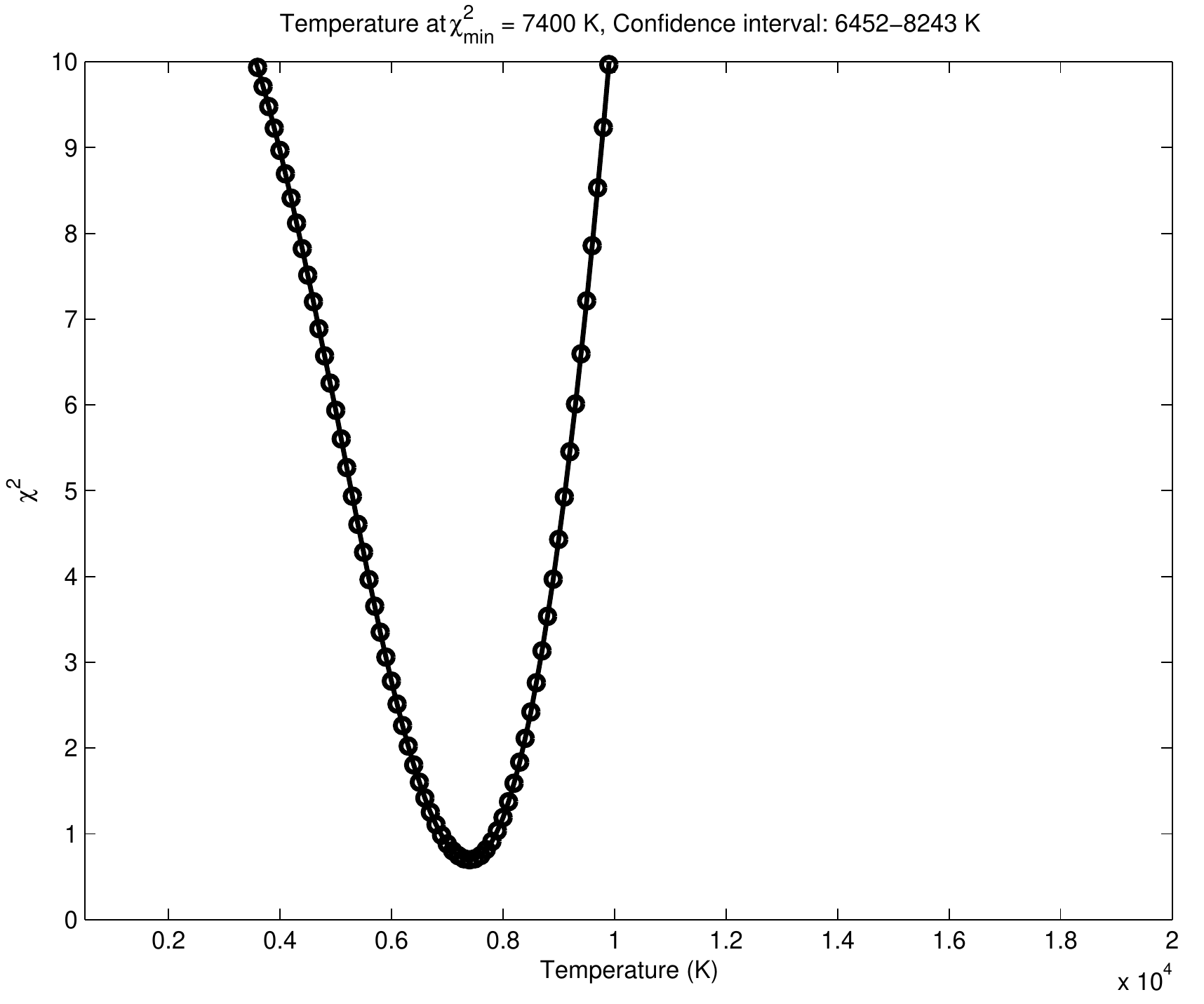}
\caption{Temperature dependence of $\chi^2$ for NGC 5315 with the inclusion of line $\WL$4267. The
temperature at $\chi^2_{min}$ and the confidence interval are shown.} \label{Chi2NGC5315-With4267}
\end{figure}

\begin{figure}[!h]
\centering{}
\includegraphics[scale=0.65]{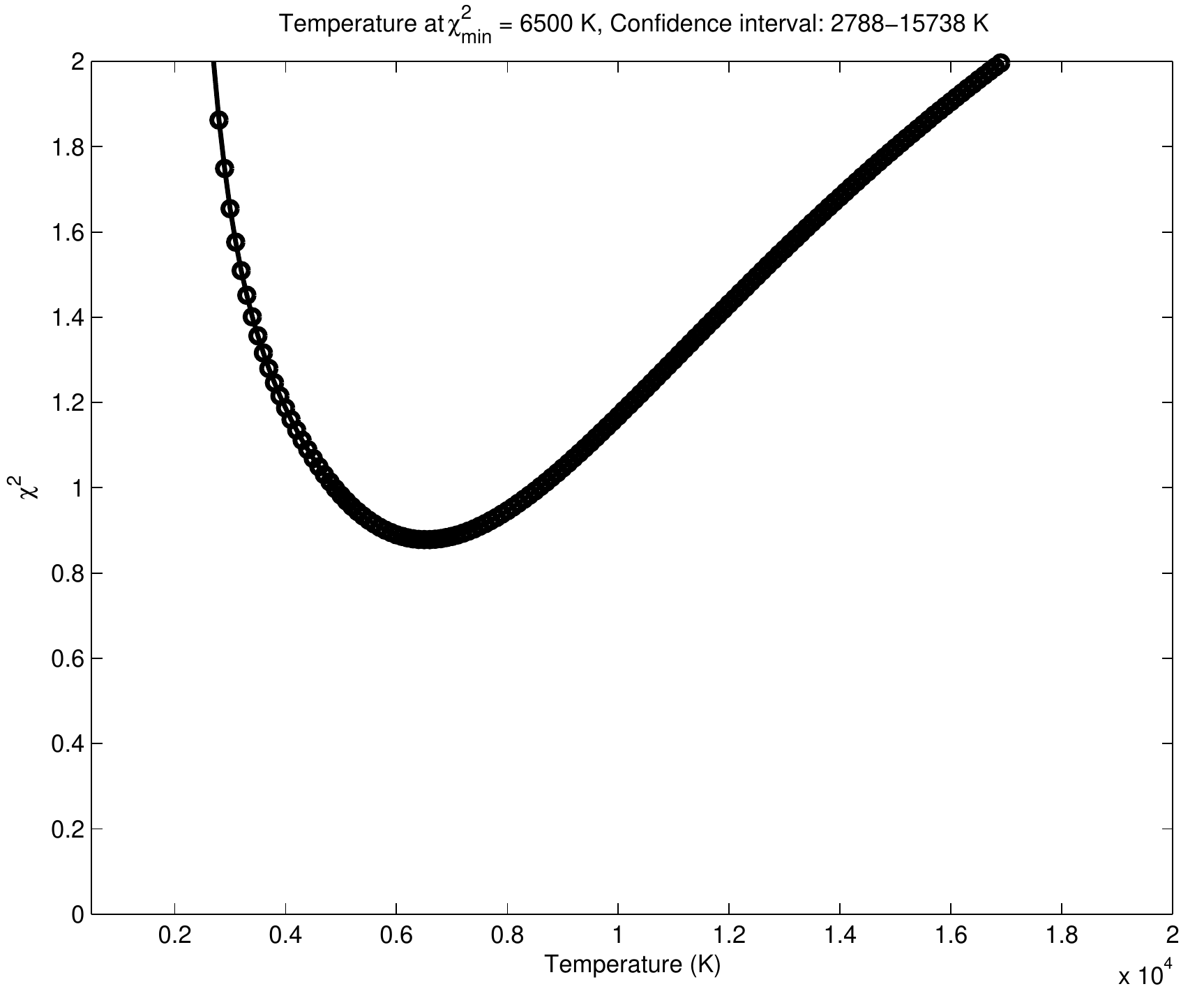}
\caption{Temperature dependence of $\chi^2$ for NGC 5315 with the exclusion of line $\WL$4267. The
temperature at $\chi^2_{min}$ and the confidence interval are shown.}
\label{Chi2NGC5315-Without4267}
\end{figure}

\begin{table} [!h]
\caption{The electron temperature in Kelvin of NGC 5315 from different atoms, ions and transitions
as obtained from the cited literature, where BJ stands for Balmer Jump. CW refers to the current
work. \label{TempTableNGC5315}}
\begin{center} 
\begin{tabular*}{\textwidth}{@{\extracolsep{\fill}}|l|c|c|c|c|c|c|c|}
\hline
    Source & \cite{MckennaKKWBA1996} & \cite{Kholtygin1998} & \cite{MathisPP1998} & \cite{TsamisBLDS2003b} & \cite{TsamisBLSD2004} & \cite{PeimbertPRE2004} &         CW \\
\hline
  H\Ii(BJ) &            &            &            &       8600 &       8600 &            &            \\
\hline
      H\II &            &            &            &            &            &       7500 &            \\
\hline
     He\Ii &            &            &            &            &      10000 &            &            \\
\hline
      C\II &            &            &            &            &            &            &       7400 \\

           &            &            &            &            &            &            &       6500 \\
\hline
      N\II &            &            &            &            &            &       9600 &            \\
\hline
    [N\II] &       9090 &            &            &      10800 &            &            &            \\
\hline
      O\Ii &            &            &            &            &            &       7730 &            \\
\hline
      O\II &            &            &            &       8100 &       5750 &            &            \\

           &            &            &            &            &       4350 &            &            \\
\hline
     O\III &            &            &            &            &            &       8850 &            \\
\hline
    O\III] &            &            &      10040 &            &            &            &            \\
\hline
   [O\III] &       8150 &            &      10330 &       9000 &       9000 &            &            \\

           &            &            &            &      18500 &            &            &            \\

           &            &            &            &       7800 &            &            &            \\
\hline
    Ne\III &            &            &            &            &            &      10700 &            \\
\hline
      S\II &            &            &            &            &            &       9000 &            \\
\hline
     S\III &            &            &            &            &            &       9150 &            \\
\hline
    [S\II] &            &            &            &      11400 &            &            &            \\
\hline
    Cl\III &            &            &            &            &            &      10500 &            \\
\hline
     Cl\IV &            &            &            &            &            &      11200 &            \\
\hline
    Ar\III &            &            &            &            &            &       8300 &            \\
\hline
      Mean &            &       7873 &            &            &            &            &            \\
\hline
\end{tabular*}
\end{center}
\end{table}

\clearpage
\subsection{NGC 7027}

NGC 7027 is a compact, bright, young, high excitation planetary nebula with one of the hottest
central stars known for a PN. Among the observed planetary nebulae, it has one of the richest lines
emission which spans most parts of the electromagnetic spectrum and has been a challenge for
observers as well as theorists. NGC 7027 is located about 0.9~kpc from the sun in the direction of
the constellation Cygnus in the Milky Way. There is evidence on the existence of substantial
quantities of heavy elements dust alongside the neutral and ionized nebular gas. Thanks to its
luminosity and compactness, it can be easily observed despite the considerable extinction caused by
interior and interstellar dust. It has been a reference object and the subject of several spectral
surveys, and probably it is the best studied PN. There are several indications that the progenitor
of NGC 7027 is a carbon-rich star \cite{AllerK1988, PequignotB1988, ZhangLLPB2005}.

Two observational data sets related to NGC 7027 were investigated: one obtained from Baluteau
\etal\ \cite{BaluteauZMP1995}, and the other obtained from Zhang \etal\ \cite{ZhangLLPB2005}.

\subsubsection{Baluteau \etal}

On surveying Baluteau \etal\ \cite{BaluteauZMP1995} for C\II\ recombination lines we obtained about
30 lines attributed to C\II\ transitions. Following the refinement process, outlined in \S\
\ref{TempMeth}, this list was reduced to just 8 lines. These lines are presented in
Table~\ref{DataBaluteauZMP1995}.

\begin{table} [!h]
\caption{The observational C\II\ transition lines of Baluteau \etal\ \cite{BaluteauZMP1995} that
were used in the least squares minimization procedure to find the electron temperature of the line
emitting region in the planetary nebula NGC 7027. The given flux is the normalized to the H$_{\beta}$ flux. \label{DataBaluteauZMP1995}}
\begin{center} 
\begin{tabular*}{\textwidth}{@{\extracolsep{\fill}}|c|c|c|c|c|c|c|}
\hline
$\WL_{lab}$ &       $I$ &      Lower &      Upper &   $\SW_l$ &   $\SW_u$ &       Type \\
\hline
   6779.93 &       11.8 &     3s$'$ \SLP4Po &      3p$'$ \SLP4De &          4 &          6 &         BB \\
\hline
   6783.91 &        2.1 &     3s$'$ \SLP4Po &      3p$'$ \SLP4De &          6 &          8 &         BB \\
\hline
   6787.22 &        3.7 &     3s$'$ \SLP4Po &      3p$'$ \SLP4De &          2 &          2 &         BB \\
\hline
   6791.47 &        4.9 &     3s$'$ \SLP4Po &      3p$'$ \SLP4De &          4 &          4 &         BB \\
\hline
   6798.10 &        0.7 &     3s$'$ \SLP4Po &      3p$'$ \SLP4De &          4 &          2 &         BB \\
\hline
   6812.28 &        0.5 &     3s$'$ \SLP4Po &      3p$'$ \SLP4De &          6 &          4 &         BB \\
\hline
   7112.48 &        4.7 &      3p$'$ \SLP4De &     3d$'$ \SLP4Fo &          6 &         10 &         BB \\
\hline
   8793.80 &       11.8 &      3p$'$ \SLP2De &     3d$'$ \SLP2Fo &          6 &          8 &         FB \\
\hline
\end{tabular*}
\end{center}
\end{table}

Standard normalization process to the total flux and emissivity was carried out, and a $\chi^2$
plot as a function of temperature over the range $T=500-20000$~K in steps of 100~K was obtained. As
seen in Figure~\ref{Chi2NGC7027-Baluteau}, the $\chi^2_{min}$ occurs at a temperature of about
11100~K, in good agreement with values obtained by other researchers using differen methods and
various transitions from different atoms and ions; a sample of which is presented in
Table~\ref{TempTableNGC7027}. There are other estimates for the electron temperature of the line
emitting regions of NGC 7027 deduced from different methods, and they generally represent average
values. For example, van Hoof \etal\ \cite{HoofBVF2000} reported average electron temperatures in
various line emitting regions of NGC 7027 obtained from a number of ions using their
photoionization model; these temperatures range between $T=10450-17990$~K with an average value of
about 14100~K. This is very close to the value derived by Aller and Keyes \cite{AllerK1988} and
Zhang \etal\ \cite{ZhangLLPB2005} of about 14000~K.

\begin{figure}[!h]
\centering{}
\includegraphics[scale=0.65]{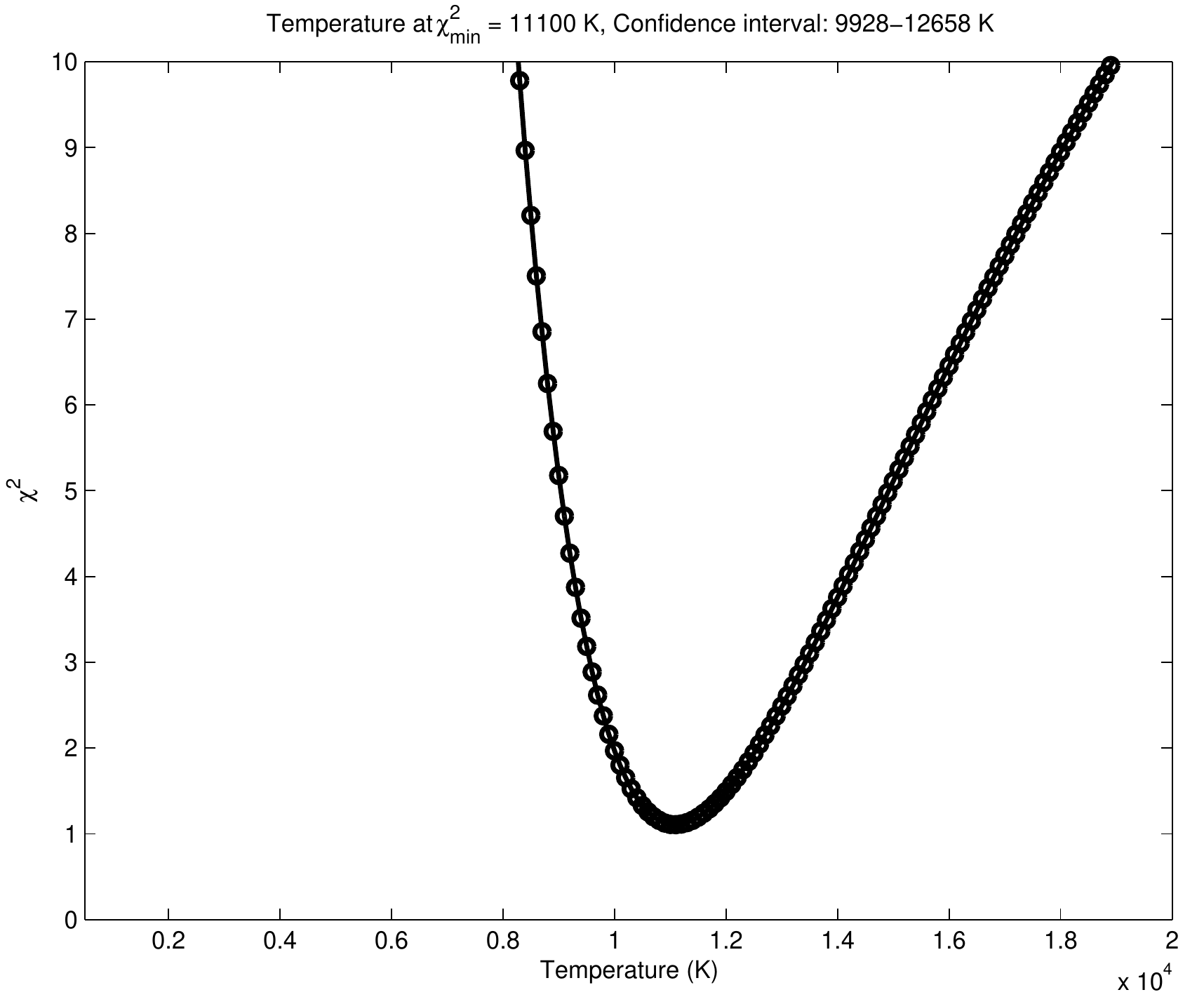}
\caption{Temperature dependence of $\chi^2$ for NGC 7027 of Baluteau \etal\ \cite{BaluteauZMP1995}.
The temperature at $\chi^2_{min}$ and the confidence interval are shown.}
\label{Chi2NGC7027-Baluteau}
\end{figure}

{\renewcommand{\tabcolsep}{1pt}
\begin{table} [!h]
\caption{The electron temperature in Kelvin for NGC 7027 derived from different atoms, ions and
transitions as obtained from the cited literature, where BJ stands for Balmer Jump and PJ for
Paschen Jump. CW refers to the current work where the first value was obtained from Baluteau \etal\
data while the other two are obtained from Zhang \etal\ data. \label{TempTableNGC7027}}
\begin{center} 
{
\begin{tabular*}{\textwidth}{@{\extracolsep{\fill}}|l|c|c|c|c|c|c|c|c|c|}
\hline
    Source & \cite{MckennaKKWBA1996} & \cite{Kholtygin1998} & \cite{MathisPP1998} & \cite{LiuLBL2004} & \cite{ZhangLWSLD2004} & \cite{TessiG2005} & \cite{PeimbertP2006} & \cite{SharpeeZWPCe2007} &         CW \\
\hline
  H\Ii(BJ) &            &            &            &      12800 &      12000 &            &      12000 &       8000 &            \\
\hline
  H\Ii(PJ) &            &            &            &            &       8000 &            &            &       8000 &            \\
\hline
     He\Ii &            &            &            &       9260 &            &            &      10000 &            &            \\

           &            &            &            &            &            &            &       8200 &            &            \\

           &            &            &            &            &            &            &      10360 &            &            \\
\hline
      C\II &            &            &            &            &            &            &            &            &      11100 \\

           &            &            &            &            &            &            &            &            &      12500 \\

           &            &            &            &            &            &            &            &            &      12000 \\
\hline
    C\III] &            &            &      12290 &            &            &            &            &            &            \\
\hline
     [N\Ii] &            &            &            &            &            &            &            &      15000 &            \\
\hline
    [N\II] &      12300 &            &            &            &            &            &            &            &            \\
\hline
      O\II &            &            &            &       7100 &            &            &            &            &            \\
\hline
     [O\Ii] &            &            &            &            &            &            &            &      11300 &            \\
\hline
    O\III] &            &            &      13300 &            &            &            &            &            &            \\
\hline
   [O\III] &       9260 &            &      14850 &      12600 &            &      14130 &      13000 &            &            \\
\hline
   [Cl\IV] &            &            &            &            &            &            &            &      13700 &            \\
\hline
  [Ar\III] &            &            &            &            &            &            &            &      12900 &            \\
\hline
   [Ar\IV] &            &            &            &            &            &      13600 &            &            &            \\
\hline
      Mean &            &      11505 &            &            &            &            &            &            &            \\

           &            &      11124 &            &            &            &            &            &            &            \\
\hline
\end{tabular*}
}
\end{center}
\end{table}
\renewcommand{\tabcolsep}{6pt}}

\subsubsection{Zhang \etal}

The observational data of this object were obtained from Zhang \etal\ \cite{ZhangLLPB2005}. On
surveying the observational list for C\II\ recombination lines we obtained about 67 lines
attributed to C\II\ transitions. Following the refinement process, outlined in \S\ \ref{TempMeth},
this list was reduced to just 20 lines. These lines are presented in Table~\ref{DataZhangLLPB2005}.

The data were normalized to the total flux and emissivity with and without line $\WL$4267, as
outlined earlier, and $\chi^2$ was calculated and plotted against temperature over the range
$T=500-30000$~K in steps of 100~K. As seen in Figures~\ref{Chi2NGC7027-Zhang-With4267} and
\ref{Chi2NGC7027-Zhang-Without4267}, the first minimizes at $T\simeq12500$~K and the second at
$T\simeq12000$~K, which reasonably agree with some of the values reported in the literature as seen
in Table~\ref{TempTableNGC7027}. However, the confidence interval in both cases poorly constrains
the temperature estimate limits especially on the upper end. It should be remarked that since no
observational error estimate was given for these lines, an error estimate based on Poisson
statistical distribution was used in the least squares procedure, as given by
Equation~\ref{EAPoisEq} in Appendix~\ref{AppErrAna}.

\begin{figure}[!h]
\centering{}
\includegraphics[scale=0.65]{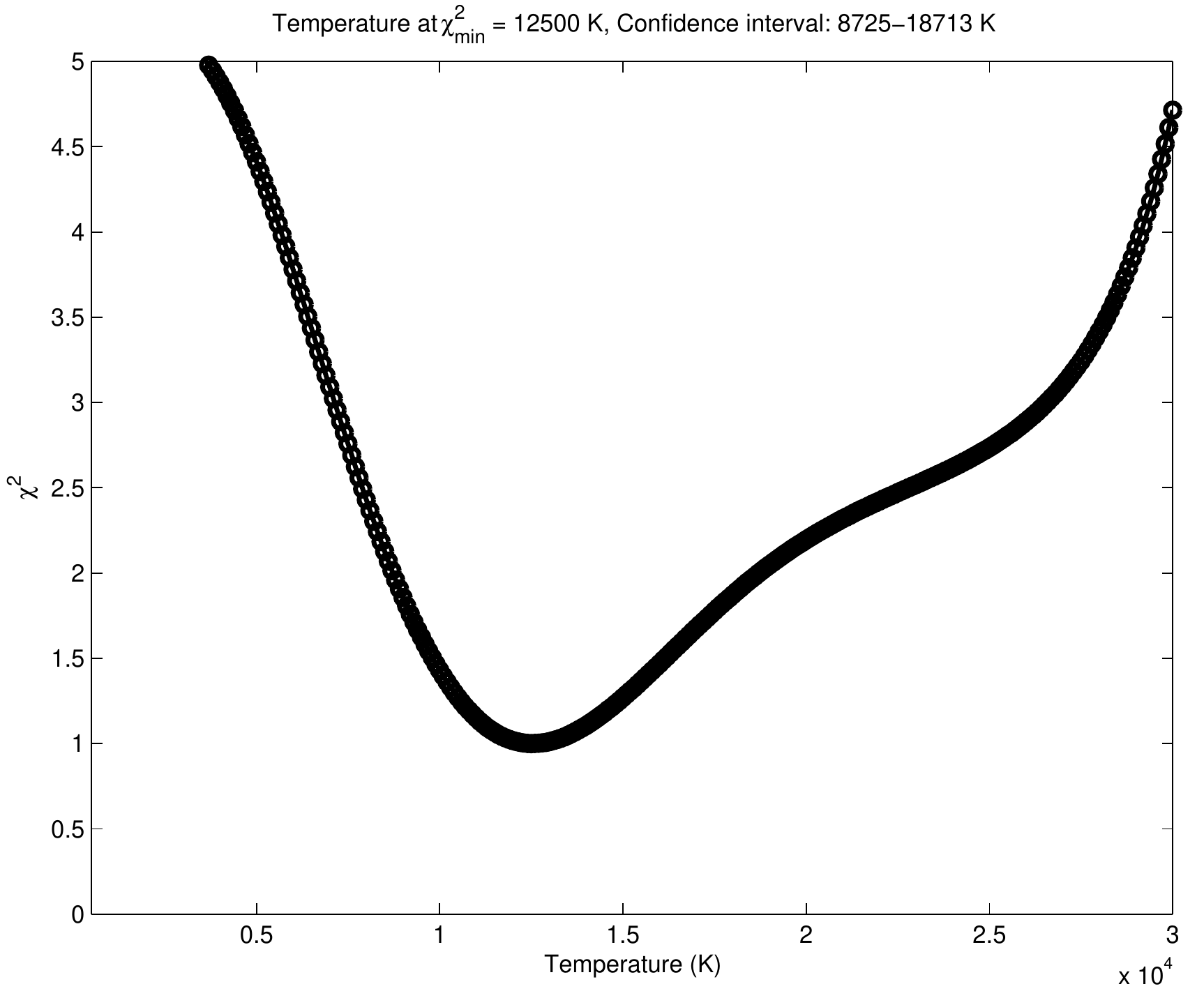}
\caption{Temperature dependence of $\chi^2$ for NGC 7027 of Zhang \etal\ \cite{ZhangLLPB2005} with
the inclusion of line $\WL$4267. The temperature at $\chi^2_{min}$ and the confidence interval are
shown.} \label{Chi2NGC7027-Zhang-With4267}
\end{figure}

\begin{figure}[!h]
\centering{}
\includegraphics[scale=0.65]{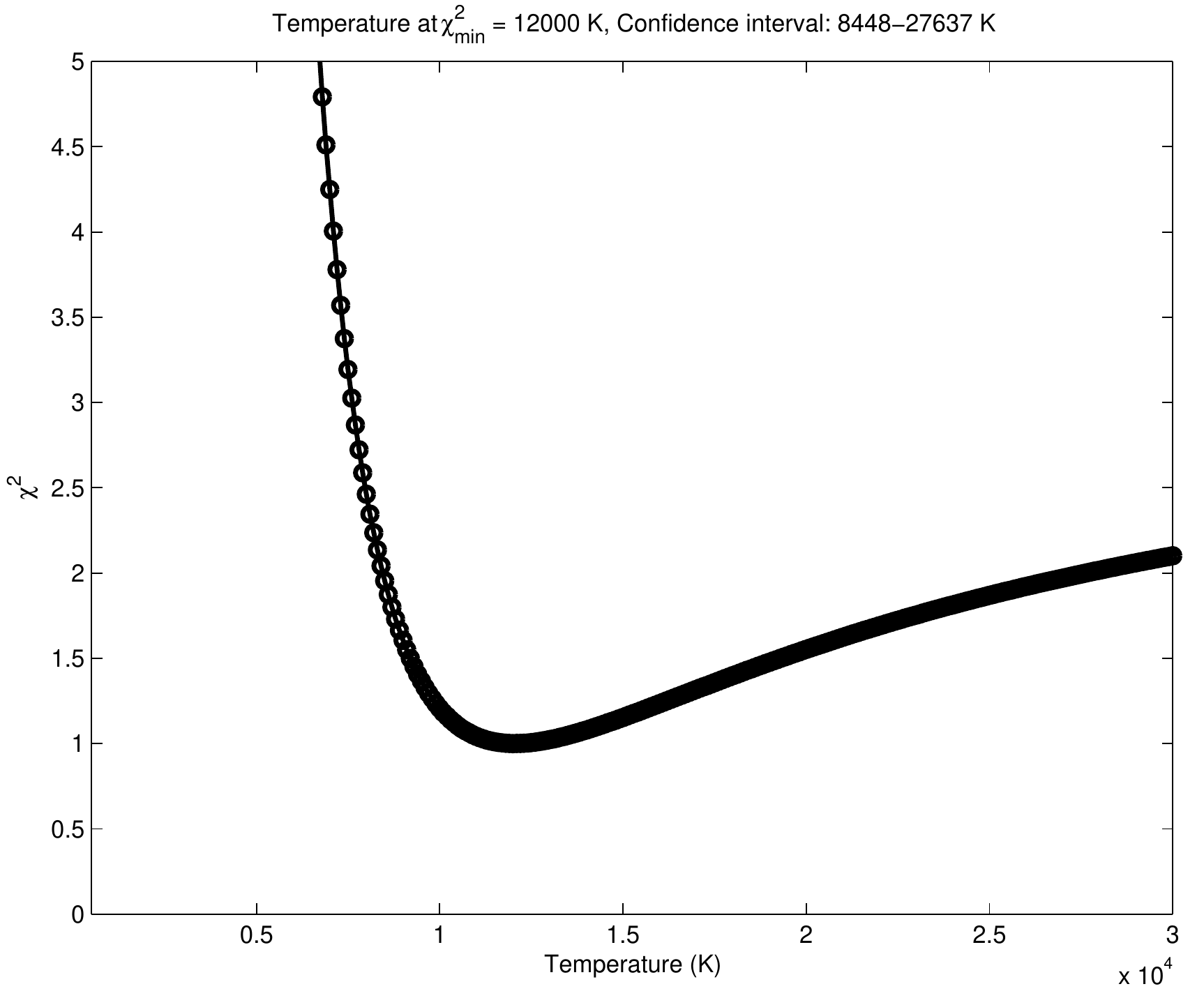}
\caption{Temperature dependence of $\chi^2$ for NGC 7027 of Zhang \etal\ \cite{ZhangLLPB2005} with
the exclusion of line $\WL$4267. The temperature at $\chi^2_{min}$ and the confidence interval are
shown.} \label{Chi2NGC7027-Zhang-Without4267}
\end{figure}

\begin{table} [!t]
\caption{The observational C\II\ transition lines of Zhang \etal\ \cite{ZhangLLPB2005} that were
used in the least squares minimization procedure to find the electron temperature of the line
emitting region in the planetary nebula NGC 7027. The given flux is the normalized to the H$_{\beta}$ flux. \label{DataZhangLLPB2005}}
\begin{center} 
\begin{tabular*}{\textwidth}{@{\extracolsep{\fill}}|c|c|c|c|c|c|c|}
\hline
$\WL_{lab}$ &       $I$ &      Lower &      Upper &   $\SW_l$ &   $\SW_u$ &       Type \\
\hline
   3588.92 &      0.018 &      3p$'$ \SLP4De &     4s$'$ \SLP4Po &          2 &          2 &         FB \\
\hline
   3590.86 &      0.059 &      3p$'$ \SLP4De &     4s$'$ \SLP4Po &         10 &          6 &         FB \\
\hline
   3876.66 &      0.026 &     3d$'$ \SLP4Fo &      4f$'$ \SLP4Ge &          6 &          8 &         FB \\
\hline
   4267.00 &      0.575 &       3d \SLP2De &      4f \SLP2Fo &         16 &         20 &         BB \\
\hline
  4323.11 &      0.004 &      3p$'$ \SLP4Pe &     4s$'$ \SLP4Po &          2 &          2 &         FB \\
\hline
   4372.35 &      0.026 &     3d$'$ \SLP4Po &      4f$'$ \SLP4De &          4 &          4 &         FF \\
\hline
   4376.56 &      0.031 &     3d$'$ \SLP4Po &      4f$'$ \SLP4De &          4 &          6 &         FF \\
\hline
   4411.16 &      0.019 &     3d$'$ \SLP2Do &      4f$'$ \SLP2Fe &          4 &          6 &         FF \\
\hline
   4618.40 &      0.009 &     3d$'$ \SLP2Fo &      4f$'$ \SLP2Ge &          6 &          8 &         FF \\
\hline
   5133.28 &      0.013 &     3s$'$ \SLP4Po &      3p$'$ \SLP4Pe &          4 &          6 &         BB \\
\hline
   5143.38 &      0.013 &     3s$'$ \SLP4Po &      3p$'$ \SLP4Pe &          4 &          2 &         BB \\
\hline
   5151.09 &      0.009 &     3s$'$ \SLP4Po &      3p$'$ \SLP4Pe &          6 &          4 &         BB \\
\hline
   5259.06 &      0.009 &     3d$'$ \SLP4Fo &      4p$'$ \SLP4De &          8 &          6 &         FB \\
\hline
   6779.93 &      0.034 &     3s$'$ \SLP4Po &      3p$'$ \SLP4De &          4 &          6 &         BB \\
\hline
   6783.90 &      0.004 &     3s$'$ \SLP4Po &      3p$'$ \SLP4De &          6 &          8 &         BB \\
\hline
   6787.22 &      0.008 &     3s$'$ \SLP4Po &      3p$'$ \SLP4De &          2 &          2 &         BB \\
\hline
   6791.47 &      0.012 &     3s$'$ \SLP4Po &      3p$'$ \SLP4De &          4 &          4 &         BB \\
\hline
   6800.68 &      0.009 &     3s$'$ \SLP4Po &      3p$'$ \SLP4De &          6 &          6 &         BB \\
\hline
   6812.29 &      0.001 &     3s$'$ \SLP4Po &      3p$'$ \SLP4De &          6 &          4 &         BB \\
\hline
   8793.80 &      0.015 &      3p$'$ \SLP2De &     3d$'$ \SLP2Fo &          6 &          8 &         FB \\
\hline
\end{tabular*}
\end{center}
\end{table}

\clearpage
\subsection{IC 418}

IC 418, or Spirograph Nebula, is a bright, young, C-enhanced, low-excitation, highly symmetric,
elliptically-shaped \planeb\ with apparent ring structure located at a distance of about 0.6~kpc in
the constellation Lepus \cite{Minkowski1953, HarringtonLSS1980, AdamsS1982, CleggSPP1983,
PottaschBRBDe1984, PhillipsRM1990, HyungAF1994, PottaschSBF2004, SharpeeBW2004}. Due to its
brightness and exceptional features, the nebula was, and is still, the subject of many theoretical
and observational studies in the last decades. The observational data about IC 418 which we used in
the current study come from Sharpee \etal\ \cite{SharpeeWBH2003}. On surveying this source of data
for C\II\ recombination lines we obtained about 93 lines attributed to C\II\ transitions. Following
the refinement process, outlined in \S\ \ref{TempMeth}, this list was reduced to just 23 lines.
These lines are presented in Table~\ref{DataSharpeeWBH2003}. In Figure~\ref{EtoIC418} the ratio of
theoretical emissivity to observational flux is plotted against electron temperature on a
linear-linear graph for the selected 23 lines. This graph, and other similar graphs for the other
objects, were used to select the suitable lines.

\begin{figure}[!h]
\centering{}
\includegraphics[scale=0.65]{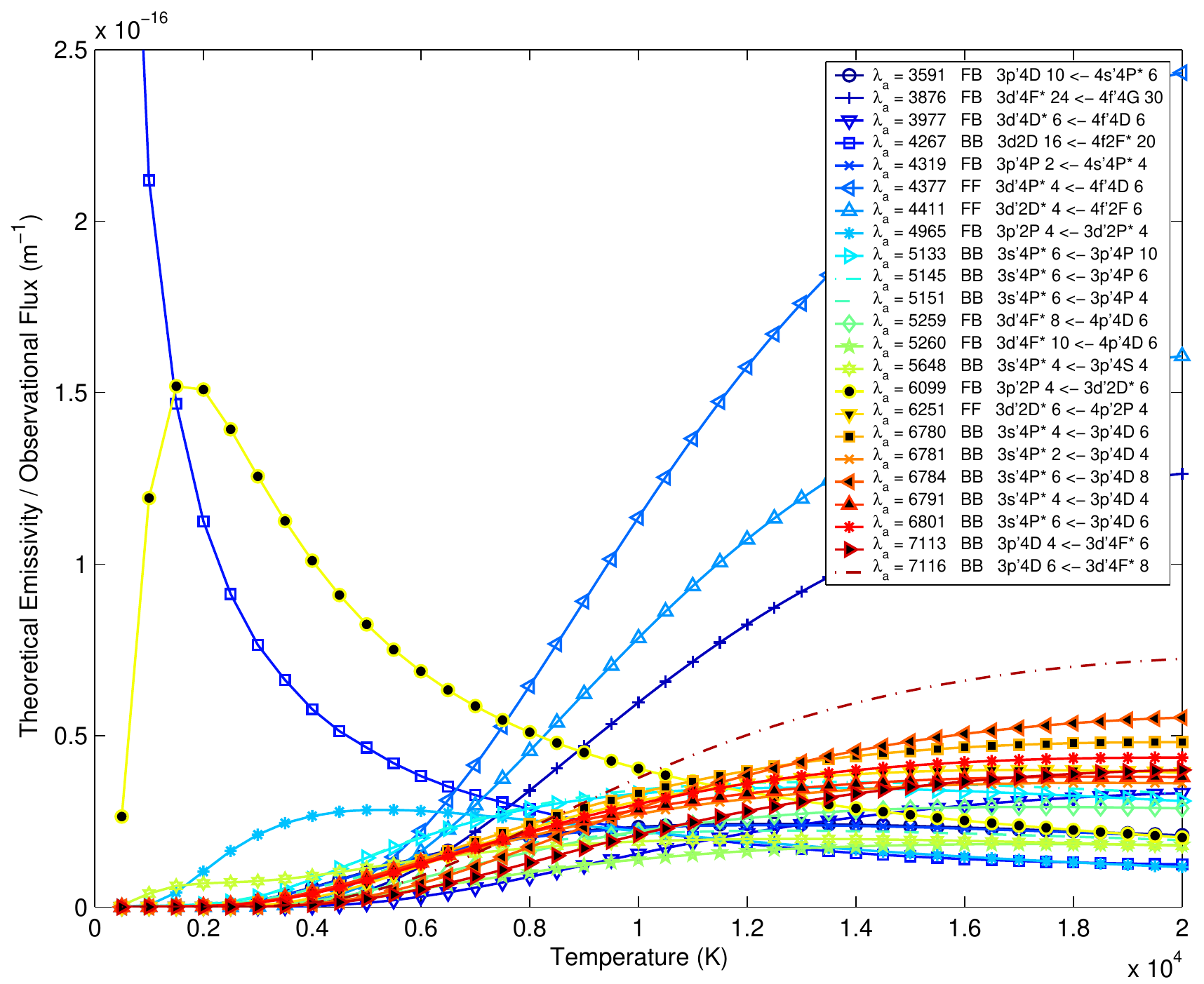}
\caption{The ratio of theoretical emissivity to observational flux as a function of temperature for
the selected C\II\ lines of IC 418 spectra of Sharpee \etal\ \cite{SharpeeWBH2003}.}
\label{EtoIC418}
\end{figure}

\begin{table} [!t]
\caption{The observational C\II\ transition lines of Sharpee \etal\ \cite{SharpeeWBH2003} that were
used in the least squares minimization procedure to find the electron temperature of the line
emitting region in the planetary nebula IC 418. The given flux is the normalized to the H$_{\beta}$ flux. \label{DataSharpeeWBH2003}}
\begin{center} 
\begin{tabular*}{\textwidth}{@{\extracolsep{\fill}}|c|c|c|c|c|c|c|}
\hline
$\WL_{lab}$ &       $I$ &      Lower &      Upper &   $\SW_l$ &   $\SW_u$ &       Type \\
\hline
  3590.757 &     0.0252 &      3p$'$ \SLP4De &     4s$'$ \SLP4Po &         10 &          6 &         FB \\
\hline
  3876.392 &     0.0069 &     3d$'$ \SLP4Fo &      4f$'$ \SLP4Ge &         24 &         30 &         FB \\
\hline
  3977.250 &     0.0026 &     3d$'$ \SLP4Do &      4f$'$ \SLP4De &          6 &          6 &         FB \\
\hline
  4267.001 &     0.5712 &       3d \SLP2De &      4f \SLP2Fo &         16 &         20 &         BB \\
\hline
  4318.606 &     0.0086 &      3p$'$ \SLP4Pe &     4s$'$ \SLP4Po &          2 &          4 &         FB \\
\hline
  4376.582 &     0.0016 &     3d$'$ \SLP4Po &      4f$'$ \SLP4De &          4 &          6 &         FF \\
\hline
  4411.152 &     0.0016 &     3d$'$ \SLP2Do &      4f$'$ \SLP2Fe &          4 &          6 &         FF \\
\hline
  4964.736 &     0.0211 &      3p$'$ \SLP2Pe &     3d$'$ \SLP2Po &          4 &          4 &         FB \\
\hline
  5132.947 &     0.0044 &     3s$'$ \SLP4Po &      3p$'$ \SLP4Pe &          6 &         10 &         BB \\
\hline
  5145.165 &     0.0040 &     3s$'$ \SLP4Po &      3p$'$ \SLP4Pe &          6 &          6 &         BB \\
\hline
  5151.085 &     0.0046 &     3s$'$ \SLP4Po &      3p$'$ \SLP4Pe &          6 &          4 &         BB \\
\hline
  5259.055 &     0.0031 &     3d$'$ \SLP4Fo &      4p$'$ \SLP4De &          8 &          6 &         FB \\
\hline
  5259.664 &     0.0032 &     3d$'$ \SLP4Fo &      4p$'$ \SLP4De &         10 &          6 &         FB \\
\hline
  5648.070 &     0.0014 &     3s$'$ \SLP4Po &      3p$'$ \SLP4Se &          4 &          4 &         BB \\
\hline
  6098.510 &     0.0011 &      3p$'$ \SLP2Pe &     3d$'$ \SLP2Do &          4 &          6 &         FB \\
\hline
  6250.760 &     0.0015 &     3d$'$ \SLP2Do &      4p$'$ \SLP2Pe &          6 &          4 &         FF \\
\hline
  6779.940 &     0.0109 &     3s$'$ \SLP4Po &      3p$'$ \SLP4De &          4 &          6 &         BB \\
\hline
  6780.600 &     0.0055 &     3s$'$ \SLP4Po &      3p$'$ \SLP4De &          2 &          4 &         BB \\
\hline
  6783.910 &     0.0022 &     3s$'$ \SLP4Po &      3p$'$ \SLP4De &          6 &          8 &         BB \\
\hline
  6791.470 &     0.0066 &     3s$'$ \SLP4Po &      3p$'$ \SLP4De &          4 &          4 &         BB \\
\hline
  6800.680 &     0.0050 &     3s$'$ \SLP4Po &      3p$'$ \SLP4De &          6 &          6 &         BB \\
\hline
  7113.040 &     0.0052 &      3p$'$ \SLP4De &     3d$'$ \SLP4Fo &          4 &          6 &         BB \\
\hline
  7115.630 &     0.0043 &      3p$'$ \SLP4De &     3d$'$ \SLP4Fo &          6 &          8 &         BB \\
\hline
\end{tabular*}
\end{center}
\end{table}

The data, with and without $\WL$4267, were normalized to the total flux and emissivity and $\chi^2$
was obtained and plotted against temperature over the range $T=500-20000$~K in steps of 100~K. As
seen in Figures~\ref{Chi2IC418-With4267} and \ref{Chi2IC418-Without4267}, the first minimizes at
$T\simeq8700$~K while the second at $T\simeq7700$~K. The electron temperature data reported in the
literature, a sample of which is presented in Table~\ref{TempTableIC418}, suffer from erratic
scattering and hence no average broad value can be safely concluded. One possibility for this
scattering, in addition to the differences arising from using different methods as well as
different atoms, ions and transitions, is the type of spectral data and the range to which they
belong as they originate from different parts in the nebula. For example, using the visible part
originating in one part of the nebula should produce a temperature different to that obtained from
radio frequency or far-infrared originating from another part. It should be remarked that Sharpee
\etal\ \cite{SharpeeWBH2003} give the signal-to-noise (S/N) ratio instead of giving observational
error; therefore the error (i.e. standard deviation) on flux was computed as the product of the
line flux times the reciprocal of S/N ratio.

\begin{figure}[!h]
\centering{}
\includegraphics[scale=0.65]{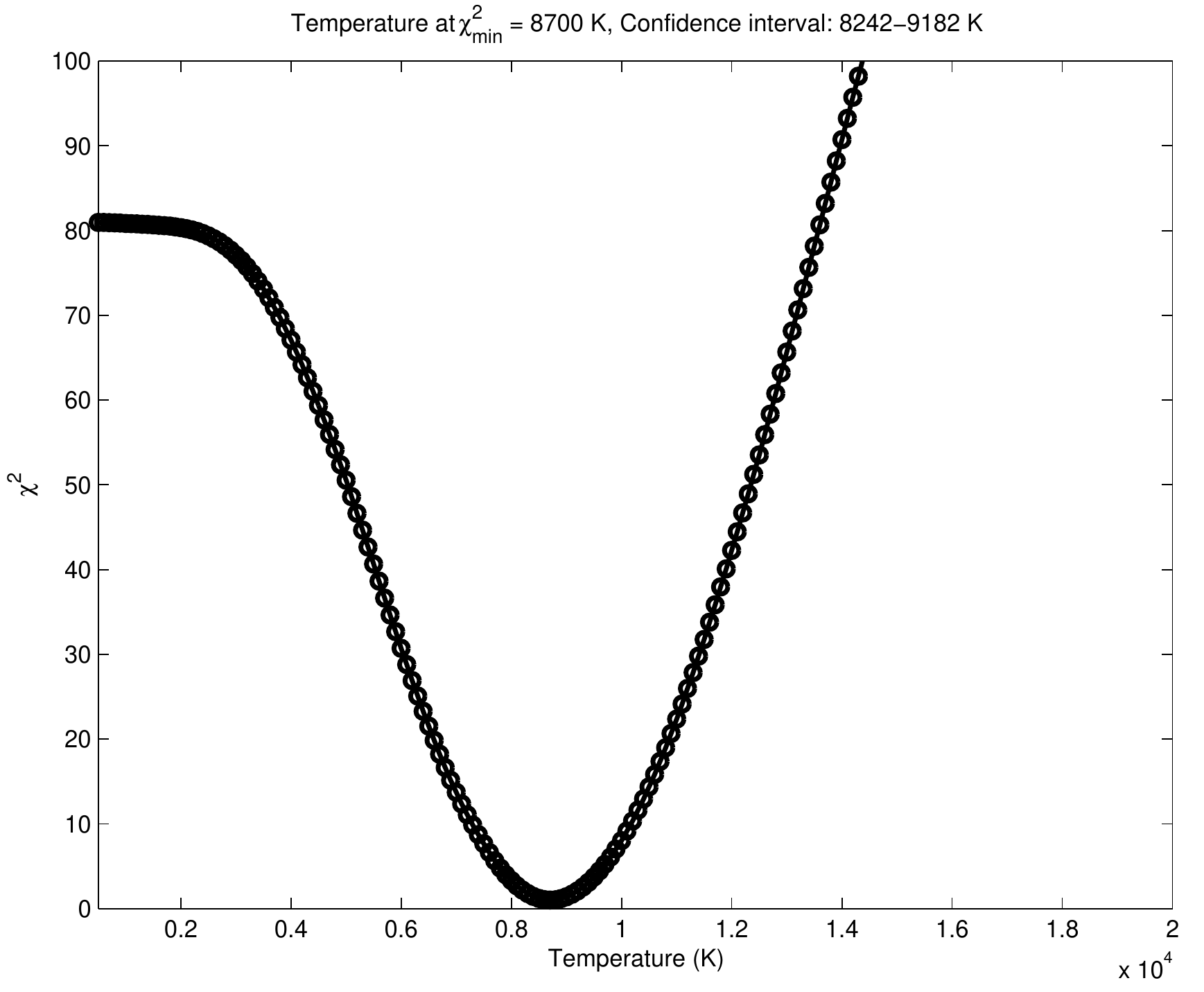}
\caption{Temperature dependence of $\chi^2$ for IC 418 with the inclusion of line $\WL$4267. The
temperature at $\chi^2_{min}$ and the confidence interval are shown.} \label{Chi2IC418-With4267}
\end{figure}

\begin{figure}[!h]
\centering{}
\includegraphics[scale=0.65]{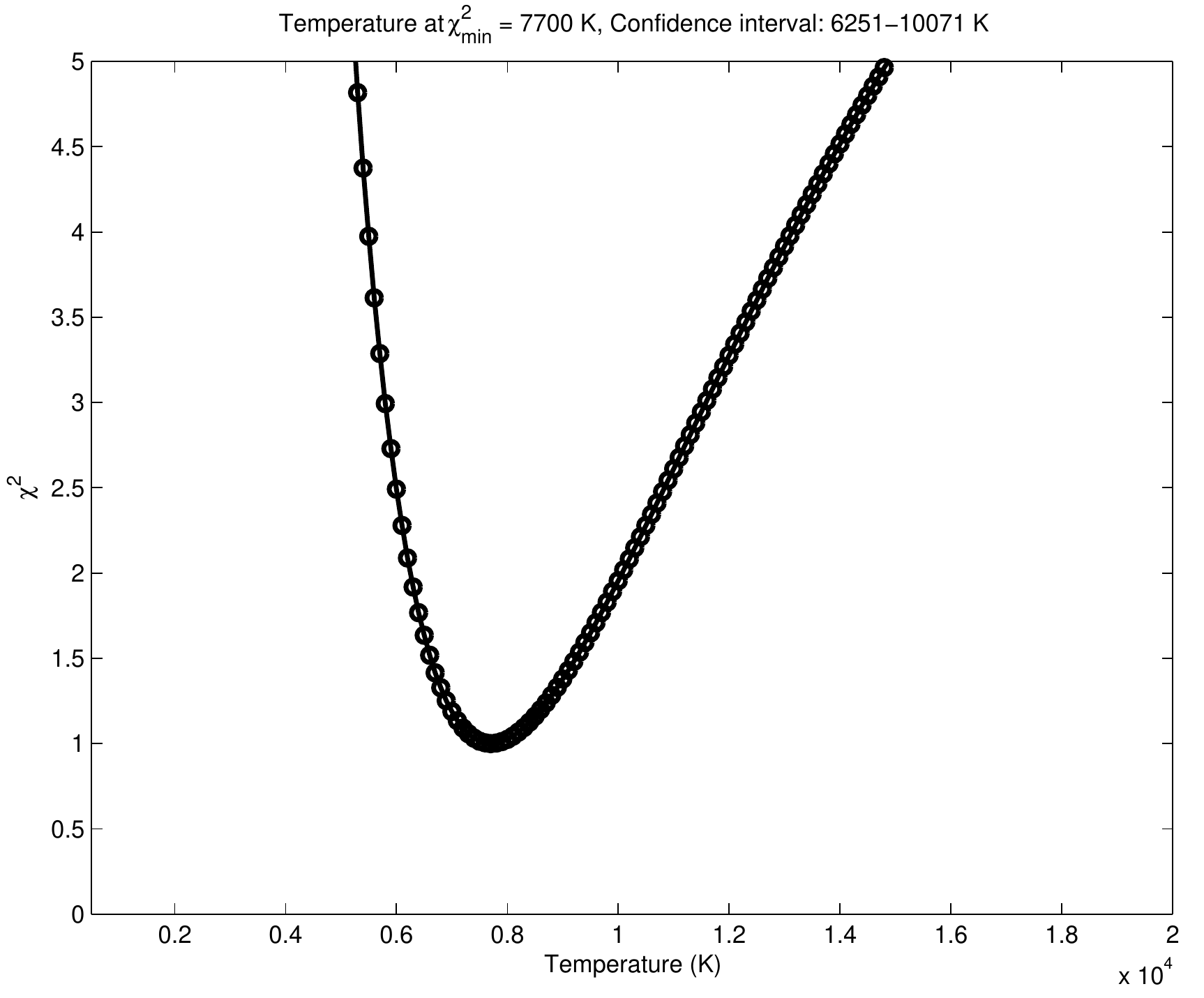}
\caption{Temperature dependence of $\chi^2$ for IC 418 with the exclusion of line $\WL$4267. The
temperature at $\chi^2_{min}$ and the confidence interval are shown.} \label{Chi2IC418-Without4267}
\end{figure}

{\renewcommand{\tabcolsep}{4pt}
\begin{table} [!h]
\caption{The electron temperature in Kelvin of IC 418 from different atoms, ions and transitions as
obtained from the cited literature, where BD stands for Balmer Discontinuity and RF for Radio
Frequency. CW refers to the current work. \label{TempTableIC418}}
\begin{center} 
{\scriptsize
\begin{tabular*}{\textwidth}{@{\extracolsep{\fill}}|l|c|c|c|c|c|c|c|c|c|c|c|c|}
\hline
    Source & \cite{Minkowski1953} & \cite{Seaton1954} & \cite{Kaler1966} & \cite{MarneS1969} & \cite{Miller1974} & \cite{HarringtonLSS1980} & \cite{AdamsS1982} & \cite{HyungAF1994} & \cite{MckennaKKWBA1996} & \cite{PottaschSBF2004} & \cite{SharpeeBW2004} &         CW \\
\hline
  H\Ii(BD) &  $>$15000 &            &            &            &            &            &            &            &            &            &            &            \\
\hline
     He\Ii &            &            &            &            &            &            &            &       9800 &            &            &            &            \\

           &            &            &            &            &            &            &            &       9800 &            &            &            &            \\
\hline
      C\II &            &            &            &            &            &            &            &       9600 &            &            &            &       8700 \\

           &            &            &            &            &            &            &            &            &            &            &            &       7700 \\
\hline
     C\III &            &            &            &            &            &            &            &       9800 &            &            &            &            \\
\hline
    C\III] &            &            &            &            &            &            &       8700 &            &            &            &            &            \\
\hline
      N\Ii &            &            &            &            &            &            &            &       8500 &            &            &            &            \\
\hline
      N\II &            &            &            &            &            &            &            &       9600 &            &            &            &            \\
\hline
    [N\II] &            &            &            &            &            &       8300 &       8200 &            &       8510 &       9400 &            &            \\

           &            &            &            &            &            &       8200 &            &            &            &            &            &            \\
\hline
   [N\III] &            &            &            &            &            &            &            &            &            &       8500 &            &            \\
\hline
      O\Ii &            &            &            &            &            &            &            &       8500 &            &            &            &            \\
\hline
      O\II &            &            &            &            &            &            &            &       9600 &            &            &            &            \\
\hline
     O\III &            &            &            &            &            &            &            &       9800 &            &            &            &            \\
\hline
    [O\II] &            &            &            &            &            &            &       8200 &            &            &            &            &            \\
\hline
   [O\III] &       7000 &            &      11200 &            &            &       9100 &       9800 &            &       9360 &       9100 &            &            \\

           &            &            &            &            &            &      10300 &       9500 &            &            &       9400 &            &            \\

           &            &            &            &            &            &            &       9600 &            &            &            &            &            \\
\hline
     Ne\II &            &            &            &            &            &            &            &       8500 &            &            &            &            \\
\hline
    Ne\III &            &            &            &            &            &            &            &       9800 &            &            &            &            \\
\hline
  [Ne\III] &            &            &            &            &            &            &            &            &            &       8000 &            &            \\
\hline
     Mg\II &            &            &            &            &            &            &            &       8500 &            &            &            &            \\
\hline
      S\II &            &            &            &            &            &            &            &       8500 &            &            &            &            \\

           &            &            &            &            &            &            &            &       8500 &            &            &            &            \\
\hline
     S\III &            &            &            &            &            &            &            &       9800 &            &            &            &            \\
\hline
   [S\III] &            &            &            &            &            &            &            &            &            &       9500 &            &            \\

           &            &            &            &            &            &            &            &            &            &       9200 &            &            \\
\hline
     Cl\II &            &            &            &            &            &            &            &       8500 &            &            &            &            \\
\hline
    Cl\III &            &            &            &            &            &            &            &       9800 &            &            &            &            \\
\hline
     Ar\II &            &            &            &            &            &            &            &       8500 &            &            &            &            \\
\hline
    Ar\III &            &            &            &            &            &            &            &       9800 &            &            &            &            \\

           &            &            &            &            &            &            &            &       9800 &            &            &            &            \\
\hline
  [Ar\III] &            &            &            &            &            &            &            &            &            &       9100 &            &            \\
\hline
        RF &            &            &            &      12500 &       6600 &            &            &            &            &            &            &            \\

           &            &            &            &      23000 &            &            &            &            &            &            &            &            \\

           &            &            &            &       7100 &            &            &            &            &            &            &            &            \\
\hline
      Mean &            &      20000 &            &            &            &            &            &            &            &            &      10000 &            \\
\hline
\end{tabular*}
}
\end{center}
\end{table}
\renewcommand{\tabcolsep}{6pt}}

\clearpage
\subsection{NGC 2867}

NGC 2867, or Caldwell 90, is a compact \planeb\ with comparatively small size and fairly strong
surface brightness located at a distance of about 0.5~kpc in the southern constellation Carina. It
has a rather warm ($\sim145000$~K) Wolf-Rayet central star of WC6 or early WO type with evidence of
being relatively rich in nitrogen. The rich spectrum of this nebula and its star has been the
subject of investigation for many studies in the last few decades with some emphasis on the optical
and ultraviolet transitions \cite{AllerKRO1981, KalerJ1989, Feibelman1998, RojasPP2009}.

The observational data of this object were obtained from Garc\'{\i}a-Rojas \etal\
\cite{RojasPP2009} where two knots have been studied: one labeled NGC 2867-1 and the other NGC
2867-2. On surveying this data source for C\II\ recombination lines we obtained about 8 lines
attributed to C\II\ transitions. Following the refinement process, outlined in \S\ \ref{TempMeth},
this list was reduced to just 2 lines. These lines are presented in Table~\ref{DataRojasPP2009}
where the first value of $I$ belongs to the first knot while the second (inside the brackets)
belongs to the second.

\begin{table} [!h]
\caption{The observational C\II\ transition lines of Garc\'{\i}a-Rojas \etal\ \cite{RojasPP2009}
that were used in the least squares minimization procedure to find the electron temperature of the
line emitting region in the planetary nebula NGC 2867. The given flux is the normalized to the H$_{\beta}$ flux. The first $I$ value belongs to knot 1 and
the second to knot 2. \label{DataRojasPP2009}}
\begin{center} 
\begin{tabular*}{\textwidth}{@{\extracolsep{\fill}}|c|c|c|c|c|c|c|}
\hline
$\WL_{lab}$ &       $I$ &      Lower &      Upper &   $\SW_l$ &   $\SW_u$ &       Type \\
\hline
   4267.15 &      0.814(1.246) &       3d \SLP2De &      4f \SLP2Fo &         16 &         20 &         BB \\
\hline
   6779.93 &      0.045(0.079) &     3s$'$ \SLP4Po &      3p$'$ \SLP4De &          4 &          6 &         BB \\
\hline
\end{tabular*}
\end{center}
\end{table}

These data sets were normalized to the total flux and emissivity and $\chi^2$ was obtained and
plotted against temperature in the range $T=500-30000$~K with a step size of 100~K. These plots are
presented in Figures~\ref{Chi2NGC2867-1} and \ref{Chi2NGC2867-2}. As seen, NGC 2867-1 graph
indicates a temperature of about 14300~K while NGC 2867-2 graph a temperature of about 16000~K. The
difference in temperature value may be caused by the difference in the physical conditions of the
two knots. Table~\ref{TempTableNGC2867} presents electron temperatures derived in previous works
from transitions of different species. As seen, our values are significantly higher than most of
the values reported in the literature. However, this may be explained by the complex structure of
this nebula and the possibility of different lines being originating from different regions with
very different physical conditions.

\begin{figure}[!h]
\centering{}
\includegraphics[scale=0.65]{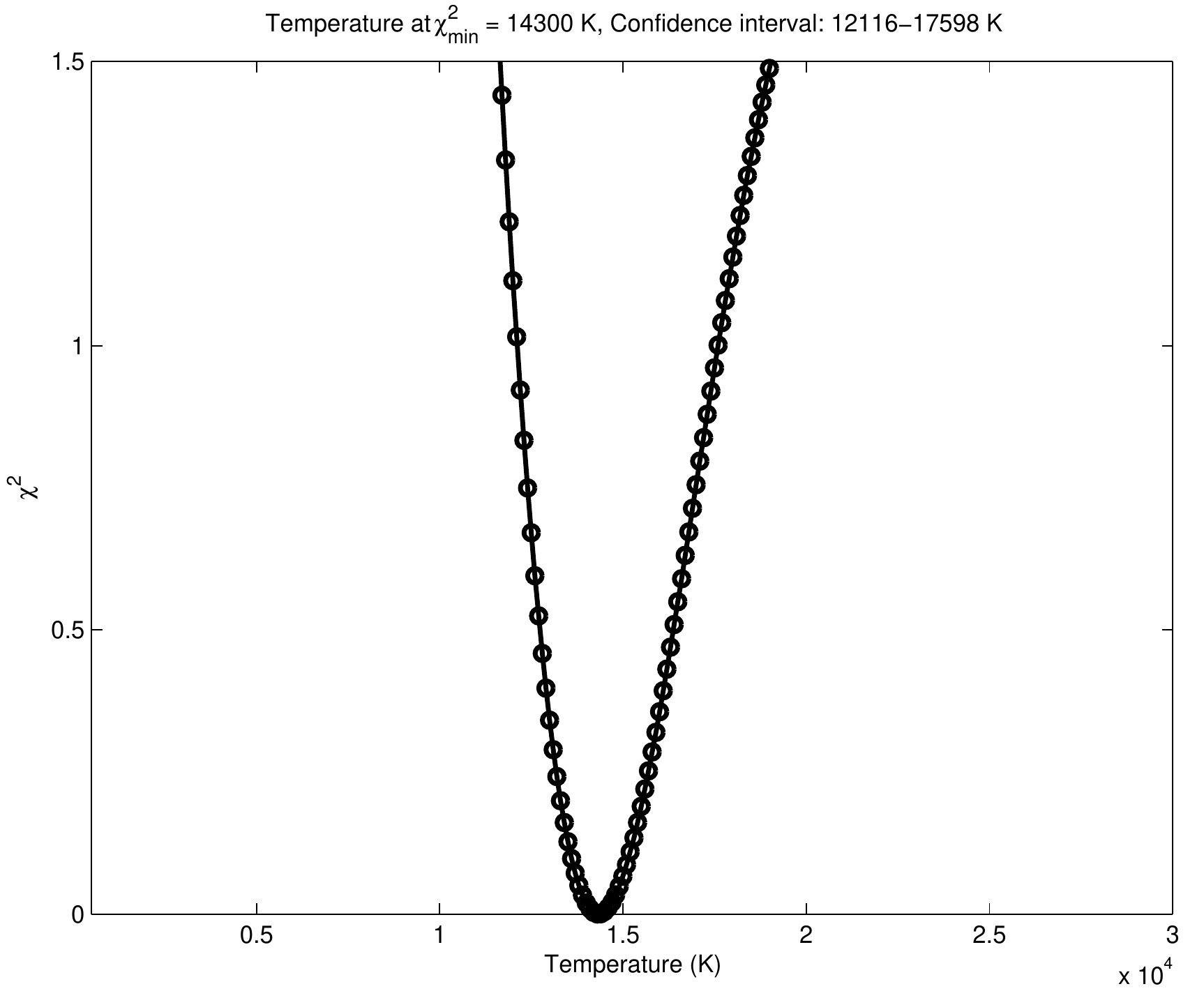}
\caption{Temperature dependence of $\chi^2$ for NGC 2867-1. The temperature at $\chi^2_{min}$ and
the confidence interval are shown.} \label{Chi2NGC2867-1}
\end{figure}

\begin{figure}[!h]
\centering{}
\includegraphics[scale=0.65]{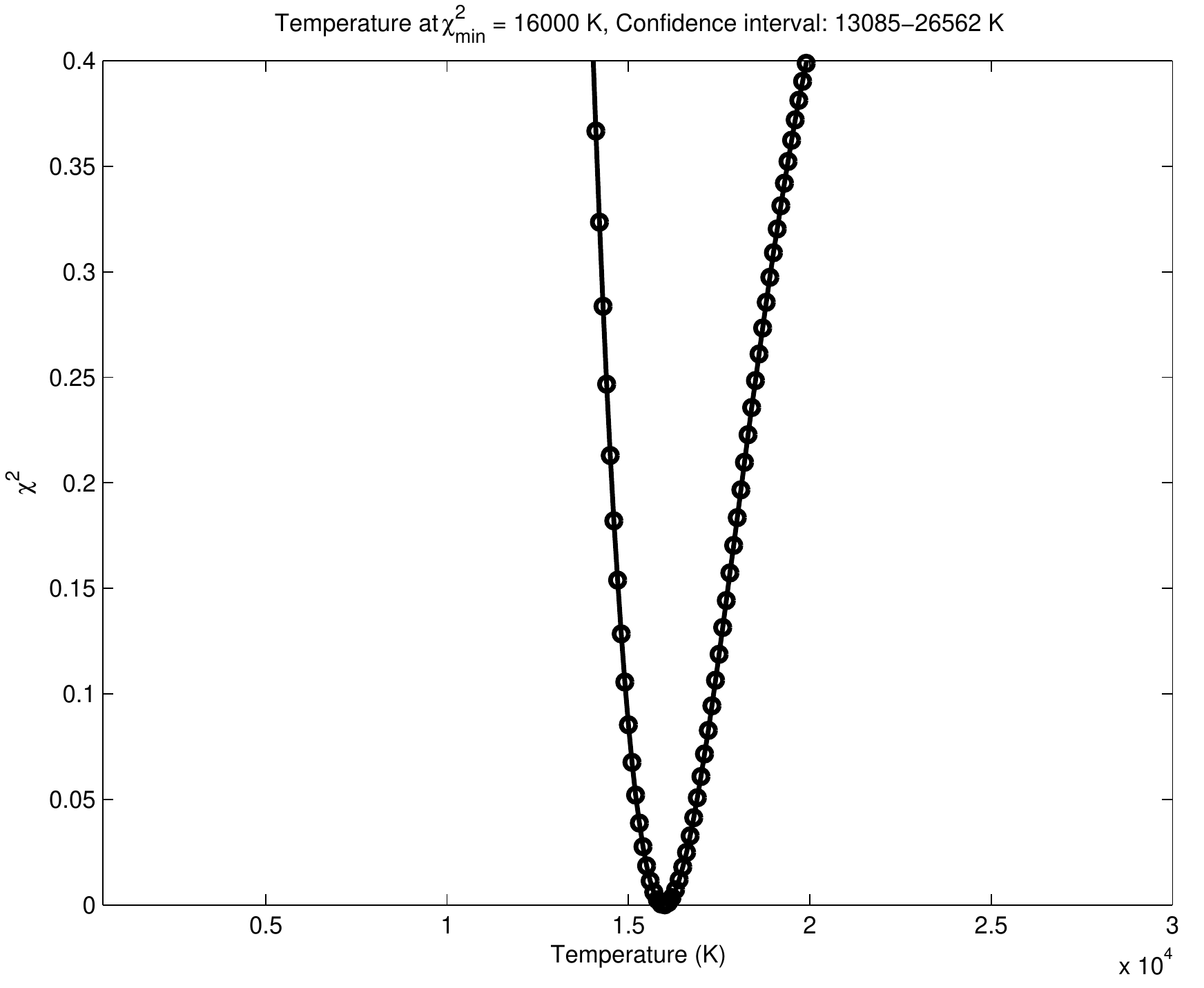}
\caption{Temperature dependence of $\chi^2$ for NGC 2867-2. The temperature at $\chi^2_{min}$ and
the confidence interval are shown.} \label{Chi2NGC2867-2}
\end{figure}

\begin{table} [!h]
\caption{The electron temperature in Kelvin of NGC 2867 from different species and transitions as
obtained from the cited literature, where BD stands for Balmer Decrement and K1 and K2 for knot 1
and knot 2. CW refers to the current work where the first value is related to knot 1 and the second
to knot 2. \label{TempTableNGC2867}}
\begin{center} 
\begin{tabular*}{\textwidth}{@{\extracolsep{\fill}}|l|c|c|c|c|c|c|c|}
\hline
    Source & \cite{AllerKRO1981} & \cite{MckennaKKWBA1996} & \cite{Kholtygin1998} & \cite{MathisPP1998} & \cite{RojasPP2009} K1 & \cite{RojasPP2009} K2 &         CW \\
\hline
  H\Ii(BD) &            &            &            &            &       8950 &       8950 &            \\
\hline
     He\Ii &            &            &            &            &      10900 &      10250 &            \\
\hline
      C\II &            &            &            &            &            &            &      14300 \\

           &            &            &            &            &            &            &      16000 \\
\hline
    C\III] &            &            &            &      10750 &            &            &            \\
\hline
    [N\II] &       8800 &       9000 &            &            &      11750 &      11750 &            \\
\hline
    O\III] &            &            &            &      10580 &            &            &            \\
\hline
   [O\III] &      11100 &      10520 &            &      11340 &      11850 &      11600 &            \\
\hline
    [S\II] &            &            &            &            &       8450 &       8250 &            \\
\hline
  [Ar\III] &            &            &            &            &      10800 &      11350 &            \\
\hline
   [Ar\IV] &            &            &            &            &      15400 &  $>$19870 &            \\
\hline
[N\II]+[O\III] &            &            &            &            &      11850 &      11600 &            \\
\hline
      Mean &            &            &      10592 &            &            &            &            \\
\hline
\end{tabular*}
\end{center}
\end{table}

\clearpage
\subsection{DQ Herculis 1934}

This is a peculiar old classical galactic nova originating from an accreting cataclysmic variable
binary system which apparently consists of a white and a red dwarfs. It is located in the Hercules
constellation at an estimated distance of about 0.2-0.4~kpc. DQ Her is characterized by its 71~s
periodically pulsating spectrum which occurs mainly in the visible, ultraviolet and X-ray regions.
The strange features of this nova include a very low electron temperature and an enhanced CNO
abundance. The former seems to correlate to the latter since low electron temperatures usually
characterize metal rich nebulae. Thanks to its reasonable brightness, pulsation, and peculiar
properties, it is one of the most observed and best studied novae \cite{BathEP1974, Nelson1976,
GallagherB1980, Ferland1980, Itoh1981, FerlandWLSSe1984, King1985, CanalleO1990, PetitjeanBP1990}.

The observational data of DQ Her which we used in the current study were obtained from Ferland
\etal\ \cite{FerlandWLSSe1984}. On surveying the data source for C\II\ recombination lines we
obtained 2 lines attributed to C\II\ transitions. These lines are presented in
Table~\ref{DataFerlandWLSSe1984}. The observational and theoretical data were then normalized to
the total flux and emissivity respectively and $\chi^2$ was obtained and plotted against
temperature in the range $T=500-5000$~K with a step size of 100~K.
The $\chi^2$ plot is shown in Figure~\ref{Chi2DQHer-2}. As seen in this figure the deduced electron
temperature is about 1600~K.

It should be remarked that the use of $\WL$1335~\AA\ line is a second exception (the first is
$\WL$4267~\AA) to our rule of using the BB transitions only if the upper state has a doubly-excited
core. The justification of this exception is that the upper state of the $\WL$1335 transition is
\nlo1s22s2p$^2$ which is connected to the C$^{2+}$ \nlo1s22s$^2$ continuum and \nlo1s22s$^2$$nl$
Rydberg states by two-electron radiative processes which are usually very weak. Another remark is
that since no observational error estimate was given for these lines, an error estimate based on
Poisson statistical distribution was used in the least squares procedure, as given by
Equation~\ref{EAPoisEq} in Appendix~\ref{AppErrAna}.

\begin{table} [!t]
\caption{The observational C\II\ transition lines of Ferland \etal\ \cite{FerlandWLSSe1984} that
were used in the least squares minimization procedure to find the electron temperature of the line
emitting region in DQ Herculis 1934. The given flux is the normalized to the H$_{\beta}$ flux. \label{DataFerlandWLSSe1984}}
\begin{center} 
\begin{tabular*}{\textwidth}{@{\extracolsep{\fill}}|c|c|c|c|c|c|c|}
\hline
$\WL_{lab}$ &       $I$ &      Lower &      Upper &   $\SW_l$ &   $\SW_u$ &       Type \\
\hline
      1335 &        270 &     2p \SLP2Po &   2s2p$^2$ \SLP2De &          10 &         14 &         BB \\
\hline
      4267 &         29 &       3d \SLP2De &      4f \SLP2Fo &         16 &         20 &         BB \\
\hline
\end{tabular*}
\end{center}
\end{table}

\begin{figure}[!h]
\centering{}
\includegraphics[scale=0.65]{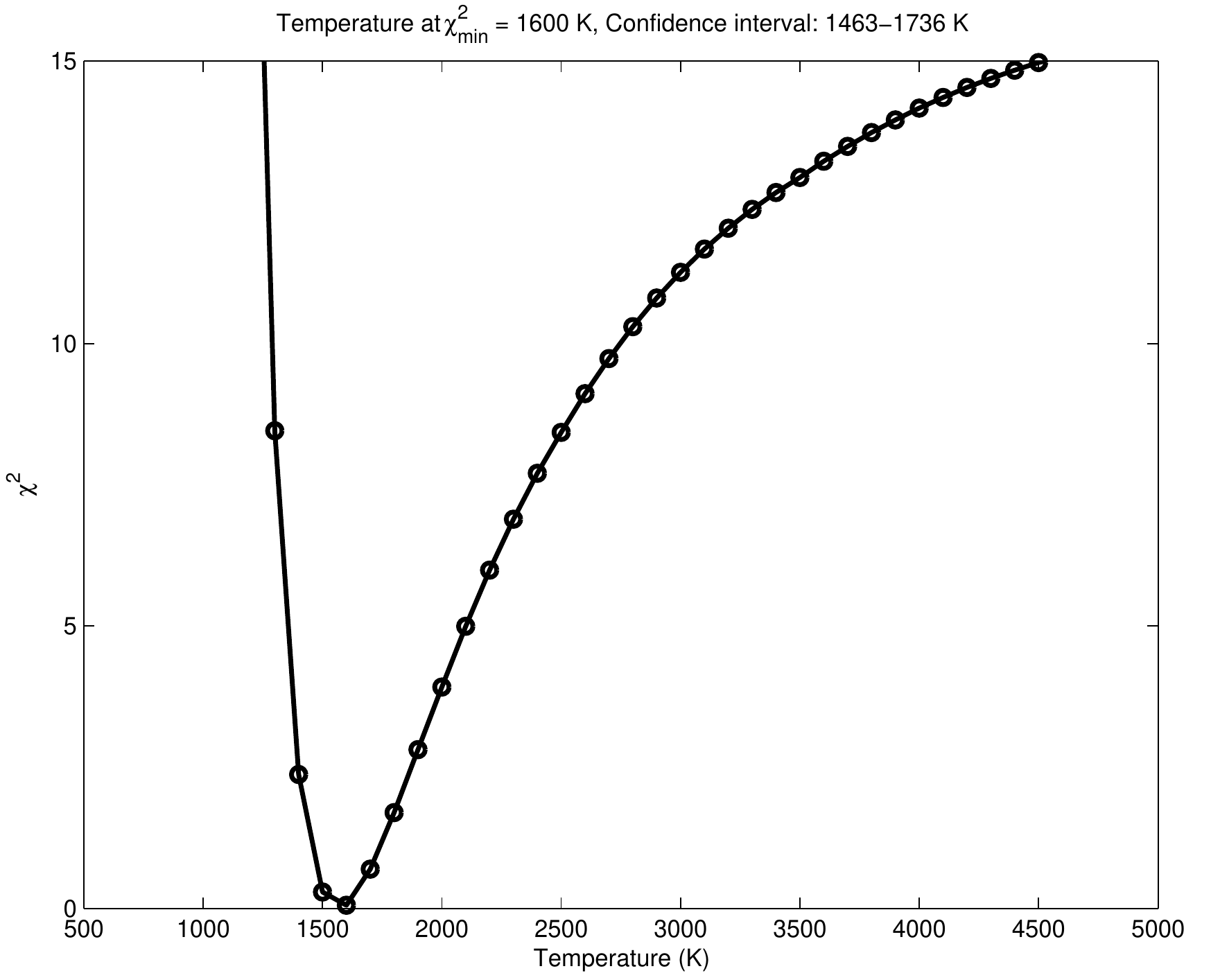}
\caption{Temperature dependence of $\chi^2$ for DQ Herculis 1934. The temperature at $\chi^2_{min}$
and the confidence interval are shown.} \label{Chi2DQHer-2}
\end{figure}

Table~\ref{TempTableDQHer} presents some of the electron temperature values for DQ Her collected
from the literature. Our value of 1600~K reasonably agrees with the values of Smits
\cite{Smits1991} and Davey \cite{ThesisDavey1995} which are also derived from C\II\ recombination
lines but with different method using the flux ratio of $\WL$1335 to $\WL$4267. Itoh
\cite{Itoh1981} seems to conclude different temperature values derived from various species and may
originate from different parts of DQ Her; some of these values are lower than 500~K while some are
much higher and may even reach 20000~K or higher although some of these derived values are based on
tentative models. The higher temperatures may also belong to the hot inner disk region, rather than
the cool outer nebula shell, where much higher temperatures have been derived for the disk
\cite{Kraft1959, Starrfield1970, SaitoBHM2010}.

\begin{table} [!h]
\caption{The electron temperature in Kelvin of DQ Herculis 1934 from different species and
transitions as obtained from the cited literature, where BC stands for Balmer Continuum. CW refers
to the current work. \label{TempTableDQHer}}
\begin{center} 
\begin{tabular*}{\textwidth}{@{\extracolsep{\fill}}|l|c|c|c|c|c|c|c|}
\hline
    Source & \cite{WilliamsWHMK1978} & \cite{GallagherS1978} & \cite{FerlandWLSSe1984} & \cite{PetitjeanBP1990} & \cite{Smits1991} & \cite{ThesisDavey1995} &         CW \\
\hline
  H\Ii(BC) &        500 &       1000 &            &        450 &            &            &            \\
\hline
      C\II &            &            &        700 &            &       1300 &       1450 &       1600 \\
\hline
    [N\II] &            &            &            &       2500 &            &            &            \\

           &            &            &            &       2400 &            &            &            \\
\hline
      Mean &            &            &        500 &            &            &            &            \\
\hline
\end{tabular*}
\end{center}
\end{table}

\clearpage
\subsection{\CPD}

\CPD\ is a cool late-type Wolf-Rayet star that is usually classified as WC10 or WC11. The star,
which is located at about 1.3-1.5~kpc, is surrounded by a young planetary nebula with complex
visible structure. The low-excitation spectrum of \CPD, which has been described as unique, shows
bright C\II\ and C\III\ emission lines indicating the star is carbon-rich. Observational studies
suggest that the star is encircled by a disk or torus of dust \cite{WebsterG1974, Thackeray1977,
HouziauxH1982, PollaccoKMWR1992, CrowtherDBS1996, CohenBSLCe1999, CohenBSLCe1999-2, CohenBLJ2002,
DemarcoJBCBH2004, Chesneau2006, ChesneauCDWLe2006}.

Here, we try to infer the electron temperature of the stellar nebular wind surrounding \CPD. The
observational data of this object were obtained from De Marco \etal\ \cite{DemarcoSB1997}. On
surveying this data source we obtained about 16 C\II\ lines. Following the refinement process,
outlined in \S\ \ref{TempMeth}, this list was reduced to 13 lines. These lines are presented in
Table~\ref{DataDemarcoSB19971}.

\begin{table} [!t]
\caption{The observational C\II\ transition lines of De Marco \etal\ \cite{DemarcoSB1997} that were
used in the least squares minimization procedure to find the electron temperature of the line
emitting region in the planetary nebula \CPD. The given flux is the absolute value in units of erg.s$^{-1}$.cm$^{-2}$. \label{DataDemarcoSB19971}}
\begin{center} 
\begin{tabular*}{\textwidth}{@{\extracolsep{\fill}}|c|c|c|c|c|c|c|}
\hline
$\WL_{lab}$ &       $I$ &      Lower &      Upper &   $\SW_l$ &   $\SW_u$ &       Type \\
\hline
   4618.73 &  4.194E-12 &     3d$'$ \SLP2Fo &      4f$'$ \SLP2Ge &          6 &          8 &         FF \\
\hline
   4619.23 &  6.167E-12 &     3d$'$ \SLP2Fo &      4f$'$ \SLP2Ge &          8 &         10 &         FF \\
\hline
   4627.63 &  1.850E-13 &     3d$'$ \SLP2Fo &      4f$'$ \SLP2Ge &          8 &          8 &         FF \\
\hline
   4953.85 &  1.437E-12 &      3p$'$ \SLP2Pe &     3d$'$ \SLP2Po &          2 &          2 &         FB \\
\hline
   4958.67 &  7.183E-13 &      3p$'$ \SLP2Pe &     3d$'$ \SLP2Po &          4 &          2 &         FB \\
\hline
   4959.92 &  7.183E-13 &      3p$'$ \SLP2Pe &     3d$'$ \SLP2Po &          2 &          4 &         FB \\
\hline
   4964.73 &  3.592E-12 &      3p$'$ \SLP2Pe &     3d$'$ \SLP2Po &          4 &          4 &         FB \\
\hline
   5107.97 &  5.906E-13 &     3d$'$ \SLP2Po &      4f$'$ \SLP2De &          4 &          4 &         FF \\
\hline
   5113.69 &  3.384E-12 &     3d$'$ \SLP2Po &      4f$'$ \SLP2De &          4 &          6 &         FF \\
\hline
   5114.26 &  2.888E-12 &     3d$'$ \SLP2Po &      4f$'$ \SLP2De &          2 &          4 &         FF \\
\hline
   8793.80 &  1.926E-12 &      3p$'$ \SLP2De &     3d$'$ \SLP2Fo &          6 &          8 &         FB \\
\hline
   8799.90 &  1.347E-12 &      3p$'$ \SLP2De &     3d$'$ \SLP2Fo &          4 &          6 &         FB \\
\hline
   8826.98 &  9.426E-14 &      3p$'$ \SLP2De &     3d$'$ \SLP2Fo &          6 &          6 &         FB \\
\hline
\end{tabular*}
\end{center}
\end{table}

The data were normalized to the total flux and emissivity and $\chi^2$ was computed and plotted
against temperature over the range $T=500-30000$~K in steps of 100~K. As seen in
Figure~\ref{Chi2CPD}, the plot indicates a temperature of about 17300~K. This agrees, within the
reported error bars, with the temperature of De Marco \etal\ \cite{DemarcoBS1996, DemarcoSB1996}
who deduced a value of 18500$\pm$1500~K for this object using a similar least squares approach.
Table~\ref{TempTableCPD} presents electron temperatures derived previously in the cited literature.
As seen there is a large scattering in the reported values. The main reason for this may be the
large temperature variations across the line emitting regions, where the high temperature lines
originate from the inner ring next to the central star while the low temperature lines come from
the relatively cold region at the perimeter of the nebula. Differences in the derivation methods
and the type of transitions used in these derivations (e.g. CEL versus ORL), as well as
observational and theoretical errors, should also contribute to the scattering. However, our value
agrees reasonably with some of these reported values.

\begin{figure}[!t]
\centering{}
\includegraphics[scale=0.65]{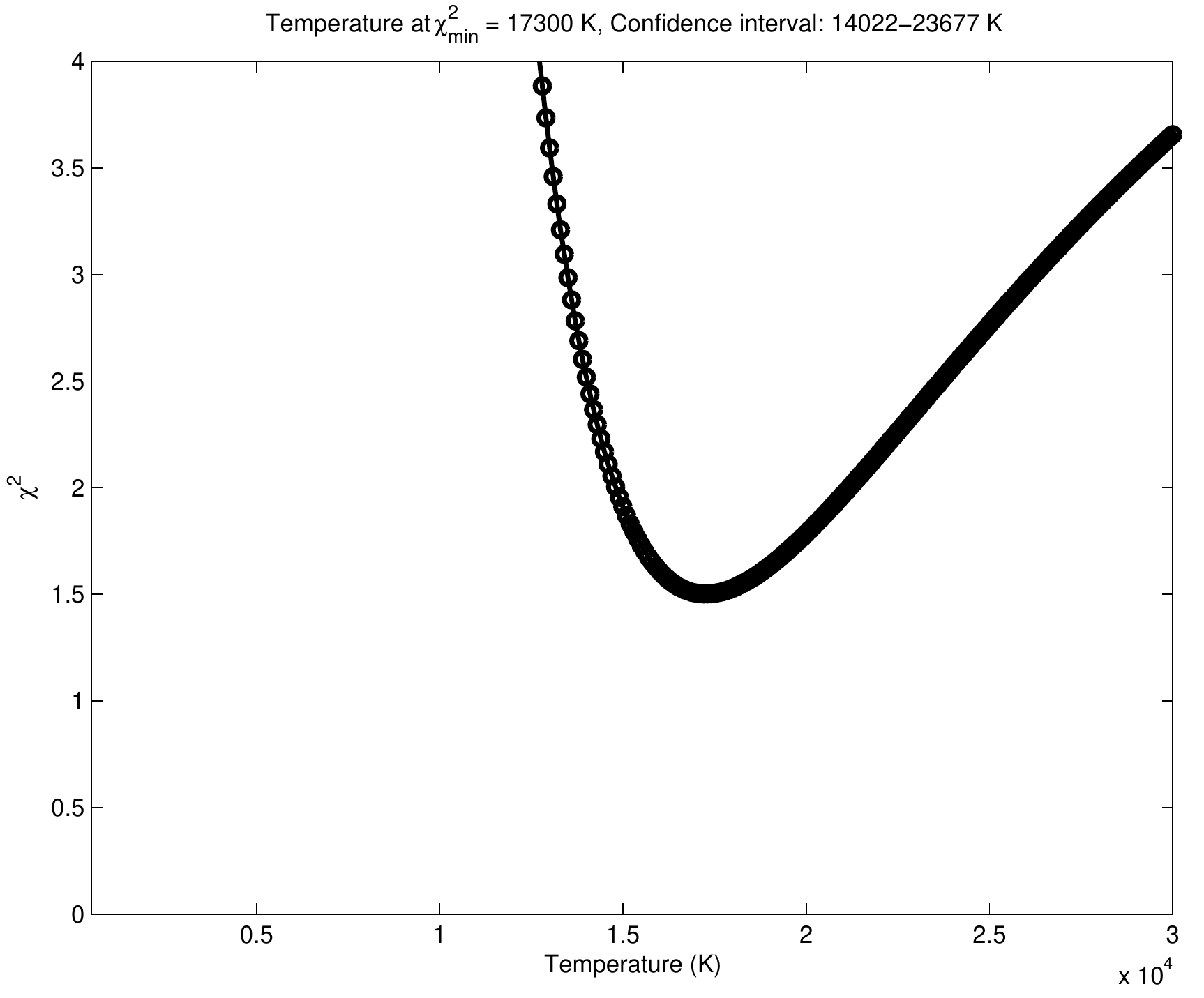}
\caption{Temperature dependence of $\chi^2$ for \CPD. The temperature at $\chi^2_{min}$ and the
confidence interval are shown.} \label{Chi2CPD}
\end{figure}

\begin{table} [!h]
\caption{The electron temperature in Kelvin of the stellar wind of the \CPD\ Wolf-Rayet star from
different species and transitions as obtained from the cited literature. CW refers to the current
work. Also Davey \cite{ThesisDavey1995} seems to support the value of 12800~K derived by Barlow and
Storey \cite{BarlowS1993}. \label{TempTableCPD}}
\begin{center} 
{\scriptsize
\begin{tabular*}{\textwidth}{@{\extracolsep{\fill}}|l|c|c|c|c|c|c|c|c|c|c|}
\hline
    Source & \cite{HouziauxH1982} & \cite{Rao1987} & \cite{RaoGN1990} & \cite{BarlowS1993} & \cite{CrowtherHS1995} & \cite{CrowtherDBS1996} & \cite{DemarcoBS1996, DemarcoSB1996} & \cite{DemarcoSB1997} & \cite{DemarcoB2001} &         CW \\
\hline
     He\Ii &            &            &            &            &      20100 &            &            &            &            &            \\
\hline
     He\II &            &            &            &            &      20800 &            &            &            &            &            \\
\hline
      C\II &            &            &            &      12800 &            &      20000 &      18500 &      21300 &      21300 &      17300 \\

           &            &            &            &            &            &            &            &      18900 &            &            \\

           &            &            &            &            &            &            &            &      21000 &            &            \\

           &            &            &            &            &            &            &            &      21700 &            &            \\
\hline
    [N\II] &            &      11000 &            &            &            &            &            &            &            &            \\
\hline
   General &      10000 &            &       7000 &            &            &            &            &            &            &            \\
\hline
\end{tabular*}
}
\end{center}
\end{table}

\clearpage
\subsection{\Het}

\Het\ is a late-type WC10 Wolf-Rayet star surrounded by a planetary nebula with an apparent ring
structure. There are many physical similarities between \CPD\ and \Het\ such as age, flux and
distance. These similarities are reflected in the strong resemblance of their observed spectra and
hence they are normally investigated jointly \cite{Cohen1975, DemarcoBS1996, DemarcoSB1997,
SahaiNW2000, Soker2002, Chesneau2006, ChesneauCDWLe2006, LagadecCMDPe2006}.

Here, we investigate the electron temperature of the stellar wind surrounding \Het\ using
observational data from De Marco \etal\ \cite{DemarcoSB1997}. On surveying this data source we
obtained about 16 C\II\ recombination lines. Following the refinement process, outlined in \S\
\ref{TempMeth}, this list was reduced to 13 lines. These lines are presented in
Table~\ref{DataDemarcoSB19972}.

\begin{table} [!h]
\caption{The observational C\II\ transition lines of De Marco \etal\ \cite{DemarcoSB1997} that were
used in the least squares minimization procedure to find the electron temperature of the line
emitting region in the planetary nebula \Het. The given flux is the absolute value in units of erg.s$^{-1}$.cm$^{-2}$. \label{DataDemarcoSB19972}}
\begin{center} 
\begin{tabular*}{\textwidth}{@{\extracolsep{\fill}}|c|c|c|c|c|c|c|}
\hline
$\WL_{lab}$ &       $I$ &      Lower &      Upper &   $\SW_l$ &   $\SW_u$ &       Type \\
\hline
   4618.73 &  2.271E-12 &     3d$'$ \SLP2Fo &      4f$'$ \SLP2Ge &          6 &          8 &         FF \\
\hline
   4619.23 &  3.339E-12 &     3d$'$ \SLP2Fo &      4f$'$ \SLP2Ge &          8 &         10 &         FF \\
\hline
   4627.63 &  1.002E-13 &     3d$'$ \SLP2Fo &      4f$'$ \SLP2Ge &          8 &          8 &         FF \\
\hline
   4953.85 &  7.141E-13 &      3p$'$ \SLP2Pe &     3d$'$ \SLP2Po &          2 &          2 &         FB \\
\hline
   4958.67 &  3.571E-13 &      3p$'$ \SLP2Pe &     3d$'$ \SLP2Po &          4 &          2 &         FB \\
\hline
   4959.92 &  3.571E-13 &      3p$'$ \SLP2Pe &     3d$'$ \SLP2Po &          2 &          4 &         FB \\
\hline
   4964.73 &  1.785E-12 &      3p$'$ \SLP2Pe &     3d$'$ \SLP2Po &          4 &          4 &         FB \\
\hline
   5107.97 &  1.973E-13 &     3d$'$ \SLP2Po &      4f$'$ \SLP2De &          4 &          4 &         FF \\
\hline
   5113.69 &  1.130E-12 &     3d$'$ \SLP2Po &      4f$'$ \SLP2De &          4 &          6 &         FF \\
\hline
   5114.26 &  9.646E-13 &     3d$'$ \SLP2Po &      4f$'$ \SLP2De &          2 &          4 &         FF \\
\hline
   8793.80 &  9.342E-13 &      3p$'$ \SLP2De &     3d$'$ \SLP2Fo &          6 &          8 &         FB \\
\hline
   8799.90 &  6.533E-13 &      3p$'$ \SLP2De &     3d$'$ \SLP2Fo &          4 &          6 &         FB \\
\hline
   8826.98 &  4.573E-14 &      3p$'$ \SLP2De &     3d$'$ \SLP2Fo &          6 &          6 &         FB \\
\hline
\end{tabular*}
\end{center}
\end{table}

The data were normalized to the total flux and emissivity, as outlined previously, and $\chi^2$ was
computed and plotted against temperature over the range $T=500-30000$~K in steps of 100~K. As seen
in Figure~\ref{Chi2He2}, the plot indicates a temperature of about 16200~K which agrees very well
with some previously-deduced values notably those of De Marco \etal\ \cite{DemarcoSB1997,
DemarcoB2001}. Table~\ref{TempTableHet} presents electron temperatures derived previously in the
cited literature. As seen there is some scattering in these values. Similar arguments to those
presented in the \CPD\ section may be given to explain this scattering in the derived electron
temperatures.

\begin{figure}[!t]
\centering{}
\includegraphics[scale=0.65]{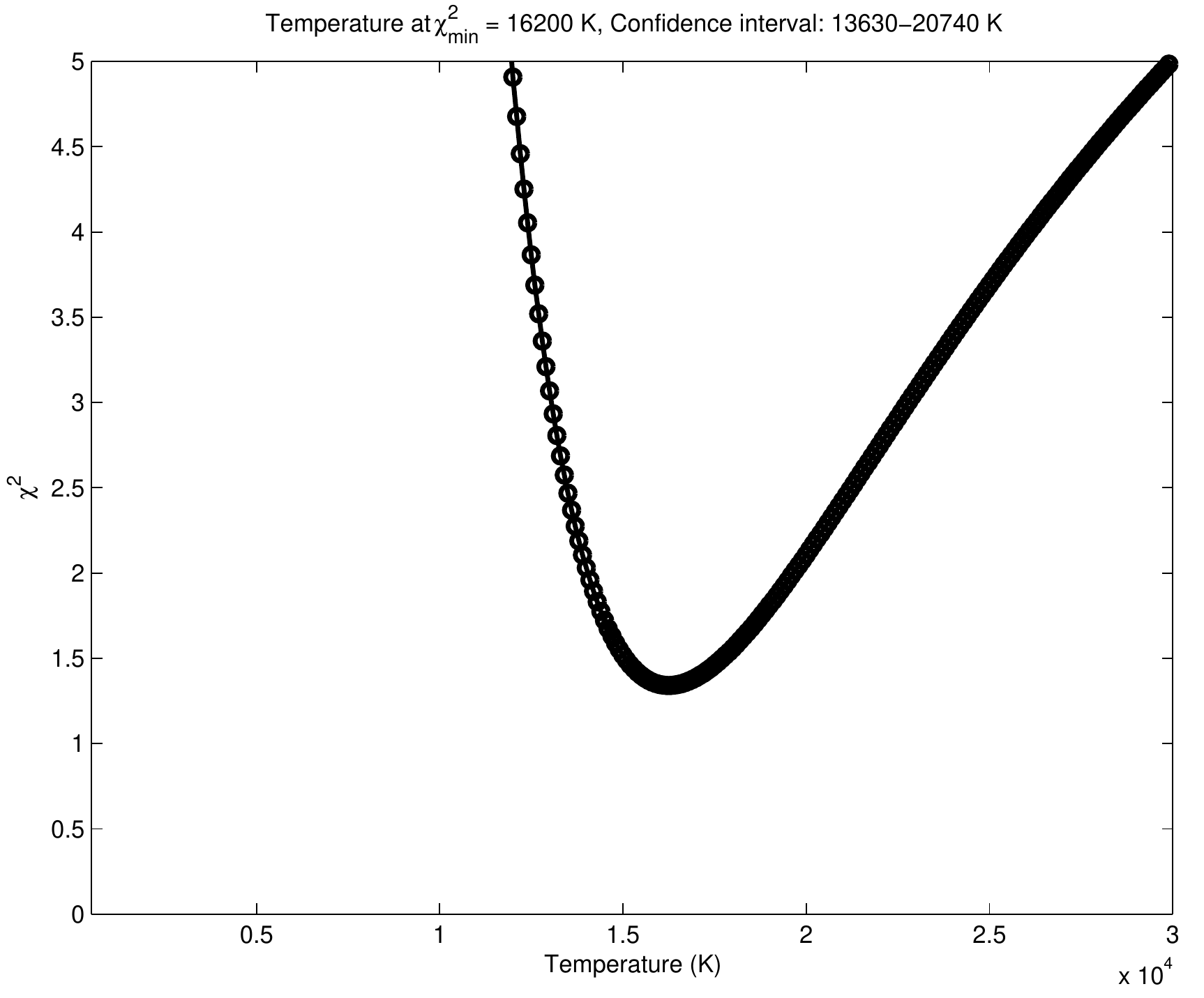}
\caption{Temperature dependence of $\chi^2$ for \Het. The temperature at $\chi^2_{min}$ and the
confidence interval are shown.} \label{Chi2He2}
\end{figure}

\begin{table} [!h]
\caption{The electron temperature in Kelvin of the stellar wind of the \Het\ Wolf-Rayet star as
obtained from the cited literature. CW refers to the current work. \label{TempTableHet}}
\begin{center} 
\begin{tabular*}{\textwidth}{@{\extracolsep{\fill}}|l|c|c|c|c|c|}
\hline
    Source & \cite{Rao1987} & \cite{DemarcoBS1996, DemarcoSB1996} & \cite{DemarcoSB1997} & \cite{DemarcoB2001} &         CW \\
\hline
      C\II &            &      13600 &      16400 &      16400 &      16200 \\

           &            &            &      17000 &            &            \\
\hline
   General &       8800 &            &            &            &            \\
\hline
\end{tabular*}
\end{center}
\end{table}

\clearpage
\section{Electron Distribution}\label{ElecDist}

As discussed in \s\ \ref{RFLines}, it has been suggested \cite{NichollsDS2012} that the discrepancy
between the results of ORLs and those of CELs is based on the assumption of a Maxwell-Boltzmann
(MB) for the electron distribution in the nebulae. Therefore by assuming a different type of
distribution, e.g. $\kappa$-distribution, the ORLs and CELs should yield very similar results for
the abundance and electron temperature. One way for testing this proposal is to obtain the electron
distribution from the transition lines obtained from the spectra of nebulae and compare to the MB
and other distributions. Since we have sufficient theoretical data for the C\II\ dielectronic
recombination with some observational data, we decided to put this proposal to test.
To obtain the electron distribution from the observed spectra, the observed flux in conjunction
with theoretical data such as recombination coefficient, are used as outlined in the following for
the cases in which the transition originates from an autoionizing upper state.

The dielectronic recombination coefficient, $\alpha(\WL)$, is related to the fraction of electrons
per unit energy, $f$, in the energy range $\epsilon$ to $\epsilon+\mathrm{d}\epsilon$ by the
following relation
\begin{equation}
 \alpha(\WL)=2\pi a_{0}^{3}\frac{\omega_{r}}{\omega_{+}}R\left(\frac{R}{\epsilon}\right)^{1/2}\Gamma^{r}bf\label{ElDiEq1}
\end{equation}
where $a_{0}$ is the Bohr radius, $\omega_{r}$ is the statistical weight of resonance, $\omega_{+}$
is the statistical weight of the recombined ion, $R$ is the Rydberg constant, $\Gamma^{r}$ is the
radiative probability of the transition, and $b$ is the departure coefficient of the autoionizing
state. Now MB distribution is given by
\begin{equation}
f_{MB}=\frac{2}{\left(kT\right)^{3/2}}\sqrt{\frac{\epsilon}{\pi}}e^{-\frac{\epsilon}{kT}}\label{ElDiEq2}
\end{equation}
where $k$ is the Boltzmann constant, and $T$ is the electron temperature. Because there is a
proportionality between the de-reddened flux, $I$, and the dielectronic recombination coefficient
multiplied by the photon energy, $\frac{hc}{\WL}$, we can write

\begin{equation}
\alpha=CI\WL\label{ElDiEq3}
\end{equation}
where $C$ is a proportionality factor. On substituting $\alpha$ from Equation \ref{ElDiEq3} into
Equation \ref{ElDiEq1} and solving for $f$ we obtain

\begin{equation}
f=DI\frac{\omega_{+}}{\omega_{r}}\left(\frac{\epsilon}{R}\right)^{1/2}\frac{\WL}{\Gamma^{r}b}\label{ElDiEq4}
\end{equation}
where $D$ is another proportionality factor. Full derivation of the electron sampling formulation
can be found in \cite{StoreyS2012}.

By obtaining the $f$ observational data points related to the FF and FB transitions for a
particular object, where $\epsilon$ in Equation \ref{ElDiEq4} stands for the energy of resonance,
and plotting them as a function of energy alongside the MB distribution, as given by
Equation~\ref{ElDiEq2}, for suitably-chosen temperatures, the observational electron distribution
can be compared to the MB theoretical distribution.

As part of the current investigation, we carried out these calculations on the data sets that
contain more than one FF or FB transition to be useful for comparison to the MB distribution. The
resonance data required for these calculations are given in Table \ref{ResTable}. In
Figures~\ref{EdNGC7009}-\ref{EdHe2-113} the results of the calculations are presented where the
$\sqrt{\E}$-scaled Maxwell-Boltzmann electron distribution for the given temperature, which is
chosen as the temperature that we obtained for that object from our least squares procedure, is
plotted against electron energy on a log-linear graph alongside the $\sqrt{\E}$-scaled observed
data points representing the electron fraction in the given energy range for the FF and FB
transitions of the selected lines from the indicated object and cited data source. The
$\sqrt{\E}$-scaled $\kappa$ and Druyvesteyn distributions are also plotted in these figures for
comparison using typical values for the parameters of these functions which are given by the
following equations \cite{BryansThesis2005}
\begin{equation}\label{kappaEq}
f_{\kappa,E_{\kappa}}\left(E\right)=\frac{2}{\sqrt{\pi}\kappa^{3/2}E_{\kappa}}\sqrt{\frac{E}{E_{\kappa}}}\frac{\Gamma\left(\kappa+1\right)}{\Gamma\left(\kappa-\frac{1}{2}\right)}\left(1+\frac{E}{\kappa
E_{\kappa}}\right)^{-\left(\kappa+1\right)}
\end{equation}
and
\begin{equation}\label{DruyvesteynEq}
f_{x,E_{x}}\left(E\right)=\frac{x}{E_{x}^{3/2}}\frac{\Gamma\left(\frac{5}{2}x\right)^{3/2}}{\Gamma\left(\frac{3}{2}x\right)^{5/2}}\sqrt{E}\exp\left(-\left[\frac{E\Gamma\left(\frac{5}{2}x\right)}{E_{x}\Gamma\left(\frac{3}{2}x\right)}\right]^{x}\right)
\end{equation}
where $\kappa$ and $x$ are parameters characterizing these distributions, while $\E_{\kappa}$ and
$E_{x}$ are characteristic energies.

As seen, there is in general a few observed data points to constrain the trend and determine
decisively if the observed electron distribution is MB or not. However, the scattering of these
data points is not far from MB, although in some cases the observational data may also broadly
resemble a Druyvesteyn distribution and this similarity may even improve for carefully chosen
Druyvesteyn parameters. An interesting point is that the fit between the observational data points
and the MB curve generally improves as the temperature of the curve approaches the broadly-known
temperature of the object.

\begin{figure}[!t]
\centering{}
\includegraphics[scale=0.65]{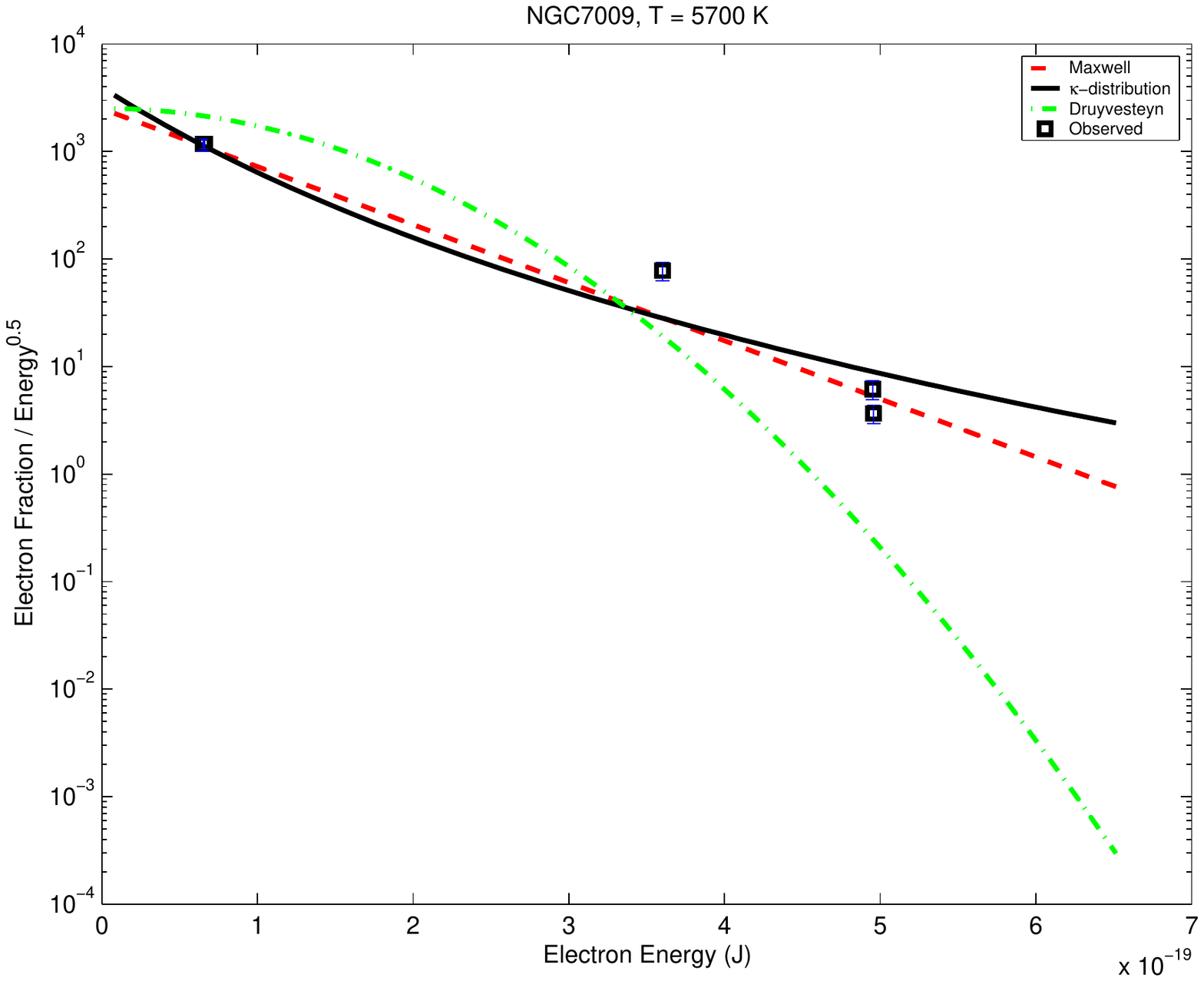}
\caption{Electron distribution plot for NGC 7009 data of Fang and Liu \cite{FangL2011}, alongside
Maxwell-Boltzmann, $\kappa$ ($=5.0$) and Druyvesteyn (with $x=2.0$) distributions for $T=5700$~K.
The $y$-axis is in arbitrary units.} \label{EdNGC7009}
\end{figure}

\begin{figure}[!t]
\centering{}
\includegraphics[scale=0.65]{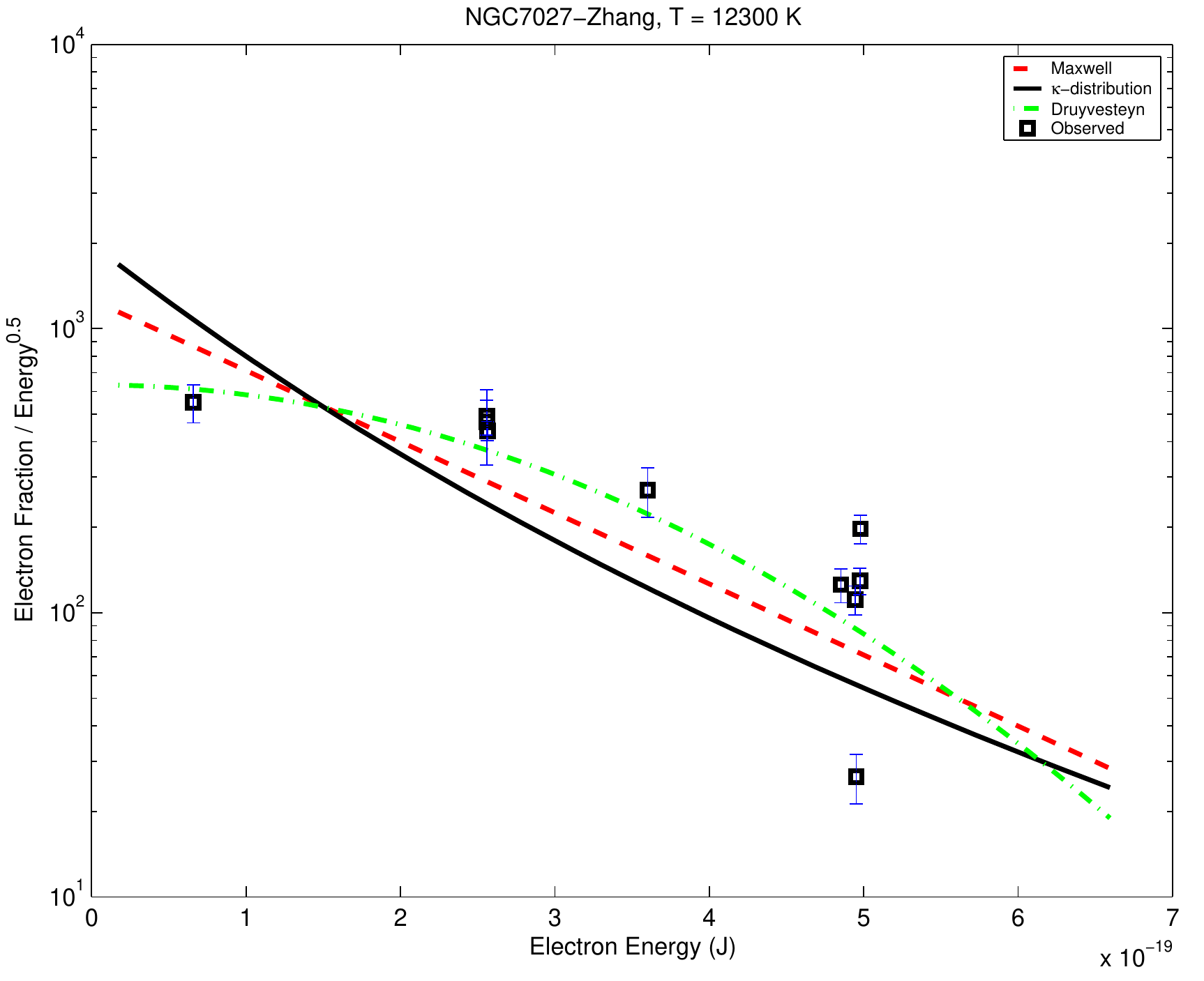}
\caption{Electron distribution plot for NGC 7027 data of Zhang \etal\ \cite{ZhangLLPB2005},
alongside Maxwell-Boltzmann, $\kappa$ ($=5.0$) and Druyvesteyn (with $x=2.0$) distributions for
$T=12300$~K. The $y$-axis is in arbitrary units.} \label{EdNGC7027-Zhang}
\end{figure}

\begin{figure}[!t]
\centering{}
\includegraphics[scale=0.65]{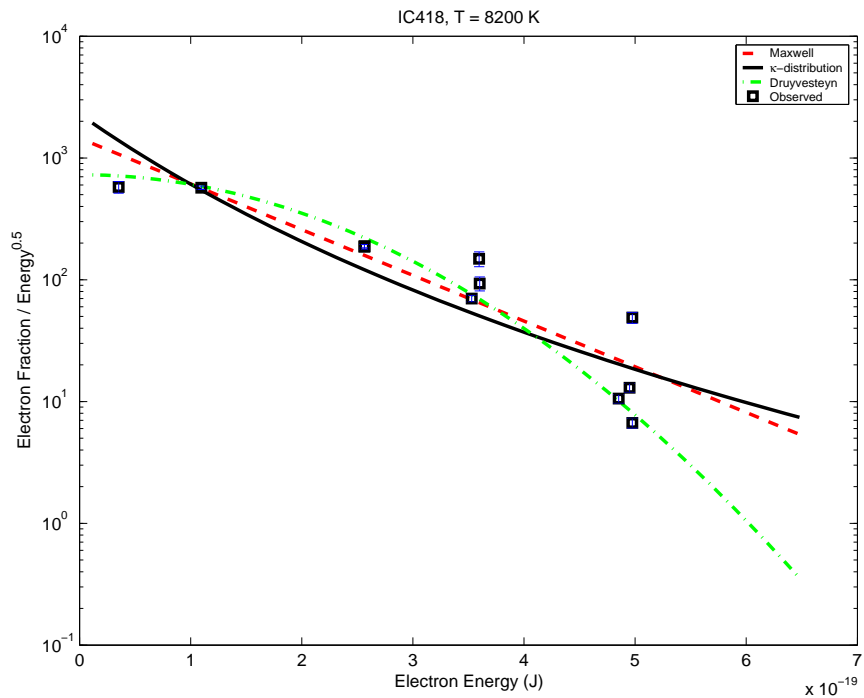}
\caption{Electron distribution plot for IC 418 data of Sharpee \etal\ \cite{SharpeeWBH2003},
alongside Maxwell-Boltzmann, $\kappa$ ($=5.0$) and Druyvesteyn (with $x=2.0$) distributions for
$T=8200$~K. The $y$-axis is in arbitrary units.} \label{EdIC418}
\end{figure}

\begin{figure}[!t]
\centering{}
\includegraphics[scale=0.65]{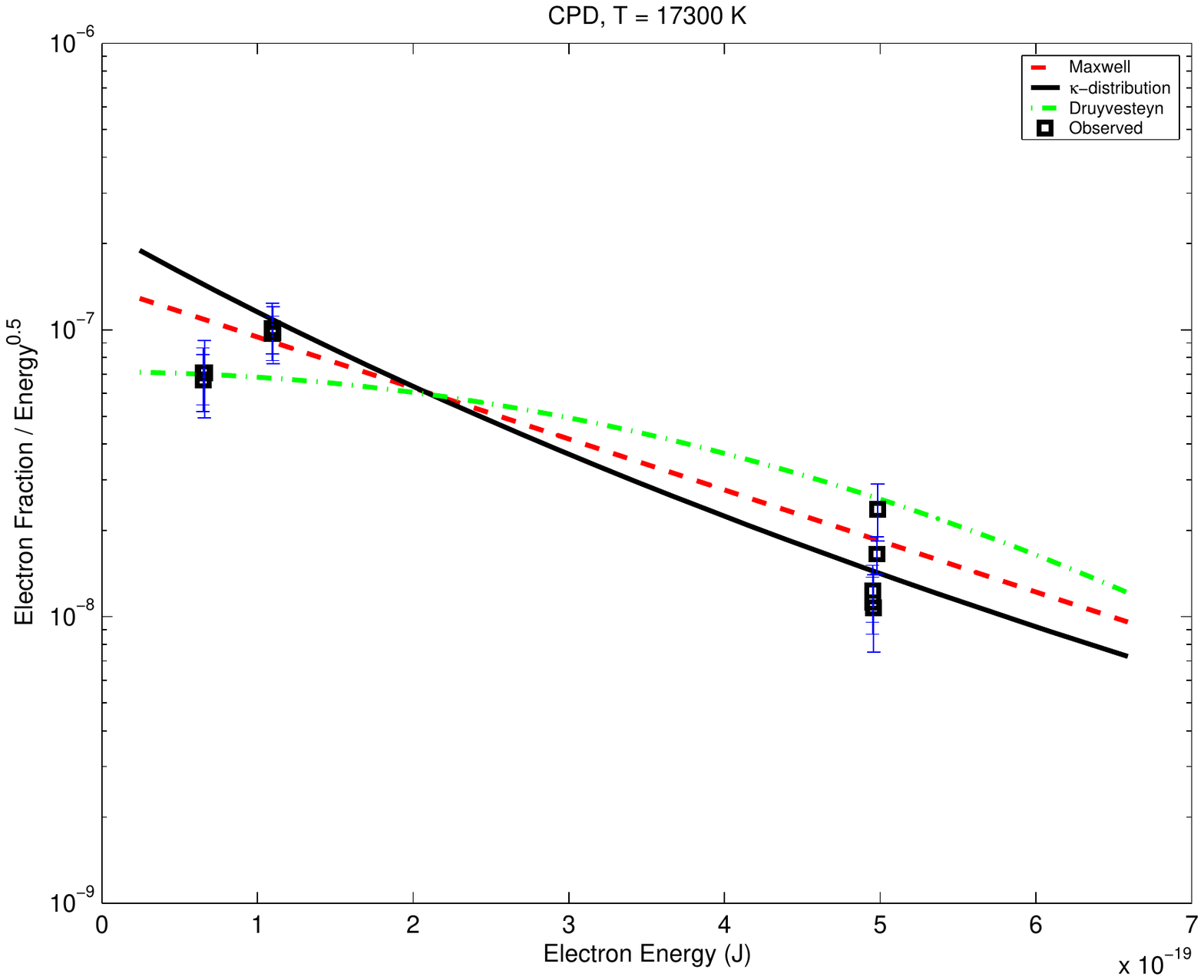}
\caption{Electron distribution plot for \CPD\ data of De Marco \etal\ \cite{DemarcoSB1997},
alongside Maxwell-Boltzmann, $\kappa$ ($=5.0$) and Druyvesteyn (with $x=2.0$) distributions for
$T=17300$~K. The $y$-axis is in arbitrary units.} \label{EdCPD}
\end{figure}

\begin{figure}[!t]
\centering{}
\includegraphics[scale=0.65]{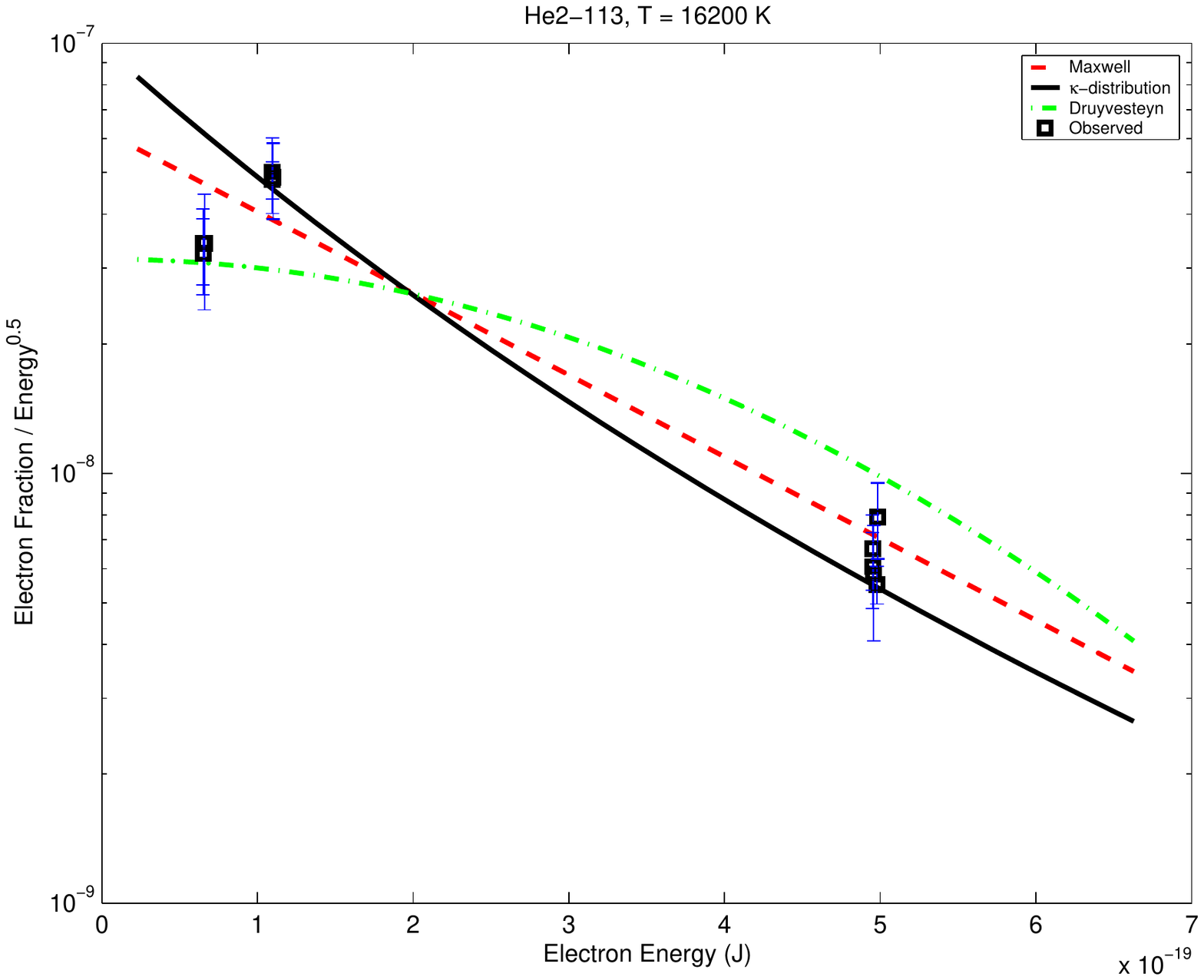}
\caption{Electron distribution plot for \Het\ data of De Marco \etal\ \cite{DemarcoSB1997},
alongside Maxwell-Boltzmann, $\kappa$ ($=5.0$) and Druyvesteyn (with $x=2.0$) distributions for
$T=16200$~K. The $y$-axis is in arbitrary units.} \label{EdHe2-113}
\end{figure}

{\singlespace
\begin{longtable}{@{\extracolsep\fill}cccccc@{}}
\caption{The 64 autoionizing states of C$^+$ used in the current study with their energy in \Ryd,
\depcoe, and autoionization probability in s$^{-1}$. The 1s$^{2}$ core is suppressed from all configurations. \label{ResTable}}\\
\hline\hline %
Index & Configuration &      Level &      $\E$ &     $\DC$ &    $\aTP$ \\ %
\hline
\endfirsthead
\caption[]{continued.}\\
\hline%
Index & Configuration &      Level &      $\E$ &     $\DC$ &    $\aTP$ \\ %
\hline
\endhead
\hline
\endfoot
\hline
         1 & 2s2p(\SLP3Po)3d & \SLPJ2Do{3/2} & 0.0160449523 & 0.0056778500 & 1.240914E+07 \\

         2 & 2s2p(\SLP3Po)3d & \SLPJ2Do{5/2} & 0.0161440982 & 0.0261058105 & 5.390253E+07 \\

         3 & 2s2p(\SLP3Po)3d & \SLPJ4Po{5/2} & 0.0198592428 & 0.0002217072 & 4.494221E+05 \\

         4 & 2s2p(\SLP3Po)3d & \SLPJ4Po{3/2} & 0.0200528871 & 0.0109357666 & 2.600220E+07 \\

         5 & 2s2p(\SLP3Po)3d & \SLPJ4Po{1/2} & 0.0201782774 & 0.0044633618 & 1.054050E+07 \\

         6 & 2s2p(\SLP3Po)3d & \SLPJ2Fo{5/2} & 0.0298595786 & 0.9981159635 & 1.159802E+12 \\

         7 & 2s2p(\SLP3Po)3d & \SLPJ2Fo{7/2} & 0.0302407616 & 0.9981660633 & 1.195966E+12 \\

         8 & 2s2p(\SLP3Po)3d & \SLPJ2Po{3/2} & 0.0502577447 & 0.9925373136 & 2.193909E+11 \\

         9 & 2s2p(\SLP3Po)3d & \SLPJ2Po{1/2} & 0.0504825543 & 0.9928976298 & 2.304847E+11 \\

        10 & 2s2p(\SLP3Po)4s & \SLPJ4Po{1/2} & 0.1174409992 & 0.9062462126 & 5.264264E+09 \\

        11 & 2s2p(\SLP3Po)4s & \SLPJ4Po{3/2} & 0.1176606145 & 0.9595276889 & 1.292497E+10 \\

        12 & 2s2p(\SLP3Po)4s & \SLPJ4Po{5/2} & 0.1180785216 & 0.0000000000 & 0.000000E+00 \\

        13 & 2s2p(\SLP3Po)4s & \SLPJ2Po{1/2} & 0.1378830490 & 0.9999959170 & 1.267096E+14 \\

        14 & 2s2p(\SLP3Po)4s & \SLPJ2Po{3/2} & 0.1383845207 & 0.9999959515 & 1.270574E+14 \\

        15 & 2s2p(\SLP3Po)4p  &  \SLPJ2Pe{1/2} & 0.1616554031 & 0.5081258069 & 4.383575E+08 \\

        16 & 2s2p(\SLP3Po)4p  &  \SLPJ2Pe{3/2} & 0.1618888697 & 0.8281132233 & 2.043629E+09 \\

        17 & 2s2p(\SLP3Po)4p  &  \SLPJ4De{1/2} & 0.1648956865 & 0.0166709970 & 1.096506E+06 \\

        18 & 2s2p(\SLP3Po)4p  &  \SLPJ4De{3/2} & 0.1650135133 & 0.9428663460 & 1.059104E+09 \\

        19 & 2s2p(\SLP3Po)4p  &  \SLPJ4De{5/2} & 0.1652179105 & 0.9575548715 & 1.436774E+09 \\

        20 & 2s2p(\SLP3Po)4p  &  \SLPJ4De{7/2} & 0.1655322976 & 0.0000000000 & 0.000000E+00 \\

        21 & 2s2p(\SLP3Po)4p  &  \SLPJ4Se{3/2} & 0.1740799826 & 0.0023540601 & 1.361289E+05 \\

        22 & 2s2p(\SLP3Po)4p  &  \SLPJ4Pe{1/2} & 0.1795026594 & 0.9721418782 & 1.500088E+09 \\

        23 & 2s2p(\SLP3Po)4p  &  \SLPJ4Pe{3/2} & 0.1796552966 & 0.9802394591 & 2.134176E+09 \\

        24 & 2s2p(\SLP3Po)4p  &  \SLPJ4Pe{5/2} & 0.1798464804 & 0.9965561757 & 1.249636E+10 \\

        25 & 2s2p(\SLP3Po)4p  &  \SLPJ2De{3/2} & 0.1854199383 & 0.9999806616 & 2.450767E+13 \\

        26 & 2s2p(\SLP3Po)4p  &  \SLPJ2De{5/2} & 0.1846436636 & 0.9999809896 & 2.436138E+13 \\

        27 & 2s2p(\SLP3Po)4p  & \SLPJ2Se{1/2} & 0.1997648585 & 0.9999637714 & 8.854266E+12 \\

        28 & 2s2p(\SLP1Po)3s &  \SLPJ2Po{1/2} & 0.2058921572 & 0.9999507406 & 6.254162E+13 \\

        29 & 2s2p(\SLP1Po)3s &  \SLPJ2Po{3/2} & 0.2059121732 & 0.9999505223 & 6.218950E+13 \\

        30 & 2s2p(\SLP3Po)4d  &  \SLPJ4Fo{3/2} & 0.2086057952 & 0.0130742833 & 1.050972E+06 \\

        31 & 2s2p(\SLP3Po)4d  &  \SLPJ4Fo{5/2} & 0.2087298186 & 0.5207186031 & 8.735439E+07 \\

        32 & 2s2p(\SLP3Po)4d  &  \SLPJ4Fo{7/2} & 0.2089176308 & 0.6046487195 & 1.231846E+08 \\

        33 & 2s2p(\SLP3Po)4d  &  \SLPJ4Fo{9/2} & 0.2091829917 & 0.0000000000 & 0.000000E+00 \\

        34 & 2s2p(\SLP3Po)4d  &  \SLPJ4Do{1/2} & 0.2137906313 & 0.0240817145 & 4.770103E+07 \\

        35 & 2s2p(\SLP3Po)4d  &  \SLPJ4Do{3/2} & 0.2138393841 & 0.0111774423 & 2.183347E+07 \\

        36 & 2s2p(\SLP3Po)4d  &  \SLPJ4Do{5/2} & 0.2139172974 & 0.0140469244 & 2.749728E+07 \\

        37 & 2s2p(\SLP3Po)4d  &  \SLPJ4Do{7/2} & 0.2140182658 & 0.0623175129 & 1.279541E+08 \\

        38 & 2s2p(\SLP3Po)4d  &  \SLPJ2Do{3/2} & 0.2181284447 & 0.3163623529 & 4.111112E+08 \\

        39 & 2s2p(\SLP3Po)4d  &  \SLPJ2Do{5/2} & 0.2182467272 & 0.1161936608 & 1.173714E+08 \\

        40 & 2s2p(\SLP3Po)4d  &  \SLPJ4Po{5/2} & 0.2200435636 & 0.0044332899 & 5.029955E+06 \\

        41 & 2s2p(\SLP3Po)4d  &  \SLPJ4Po{3/2} & 0.2202300088 & 0.0095968218 & 1.099796E+07 \\

        42 & 2s2p(\SLP3Po)4d  &  \SLPJ4Po{1/2} & 0.2203477445 & 0.0066028492 & 7.552275E+06 \\

        43 & 2s2p(\SLP3Po)4f  &  \SLPJ2Fe{5/2} & 0.2225695047 & 0.4062779057 & 1.887721E+08 \\

        44 & 2s2p(\SLP3Po)4f  &  \SLPJ2Fe{7/2} & 0.2226518832 & 0.0014994709 & 4.083866E+05 \\

        45 & 2s2p(\SLP3Po)4f  &  \SLPJ4Fe{3/2} & 0.2226157059 & 0.3253636102 & 1.231093E+08 \\

        46 & 2s2p(\SLP3Po)4f  &  \SLPJ4Fe{5/2} & 0.2226627273 & 0.3227642510 & 1.220007E+08 \\

        47 & 2s2p(\SLP3Po)4f  &  \SLPJ4Fe{7/2} & 0.2227230532 & 0.0414199477 & 1.122416E+07 \\

        48 & 2s2p(\SLP3Po)4f  &  \SLPJ4Fe{9/2} & 0.2227599595 & 0.0926167497 & 2.606316E+07 \\

        49 & 2s2p(\SLP3Po)4d  &  \SLPJ2Fo{5/2} & 0.2259594181 & 0.9967124124 & 3.003092E+11 \\

        50 & 2s2p(\SLP3Po)4d  &  \SLPJ2Fo{7/2} & 0.2263462510 & 0.9968281849 & 3.121474E+11 \\

        51 & 2s2p(\SLP3Po)4f  &  \SLPJ4Ge{5/2} & 0.2267242446 & 0.0410695674 & 1.127515E+07 \\

        52 & 2s2p(\SLP3Po)4f  &  \SLPJ4Ge{7/2} & 0.2268078989 & 0.3627478405 & 1.492546E+08 \\

        53 & 2s2p(\SLP3Po)4f  &  \SLPJ4Ge{9/2} & 0.2270049148 & 0.3399201973 & 1.350812E+08 \\

        54 & 2s2p(\SLP3Po)4f  &  \SLPJ4Ge{11/2} & 0.2272718249 & 0.0000000000 & 0.000000E+00 \\

        55 & 2s2p(\SLP3Po)4f  &  \SLPJ2Ge{7/2} & 0.2271098017 & 0.6893371371 & 5.331860E+08 \\

        56 & 2s2p(\SLP3Po)4f  &  \SLPJ2Ge{9/2} & 0.2274615507 & 0.7843011811 & 8.601144E+08 \\

        57 & 2s2p(\SLP3Po)4f  &  \SLPJ4De{7/2} & 0.2281245886 & 0.0002600797 & 6.139194E+04 \\

        58 & 2s2p(\SLP3Po)4f  &  \SLPJ4De{5/2} & 0.2282086986 & 0.9951533708 & 4.934209E+10 \\

        59 & 2s2p(\SLP3Po)4f  &  \SLPJ4De{3/2} & 0.2284089951 & 0.9846397825 & 1.519940E+10 \\

        60 & 2s2p(\SLP3Po)4f  &  \SLPJ4De{1/2} & 0.2285283711 & 0.0000000000 & 0.000000E+00 \\

        61 & 2s2p(\SLP3Po)4f  &  \SLPJ2De{5/2} & 0.2284110910 & 0.9979488054 & 9.860173E+10 \\

        62 & 2s2p(\SLP3Po)4f  &  \SLPJ2De{3/2} & 0.2286146681 & 0.9984868233 & 1.359512E+11 \\

        63 & 2s2p(\SLP3Po)4d  &  \SLPJ2Po{3/2} & 0.2332306002 & 0.9986068098 & 5.007744E+11 \\

        64 & 2s2p(\SLP3Po)4d  &  \SLPJ2Po{1/2} & 0.2334757310 & 0.9986193975 & 5.083958E+11 \\
\hline
\end{longtable}
}

\section{Epilogue}\label{TempEpilogue}

In the end of this chapter, an important issue should be addressed that is how to justify the large
discrepancy in the temperature values obtained by different investigators from different atoms,
ions, and types of transition. In the following points we propose some possible reasons for this
discrepancy.

\begin{itemize}

\item
Differences in theoretical, observational and numerical techniques used to infer electron
temperatures. Since the frameworks of these techniques are different the results may also differ.

\item
The astronomical objects in general and \planeb e in particular are extensive objects with complex
structure of varying physical conditions. Hence, the collected spectrum normally originates from
different regions in the object which are subject to different physical conditions, including
temperature. Therefore, there is no contradiction in some cases of varying electron temperatures as
they can be deduced from transitions originating from different regions.

\item
Selectivity of electron temperature by different types of transition (e.g. forbidden or
recombination) from the same or different species. Because each type of transition of a particular
species is optimal at certain ranges of electron energy, the deduced temperature is dependent on
these factors \cite{HoofBVF2000}.

\item
Finally, observational and theoretical errors (gross, systematic, random, etc.) should also be
considered as a possible reason in some cases especially for some eccentric results.

\end{itemize}

%

\newpage
\thispagestyle{empty} \vspace*{5.0cm} \phantomsection \addcontentsline{toc}{chapter}{\protect
\numberline{} Part II: Molecular Physics}

{\center

\LARGE{\bf Part II

\vspace*{1.0cm}

Molecular Physics:  \vspace{0.3cm} \\
A Computed Line List for Molecular Ion \htdp} \vspace*{2.0cm} \\
}

\chapter{Computed Line List for \htdp} 

In this part of the thesis, we present our comprehensive, highly-accurate calculated line list of
frequencies and transition probabilities for the singly deuterated isotopologue of \htp, \htdp. The
list, called \STo, contains over 22 million rotational-vibrational transitions occurring between
more than 33 thousand energy levels and covers frequencies up to 18500 \wn. Due to its simplicity
and cosmological and astrophysical importance, \htdp\ has been the subject of a substantial number
of spectroscopic and astronomical studies in the last three decades. We hope that our line list and
the associated conclusions presented in this thesis will be a valuable contribution in this
direction that other researchers will benefit from \cite{SochiT2010}.

\section{Introduction}

\htdp\ is one of the simplest polyatomic quantum systems and the lightest asymmetric top molecular
ion. It has a substantial permanent electric dipole moment of about 0.6~Debye due to the
displacement of the center of charge from the center of mass. It consists of two electrons bound to
three nuclei (two protons and one deuteron) forming a triangular shape at equilibrium. The molecule
is a prolate top with three normal vibrational modes: a symmetric breathing mode $\nu_1$, a bending
mode $\nu_2$, and an asymmetric stretch $\nu_3$. All these modes are infrared active. Depending on
the spin alignment of its protons, it has two different species, ortho and para, with a spin weight
ratio of 3:1. Because \htdp\ has a permanent dipole moment, it possesses pure rotational
transitions which mainly occur in the sub-millimeter region (far infrared and microwave) and hence
it can be observed in emission or absorption from ground-based detectors. This all contrasts
strongly with the non-deuterated \htp\ molecular ion which has no allowed rotational spectrum and
only one infrared active vibrational mode \cite{DalgarnoHNK1973, AmanoW1984, BogeyDDDL1984,
FarnikDKPTN2002}.

In cool astrophysical environments, where this molecule is mostly found and where it plays very
important role, the molecule is formed by several reactions; the main one is
\begin{equation}\label{htdpEq}
\textrm{H}_3^+ + \textrm{HD} \rightarrow \textrm{H}_2\textrm{D}^+ + \textrm{H}_2
\end{equation}
Because this and other similar formation reactions are exothermic, the formation of \htdp\ in the
cold interstellar environment is favored. Consequently, the abundance of this species is enhanced
relative to the D/H ratio.

\htdp\ possesses very rich rotational and roto-vibrational spectra. As an isotopomer of \htp,
\htdp\ is a major participant in chemical reactions taking place in the interstellar medium. In
particular, it plays a key role in the deuteration processes in these environments
\cite{Millar2003}. Although the existence of \htdp\ in the interstellar medium (ISM) and
astrophysical objects was contemplated decades ago \cite{DalgarnoHNK1973, DishoeckPKB1992} with
some reported tentative sighting \cite{PhillipsBKWC1985, PaganiWFKGe1992, BoreikoB1993}, it is only
relatively recently that the molecule was firmly detected in the ISM via one of its rotational
transition lines \cite{StarkTD1999}. There have been many subsequent astronomical studies related
to \htdp\ spectra \cite{CaselliTCB2003, VastelPY2004,CeccarelliDLCC2004, StarkSBHDe2004,
HogerheijdeCETAe2006, HarjuHLJMe2006, CernicharoPG2007, CaselliVCTCB2008}. In particular these
spectra have been used to investigate the mid-plane of proto-planetary disks \cite{RamosCE2007},
the kinematics of the centers of pre-stellar cores \cite{TakCC2005}, and the suggestion of possible
use as a probe for the presence of the hypothetical cloudlets forming the baryonic dark matter
\cite{CeccarelliD2006}. Vibrational transitions of \htdp\ have yet to be observed astronomically.

The important role of \htdp\ in the astrophysical molecular chemistry, especially in the deuterium
fractionation processes, was recognized since the early stages of radio astronomy development.
Because \htp\ lacks a permanent dipole moment and hence cannot be detected via radio astronomy, its
asymmetrical isotopomers, namely \htdp\ and \dthp, are the most promising tracers of the extremely
cold and highly dense interstellar clouds. These species are the last to remain in gaseous state
under these extreme conditions with observable pure rotational spectra \cite{DalgarnoHNK1973,
HerbstM2008Book}. Of these two isotopomers, \htdp\ is the more abundant and easier to observe and
hence it is the main species to utilize in such astrophysical investigations.

\htdp\ is the key molecular ion that drives deuterium fractionation of various molecular species by
isotope exchange processes in the interstellar medium especially at low temperatures. This results
in considerable enhancement of deuterated molecules that may also participate in other deuteration
processes by passing their deuterium content to other species. At very low temperatures
($\sim$10~K), where abundant species, such as CO, N$_2$ and O$_2$, that consume \htdp\ freeze out
onto the surface of interstellar dust grains, and where the destruction channels of \htdp\ in the
H/D exchange reactions are blocked due to zero-point energy effects, the formation of \htdp\ is
strongly enhanced. This can lead to \htdp\ becoming more abundant than \htp\ although the cosmic
abundance of deuterium is only about $2 \times 10^{-5}$ that of hydrogen. In these conditions
\htdp\ becomes the main source of deuterium fractionation as it is more abundant than any other
deuterated species, with subsequent effective deuteration of other species \cite{ShyFW1981,
WarnerCPW1984, StarkTD1999, Millar2003, HerbstM2008Book, TennysonBook2005}.

The potential importance of \htdp\ spectroscopy for cosmology is obvious as it is a primordial
molecular species that could have a considerable abundance in the early universe. In particular, it
may have played a role in the cooling of primordial proto-objects \cite{Dubrovich1993,
DubrovichL1995, DubrovichP2000, GalliP1998, SchleicherGPCKe20008}. Its significance is highlighted
by a number of studies which consider the role of \htdp\ in spectral distortions of Cosmic
Microwave Background Radiation and whether this can be used to determine the \htdp\ and deuterium
abundances at different epochs \cite{DubrovichL1995}. Finally the \htdp\ 372 GHz line has been
considered as a probe for the presence of dark matter \cite{CeccarelliD2006}.

In the following we present a short literature review summarizing some landmarks in the
investigation of this molecule.

\subsection{Literature Review}

\htdp\ is a highly important molecule and hence it was the subject of extensive studies both at
laboratory and astrophysical levels in the last few decades. It is regarded as one of the benchmark
systems for theoretical and experimental molecular investigations \cite{DalgarnoHNK1973,
WarnerCPW1984, MillerTS1989, DinelliNPT1997}.

The first successful spectroscopic investigation of \htdp\ was carried out by Shy \etal\
\cite{ShyFW1981} where nine rotational-vibrational transitions were measured in the infrared region
between 1800 and 2000~\wn\ using Doppler-tuned fast-ion laser technique, but no specific
spectroscopic assignments were made. This was followed by other spectroscopic investigations which
include the observation and identification of the strong and highly-important rotational
$1_{10}-1_{11}$ transition line of ortho-\htdp\ at 372 GHz in 1984 by Bogey \etal\
\cite{BogeyDDDL1984} and Warner \etal\ \cite{WarnerCPW1984} using discharge techniques. The three
fundamental vibrational bands of the \htdp\ ion were also detected and identified by Amano and
Watson \cite{AmanoW1984}, Amano \cite{Amano1985} and Foster \etal\ \cite{FosterMPWPe1986} using
laser spectroscopy.

F\'{a}rn\'{\i}k \etal\ \cite{FarnikDKPTN2002} measured transitions to overtones 2$\nu_2$ and
2$\nu_3$ and to combination $\nu_2+\nu_3$ in jet-cooled \htdp\ ions using high-resolution infrared
spectroscopic technique. Amano and Hirao \cite{AmanoH2005} also measured the transition frequency
of the $1_{10}-1_{11}$ line of \htdp\ at 372 GHz with improved accuracy [372.421385(10) GHz].
Moreover, the transition frequency of the $3_{21}-3_{22}$ line of \htdp\ was observed and measured
[646.430293(50) GHz] for the first time in this investigation. Hlavenka \etal\
\cite{HlavenkaKPVKG2006} measured second overtone transition frequencies of \htdp\ in the infrared
region using cavity ringdown absorption spectroscopy. Other examples of spectroscopic
investigations include those of Asvany \etal\ \cite{AsvanyHMKSTS2007}, who detected 20 lines with
full spectroscopic assignment using laser induced reaction techniques, and the relatively recent
work of Yonezu \etal\ \cite{YonezuMMTA2009}, who reported precise measurement of the transition
frequency of the \htdp\ $2_{12}-1_{11}$ line at 2.363 THz, alongside three more far-infrared lines
of \htdp, using tunable far-infrared spectrometry. However, none of these studies measured absolute
line intensities, although the work of F\'{a}rn\'{\i}k \etal\ \cite{FarnikDKPTN2002} and Asvany
\etal\ \cite{AsvanyHMKSTS2007} give relative intensities.

On the astronomical side, a tentative observation of the $1_{10}-1_{11}$ transition line of
ortho-\htdp\ at 372 GHz from NGC 2264 has been reported by Phillips \etal\ \cite{PhillipsBKWC1985}.
This line is the strongest and the most important of the \htdp\ lines as it involves some of the
lowest rotational levels and is the main one that can be observed from ground-based telescopes.
Pagani \etal\ \cite{PaganiWFKGe1992} have also searched for this line in several dark clouds with
no success; nonetheless they derived upper limits for the ortho-\htdp\ column density. A similar
failed attempt to observe this line in a number of interstellar clouds including NGC 2264 was made
by van Dishoeck \etal\ \cite{DishoeckPKB1992}. In their search for the rotational transitions of
\htdp, Boreiko and Betz \cite{BoreikoB1993} detected an absorption feature toward the IRc2 region
of M42. The feature was suspected to be from the lowest rotational $1_{01}-0_{00}$ transition at
1370 GHz of para-\htdp.

The first confirmed astronomical observation of \htdp\ was made in 1999 by Stark \etal\
\cite{StarkTD1999} who reported the detection of the $1_{10}-1_{11}$ ground-state transition line
of ortho-\htdp\ at 372 GHz in emission from the young stellar object NGC 1333 IRAS 4A. This was
followed by the detection of this line towards the pre-stellar core L1544 by Caselli \etal\
\cite{CaselliTCB2003} and the detection of the same line toward another pre-stellar core 16293E by
Vastel \etal\ \cite{VastelPY2004}. The 372 GHz line of \htdp\ was also detected toward a disk
source DM Tau, and tentatively detected toward TW Hya by Ceccarelli \etal\
\cite{CeccarelliDLCC2004}. Other confirmed or tentative sightings of this line include the
observation toward I16293A \cite{StarkSBHDe2004}, toward L1544 and L183 \cite{VastelCCPWe2006,
VastelPCCP2006}, in the starless core Barnard 68 \cite{HogerheijdeCETAe2006}, and in a massive
pre-stellar core in Orion B \cite{HarjuHLJMe2006}. \htdp\ has also been detected in absorption
toward Sgr B2 in the galactic center through the far-infrared $2_{12}-1_{11}$ transition of
ortho-\htdp\ at 2363.325 GHz \cite{CernicharoPG2007}. Recently, \htdp\ has been observed through
the 372 GHz $1_{10}-1_{11}$ line in seven starless cores and four protostellar cores by Caselli
\etal\ \cite{CaselliVCTCB2008}.

\subsection{Line Lists}

Several synthetic \htdp\ line lists have been generated from \abin\ calculations in the last few
decades. Prominent examples include the line list of Miller \etal\ \cite{MillerTS1989} and another
one generated by Tennyson and co-workers as part of the work reported by Asvany \etal\
\cite{AsvanyHMKSTS2007}. Miller \etal 's list extends to rotational quantum number $J=30$ and
covers all levels up to 5500~\wn\ above the ground state. These lists were used in a number of
studies for various purposes such as spectroscopic assignment of energy levels and transition lines
from astronomical observations, and for computation of low-temperature partition functions
\cite{SidhuMT1992}. The line lists also played an important role in motivating and steering the
experimental and observational work in this field \cite{MillerTS1989, AsvanyHMKSTS2007}.
Nonetheless, the previous \htdp\ line lists suffered from limitations that include low energy
cut-off and the inclusion of a limited number of levels especially at high $J$. These limitations,
alongside the recent developments in computing technology and the availability of high accuracy
\abin\ models of \htp\ and its isotopologues, including \htdp, represented by the potential and
dipole surfaces of Polyansky and Tennyson \cite{PolyanskyT1999}, provided the motivation to
generate a more comprehensive and accurate line list. The intention is that the new list will both
fill the previous gaps and provide data of better quality.

The \htdp\ line list presented in this thesis, called \STo, consists of 22164810 transition lines
occurring between 33330 rotational-vibrational levels. These are all the energy levels with
rotational quantum number $J\leq 20$ and frequencies below 18500~\wn\ as an upper cut-off limit for
both levels and transitions. This line list can be seen as a companion to the \htp\ line list of
Neale \etal\ \cite{NealeMT1996} which has been extensively used for astrophysical studies; although
for reasons explained below, the \STo\ line list is actually expected to be more accurate. More
details about \STo\ will follow.

%

\section{Method}

As indicated already, the main code used to generate the \htdp\ line list is the DVR3D code of
Tennyson \etal\ \cite{TennysonKBHPRZ2004} which is a suite of programs for performing
rotational-vibrational calculations on triatomic molecules. The code calculates, among other
things, wavefunctions, energy levels, transition lines, dipole moments, and transition
probabilities. DVR3D is based on a discrete variable representation (DVR) approach by using
Gauss-Jacobi and Gauss-Laguerre quadrature schemes for representing the internal coordinates. The
code uses an exact Hamiltonian, within the Born-Oppenheimer approximation, and requires potential
energy and dipole surfaces to be supplied as an input. In general, it is these surfaces which
largely determine the accuracy of the resulting calculations \cite{PolyanskyCSZBe2003} assuming a
good choice of DVR3D input parameters. The DVR3D code consists of four main stages: vibration,
rotation, dipole and spectra. These stages are fully documented by Tennyson \etal\
\cite{TennysonKBHPRZ2004}, and hence the following discussion about these stages is restricted to
what is necessary to clarify the method that we followed in generating the \STo\ list. In general,
the parameters of these stages were chosen following extensive consistency and convergence tests,
some of which are outlined. The results were regularly compared to the available experimental and
observational data and the theoretical results from previous investigations. A sample of these
comparisons, validations and preliminary results will follow in the subsequent sections.


The vibrational stage of the DVR3D suite requires an accurate model for the variation of electronic
potential as a function of nuclear geometry. This potential energy surface, which is supplied
through a fortran subroutine, is usually obtained from \abin\ calculations enhanced by empirical
adjustment to improve the fit to experimental data. For the generation of \STo\ list, we used the
\htp\ global potential surface of Polyansky \etal\ \cite{PolyanskyPKT2000}, who used the ultra-high
accurate \abin\ data of Cencek \etal\ \cite{CencekRJK1998} supplemented by extra data points to
reach about 200 data points in total. The surface was constrained at high energy by the data of
Schinke \etal\ \cite{SchinkeDL1980}. We complemented this surface by adding the \htdp\ adiabatic
correction term fitted by Polyansky and Tennyson \cite{PolyanskyT1999}. To allow for non-adiabatic
corrections to the Born-Oppenheimer approximation, we employed Polyansky and Tennyson's
\cite{PolyanskyT1999} vibrational mass scaling where different values for the reduced nuclear mass
were used for modeling vibrational and rotational motions. We used $\mu_\textrm{H}=1.0075372$~u and
$\mu_\textrm{D}=2.0138140$~u for the vibrational atomic masses, and $\mu_\textrm{H}=1.00727647$~u
and $\mu_\textrm{D}=2.01355320$~u for the rotational atomic masses. The accuracy of this model will
be assessed below.

Calculations were performed in Jacobi scattering coordinate system ($r_1, r_2, \theta$),
schematically presented in Figure~\ref{Jacobi}, to depict the triatomic \htdp\ molecule, where
$r_1$ represents the diatomic distance (H-H), $r_2$ is the separation of the D atom from the center
of mass of the diatom, and $\theta$ is the angle between $r_1$ and $r_2$. This coordinate system is
believed to be more appropriate for an asymmetric molecule like \htdp. Radial basis functions of
Morse oscillator type were used to model $r_1$ \cite{TennysonS1982}, while spherical oscillators
were used to model $r_2$ \cite{TennysonS1983}. Following Polyansky and Tennyson
\cite{PolyanskyT1999}, the Morse parameters for $r_1$ were set to $r_e = 1.71$, $D_e = 0.10$ and
$\omega_e = 0.0108$, with 20 Gauss-Laguerre grid points. Parameters of the spherical oscillator
functions were set to $\alpha = 0.0$ and $\omega_e = 0.0075$ with 44 Gauss-Laguerre grid points,
following extensive tests on convergence of the vibrational band origins. The essence of these
tests is to vary these parameters systematically to minimize the sum of the energy of the
rotation-less states. 36 Gauss-Legendre grid points were used to represent the angular motion. The
final vibrational Hamiltonian matrix used was of dimension 2000. In this investigation six
combinations of $\alpha$ and $\omega_e$ ($\alpha=0.0,0.5$ and 1.0 with $\omega_e=0.0075$ and
0.0095) were used for test in conjunction with 16 different values ($30,32,\dots,60$) for the
number of $r_2$ mesh points (NPNT2).

\begin{figure}[!t]
\centering{}
\includegraphics[scale=1]{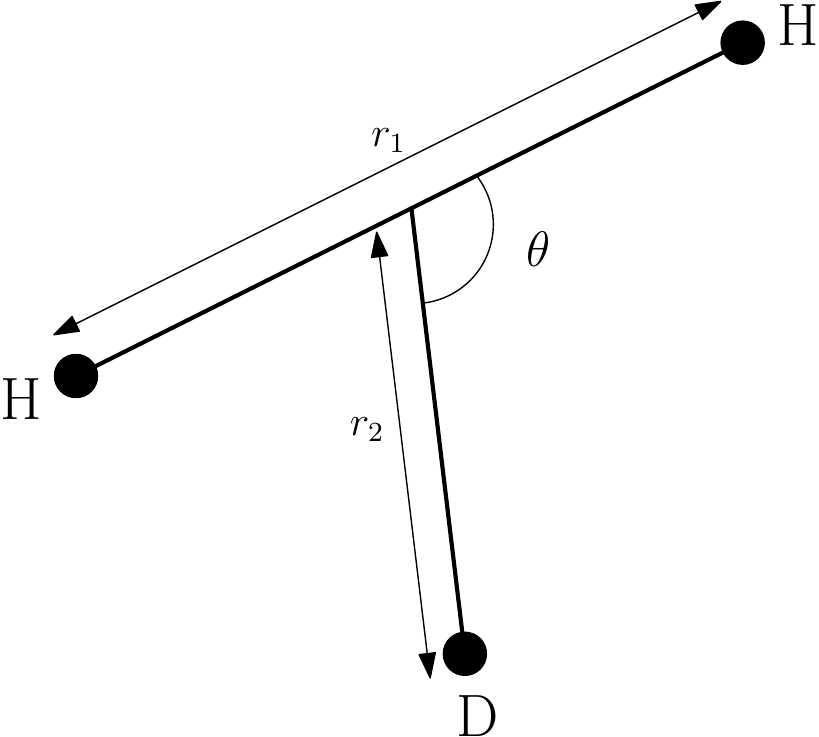}
\caption{Jacobi scattering coordinate system.} \label{Jacobi}
\end{figure}

The vibrational stage requires the maximum dimension of the largest intermediate 2D Hamiltonian,
MAX2D, and the maximum dimension of the final Hamiltonian, MAX3D, as an input. For these parameters
we used values of 1000 and 2000 respectively. The vibrational stage also requires the number of
eigenvalues and eigenvectors to be archived for use in the rotational stage (NEVAL). A value of
4000 were chosen for this parameter based on the results of convergence tests. For the number of
DVR mesh points in $r_2$ from Gauss-associated Laguerre quadrature (NPNT2) a value of 44 were
chosen, while values of 36 and 20 were used for the number of mesh points in $\theta$ (NALF) and in
$r_1$ (NPNT1) respectively. The selection of these values was mainly based on the results obtained
from previous investigations and our own convergence tests.


In the rotational stage, a value of 2000 was used for the number of vibrational levels for each $k$
to be read and used (NVIB). The size of the Hamiltonian matrix in this stage (IBASS), which is a
function of $J$, was set to $1800(J+1)$ following a series of tests on $J=3$ and $J=15$ in which
this parameter was varied incrementally until convergence within an error margin of 0.01~\wn\ was
reached. These tests demonstrated that choosing sufficiently large values for the size of
rotational Hamiltonian (IBASS), although computationally very expensive, is crucial for obtaining
reliable results. Our aim was to obtain convergence to within 0.01~\wn\ for all rotation-vibration
levels considered. Our tests suggest we achieved this except, possibly, for some of the highest
lying levels. For these levels our basic model, and in particular our corrections to the
Born-Oppenheimer approximation, are not expected to be reliable to this accuracy. Recently, an
improved \abin\ adiabatic potential energy surfaces for \htp\ and its isotopomers have been
developed by Pavanello \etal\ \cite{PavanelloAAZMe2012}. These surfaces include diagonal
Born-Oppenheimer corrections calculated from precise wavefunctions. The use of these surfaces in
the future should improve the accuracy of these levels.


To compute the intensity of the vibration-rotation transitions, the transition dipole moments
between the states concerned must be known. Therefore, a dipole moment surface, supplied to the
dipole stage in the form of an independent fortran subroutine, is required. This surface is usually
obtained from \abin\ calculations without empirical adjustment due to limitation on the quality of
experimental data if such data are available at all. For \STo\ generation, we used the \abin\
dipole surface of R\"{o}hse \etal\ \cite{RohseKJK1994} to calculate the components of the \htdp\
dipole. In the DIPOLE3 module of DVR3D we set the number of Gauss-Legendre quadrature points used
for evaluating the wavefunctions and dipole surface (NPOT) to 50. This choice is consistent with
the requirement that this parameter should be slightly larger than the number of DVR points used to
calculate the underlying wavefunctions.


Finally, the last stage, spectra, is for producing the list of transition lines. The data of each
line in this list include the attributes of the upper and lower levels, the frequency of the line
in \wn, the line strength, the temperature-dependent absolute and relative intensities, and the
Einstein $\EAC$ coefficient. The actual transitions file of the \STo\ list represents a summary of
the transitions file produced by spectra stage, where the upper and lower levels in the \STo\ file
are identified by a unique index from the levels file alongside the Einstein $\EAC$ coefficient.
These data augmented with the data in the levels file are sufficient to find the intensity and
other temperature-dependent quantities. The purpose of this compact form presentation of the
transition data is to reduce file size and avoid possible errors and confusion.

The final \STo\ line list consists of two main files: one for the levels and the other for the
transitions. These two files are constructed and formatted according to the method and style of the
BT2 water line list of Barber \etal\ \cite{BarberTHT2006}. The total amount of CPU time spent in
producing the \STo\ list including preparation, convergence tests, and verifying the final results
was about 8000 hours. We used the serial version of the DVR3D suite on PC platforms running Linux
(Red Hat) operating system, as part of the computing network of the Physics Department, University
College London. Both 32- and 64-bit machines were used in this work although the final data were
produced mainly on 64-bit machines due to the huge memory requirement of the DVR3D code especially
for the high-$J$ calculations. Large samples of data produced by these machines were compared and
found to be identical.

It should be remarked that the outlined variational nuclear motion procedure used for these
calculations provides rigorous quantum numbers: $J$, ortho/para and parity $p$, but not the
standard approximate quantum numbers in normal mode, rigid rotor notation. We therefore hand
labeled those levels for which such quantum numbers could be assigned in a fairly straightforward
fashion. 5000 of these levels are fully designated with 3 rotational ($J,K_a,K_c$) and 3
vibrational ($v_1,v_2,v_3$) quantum numbers, while 341 levels are identified with rotational
quantum numbers only. The assignments were made by first identifying 176 experimentally-assigned
levels. The rotation-less levels are then assigned by taking $K_a=K_c=0$ when $J=0$. This is
followed by the assignment of the vibrational ground state levels which are obtained mainly from
Polyansky \etal\ \cite{PolyanskyMT1993}. All these assignments can then provide a guess for many
other levels. For example, the energy of the ($v_1,v_2,v_3,J,K_a,K_c$) level can be estimated from
the energy of the ($0,0,0,J,K_a,K_c$) and ($v_1,v_2,v_3,0,0,0$) levels. Many gaps can then be
filled as the missing levels can be uniquely identified from the identified ones in their
neighborhood. Some of these assignments are made as a tentative guess and hence the assignments in
general should be treated with caution. Tables \ref{sampleLevTable} and \ref{sampleTraTable}
present samples of the ST1 levels and transitions files respectively. A sample of the input data
files used as an input to the DVR3D code in various stages of production of \STo, as well as the
input files of Spectra-BT2 code used to generate the synthetic spectra, is given in
Appendix~\ref{AppInDataM}.

\begin{table}
\centering %
\caption{Sample of the \STo\ levels file. The columns from left to right are for: index of level in
file, $J$, symmetry, index of level in block, frequency in \wn, $v_1$, $v_2$, $v_3$, $J$, $K_a$,
$K_c$. We used -2 to mark unassigned quantum numbers.} \vspace{0.5cm}
\label{sampleLevTable} %
\begin{tabular*}{\textwidth}{c@{\extracolsep{\fill}}rrrrrrrrrrr}
\hline
720 &          1 &          4 &         89 & 17645.80538 &         -2 &         -2 &         -2 &          1 &          1 &          0 \\
721 &          1 &          4 &         90 & 17699.90283 &         -2 &         -2 &         -2 &          1 &          1 &          0 \\
722 &          1 &          4 &         91 & 17712.30089 &         -2 &         -2 &         -2 &          1 &          1 &          0 \\
723 &          1 &          4 &         92 & 17778.46381 &         -2 &         -2 &         -2 &          1 &          1 &          0 \\
724 &          1 &          4 &         93 & 17840.99975 &         -2 &         -2 &         -2 &          1 &          1 &          0 \\
725 &          1 &          4 &         94 & 17902.76377 &         -2 &         -2 &         -2 &          1 &          1 &          0 \\
726 &          1 &          4 &         95 & 18005.04261 &         -2 &         -2 &         -2 &          1 &          1 &          0 \\
727 &          1 &          4 &         96 & 18113.51953 &         -2 &         -2 &         -2 &          1 &          1 &          0 \\
728 &          1 &          4 &         97 & 18163.92137 &         -2 &         -2 &         -2 &          1 &          1 &          0 \\
729 &          1 &          4 &         98 & 18226.60371 &         -2 &         -2 &         -2 &          1 &          1 &          0 \\
730 &          1 &          4 &         99 & 18265.61525 &         -2 &         -2 &         -2 &          1 &          1 &          0 \\
731 &          1 &          4 &        100 & 18379.99989 &         -2 &         -2 &         -2 &          1 &          1 &          0 \\
732 &          1 &          4 &        101 & 18499.05736 &         -2 &         -2 &         -2 &          1 &          1 &          0 \\
733 &          2 &          1 &          1 &  131.63473 &          0 &          0 &          0 &          2 &          0 &          2 \\
734 &          2 &          1 &          2 &  223.86306 &          0 &          0 &          0 &          2 &          2 &          0 \\
735 &          2 &          1 &          3 & 2318.35091 &          0 &          1 &          0 &          2 &          0 &          2 \\
736 &          2 &          1 &          4 & 2427.09231 &          0 &          1 &          0 &          2 &          2 &          0 \\
737 &          2 &          1 &          5 & 2490.93374 &          0 &          0 &          1 &          2 &          1 &          2 \\
738 &          2 &          1 &          6 & 3123.27957 &          1 &          0 &          0 &          2 &          0 &          2 \\
739 &          2 &          1 &          7 & 3209.80678 &          1 &          0 &          0 &          2 &          2 &          0 \\
740 &          2 &          1 &          8 & 4407.78525 &          0 &          2 &          0 &          2 &          0 &          2 \\
741 &          2 &          1 &          9 & 4512.34757 &          0 &          2 &          0 &          2 &          2 &          0 \\
742 &          2 &          1 &         10 & 4563.25961 &          0 &          1 &          1 &          2 &          1 &          2 \\
743 &          2 &          1 &         11 & 4761.25784 &          0 &          0 &          2 &          2 &          0 &          2 \\
744 &          2 &          1 &         12 & 4845.07742 &          0 &          0 &          2 &          2 &          2 &          0 \\
745 &          2 &          1 &         13 & 5157.02577 &          1 &          1 &          0 &          2 &          0 &          2 \\
746 &          2 &          1 &         14 & 5256.75108 &          1 &          1 &          0 &          2 &          2 &          0 \\
747 &          2 &          1 &         15 & 5392.06999 &          1 &          0 &          1 &          2 &          1 &          2 \\
748 &          2 &          1 &         16 & 6006.36278 &          2 &          0 &          0 &          2 &          0 &          2 \\
749 &          2 &          1 &         17 & 6088.92801 &          2 &          0 &          0 &          2 &          2 &          0 \\
750 &          2 &          1 &         18 & 6404.39091 &          0 &          3 &          0 &          2 &          0 &          2 \\
751 &          2 &          1 &         19 & 6514.12841 &          0 &          3 &          0 &          2 &          2 &          0 \\
752 &          2 &          1 &         20 & 6536.82044 &          0 &          2 &          1 &          2 &          1 &          2 \\
753 &          2 &          1 &         21 & 6730.08074 &          0 &          1 &          2 &          2 &          0 &          2 \\
754 &          2 &          1 &         22 & 6881.05077 &          0 &          1 &          2 &          2 &          2 &          0 \\
755 &          2 &          1 &         23 & 6988.69291 &         -2 &         -2 &         -2 &          2 &         -2 &         -2 \\
756 &          2 &          1 &         24 & 7123.16702 &          1 &          2 &          0 &          2 &          0 &          2 \\
757 &          2 &          1 &         25 & 7211.77759 &          1 &          2 &          0 &          2 &          2 &          0 \\
758 &          2 &          1 &         26 & 7333.23873 &         -2 &         -2 &         -2 &          2 &         -2 &         -2 \\
759 &          2 &          1 &         27 & 7564.78576 &         -2 &         -2 &         -2 &          2 &         -2 &         -2 \\
\hline
\end{tabular*}
\end{table}

\begin{table}[!t]
\centering %
\caption{Sample of the \STo\ transitions file. The first two columns are for the indices of the two
levels in the levels file, while the third column is for the $\EAC$ coefficients in s$^{-1}$.}
\vspace{0.5cm}
\label{sampleTraTable} %
\begin{tabular}{rrr}
\hline
30589 &\hspace{0.75cm}      29553 &\hspace{0.75cm}   7.99E-04 \\
19648 &      18049 &   8.37E-03 \\
8943 &       7423 &   5.55E-01 \\
8490 &       7981 &   2.18E-03 \\
20620 &      22169 &   6.91E-04 \\
17613 &      15937 &   5.62E-03 \\
13046 &      11400 &   1.15E+00 \\
20639 &      20054 &   7.26E-03 \\
14433 &      17117 &   2.49E-03 \\
25960 &      28074 &   1.92E-03 \\
10371 &       8748 &   3.80E-02 \\
10380 &      12978 &   2.61E-04 \\
6353 &       6778 &   4.34E-04 \\
12834 &      14507 &   8.68E-04 \\
19391 &      18843 &   2.05E-03 \\
29648 &      27652 &   9.58E-05 \\
25891 &      26331 &   3.67E-04 \\
16727 &      19473 &   1.23E-03 \\
17387 &      14665 &   1.63E-03 \\
9227 &      11797 &   9.18E-06 \\
\hline
\end{tabular}
\end{table}

\section{Validation and Primary Results}

In this section we present a number of comparisons that we made between the \STo\ list and data
sets found in the literature to validate our list. We also present some preliminary results that we
obtained from \STo.

\subsection{Comparison to Experimental and Theoretical Data}

To validate our results, a number of comparisons between the \STo\ line list and experimental data
found in the literature were carried out. The main data sources of laboratory measurements that
have been used in these comparisons are presented in Table~\ref{ComparisonTable}. The table also
gives statistical information about the discrepancy between the calculated line frequencies and
their experimental counterparts. Table~\ref{asvanyTable} presents a rather detailed account of this
comparison for a sample data extracted from one of these data sets, specifically that of Asvany
\etal\ \cite{AsvanyHMKSTS2007}. This table also contains a comparison of relative Einstein \EBC\
coefficients between theoretical values from \STo\ and measured values from Asvany \etal. The
theoretical values in this table are obtained from the calculated \EAC\ coefficients using the
relation \cite{TennysonBook2005}
\begin{equation}\label{ebcEq}
    \EBC_{lu}=\frac{(2J'+1)\C^{3}\EAC_{ul}}{(2J''+1)8\pi \h\F^{3}}
\end{equation}
where $\EAC_{ul}$ and $\EBC_{lu}$ are the Einstein A and B coefficients respectively for transition
between upper ($u$) and lower ($l$) states, $J'$ and $J''$ are the rotational quantum numbers for
upper and lower states, $\C$ is the speed of light, $\h$ is \Planck's constant, and $\F$ is the
transition frequency.

Tables \ref{asvanyTable1} and \ref{asvanyTable2} present more comparisons to Asvany \etal\
\cite{AsvanyHMKSTS2007} data. As seen, in all cases the \STo\ values agree very well with the
measured values of Asvany \etal\ within acceptable experimental errors. Other comparisons to
previous theoretical data, such as that of Polyansky and Tennyson \cite{PolyanskyT1999}, were also
made and the outcome was satisfactory in all cases.

\begin{table}[!t]
\centering %
\caption{The main experimental data sources used to validate the \STo\ list. Columns 2 and 3 give
the number of data points and the frequency range of the experimental data respectively, while the
last four columns represent the minimum, maximum, average, and standard deviation of discrepancies
(i.e. observed minus calculated) in \wn\ between the \STo\ and the experimental data sets.}
\vspace{0.2cm}
\label{ComparisonTable} %
{\normalsize
\begin{tabular*}{\textwidth}{l@{\extracolsep{\fill}}ccrrrc}
\hline
    Source &        $N$ & Range (\wn) &       Min. &       Max. & Ave. & SD \\
\hline
Shy \etal\ \cite{ShyFW1981}&          9 & 1837 -- 1953 &     $-$0.014 &      0.116 & 0.033 & 0.052 \\
Amano and Watson \cite{AmanoW1984}&         27 & 2839 -- 3179 &     $-$0.315 &      0.054 & 0.033 & 0.065 \\
Amano \cite{Amano1985}&         37 & 2839 -- 3209 &     $-$0.024 &      0.054 & 0.022 & 0.019 \\
Foster \etal\ \cite{FosterMPWPe1986}&   73 & 1837 -- 2603 &     $-$0.134 &      0.213 & 0.072 & 0.067 \\
F\'{a}rn\'{\i}k \etal\ \cite{FarnikDKPTN2002}&   8 & 4271 -- 4539 &      0.046 &      0.172 & 0.104 & 0.050 \\
Asvany \etal\ \cite{AsvanyHMKSTS2007}& 25 & 2946 -- 7106 &      0.008 &      0.242 & 0.120 & 0.088 \\
\hline
\end{tabular*}
}
\end{table}


\begin{table}[!t]
\centering %
\caption{Comparison between measured (Asvany \etal\ \cite{AsvanyHMKSTS2007}) and calculated (ST1)
frequencies and relative Einstein $\EBC$ coefficients for a number of transition lines of \htdp.
These coefficients are normalized to the last line in the table. The absolute B coefficients, as
obtained from the A coefficients of \STo\ list using Equation~\ref{ebcEq}, are also shown in column
5 as multiples of $10^{14}$ and in units of m.kg$^{-1}$.}
\label{asvanyTable} %
\vspace{0.2cm} %
{\normalsize
\begin{tabular*}{\textwidth}{c@{\extracolsep{\fill}}c|rrc|lr}
\hline
\multicolumn{ 2}{c|}{{\bf Transition}} & \multicolumn{ 2}{c}{{\bf Freq. (\wn)}} & {\bf $\EBC$} & \multicolumn{ 2}{c}{{Relative \bf $\EBC$}} \\

      Vib. &       Rot. &        Obs. &        \STo\ &     \STo\   &        Obs. &        \STo\ \\
\hline
   (0,3,0) & \rotn 110 $\leftarrow$ \rotn 111 &   6303.784 &   6303.676 &       8.05 &       0.29 &       0.29 \\
   (0,3,0) & \rotn 101 $\leftarrow$ \rotn 000 &   6330.973 &   6330.850 &       8.59 & 0.32$\pm$0.02 &       0.31 \\
   (0,2,1) & \rotn 000 $\leftarrow$ \rotn 111 &   6340.688 &   6340.456 &       7.36 & 0.27$\pm$0.03 &       0.27 \\
   (0,2,1) & \rotn 202 $\leftarrow$ \rotn 111 &   6459.036 &   6458.794 &       9.17 & 0.35$\pm$0.04 &       0.34 \\
   (0,2,1) & \rotn 111 $\leftarrow$ \rotn 000 &   6466.532 &   6466.300 &       27.3 &          1.00 &       1.00 \\
\hline
\end{tabular*}
}
\end{table}


\begin{table}[!t]
\centering %
\caption{Asvany \cite{AsvanyHMKSTS2007} data for the second overtone and combination transitions of
\htdp\ compared to the corresponding data as calculated from the \STo\ list.}
\label{asvanyTable1} %
\vspace{0.2cm} %
{\normalsize
\begin{tabular*}{\textwidth}{r@{\extracolsep{\fill}}r|rr|rr|rr}
\hline
\multicolumn{ 2}{c|}{{\bf Transition}} & \multicolumn{ 2}{c|}{{\bf $\F$ (\wn)}} & \multicolumn{ 2}{c|}{{\bf $\EAC$ (s$^{-1})$}} & \multicolumn{ 2}{c}{{\bf $\EBC$ (s$^{-1})$}} \\
      Vib. &       Rot. &        Asv &        \STo\ &        Asv &        \STo\ &        Asv &        \STo\ \\
\hline
   (0,3,0) & \rotn 000 $\leftarrow$ \rotn 101 &   6241.966 &   6241.854 &       7.08 &       7.08 &       0.21 &       0.21 \\
   (0,3,0) & \rotn 111 $\leftarrow$ \rotn 110 &   6270.392 &   6270.272 &       2.13 &       2.13 &       0.19 &       0.19 \\
   (0,3,0) & \rotn 110 $\leftarrow$ \rotn 111 &   6303.784 &   6303.676 &       3.36 &       3.36 &       0.29 &       0.29 \\
   (0,3,0) & \rotn 101 $\leftarrow$ \rotn 000 &   6330.973 &   6330.850 &       1.21 &       1.21 &       0.31 &       0.31 \\
           &            &            &            &            &            &            &            \\
   (0,2,1) & \rotn 000 $\leftarrow$ \rotn 111 &   6340.688 &   6340.456 &       9.36 &       9.37 &       0.27 &       0.27 \\
   (0,2,1) & \rotn 101 $\leftarrow$ \rotn 110 &   6369.460 &   6369.219 &       6.04 &       6.04 &       0.51 &       0.51 \\
   (0,2,1) & \rotn 110 $\leftarrow$ \rotn 101 &   6433.742 &   6433.514 &       4.64 &       4.64 &       0.38 &       0.38 \\
   (0,2,1) & \rotn 202 $\leftarrow$ \rotn 111 &   6459.036 &   6458.794 &       2.47 &       2.47 &       0.34 &       0.34 \\
   (0,2,1) & \rotn 111 $\leftarrow$ \rotn 000 &   6466.532 &   6466.300 &       4.10 &       4.10 &       1.00 &       1.00 \\
   (0,2,1) & \rotn 303 $\leftarrow$ \rotn 212 &   6483.576 &   6483.335 &       3.86 &       3.86 &       0.44 &       0.44 \\
   (0,2,1) & \rotn 212 $\leftarrow$ \rotn 101 &   6491.349 &   6491.124 &       4.49 &       4.49 &       0.60 &       0.60 \\
   (0,2,1) & \rotn 221 $\leftarrow$ \rotn 110 &   6573.837 &   6573.616 &       3.64 &       3.64 &       0.47 &       0.47 \\
   (0,2,1) & \rotn 220 $\leftarrow$ \rotn 111 &   6589.412 &   6589.193 &       2.49 &       2.49 &       0.32 &       0.32 \\
           &            &            &            &            &            &            &            \\
   (1,2,0) & \rotn 000 $\leftarrow$ \rotn 101 &   6945.877 &   6945.831 &      10.25 &      10.30 &       0.22 &       0.23 \\
   (1,2,0) & \rotn 111 $\leftarrow$ \rotn 110 &   6974.252 &   6974.199 &       5.54 &       5.54 &       0.36 &       0.36 \\
   (1,2,0) & \rotn 110 $\leftarrow$ \rotn 111 &   7004.803 &   7004.759 &       5.10 &       5.10 &       0.33 &       0.33 \\
   (1,2,0) & \rotn 101 $\leftarrow$ \rotn 000 &   7039.362 &   7039.309 &       3.72 &       3.72 &       0.70 &       0.70 \\
   (1,2,0) & \rotn 212 $\leftarrow$ \rotn 111 &   7066.839 &   7066.777 &       3.60 &       3.60 &       0.37 &       0.37 \\
   (1,2,0) & \rotn 202 $\leftarrow$ \rotn 101 &   7077.529 &   7077.470 &       4.05 &       4.05 &       0.42 &       0.42 \\
   (1,2,0) & \rotn 211 $\leftarrow$ \rotn 110 &   7105.518 &   7105.464 &       3.38 &       3.38 &       0.35 &       0.35 \\
\hline
\end{tabular*}
}
\end{table}


\begin{table}[!t]
\centering %
\caption{Asvany \cite{AsvanyHMKSTS2007} data for the transitions of the (1,0,0) band of \htdp\
detected with an optical parametric oscillator systems compared to the corresponding data as
calculated from the \STo\ list.}
\label{asvanyTable2} %
\vspace{0.2cm} %
{\normalsize
\begin{tabular*}{\textwidth}{r|@{\extracolsep{\fill}}rr|rr|rr}
\hline
{\bf Transition} & \multicolumn{ 2}{c|}{{\bf $\F$ (\wn)}} & \multicolumn{ 2}{c|}{{\bf $\EAC$ (s$^{-1}$)}} & \multicolumn{ 2}{c}{{\bf $\EBC$ (s$^{-1}$)}} \\
           &        Asv &        \STo\ &        Asv &        \STo\ &        Asv &        \STo\ \\
\hline
\rotn 000 $\leftarrow$ \rotn 101 &   2946.805 &   2946.790 &     53.167 &     53.200 &      0.318 &      0.318 \\
\rotn 110 $\leftarrow$ \rotn 111 &   3003.279 &   3003.267 &     27.509 &     27.500 &      0.466 &      0.465 \\
\rotn 101 $\leftarrow$ \rotn 000 &   3038.182 &   3038.158 &     20.353 &     20.400 &      1.000 &      1.000 \\
\rotn 212 $\leftarrow$ \rotn 111 &   3068.850 &   3068.817 &     20.088 &     20.100 &      0.532 &      0.531 \\
\rotn 220 $\leftarrow$ \rotn 101 &   3164.118 &   3164.110 &     1.5976 &     1.6000 &      0.040 &      0.039 \\
\hline
\end{tabular*}
}
\end{table}

\subsection{Partition Function}

The partition function of a system consisting of an ensemble of particles in thermodynamic
equilibrium is given by
\begin{equation}\label{partEq}
    \pf(\T) = \sum_{i}(2J_i+1)\nsdf_{i}e^{(\frac{-\E_{i}}{\BOLTZ\T})}
\end{equation}
where $i$ is an index running over all energy states of the ensemble, $\E_i$ is the energy of state
$i$ above the ground level which has rotational angular momentum $J_i$, $k$ is Boltzmann's
constant, $T$ is temperature, and $g_i$ is the nuclear spin degeneracy factor which is 1 for para
states and 3 for ortho ones.

Using the \STo\ energy levels we calculated the partition function of \htdp\ for a range of
temperatures and compared the results to those obtained by Sidhu \etal\ \cite{SidhuMT1992}.
Table~\ref{sidhuTable} presents these results for a temperature range of 5 -- 4000~K. The results
are also graphically presented in Figure~\ref{Partition}. The table and figure reveal that although
our results and those of Sidhu \etal\ agree very well at low temperatures (below 1200~K), they
differ significantly at high temperatures and the deviation increases as the temperature rises.
This can be explained by the fact that \STo\ contains more energy levels, particularly at high
energy, which contribute increasingly at high temperatures. Moreover, the Sidhu list has a lower
limit for the upper cut-off energy and hence the levels at the top cannot contribute to the
high-energy transitions beyond a certain limit. We therefore expect our partition function to be
the more reliable one for $T > 1200$~K.

Using a Levenberg-Marquardt nonlinear curve-fitting routine, we fitted the \STo\ partition function
curve in the indicated range to a fifth order polynomial in temperature
\begin{equation}\label{partfit}
    \pf(\T) = \sum_{i=0}^5 a_i \T^i
\end{equation}
and obtained the following coefficients for $\T$ in Kelvin:
{
\begin{eqnarray}
\nonumber  a_0 &=& -0.300315 \\
\nonumber  a_1 &=& +0.094722 \\
\nonumber  a_2 &=& +0.000571 \\
\nonumber  a_3 &=& -3.24415 \times 10^{-7} \\
\nonumber  a_4 &=& +2.01240 \times 10^{-10} \\
           a_5 &=& -1.94176 \times 10^{-14}
\end{eqnarray}}
The fitting curve is not shown on Figure~\ref{Partition} because it is virtually identical to the
\STo\ curve within the graph resolution. The root-mean-square of the percentage error between the
polynomial fit and the line list values, as given in Table~\ref{sidhuTable}, is about 0.18. It
should be remarked that the fitting polynomial is valid only within the fitting temperature range
(5 -- 4000~K) and hence cannot be extrapolated beyond this range. Moreover, we recommend using the
line list directly to obtain more reliable values for the partition function because the purpose of
the fitting polynomial is to show the trend rather than to be a substitute for the line list.

\begin{table}[!t]
\centering %
\caption{The partition function of \htdp\ for a number of temperatures as obtained from Sidhu
\etal\ \cite{SidhuMT1992} and \STo\ line lists.}
\label{sidhuTable} %
\vspace{0.2cm} {\normalsize
\begin{tabular}{|r|rr|r|rr|}
\hline
  $\T$(K) & \multicolumn{ 2}{r|}{Partition Function} &   $\T$(K) & \multicolumn{ 2}{r|}{Partition Function} \\

           &      Sidhu &        \STo\ &            &      Sidhu &        \STo\ \\
\hline
         5 &       1.00 &       1.00 &        800 &     347.58 &     347.53 \\
        10 &       1.01 &       1.01 &        900 &     426.24 &     426.24 \\
        20 &       1.07 &       1.28 &       1000 &     515.43 &     516.31 \\
        30 &       2.15 &       2.15 &       1200 &     731.43 &     737.61 \\
        40 &       3.46 &       3.46 &       1400 &    1004.25 &    1026.84 \\
        50 &       5.05 &       5.05 &       1600 &    1339.43 &    1401.52 \\
        60 &       6.82 &       6.82 &       1800 &    1738.89 &    1881.33 \\
        70 &       8.73 &       8.73 &       2000 &    2203.94 &    2487.98 \\
        80 &      10.76 &      10.76 &       2200 &    2735.31 &    3244.83 \\
        90 &      12.90 &      12.89 &       2400 &    3327.58 &    4176.19 \\
       100 &      15.12 &      15.12 &       2600 &    3983.83 &    5306.36 \\
       150 &      27.56 &      27.55 &       2800 &    4698.51 &    6658.69 \\
       200 &      42.03 &      42.02 &       3000 &            &    8254.62 \\
       300 &      76.49 &      76.46 &       3200 &            &   10112.90 \\
       400 &     117.48 &     117.44 &       3400 &            &   12249.00 \\
       500 &     164.51 &     164.53 &       3600 &            &   14674.90 \\
       600 &     218.11 &     217.98 &       3800 &            &   17398.90 \\
       700 &     278.66 &     278.58 &       4000 &            &   20425.60 \\
\hline
\end{tabular}
}
\end{table}

\begin{figure}[!t]
\centering
\includegraphics[scale=0.6]{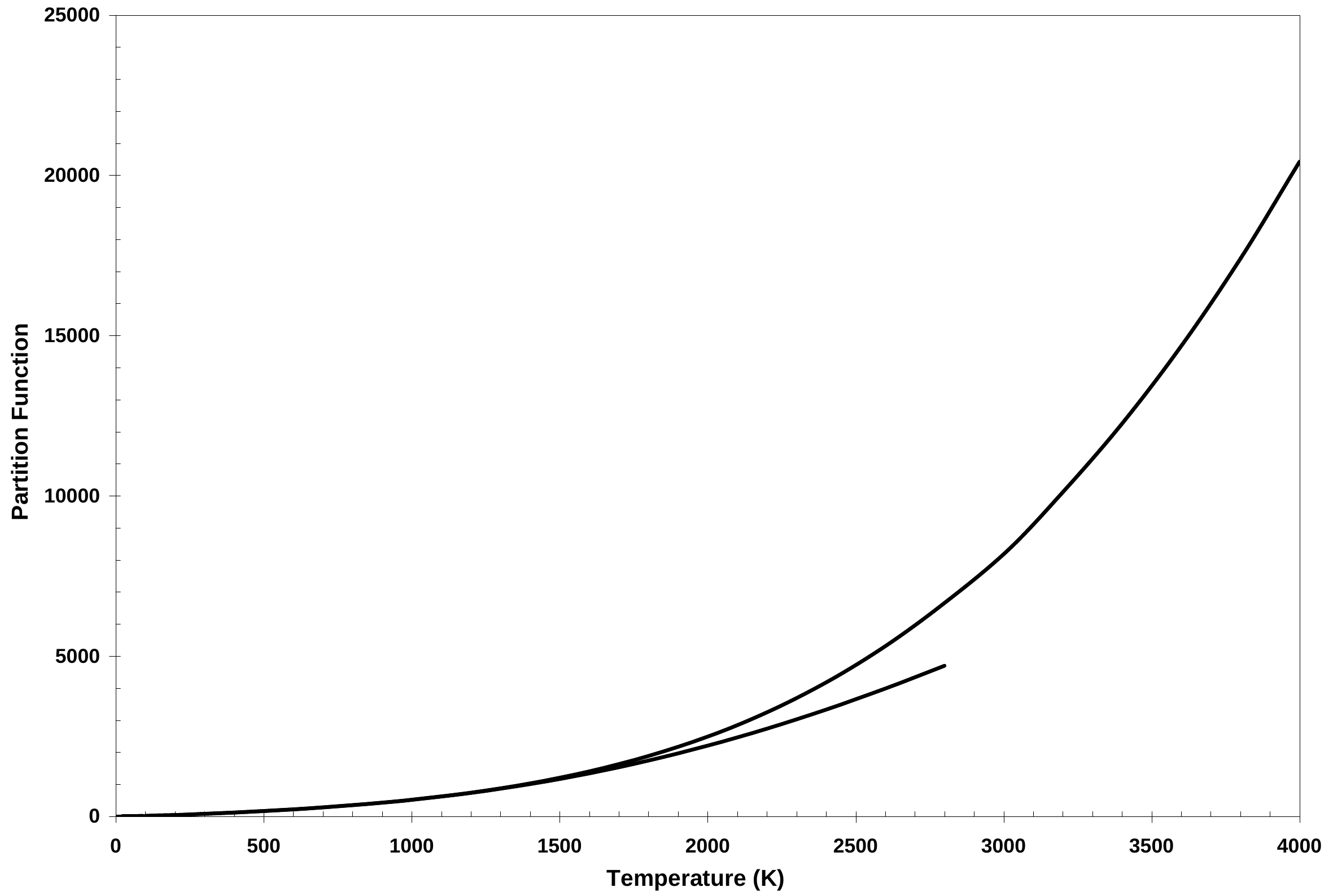}
\caption{The \htdp\ partition functions of \STo\ (upper curve) and Sidhu \etal\ \cite{SidhuMT1992}
(lower curve).} \label{Partition}
\end{figure}

\subsection{Cooling Function}

ST1 was also used to compute the cooling function of \htdp\ as a function of temperature. The
cooling function $W$ which quantifies the rate of cooling per molecule for dense gas in local
thermodynamic equilibrium is given by
\begin{equation} \label{coolEq}
    \cf(\T) = \frac{1}{\pf} \left( \sum_{u,l} \EAC_{ul} (\E_{u}-\E_{l}) (2J_{u}+1)\nsdf_{u}e^{(\frac{-\E_{u}}{\BOLTZ\T})} \right)
\end{equation}
where $\pf$ is the partition function as given by Equation~\ref{partEq}, $\EAC$ is the spontaneous
transition probability, $\E$ is the energy of the state above the ground level, $J$ is the
rotational angular momentum quantum number, $k$ is Boltzmann's constant, $T$ is the temperature,
$g$ is the nuclear spin degeneracy factor which, for \htdp, is 1 for para states and 3 for ortho
states, and $u$ and $l$ stand for the upper and lower levels respectively.

Figure~\ref{Cooling} graphically presents \STo\ cooling function (upper curve) as a function of
temperature for a certain temperature range alongside the cooling function of Neale \etal\
\cite{NealeMT1996} for \htp\ ion as obtained from a digitized image. As seen, the two curves match
very well on the main part of the common temperature range ($T > 600$~K). However the \htdp\
cooling function continues to be significant at lower temperatures, whereas at these temperatures
the cooling curve of \htp\ was not given by Neale \etal\ \cite{NealeMT1996} because they considered
it too small to be of significance. This shows why the cooling properties of \htdp\ could be
important in the cold interstellar clouds. The larger values of \STo\ can be explained by the
inclusion of more energy states and the higher upper cut-off energy of this list. It should be
remarked that Miller \etal\ \cite{MillerSMT2012} have recently improved significantly the \htp\
cooling function of Neale \etal\ \cite{NealeMT1996} by improving the fit to the calculated values
of the partition function used in the original computations.

Using a Levenberg-Marquardt curve-fitting routine, we fitted our \htdp\ cooling function curve to a
fourth order polynomial in temperature over the range 10 -- 4000~K and obtained
\begin{equation} \label{coolfit}
    \cf(\T) =  \sum_{i=0}^4 b_i \T^i
\end{equation}
where
\begin{eqnarray}
\nonumber  b_0 &=& -1.34302 \times 10^{-16} \\
\nonumber  b_1 &=& +1.56314 \times 10^{-17} \\
\nonumber  b_2 &=& -2.69764 \times 10^{-19} \\
\nonumber  b_3 &=& +7.03602 \times 10^{-22} \\
           b_4 &=& -1.10821 \times 10^{-25}
\end{eqnarray}
The fit was essentially perfect on a linear scale.

\begin{figure}[!t]
\centering
\includegraphics[scale=0.8]{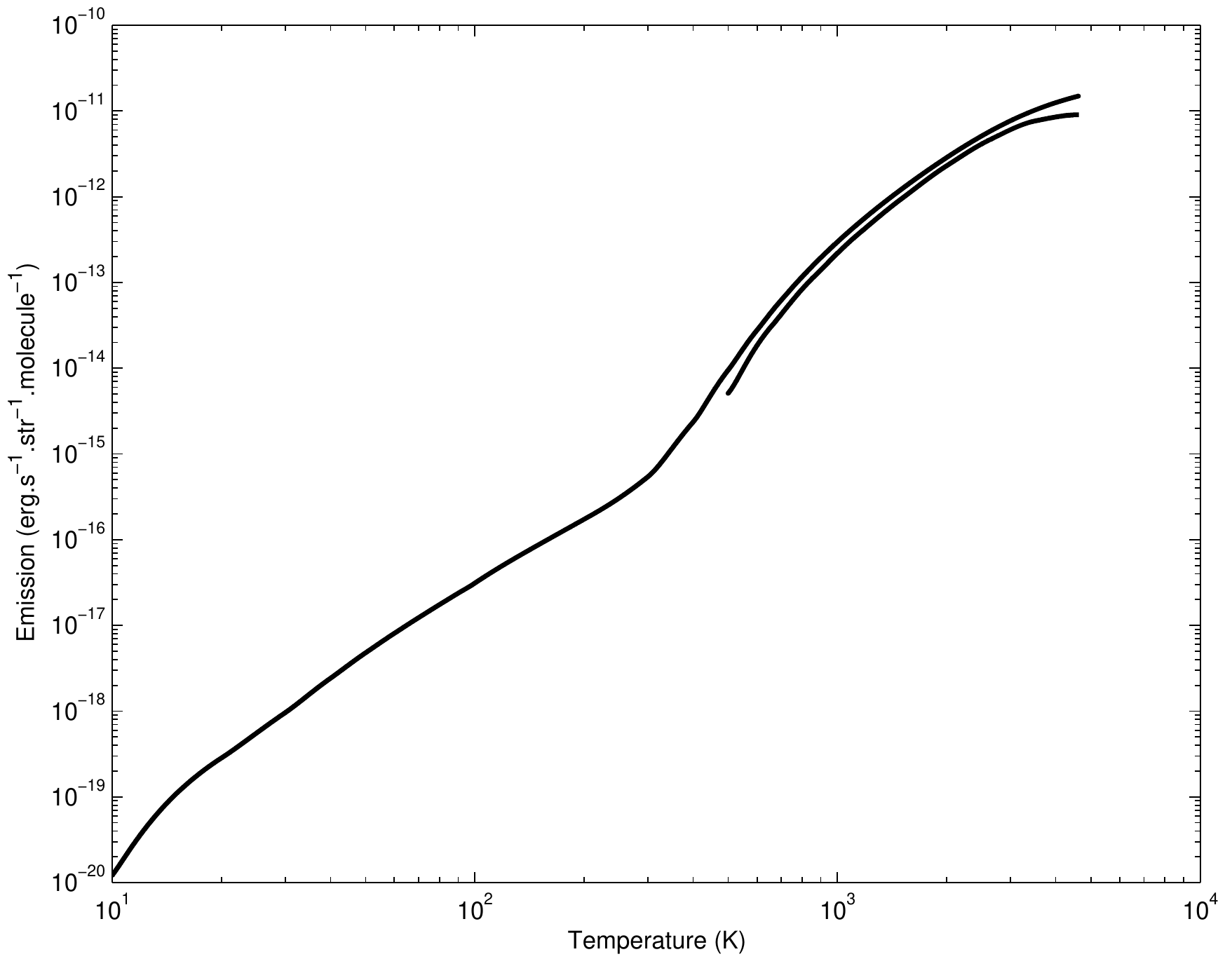}
\caption{A graph of per molecule emission of \htdp\ (upper curve) and \htp\ (lower curve) as a
function of temperature on a log-log scale. The \htdp\ curve is obtained from \STo\ while the \htp\
curve is obtained from a digitized image from Neale \etal\ \cite{NealeMT1996}.} \label{Cooling}
\end{figure}

\subsection{Synthetic Spectra}

One of the main uses of line lists such as \STo\ is to generate temperature-dependent synthetic
spectra. These spectra were generated from \STo\ using the Spectra-BT2 code, which is described by
Barber \etal\ \cite{BarberTHT2006} and is a modified version of the original spectra code of
Tennyson \etal\ \cite{TennysonMS1993}. A Gaussian line profile with 10 data points per \wn\ was
used in these computations. In Figure~\ref{SynthSpec} a sample of these spectra for the
temperatures $T=100, 500, 1000$ and 2000~K over the frequency range $0-10000$~\wn\ is presented.
Figure~\ref{SynthSpec} (a) shows a temperature-dependent \htdp\ absorbtion spectrum for the region
largely associated with pure rotational transitions, while Figures~\ref{SynthSpec} (b), (c) and (d)
show the corresponding spectrum for the vibrational region. As is usual with rotation-vibration
spectra, \htdp\ spectra show a strong dependence on temperature.

\begin{figure}[!h]
\centering
\includegraphics[width=1.0\textwidth]{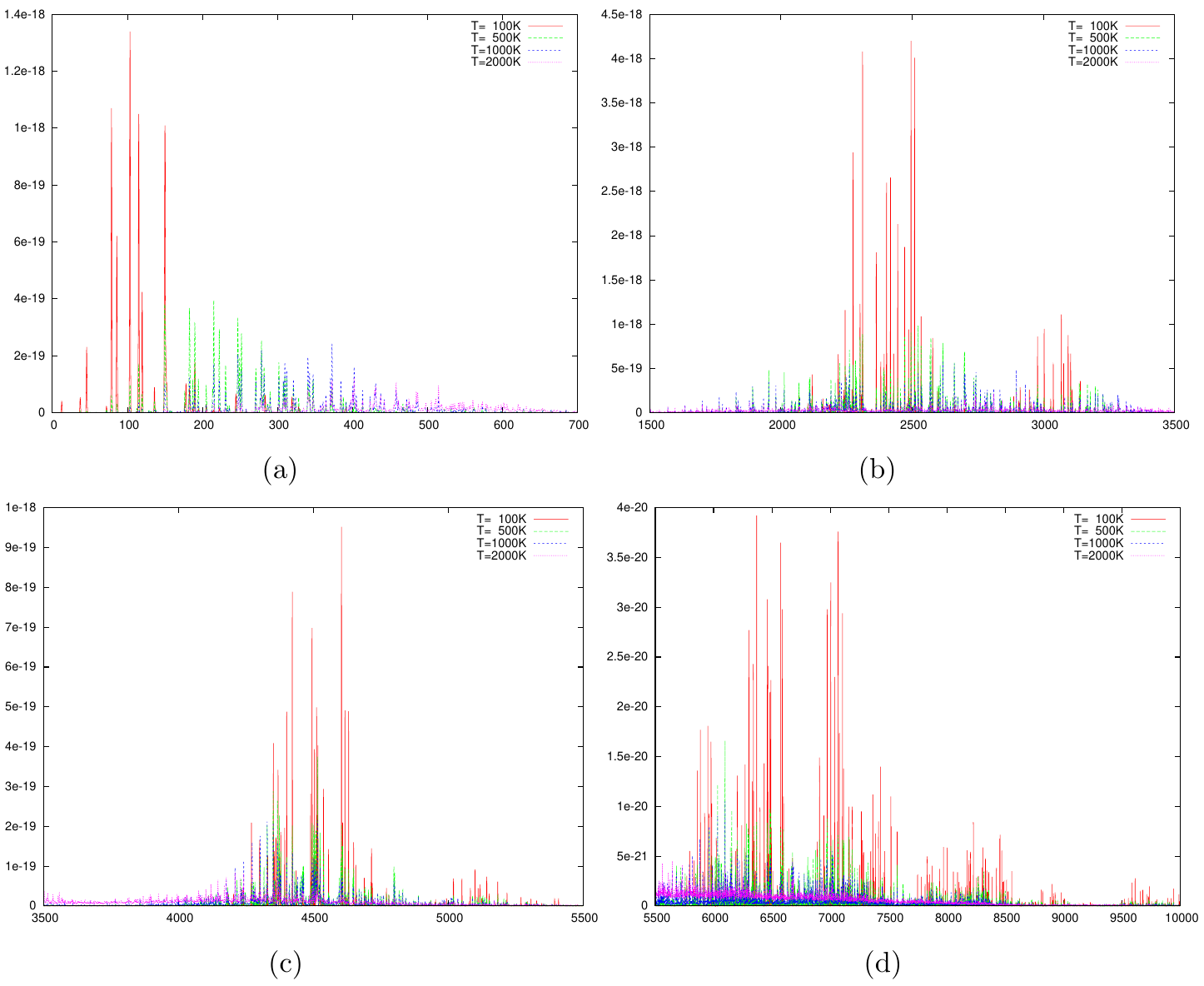}
\caption{A graph of integrated absorption coefficient in cm.molecule$^{-1}$ (on $y$ axis) as a
function of wave number in \wn\ (on $x$ axis) within the range $0-10000$~\wn\ for $\T=$ 100, 500,
1000 and 2000~K.} \label{SynthSpec}
\end{figure}

\chapter{Conclusions and Future Work} \label{Conclusions} 
%
\section{Conclusions} \label{Conclusions}

In the first part of our atomic investigation, a list of line effective recombination coefficients
is generated for the atomic ion C$^+$ using the \Rm-matrix \cite{BerringtonEN1995}, \AS\
\cite{EissnerJN1974, NussbaumerS1978, BadnellAS2008} and Emissivity \cite{SochiEmis2010} codes.
These lines are produced by dielectronic capture and subsequent radiative decays of the low-lying
autoionizing states above the threshold of C$^{2+}$ \nlo1s2\nlo2s2 \SLP1Se with a principal quantum
number $n<5$ for the captured electron. The line list, called \SSo, contains 6187 optically-allowed
transitions which include many C~{\sc ii} lines observed in astronomical spectra. Beside the
effective recombination coefficients, the list also includes detailed data of level energies for
bound and resonance states, and oscillator strengths. The theoretical results for energy and
\finstr\ splitting agree very well with the available experimental data for both resonances and
bound states. In the course of this investigation, a novel method, namely the \Km-matrix method,
for finding and analyzing resonances was developed as an alternative to the \QB\
\cite{QuigleyB1996, QuigleyBP1998} and Time-Delay \cite{StibbeT1998} methods. The \Km-matrix and
\QB\ methods produce virtually identical results; however, as far as the one-channel resonances
related to the search for the low-lying autoionizing states is concerned, the \Km-matrix method
offers a superior alternative to the \QB\ method in terms of numerical stability, computational
viability and comprehensiveness. The big advantage of \QB\ is that it is more general and can be
used for multi-channel as well as single-channel resonances. The \Km-matrix method was implemented,
elaborated and used in the production of the atomic data leading to this investigation. The
theoretical results of our line list were validated by a number of tests such as comparing the
energy and \finstr\ splitting to the available experimental data \cite{NIST2010} of autoionizing
and bound states. A conclusive comparison to theoretical data previously reported in the
literature; which include radiative transition probabilities, effective dielectronic recombination
rate coefficients and autoionization probabilities; was carried out to verify our data. All these
tests indicate the reliability of our line list. We expect \SSo\ to fill some of the existing gaps
in the theoretical data needed for spectroscopic and astronomical applications.

In the second part of our atomic investigation, we studied the electron temperature of the emitting
region of recombination lines in a number of astronomical objects, mainly planetary nebulae, by
analyzing the transition lines of C\II\ which originate from dielectronic recombination processes
from the low-lying autoionizing states and subsequent cascade decays. The investigation is based on
the use of theoretical data from SS1 list, and observational data from several sources in the
refereed literature. In the emissivity analysis we used a least squares minimization method to find
the electron temperature of the line emitting region in the investigated objects. Our results were
broadly consistent with the reported findings of other studies in the literature, which are mainly
obtained from other spectral lines with the use of different analysis techniques. The investigation
was concluded by proposing and applying a method for determining electron energy distribution from
observational data supplemented by theoretical data from SS1 to investigate the long-standing
problem in the nebular studies related to the dichotomy between the results of abundance and
temperature obtained from ORLs and CELs. Although no definite conclusion has been reached, due to
the limited quantity of the available pertinent observational data, the observational data seem to
suggest a Maxwell-Boltzmann electron distribution opposite to what has been suggested recently in
the literature of a $\kappa$-distribution.

In the molecular part of this study we investigated the generation of a new line list, called \STo,
for the highly-important triatomic ion \htdp. The list, which can be obtained from the Centre de
Donn\'{e}es astronomiques de Strasbourg (CDS) database, comprises over 33 thousand
rotational-vibrational energy levels and more than 22 million transition lines archived within two
files. Various rigorous tests were performed during all stages of production and after. All these
tests confirmed that, although ST1 is based entirely on \abin\ quantum mechanics, it is
sufficiently accurate for almost all astronomical purposes. The one possible exception is for
predicting the frequency of pure rotational transitions which are often needed to high accuracy and
which are therefore better obtained from measured frequencies using combination differences. A
number of comparisons to the available experimental and theoretical data were made. Several primary
results, which include partition and cooling functions as well as temperature-dependent synthetic
spectra, were also drawn from this list and presented in this thesis. These comparisons and results
also confirmed the reliability of the list. We expect \STo\ to be a helpful tool in analyzing and
understanding physical processes in various \htdp\ systems at spectroscopic and astronomical
levels.

\section{Recommendations for Future Work} \label{Recommendations}
There are many aspects of the current work that can be elaborated and extended in the future; some
of these prospectives are outlined in the following points

\begin{itemize}

\item
Extending the theoretical and computational treatments of the atomic investigation by employing
more comprehensive theory to incorporate processes and phenomena other than dielectronic
recombination, such as \radrec\ and collisional processes. Some of these treatments are already
developed and implemented within existing codes while others require developing novel tools.

\item
Extending the \Km-matrix method to include multi-channel resonances and implementing it within the
\Rm-matrix code or as a stand alone code.

\item
Pushing the current limit on the \priquanum\ of the active electron in the autoionizing states
beyond the $n<5$ condition which is adopted in the present work. This extension becomes more
important at higher temperatures, as the contribution of resonances arising from the $n \geq 5$
levels increases.

\item
Extending the electron temperature analysis to more data sets and including even the bound-bound
transition lines. This depends on the availability of a more comprehensive C\II\ list generated on
the base of a more comprehensive theoretical foundation which includes processes other than
dielectronic recombination, as indicated earlier.

\item
Improving the least squares method for electron temperature analysis and performing more tests and
validations by application to more data sets to identify and fix any possible weaknesses. One of
the objectives is to reduce the temperature dependency of the $\chi^2$ curve shape which affects
the confidence interval.

\item
Extending the \htdp\ line list beyond its current limits of $J\leq 20$ for rotational quantum
number and $\F<18500$~\wn\ for frequency.

\end{itemize}

%


} 

\newpage

{\singlespace
\phantomsection \addcontentsline{toc}{chapter}{\protect \numberline{} Bibliography} %
\bibliographystyle{unsrt}

}

\appendix

\chapter[Using Lifetime Matrix to Investigate Single-Channel Resonances]{} \label{AppKmatrix}

\begin{spacing}{2}
{\LARGE \bf Using Lifetime Matrix to Investigate Single-Channel
Resonances}
\end{spacing}
\vspace{1.0cm}
\noindent In this Appendix we present the \Km-matrix method of P.J. Storey which is based on using
the lifetime matrix $\Mm$ expressed in terms of the reactance matrix $\Km$ to investigate
single-channel resonances. This method is used in this study to investigate the low-lying
autoionizing states of C$^+$.

According to Smith \cite{Smith1960}, the \Mm-matrix is given, in terms of the scattering matrix
\Sm, by
\begin{equation}\label{QmatrixS1}
    \Mm = -\iu \, \Dirac \, \Sm^{*} \frac{d \Sm}{d \E}
\end{equation}
where \Sm-matrix is defined, in terms of \Km-matrix, by
\begin{equation}\label{SmatrixEq3}
    \Sm = \frac{\IM + \iu \Km}{\IM - \iu \Km}
\end{equation}
with \IM\ being the identity matrix. In the case of single-channel states, \Mm, \Sm\ and \Km\ are
one-element matrices. To indicate this fact we annotate them with $\M$, $\Ss$ and $\K$. From
\ref{QmatrixS1} and \ref{SmatrixEq3}, the following relation may be derived
\begin{equation}\label{QmatrixK}
    \M = \frac{2}{1 + \K^{2}}  \D\K\E
\end{equation}
where $\Dirac = 1$ in atomic units. It is noteworthy that since $\K$
is real, $\M$ is real as it should be.

Smith \cite{Smith1960} has demonstrated that the expectation value of $\M$ is the lifetime of the
state, $\LTS$. Now if $\K$ has a simple pole at energy $\Eo$, given by
\begin{equation}\label{KgEEo}
    \K = \frac{\PKF}{\E - \Eo}
\end{equation}
where $\PKF$ is a parameter with dimensions of energy, then from \ref{QmatrixK} we obtain
\begin{eqnarray}\label{QE}
    \M(\E) &=& \frac{-2 \PKF}{(1 + \K^{2}) (\E - \Eo)^{2}} \nonumber \\
           &=& \frac{-2 \PKF}{(\E - \Eo)^{2} + \PKF^{2}}
\end{eqnarray}
and hence the maximum value of $\M(\E)$ is $\M(\Eo) = -2 / \PKF$. On substituting the half maximum,
$-1/\PKF$, into Equation~\ref{QE} we find
\begin{equation}\label{EEo}
    \E - \Eo = \pm \PKF
\end{equation}
and therefore, the full width at half maximum is
\begin{equation}\label{FWHMe}
    \FWHMe = | 2 \PKF |
\end{equation}

According to Smith \cite{Smith1960}, the lifetime of the state $\LTS$, and the \fwhm\ $\FWHMe$ are
related by
\begin{equation}\label{LTSFWHM}
    \LTS = \frac 1 \FWHMe  \verb|       |(\AU)
\end{equation}
Hence, for a \Km-matrix pole given by \ref{KgEEo} the autoionization probability, $\aTP$, which by
definition equals the reciprocal of the lifetime, is given in atomic units by
\begin{equation}\label{DPLTS}
    \aTP = \frac 1 \LTS = \FWHMe = | 2 \PKF |  \verb|       |(\AU)
\end{equation}
Now if we consider a \Km-matrix with a pole superimposed on a
background $\BGCK$
\begin{equation}\label{KwithBG}
    \K = \BGCK + \frac{\PKF}{\E - \Eo}
\end{equation}
then from \ref{QmatrixK} we find
\begin{eqnarray}\label{QwithBG}
    \M(\E) &=& \frac{-2 \PKF}{(1 + \K^{2}) (\E - \Eo)^{2}} \nonumber \\
           &=& \frac {-2 \PKF} {(1 + \BGCK^{2}) (\E - \Eo)^{2} + 2 \BGCK \PKF (\E - \Eo) + \PKF^{2}}
\end{eqnarray}

The maximum value of $\M(\E)$ occurs when the denominator has a
minimum, that is when
\begin{equation}\label{Emax}
    \E = \Eo - \frac {\BGCK \PKF} {1 + \BGCK^{2}}
\end{equation}
and hence
\begin{equation}\label{Qmax}
    \Mmax = - \frac {2 (1 + \BGCK^{2})} {\PKF}
\end{equation}
This reveals that by including a non-vanishing background the peak
of $\M$ is shifted relative to the pole position, $\E = \Eo$, and
the peak value is modified. If we now calculate the \fwhm, $\FWHMe$,
by locating the energies where $\M = \frac{1}{2} \Mmax$ from solving
the quadratic
\begin{equation}\label{quadratic}
    (1 + \BGCK^{2}) (\E - \Eo)^{2}
    + 2 \BGCK \PKF (\E - \Eo)
    - \frac{\PKF^{2} (1 - \BGCK^{2})}{(1 + \BGCK^{2})}
    = 0
\end{equation}
we find
\begin{equation}\label{FWHMwithBG}
    \FWHMe = \frac {|2 \PKF|} {1 + \BGCK^{2}}
\end{equation}
As \fwhm\ and autoionization probability are numerically equal in
atomic units, we obtain
\begin{equation}\label{DPwithBG}
    \aTP = \frac {|2 \PKF|} {1 + \BGCK^{2}} \verb|       |({\AU})
\end{equation}

The two parameters of primary interest are the resonance position $\Er$ and the resonance width
$\RW$. However, for an energy point $\Ei$ with a \Km-matrix value $\SVKi$, Equation~\ref{KwithBG}
has three unknowns, $\BGCK$, $\PKF$ and $\Eo$ which are needed to find $\Er$ and $\RW$, and hence
three energy points at the immediate neighborhood of $\Eo$ are required to identify the unknowns.
Since the \Km-matrix changes sign at the pole, the neighborhood of $\Eo$ is located by testing the
\Km-matrix value at each point of the energy mesh for sign change and hence the three points are
obtained accordingly. Now, if we take the three consecutive values of $\K$
\begin{equation}\label{K123}
    \K_{i} = \BGCK + \frac{\PKF}{\E_{i} - \Eo} \verb|           | (i=1,2,3)
\end{equation}
and define
\begin{equation}\label{DELEKjk}
    \DELE jk = \E_{j} - \E_{k}
    \verb|       |  \&  \verb|       |
    \DELK jk = \K_{j} - \K_{k}
\end{equation}
then from \ref{K123} we obtain
\begin{equation}\label{KDiff12}
    \DELK 12
    = \PKF \left( \frac{1}{\DELE 10} - \frac{1}{\DELE 20} \right)
    = \PKF \left( \frac{\DELE 21}{\DELE 10 \DELE 20} \right)
\end{equation}
\begin{equation}\label{KDiff23}
    \DELK 23
    = \PKF \left( \frac{1}{\DELE 20} - \frac{1}{\DELE 30} \right)
    = \PKF \left( \frac{\DELE 32}{\DELE 20 \DELE 30} \right)
\end{equation}
and
\begin{equation}\label{DK12byDK23}
    \frac{\DELK 12}{\DELK 23}
    = \frac{\DELE 30 \DELE 21}{\DELE 10 \DELE 32}
\end{equation}
Therefore
\begin{equation}\label{DE30byDE10}
    \frac{\DELE 30}{\DELE 10}
    = \frac{\E_{3} - \Eo}{\E_{1} - \Eo}
    = \frac{\DELK 12 \DELE 32}{\DELK 23 \DELE 21}
\end{equation}

On algebraically manipulating \ref{DE30byDE10} we find
\begin{equation}\label{E0}
\boxed{
    \Eo = \frac{ \E_{1} \DELK 12 \DELE 32 - \E_{3} \DELK 23 \DELE 21}
             {\DELK 12 \DELE 32 - \DELK 23 \DELE 21}
}
\end{equation}
Having located the pole position, $\Eo$, the following relation can be obtained from \ref{KDiff12}
\begin{equation}\label{PKF}
\boxed{
    \PKF
    = \frac{\DELK 12 \DELE 10 \DELE 20}{\DELE 21}
}
\end{equation}
Similarly, from \ref{K123} we obtain
\begin{equation}\label{BGCK}
\boxed{
    \BGCK = \K_{1} - \frac{\PKF}{\DELE 10}
}
\end{equation}
Finally, $\Er$ and $\RW$ can be computed from Equation~\ref{Emax} and Equation~\ref{FWHMwithBG}
respectively.

\chapter[Error Analysis]{} \label{AppErrAna}

\begin{spacing}{2}
{\LARGE \bf Error Analysis}
\end{spacing}
\vspace{1.0cm}
\noindent In this Appendix we present the formulae of P.J. Storey to account for the errors in the
observational data used in the least squares optimization method where the flux of each line is
normalized to the total flux of all lines involved in the optimization process.

In the following we define the symbols used in these formulae

$f_{i}$ is the observed flux of line $i$

$f$ is the sum of observed flux of all $N$ lines involved in the least squares process

$y_{i}^{O}=\frac{f_{i}}{f}$

$y_{i}^{C}$ is the ratio of the theoretical emissivity of line $i$ to the sum of the theoretical
emissivity of all $N$ lines

$\sigma_{y_{i}^{O}}^{2}$ is the variance of the observed ratio $y_{i}^{O}$

$\sigma_{f_{i}}$ is the standard deviation of the observed flux of line $i$

In the following, we adopt a basic definition of $\chi^{2}$ based on a least squares difference and
a statistical weight given by the reciprocal of variance, and hence other factors that may enter in
the definition of $\chi^{2}$, such as scaling factors like degrees of freedom, are not included for
the sake of simplicity. The method therefore is based on minimizing $\chi^{2}$ given by

\begin{equation}
\chi^{2}=\sum_{i=1}^{N}\frac{\left(y_{i}^{O}-y_{i}^{C}\right)^{2}}{\sigma_{y_{i}^{O}}^{2}}
\end{equation}

From the definition of $y_{i}^{O}$ it can be shown that

\begin{equation}
\sigma_{y_{i}^{O}}^{2}=\left(\frac{\partial y_{i}^{O}}{\partial
f_{i}}\right)^{2}\sigma_{f_{i}}^{2}+\left(\frac{\partial y_{i}^{O}}{\partial
f}\right)^{2}\sigma_{f}^{2}=\left(\frac{1}{f}\right)^{2}\sigma_{f_{i}}^{2}+\left(-\frac{f_{i}}{f^{2}}\right)^{2}\sigma_{f}^{2}
\end{equation}

Dividing through by $\left(y_{i}^{O}\right)^{2}$ we get

\begin{equation}
\frac{\sigma_{y_{i}^{O}}^{2}}{\left(y_{i}^{O}\right)^{2}}=\frac{\left(\frac{1}{f}\right)^{2}\sigma_{f_{i}}^{2}}{\left(\frac{f_{i}}{f}\right)^{2}}+\frac{\left(\frac{f_{i}}{f^{2}}\right)^{2}\sigma_{f}^{2}}{\left(\frac{f_{i}}{f}\right)^{2}}
\end{equation}

that is

\begin{equation}
\sigma_{y_{i}^{O}}^{2}=\left(y_{i}^{O}\right)^{2}\left(\frac{\sigma_{f_{i}}^{2}}{f_{i}^{2}}+\frac{\sigma_{f}^{2}}{f^{2}}\right)\end{equation}

Now, since $f_{i}$'s are mutually independent

\begin{equation}
\sigma_{f}^{2}=\sum_{j}^{N}\left(\frac{\partial f}{\partial
f_{j}}\right)^{2}\sigma_{f_{j}}^{2}=\sum_{j}^{N}\sigma_{f_{j}}^{2}
\end{equation}

On substituting, we obtain

\begin{equation}\label{EAf1}
\boxed{\sigma_{y_{i}^{O}}^{2}=\left(y_{i}^{O}\right)^{2}\left(\frac{\sigma_{f_{i}}^{2}}{f_{i}^{2}}+\frac{\sum_{j}^{N}\sigma_{f_{j}}^{2}}{f^{2}}\right)}
\end{equation}

In the following, we investigate three types of error.

\section{Observed Error}

In this case, Equation~\ref{EAf1} should be used as it stands where $\sigma_{f_{i}}$ and
$\sigma_{f_{j}}$ are the errors given by the observer as standard deviations on the observed flux
of lines $i$ and $j$ respectively.

\section{Fixed Percentage Error}

If we assume a constant percentage error on the flux of all lines, that is

\begin{equation}
\sigma_{f_{i}}=\gamma f_{i}\end{equation}

then from Equation~\ref{EAf1} we obtain

\begin{equation}
\sigma_{y_{i}^{O}}^{2}=\left(y_{i}^{O}\right)^{2}\left(\gamma^{2}\frac{f_{i}^{2}}{f_{i}^{2}}+\gamma^{2}\frac{\sum_{j}^{N}f_{j}^{2}}{f^{2}}\right)
\end{equation}

that is

\begin{equation}
\boxed{\sigma_{y_{i}^{O}}^{2}=\left(y_{i}^{O}\right)^{2}\gamma^{2}\left(1+\frac{\sum_{j}^{N}f_{j}^{2}}{f^{2}}\right)}\end{equation}

\section{Poisson Error}

Now if we assume, based on Poisson counting statistics, that the error on the observed flux is
proportional to the square root of flux, i.e.

\begin{equation}
\sigma_{f_{i}}^{2}=\beta f_{i}\end{equation}

where $\beta$ is an unknown proportionality factor with the same dimensionality as $f_{i}$, then
equation \ref{EAf1} becomes

\begin{equation}
\sigma_{y_{i}^{O}}^{2}=\beta\frac{f_{i}^{2}}{f^{2}}\left(\frac{1}{f_{i}}+\frac{1}{f}\right)\end{equation}

that is

\begin{equation}\label{EAPoisEq}
\boxed{ \sigma_{y_{i}^{O}}^{2}=\frac{\beta f_{i}}{f^{2}}\left(1+\frac{f_{i}}{f}\right)}
\end{equation}

Since $\beta$ is unknown, $\chi^{2}$ may be normalized to its value at the minimum to obtain a more
realistic error estimate.

\chapter[Emissivity Program Documentation]{} \label{AppEmissMan} 

\begin{spacing}{2}
{\LARGE \bf Emissivity Program Documentation}
\end{spacing}
\vspace{1.0cm}
\noindent
\EmiCod\ code is a command line program that was developed during this investigation to implement
the atomic transition and emissivity model. Its main functionality is to find all possible
transitions and calculate the emissivity and recombination coefficients of the transition lines
from \dierec\ and cascade decay all the way down to the ground or a metastable state, that is all
free-free, free-bound and bound-bound transitions. \EmiCod\ is written in C++ computer language and
mixes procedural with object oriented programming. The program was compiled successfully with no
errors or warnings using Dev-C++ compiler on Windows, and g++ compiler on Cygwin and Red Hat Linux.
Sample results produced by these three versions were compared and found to be identical. Elaborate
checks were carried out in all stages of writing and debugging the program and the output was
verified. Thorough independent checks on sample emissivity data produced by \EmiCod\ were performed
and found to be consistent.

\section{Theoretical background}

In thermodynamic environments, an excited atomic state is populated by recombination and radiative
decay from accessible higher states, and depopulated by autoionization and radiative decay to lower
states. The population of a resonance state is, therefore, determined by the balance between
recombination, autoionization and radiative decay. In thermodynamic equilibrium situations,
autoionization dominates and the population is described by \Saha\ equation. In non-thermodynamic
equilibrium situations, processes other than autoionization have significant contribution to the
population and hence detailed equilibrium calculations are required to account for these processes.
In these situations, a departure coefficient, defined as the ratio of autoionization probability to
the sum of autoionization and total radiative probabilities, is used to measure the departure from
thermodynamic equilibrium and quantify the contribution of various processes.

The first step in the transition lines calculations is to find the radiative transition
probabilities of all types of transitions (free-free, free-bound and bound-bound) which, for a
given photon energy and statistical weights of the upper and lower states, can be obtained from the
oscillator strengths. The total radiative transition probability of each state can then be obtained
by summing up the transition probabilities of the given state to all optically-accessible lower
states. The population of all states can then be computed. For autoionizing states, the population
is obtained by summing up the \Saha\ capture term, given by Saha population times the departure
coefficient, and the radiative decay term given by the sum over all accessible upper states of the
population of these states times the ratio of the radiative probability of the transition to the
sum of total radiative and autoionizing probabilities of the lower state. For bound states, the
population is simply computed by summing up over all accessible upper states the population of
these states times the ratio of the radiative probability of the transition to the total radiative
probability of the lower state.

On finding all optically-allowed transitions, the emissivity of lines arising from these
transitions can be obtained from
\begin{equation}\label{emissivity1}
    \EMISS_{ul} = \NDu \rTPul E_p
\end{equation}
while the equivalent effective recombination coefficient is computed from
\begin{equation}\label{EmissRecCoeff}
    \RCf = \frac{\EMISS_{ul}} {\NDe \NDi E_p}
\end{equation}
In the last equations, $\NDu$ is the population of upper state, $\rTPul$ is the radiative
transition probability from upper to lower state, $E_p$ is the photon energy, and $\NDe$ and $\NDi$
are the number density of electrons and ions respectively.

This theoretical model, with some other extensions and elaborations, represent the basic framework
of \EmiCod\ code \cite{SochiEmis2010} which was the main tool used by the author of this thesis to
perform the calculations of \dierec\ transition lines using the raw atomic data generated mainly by
\Rm-matrix and \AS\ programs. The \EmiCod\ calculations include identifying the allowed electric
dipole transitions, vacuum and air wavelengths, radiative transition probabilities, dimensional and
normalized emissivities, effective recombination coefficients, decay routes, and so on. The code
can also perform comparison to observational data and analyze the results using a least squares
optimization procedure. Moreover, it can produce electron energy distribution for the FF and FB
transitions of the observed spectra. Some practical details about these calculations and how to be
performed by \EmiCod\ will be highlighted in the subsequent sections of this appendix.

\section{Input and Output Files}

\EmiCod\ code reads from plain text input files and writes the results to a main plain text output
file called `Transitions'. The structure of this output file is explained in section
\ref{TransitionsFile}. Other secondary output data files are also produced for particular purposes.
The required input text files are `Input', `ResEmis', `ELEVEmis' and `FVALUE'. The structure of
these files, apart from the `FVALUE' file which is produced by the \STGBB\ stage of the \Rm-matrix
code and is used by \EmiCod\ program with no modification, apart from possible inclusion of
additional data, is outlined below. Two other input data files are also required if comparison and
analysis to observational data are needed. One of these is a data file that contains astronomical
observations while the other includes mapping information of the observational lines to their
theoretical counterparts using their indices. In the following subsections we outline the files
used by the \EmiCod\ code for input and output.

\subsection{`Input' Data File} \label{InputFile}
The program uses a keyword based input file called `Input' which controls the program flow and its
main parameters. All keywords are optional with no order required. If keywords are omitted, default
values will be used. Comments can be inserted in any part of the input file as long as they do not
intrude between a keyword and its parameter(s) or between the parameters. No commentary line should
be initiated with a keyword. Each keyword must occur first on its line of text followed by its
parameter(s) in the order that will be given. There is no restriction on using spaces and new lines
as long as they obey the aforementioned rules.

\vspace{0.5cm}

\noindent {\bf \large 1. `Temperature(K)'} \vspace{0.3cm}

\noindent The first entry after this keyword is a boolean flag to choose one of the two available
schemes, followed by the number of temperature entries in K. If the flag is on (i.e. `1'), this
must be followed by the initial temperature and the interval between the temperature points in this
order; and if it is off, it should be followed by the temperature points. The first scheme is most
suitable for a large and regularly spaced temperature vector, while the second is more suitable for
small or irregularly spaced temperature vector. If this keyword is omitted, the default is a single
temperature entry of 10000.

\vspace{0.5cm}

\noindent {\bf \large 2. `Ni(m\^\,-3)'} \vspace{0.3cm}

\noindent This is the number density of the ions in m$^{-3}$. If this keyword is omitted, the
default value is $10^{10}$.

\vspace{0.5cm}

\noindent {\bf \large 3. `Ne(m\^\,-3)'} \vspace{0.3cm}

\noindent This is the number density of the electrons in m$^{-3}$. If this keyword is omitted, the
default value is $10^{10}$.

\vspace{0.5cm}

\noindent {\bf \large 4. `ResidualCharge'} \vspace{0.3cm}

\noindent This is the residual charge of the ion. It is required for scaling some of the data
obtained from the \Rm-matrix code output files. If this keyword is omitted, the default value is 2.

\vspace{0.5cm}

\noindent {\bf \large 5. `AstronomicalData'} \vspace{0.3cm}

\noindent This keyword requires a boolean flag to read astronomical data from a data file (`0' for
`No' and `1' for `Yes'). If the flag is on, a second entry, which is the name of the data file to
read the data from, will be required. The structure of the astronomical data file requires the file
to start with a comment that can occupy any number of text lines but it should be terminated with
`EndOfComment' string to mark the end. This should be followed by the transition lines data
organized in sets preferably each set on a single line of text. When the normalization is on, as
will be explained later on, the program will normalize the astronomical data to the first data set.
This facilitates the normalization as the user can simply normalize to any line in the astronomical
data by copying the data of that line and paste at the front. The data for each line usually
consist of the observed and laboratory wavelengths, the observed flux (normally after correcting
for dust extinction and normalizing such that $I({\rm H}_{\beta}) = 100$), the observed error
(standard deviation as percentage), the ionic identification, the type of transition (FF, FB or
BB), the lower spectral term of the transition, the upper spectral term of the transition, the
statistical weight of the lower level, and the statistical weight of the upper level. There is
another option for normalization that is normalizing the observational fluxes to the total flux and
the theoretical emissivities to the total emissivity.

If the user chooses to read astronomical data, the data lines will be inserted between the
theoretical lines in the `Transitions' output file according to their observed wavelength. If this
keyword is omitted, the default value is 0.

\vspace{0.5cm}

\noindent {\bf \large 6. `TestsFlag'} \vspace{0.3cm}

\noindent This keyword requires one entry which is a flag for carrying the first (in-out) and
second (ground-metastable) consistency tests. These tests are described in section \ref{Tests} of
this appendix. The options for this flag are: `1' for running the first test only, `2' for running
the second test only and `3' for running both tests. Any other value means that no test will be
carried out. If this keyword is omitted, the default value is 0.

\vspace{0.5cm}

\noindent {\bf \large 7. `NormalizationChoice'} \vspace{0.3cm}

\noindent This keyword controls the generation and writing of normalization data for theoretical
and observational lines to the `Transitions' output file alongside the original emissivity data.
The options for this keyword are

\begin{itemize}

\item `0' for no normalization. In this case no normalization data
will be written to the `Transitions' file. Any invalid normalization choice will also be treated
like zero.

\item `1' for internal normalization with the theoretical data of one of the
transition lines produced by the program. In this case this should be followed by the index of the
normalization line.

\item `2' for normalizing with respect to an outside set of emissivity
values corresponding to the temperatures included under the `Temperature(K)' keyword. The
emissivity values should be on the next line(s) of text and in the SI units (i.e. in
J.s$^{-1}$.m$^{-3}$).

\item `3' for normalizing with respect to an outside set of emissivity
data in the form of effective \reccoe s corresponding to the specified temperatures. In this case
the \reccoe s should be followed by the wavelength of the line to be normalized to.
%
The \reccoe s in the input data file should be in cm$^{3}$.s$^{-1}$ while the wavelength, which
follows the \reccoe s, should be in nanometer. The reason for this choice is to avoid potential
error in conversion as the available data in the literature are mostly given in these units rather
than SI units. These values should be entered on the next line(s) of text in the data file.

\item `4' for normalizing with respect to an outside set of emissivity
data in the form of effective \reccoe s corresponding to a set of temperatures that may be
different from those included under the `Temperature(K)' keyword. In this case, the \reccoe s
corresponding to the temperature values of the `Temperature(K)' keyword are obtained by
interpolation or extrapolation using `polint' algorithm \cite{PressTVF2002}. This algorithm is a
polynomial interpolation routine which implements Neville's interpolating method and is regarded as
an improvement to the classical Lagrange's formula for polynomial interpolation. For this option,
the normalization choice should be followed by the number of points which identify the order of
interpolation (e.g. `3' for 3-point second order interpolation and so on). The recommended value
for the number of points is `3' or `4'. Going beyond this should be for a good reason, otherwise
the interpolation may fail to produce reasonable results. The details should be sought in the
literature of numerical methods. An initial investigation suggests that the optimal number of
points may depend on the proximity to the tabulated data points and the nature of the process, i.e.
interpolation vs. extrapolation. It may be a good idea to have a trial run inspecting the estimated
errors which can be found in the `Transitions' output file before choosing the order of
interpolation. The required data entries, which should follow the number of points, are the number
of data pairs (i.e. temperature in K and \reccoe\ in cm$^{3}$.s$^{-1}$) followed by the data pairs
followed by the wavelength of the line of normalization in nanometer. These values should be
included on the next line(s) of text in the data file.

\end{itemize}

If this keyword is omitted, the default value is 0.

\vspace{0.5cm}

\noindent {\bf \large 8. `RecCoefficient'} \vspace{0.3cm}

\noindent This keyword controls the production and writing of the effective \reccoe s, $\RCf$,
which are equivalent to the given emissivities. Vacuum wavelength of the transition line will be
used in these calculations as the air wavelength is not available for $\WL < 2000$~\AA. The keyword
should be followed by a boolean flag, i.e. `1' for `Yes' and `0' for `No'. Any other choice will be
treated as zero. The written \reccoe s to the `Transitions' file will be in the SI units
(m$^{3}$.s$^{-1}$). If this keyword is omitted, the default value is 0.

\vspace{0.5cm}

\noindent {\bf \large 9. `TheoAstroLS'} \vspace{0.3cm}

\noindent This keyword controls the algorithm to obtain the sum of weighted square differences
between the observational lines and their theoretical counterparts over the input temperature
range. This sum, in its basic form, is given by
\begin{equation}\label{sumLSD}
    \chi^2 = \sum_{i=1}^N w_i \left(I_i^{no} - \varepsilon_i^{nt} \right)^{2}
\end{equation}
where $I_i^{no}$ is the normalized observational flux of line $i$, $\varepsilon_i^{nt}$ is the
normalized theoretical emissivity, $w_i$ is a statistical weight which mainly accounts for error,
and $N$ is the number of lines involved in the least squares procedure. This algorithm also offers
the possibility of computing the temperature confidence interval.

The keyword requires a boolean flag to compute $\chi^2$ or not. If the flag is on, the name of the
file that maps the indices of the theoretical and observational lines should follow. 
The structure of the mapping file requires that the index of the observational line should be
followed by the index of its theoretical counterpart. If an observational line corresponds to more
than one theoretical line, which occurs when observational lines from the same multiplet are
blended or when the lines are too close to be entirely resolved observationally, the mapping should
be repeated independently for each theoretical line. The rest of the file can be used for
commenting or storing other data if the index mapping is terminated with `-999'.

The results of the least squares minimization algorithm are written to a file called `TNV'. This
file contains information on $\GoF$, temperature at minimum $\GoF$, and the temperature limits for
the confidence interval. This is followed by the temperatures with the corresponding least squares
residuals.

\vspace{0.5cm}

\noindent {\bf \large 10. `DecayRoutes'} \vspace{0.3cm}

\noindent This keyword is for finding the decay routes to a particular state, bound or free, from
all upper states. The parameters required are a boolean flag to run this algorithm or not (`1' for
`Yes' and `0' for `No'). In the first case, the configuration, term and 2$J$ of the state should
follow. As the number of decay routes can be very large (millions and even billions) especially for
the low bound states, another parameter is used to control the maximum number of detected routes,
and this parameter should follow. The results (number of decay routes and the routes themselves
grouped in complete and non-complete) are output to `DecayRoutes' file. If the number of detected
routes exceeds the maximum number, say $n$, only the first $n$ routes will be written to the
`DecayRoutes' file. The scheme is that the program keeps detecting the routes as long as the total
number of complete and uncomplete routes detected at any point during the program execution does
not exceed the maximum number. The algorithm follows the routes from the given state upwards, and
hence the number of detected routes, complete and uncomplete, steadily increases as iteration
progresses. Consequently, to find a substantial number of complete routes or all of them if this is
feasible, the maximum number should be set to a sufficiently large value. It is recommended that
for the bound states and low resonances the maximum number should be increased in steps until the
output satisfies the user need. If the maximum number was set to a very large value the program may
fail or take very long time to complete; moreover the data produced will become virtually useless
thanks to its massive size.

The states of each decay route are identified by their configuration, term and 2$J$. For
resonances, the \Saha\ capture term and the radiative decay term are also given when relevant.
These temperature-dependent data are given for a single temperature, that is the first value in the
temperature array.

\subsection{`ResEmis' Input Data File} \label{InputFile}
This is the file containing the data for resonances. The file contains four main sections:

\begin{enumerate}

\item The first line in the file is a comment line. This is followed
by the number of resonances followed by a number of text lines matching the number of resonances.
Each one of these lines contains an index identifying the resonance, the energy position of the
resonance in $\Rc$-scaled \Ryd, the width in $\Rc$-scaled \Ryd, the configuration, term, 2$J$,
parity, a flag for marking the energy position data as experimental (1) or theoretical (0), and a
flag marking the resonances, according to their departure coefficients $\DC$, as `good' (1) or not
(0) so that the bad ones with low $\DC$ can be excluded from computations if required. Each of the
configuration and term should be a single string with no space in between.

\item After the resonances data, a comment is expected. This comment,
which can span any number of text lines, should be terminated by the string `EndOfComment' to mark
the end. Next, the number of the bound-state symmetries is expected. This is followed, possibly, by
a comment to the end of the line. The \oscstr s ($\OS$-values) for the free-bound transitions then
follow in sections according to the symmetries of the bound states, so the number of the
$\OS$-values sections is the same as the number of bound state symmetries. Each section of the
$\OS$-values is headed by a line containing 2$J$, parity and the number of bound states in that
symmetry. This is followed by a two-dimensional array of $\OS$-values where the columns stand for
the resonances as ordered previously, whereas the rows are for the bound states of the given
symmetry. For the transitions which are forbidden by the \eledip\ rules for $J$ and parity, the
$\OS$-values should be set to zero. This convention is adopted to accommodate possible future
extension to include optically-forbidden transitions.

\item Next a comment is expected which should be terminated by the
`EndOfComment' string to mark the end. The data for the photon energy at resonance position which
correspond to the $\OS$-value data in the previous section, then follow, sectioned and formatted as
for the $\OS$-value data. The photon energy data, which are obtained from the `XSECTN' file
produced by stage \STGBF\ of the \Rm-matrix code, are in \Ryd.

\item The last section starts with a comment terminated by
the `EndOfComment' string. The $\OS$-value data for the free-free transitions are then included in
a two-dimensional array where the columns stand for the odd resonances while the rows stand for the
even ones. This section is required only if the resonances are of mixed parity.

\end{enumerate}

\subsection{`ELEVEmis' Input Data File} \label{InputFile}
This file contains the bound states data. The file starts with a comment terminated by the
`EndOfComment' marker. Next, the number of bound states symmetries are expected. This is followed
by a data block for each symmetry. Each block is headed by a line containing 2$J$, parity and the
number of bound states in that symmetry. Next, a number of text lines as the number of bound states
in that symmetry are expected. Each line contains (in the following order) an index identifying the
state, the energy of the bound state in $\Rc$-scaled \Ryd, the \effquanum, the configuration, term
and a flag for marking the energy data as experimental (1) or theoretical (0). Each of the
configuration and term should be a single string with no space in between. The \effquanum, though
not used in the program, is included because the original data come from the `ELEV' data file
generated by the \STGB\ stage of the \Rm-matrix code. It is kept for possible use in the future.

\subsection{`Transitions' Output Data File} \label{TransitionsFile}
This is the main output file of \EmiCod\ program. The file starts with a number of text lines
summarizing the input data used and the output results, followed by a few commentary lines
explaining the symbols and units. This is followed by a number of data lines matching the number of
transitions. The data for each transition includes (in the following order) an index identifying
the transition, status (FF, FB or BB transition), two joined boolean flags describing the
experimental state of the energy data of the upper and lower levels (`0' for theoretical data and
`1' for experimental), the attributes of the upper and lower levels (configuration, term, 2$J$ and
parity), wavelength in vacuum, wavelength in air (only for $\WL > 2000$~\AA), the radiative
transition probability, the emissivities corresponding to the given temperatures, the normalized
emissivities and the effective \reccoe s corresponding to the given emissivities. Writing the
normalized emissivities and the effective \reccoe s is optional and can be turned off, as described
under `NormalizationChoice' and `RecCoefficient' keywords. As mentioned earlier, if the user
chooses to read the astronomical data, the observational lines will be inserted in between the
theoretical lines according to their lab wavelength.

\section{Tests} \label{Tests}
Apart from the normal debugging and testing of the program components to check that they do what
they are supposed to do, two physical tests are incorporated within the program to validate its
functionality and verify that no serious errors have occurred in processing and computing the data.
These tests are the population-balance test and the ground-metastable test. The first test relies
on the fact that the population of each state should equal the depopulation. This balance is given
by the relation
\begin{equation}\label{firstTest}
    \sum_{j>i} \ND_{j} \rTP_{ji} = \ND_{i} \sum_{k<i} \rTP_{ik}
\end{equation}
where $\ND$ is the population of the indicated state, $\rTP$ is the radiative transition
probability, $i$ is the index for the state of concern, $j$ is an index for the states above state
$i$, and $k$ is an index for the states below state $i$.

The second test is based on the fact that the total number of the electrons leaving the resonances
in radiative decay should equal the total number arriving at the ground and metastable states. This
balance is given by the relation
\begin{equation}\label{secondTest}
    \sum_{\forall j} \ND_{j} \rTP_{j} = \sum_{\forall i, k>i} \ND_{k} \rTP_{ki}
\end{equation}
where $i$ is an index for ground and metastable states and $j$ is an index for resonances.

%
%
%
%

%

\chapter{Input Data for \Rm-matrix and \AS} \label{AppInData}
In this appendix we include the input data files that we used to generate our theoretical results
from \Rm-matrix and \AS. This is for the purpose of thoroughness and to enable the interested
researcher to regenerate and check our results. It should be remarked that the structure of the
\Rm-matrix and \AS\ data files is fully explained in the write-up of these codes as given by
Badnell \cite{BadnellRmax2002, BadnellAS2008}.

\section{\Rm-matrix Input Data} \label{RmaxData}
In this section we present our input data files for each stage of
\Rm-matrix.

\newcommand{\CS}      {\vspace{-0.5cm}}

\CS
\subsection{\STGO} \label{RmaxDataSTGO}
The input data file for this stage is:

{\tiny
\begin{spacing}{1}
\begin{verbatim}
STO-
 &STG1A RAD='YES' RELOP='YES' ISMITN=1 &END
 &STG1B NZED=6 NELC=4 MAXORB=7 MAXLA=3 MAXLT=10 MAXC=16 ISMIT(1)=30 ISMIT(2)=31 ISMIT(3)=32 ISMIT(4)=43 MAXE=8 IBC=1 &END
  1 0 2 0 2 1 3 0 3 1 3 2 4 3
    5
    1    1    2    2    2
     5.13180       8.51900       2.01880       4.73790       1.57130
    21.28251       6.37632       0.08158      -2.61339      -0.00733
    5
    1    1    2    2    2
     5.13180       8.51900       2.01880       4.73790       1.57130
    -5.39193      -1.49036       5.57151      -5.25090       0.94247
    3
    1    2    3
     1.75917       1.75917       1.75917
     5.69321     -19.54864      10.39428
    4
    2    2    2    2
     1.47510       3.19410       1.83070       9.48450
     1.01509       3.80119       2.75006       0.89571
    2
    2    3
     1.98138       1.96954
    14.41203     -10.88586
    1
    3
     2.11997
     5.84915
    1
    4
     2.69086
     9.69136
     10.0000000     0.0
    5
    1    1    2    2    2
     5.13180       8.51900       2.01880       4.73790       1.57130
    21.28251       6.37632       0.08158      -2.61339      -0.00733
    5
    1    1    2    2    2
     5.13180       8.51900       2.01880       4.73790       1.57130
    -5.39193      -1.49036       5.57151      -5.25090       0.94247
    3
    1    2    3
     1.75917       1.75917       1.75917
     5.69321     -19.54864      10.39428
    4
    2    2    2    2
     1.47510       3.19410       1.83070       9.48450
     1.01509       3.80119       2.75006       0.89571
    2
    2    3
     1.98138       1.96954
    14.41203     -10.88586
    1
    3
     2.11997
     5.84915
    1
    4
     2.69086
     9.69136
     10.0000000     0.0
\end{verbatim}
\end{spacing}
}

\CS
\subsection{\STGT}
The input data file for this stage is:

{\tiny
\begin{spacing}{1}
\begin{verbatim}
STO-
 &STG2A RAD='YES' RELOP='YES' &END
 &STG2B MAXORB=7 NELC=4 NAST=26 INAST=0 MINST=2 MAXST=4 MINLT=0 MAXLT=8 &END
  1 0 2 0 2 1 3 0 3 1 3 2 4 3
  18
   2 0 0 0 0 0 0
   2 2 2 2 2 2 2
   2 2 0 0 0 0 0 0
   2 0 2 0 0 0 0 0
   2 1 0 1 0 0 0 0
   2 0 0 2 0 0 0 0
   2 0 1 0 1 0 0 0
   2 0 0 0 2 0 0 0
   2 0 0 0 0 2 0 0
   2 1 1 0 0 0 0 0
   2 1 0 0 1 0 0 0
   2 0 0 1 1 0 0 0
   2 0 1 1 0 0 0 0
   2 0 1 0 0 1 0 0
   2 0 0 0 1 1 0 0
   2 0 0 0 0 1 1 0
   2 1 0 0 0 1 0 0
   2 0 0 1 0 1 0 0
   2 0 1 0 0 0 1 0
   2 0 0 0 1 0 1 0
    0   1   0
    1   3   1
    1   1   1
    1   3   0
    2   1   0
    0   1   0
    0   3   0
    0   1   0
    1   1   1
    1   3   1
    1   3   1
    2   3   0
    1   1   1
    1   1   0
    2   1   0
    2   3   0
    0   3   0
    3   3   1
    2   1   1
    1   3   1
    1   3   0
    2   3   1
    2   1   0
    3   1   1
    1   1   1
    0   1   0
  48
   2 0 0 0 0 0 0
   2 2 3 2 3 3 1
   2 2 1 0 0 0 0 0
   2 2 0 1 0 0 0 0
   2 2 0 0 1 0 0 0
   2 2 0 0 0 1 0 0
   2 2 0 0 0 0 1 0
   2 1 2 0 0 0 0 0
   2 0 3 0 0 0 0 0
   2 0 2 1 0 0 0 0
   2 0 2 0 1 0 0 0
   2 0 2 0 0 1 0 0
   2 0 2 0 0 0 1 0
   2 1 1 1 0 0 0 0
   2 1 0 2 0 0 0 0
   2 1 0 1 1 0 0 0
   2 1 0 1 0 1 0 0
   2 1 0 1 0 0 1 0
   2 0 1 2 0 0 0 0
   2 0 0 2 1 0 0 0
   2 0 0 2 0 1 0 0
   2 0 0 2 0 1 1 0
   2 1 1 0 1 0 0 0
   2 0 1 1 1 0 0 0
   2 0 1 0 2 0 0 0
   2 0 1 0 1 1 0 0
   2 0 1 0 1 0 1 0
   2 1 0 0 2 0 0 0
   2 0 0 1 2 0 0 0
   2 0 0 0 3 0 0 0
   2 0 0 0 2 1 0 0
   2 0 0 0 2 0 1 0
   2 1 0 0 0 2 0 0
   2 0 1 0 0 2 0 0
   2 0 0 1 0 2 0 0
   2 0 0 0 1 2 0 0
   2 0 0 0 0 3 0 0
   2 0 0 0 0 2 1 0
   2 1 1 0 0 1 0 0
   2 1 1 0 0 0 1 0
   2 1 0 0 1 1 0 0
   2 1 0 0 1 0 1 0
   2 0 0 1 1 1 0 0
   2 0 0 1 1 0 1 0
   2 0 1 1 0 1 0 0
   2 0 1 1 0 0 1 0
   2 0 1 0 0 1 1 0
   2 0 0 0 1 1 1 0
   2 1 0 0 0 1 1 0
   2 0 0 1 0 1 1 0
\end{verbatim}
\end{spacing}
}

\CS
\subsection{\STGJK}
The input data file for this stage is:

{\tiny
\begin{spacing}{1}
\begin{verbatim}
S.S.
 &STGJA RAD='YES' &END
 &STGJB JNAST=46 IJNAST=12 &END
 0  0
 0  1
 2  1
 4  1
 2  1
 0  0
 2  0
 4  0
 4  0
 0  0
 2  0
 0  0
 2  1
 0  1
 2  1
 4  1
 2  0
 4  0
 6  0
 0  1
 2  1
 4  1
 4  0
 2  1
 2  0
 4  1
 6  1
 8  1
 2  0
 4  0
 6  0
 4  1
 2  0
 0  1
 2  1
 4  1
 2  1
 4  1
 6  1
 0  0
 2  0
 4  0
 4  0
 2  1
 6  1
 0  0
   1  0
   3  0
   5  0
   7  0
   9  0
   11 0
   1  1
   3  1
   5  1
   7  1
   9  1
   11 1
\end{verbatim}
\end{spacing}
}

\CS
\subsection{\STGTH}
The input data file for this stage is:

{\tiny
\begin{spacing}{1}
\begin{verbatim}
S.S.
 &STG3A RAD='YES'  &END
 &STG3B &END
\end{verbatim}
\end{spacing}
}

\CS
\subsection{\STGF}
The parameters in the input data file with a sample data is:

{\tiny
\begin{spacing}{1}
\begin{verbatim}
&STGF IMODE = 0 IPRKM = 1 IQDT = 0 IRD0 = 99 IMESH = 1 IEQ = -1 PERT = 'NO'
       LRGLAM = -1 IBIGE = 0 IPRINT = 3 IRAD = 1 IOPT1 = 2 &END
&MESH1 MXE = 1000 E0 = 0.0298042  EINCR = 5.0E-8 &END
  0   3   1
-1 -1 -1
\end{verbatim}
\end{spacing}
}

\CS
\subsection{\STGB}
The input file for this stage is:

{\tiny
\begin{spacing}{1}
\begin{verbatim}
&STGB IPERT=0 IRAD=1 &END
  0 1 0
0.1 13.0 0.001
  0 3 0
0.1 13.0 0.001
  0 5 0
0.1 13.0 0.001
  0 7 0
0.1 13.0 0.001
  0 9 0
0.1 13.0 0.001
  0 1 1
0.1 13.0 0.001
  0 3 1
0.1 13.0 0.001
  0 5 1
0.1 13.0 0.001
  0 7 1
0.1 13.0 0.001
  0 9 1
0.1 13.0 0.001
  0 11 1
0.1 13.0 0.001
 -1 -1 -1
\end{verbatim}
\end{spacing}
}

\CS
\subsection{\STGBF}
For this stage, default values are used, that is:

{\tiny
\begin{spacing}{1}
\begin{verbatim}
&STGBF  &END
\end{verbatim}
\end{spacing}
}

\CS
\subsection{\STGBB}
The input data file for this stage is:

{\tiny
\begin{spacing}{1}
\begin{verbatim}
&STGBB IPRINT=0 IBUT=0  &END
  0 1 0   0 1 1
  0 1 0   0 3 1
  0 3 0   0 1 1
  0 3 0   0 3 1
  0 3 0   0 5 1
  0 5 0   0 3 1
  0 5 0   0 5 1
  0 5 0   0 7 1
  0 7 0   0 5 1
  0 7 0   0 7 1
  0 7 0   0 9 1
  0 9 0   0 7 1
  0 9 0   0 9 1
  0 9 0   0 11 1
  0 11 0  0 9 1
  0 11 0  0 11 1
 -1 -1 -1 -1 -1 -1
\end{verbatim}
\end{spacing}
}

\CS
\subsection{\STGQB}
The parameters in the input data file with a sample data is:

{\tiny
\begin{spacing}{1}
\begin{verbatim}
0 0 0
  1
0 3 1
2.945  4.1  1.0E-8
\end{verbatim}
\end{spacing}
}

\section{\AS\ Input Data} \label{ASData}
\AS\ code was used in various stages to produce required theoretical atomic data or to check the
results of \Rm-matrix. In the next sections we present some of these input data files.

\CS
\subsection{Polarizability} \label{ASPloar}
The following input data file was used to decide which terms of the target are the most important
ones by having the largest polarizability, as explained in \S\ \ref{Target}.

{\tiny
\begin{spacing}{1}
\begin{verbatim}
S.S.
123456789 22 12513 23 12514 12515 12516 13514 13515 13516
                10 20 21
 &SALGEB  RUN='  ' RAD='YES' CUP='IC' TITLE='C III_9'
          KORB1=1 KORB2=1 KUTSO=0  &END
 &SMINIM  NZION=6 INCLUD=6 PRINT='FORM' NLAM=6 RADOUT='YES' &END
  1.00000 1.00000 1.00000 -1.0000 -1.0000 -1.0000
\end{verbatim}
\end{spacing}
}

\CS
\subsection{Transition Probabilities} \label{AsTraPro}
The following input data were used to compute the radiative transition probabilities, as discussed
in \S\ \ref{PracAsp}.

{\tiny
\begin{spacing}{1}
\begin{verbatim}
S.S.
123456789 12513516 12513515 12523 22516 12513517
                10 20 21 30 31 32
 &SALGEB  RUN='  ' RAD='YES' CUP='IC' TITLE='Resonance'
          KORB1=1 KORB2=1 KUTSO=0  &END
 &SMINIM  NZION=6 INCLUD=6 PRINT='FORM' NLAM=6 RADOUT='YES' &END
  1.00000 1.00000 1.00000 1.0000 1.0000 1.0000
\end{verbatim}
\end{spacing}
}

\CS
\subsection{$\OS$-values} \label{AsfVal}
The following input data file was used to generate $\OS$-values for all FF transitions. It was also
used to generate $\OS$-values for the free-bound and bound-bound transitions for the topmost bound
states, namely the \nlo1s22s2p(\SLP3Po)3d~\SLP4Fo and \SLP4Do levels, as these quartets with their
large \effquanum\ are out of range of \RMAT\ validity.

{\tiny
\begin{spacing}{1}
\begin{verbatim}
 S.S. 2s2 nl (2s<nl<7p); 2s 2p nl (2s<nl<7p); 2p3; 2p2 nl (2p<nl<7p)
 123456789 22513 22514 22515 22516 22517 22518 22519 2251A 2251B 2251C
           2251D 2251E 2251F 2251G 2251H 2251I 2251J 2251K 2251L 2251M
           12523 12513514 12513515 12513516 12513517 12513518 12513519
           1251351A 1251351B 1251351C 1251351D 1251351E 1251351F 1251351G
           1251351H 1251351I 1251351J 1251351K 1251351L 1251351M 33
           23514 23515 23516 23517 23518 23519 2351A 2351B 2351C 2351D
           2351E 2351F 2351G 2351H 2351I 2351J 2351K 2351L 2351M

                10
                20 21
                30 31 32
                40 41 42 43
                50 51 52 53 54
                60 61 62 63 64 65
                70

 &SALGEB  RUN='  ' RAD='YES' CUP='IC' TITLE='Resonance'
          KORB1=1 KORB2=1 KUTSO=0  &END
 &SMINIM  NZION=6 INCLUD=500 PRINT='FORM' NLAM=22 NVAR=22 RADOUT='YES' &END
 1.43240 1.43380 1.39690 1.25760 1.20290 1.35930 1.25830 1.19950 1.35610 1.41460 1.26080
 1.20020 1.35770 1.41420 1.32960 1.26370 1.20250 1.36210 1.41520 1.41420 2.34460 1.26790
 1 2 3 4 5 6 7 8 9 10 11 12 13 14 15 16 17 18 19 20 21 22
\end{verbatim}
\end{spacing}
}

\chapter[Tables]{} \label{Tables} 

\begin{spacing}{2}
{\LARGE \bf Tables}
\end{spacing}
\vspace{1.0cm}
\noindent

In this appendix, we present a sample of the data produced during this investigation. In
Table~\ref{BTable} the theoretical results for the energies of the bound states of C$^+$ below the
C$^{2+}$ \SLPJ1Se0\ threshold are given alongside the available experimental data from the \NIST\
database \cite{NIST2010}. Similarly, Table~\ref{RTableKQ} presents the energy and autoionization
width data for the resonances as obtained by the \Km-matrix and \QB\ methods. In these tables, a
negative \finstr\ splitting indicates that the theoretical levels are in reverse order compared to
their experimental counterparts. It is noteworthy that due to limited precision of figures in these
tables, some data may appear inconsistent, e.g. a zero \finstr\ splitting from two levels with
different energies. Full-precision data in electronic format are available from the author of this
thesis on request.

Regarding the bound states, all levels with \effquanum\ $\EQN$ between 0.1-13 for the outer
electron and $0\le l \le 5$ (142 states) are sought and found by \Rm-matrix. The 8 uppermost bound
states in Table~\ref{BTable}, i.e. the levels of \nlo1s22s2p(\SLP3Po)3d~\SLP4Fo and \SLP4Do, have
quantum numbers higher than 13 and hence are out of range of the \Rm-matrix approximation validity;
therefore only experimental data are included for these states. Concerning the resonances, we
searched for all states with $n<5$ where $n$ is the \priquanum\ of the active electron. 61 levels
were found by the \Km-matrix method and 55 by the \QB\ method.

Tables \ref{fValues261o}-\ref{fValues265o} present a sample of the $\OS$-values of the free-bound
transitions for some bound symmetries as obtained by integrating \pcs\ over photon energy where
these data are obtained from stage STGBF of the \Rm-matrix code. The columns in these tables stand
for the bound states identified by their indices as given in Table~\ref{BTable} while the rows
stand for the resonances represented by their indices as given in Table~\ref{RTableKQ}. The
superscript denotes the power of 10 by which the number is to be multiplied. An entry of `0' in the
$\OS$-value tables indicates that no peak was observed in the \pcs\ data.

The reason for presenting a sample of the $\OS$-values for the free-bound transitions only is that
they require considerable effort to produce.
No data belonging to $\OS$-values for the bound-bound transitions are given here because they can
be easily obtained from the `FVALUE' file generated by the \Rm-matrix code using the input data
files given in \S\ \ref{RmaxData} in Appendix \ref{AppInData}. Similarly, no data related to the
$\OS$-values for the free-free transitions are provided in this appendix because they can be easily
generated by \AS\ using the input data file supplied in \S\ \ref{AsfVal} in Appendix
\ref{AppInData}.

Finally, Table~\ref{ListSample} contains a sample of the effective recombination coefficients for
transitions extracted from our list, \SSo, in a wavelength range where several lines have been
observed in the spectra of planetary nebulae. It should be remarked that the data of \SSo\ are
generated assuming electron and ion number density of 10$^{10}$~m$^{-3}$.

\onehalfspace


\newpage

\phantomsection \addcontentsline{toc}{section}{\protect \numberline{} Bound States} %

{\footnotesize

\begin{longtable}{@{\extracolsep\fill}lllllllr@{}}
\caption{The available experimental data from NIST for the bound states of C$^+$ below the C$^{2+}$
\SLPJ1Se0\ threshold alongside the theoretical results from \Rm-matrix\ calculations.
\label{BTable}} \\
\hline\hline %
{In.$^a$} & {Config.$^b$} & {Level} & {NEEW$^c$} & {NEER$^d$} & {FSS$^e$} & {TER$^f$} & {FSSR$^g$} \\ %
\hline
\endfirsthead
\caption[]{continued.}\\
\hline\hline %
{In.} & {Config.} & {Level} & {NEEW} & {NEER} & {FSS} & {TER} & {FSSR} \\ %
\hline
\endhead
\hline
\endfoot

1   &   2s$^2$2p    &   \SLPJ2Po{1/2}   &   0   &   -1.792141   &       &   -1.792571   &       \\
2   &   2s$^2$2p    &    \SLPJ2Po{3/2}  &   63.42   &   -1.791563   &   63.4    &   -1.791994   &   63.3    \VS\\

3   &   2s2p$^2$    &    \SLPJ4Pe{1/2}  &   43003.3 &   -1.400266   &       &   -1.401292   &       \\
4   &   2s2p$^2$    &    \SLPJ4Pe{3/2}  &   43025.3 &   -1.400065   &       &   -1.401090   &       \\
5   &   2s2p$^2$    &    \SLPJ4Pe{5/2}  &   43053.6 &   -1.399807   &   50.3    &   -1.400755   &   59.0    \VS\\

6   &   2s2p$^2$    &    \SLPJ2De{5/2}  &   74930.1 &   -1.109327   &       &   -1.105247   &       \\
7   &   2s2p$^2$    &    \SLPJ2De{3/2}  &   74932.62    &   -1.109304   &   2.5 &   -1.105266   &   -2.1    \VS\\

8   &   2s2p$^2$    &    \SLPJ2Se{1/2}  &   96493.74    &   -0.912825   &       &   -0.899439   &       \VS\\

9   &   2s2p$^2$    &    \SLPJ2Pe{1/2}  &   110624.17   &   -0.784059   &       &  -0.772978   &       \\
10  &   2s2p$^2$    &    \SLPJ2Pe{3/2}  &   110665.56   &   -0.783682   &   41.4    &   -0.772570   &   44.8    \VS\\

11  &   2s$^2$3s    &    \SLPJ2Se{1/2}  &   116537.65   &   -0.730171   &       &   -0.727698   &       \VS\\

12  &   2s$^2$3p    &    \SLPJ2Po{1/2}  &   131724.37   &   -0.591780   &       &   -0.592081   &       \\
13  &   2s$^2$3p    &    \SLPJ2Po{3/2}  &   131735.52   &   -0.591678   &   11.1    &   -0.591980   &   11.1    \VS\\

14  &   2p$^3$  &    \SLPJ4So{3/2}  &   142027.1    &   -0.497894   &       &   -0.487491   &       \VS\\

15  &   2s$^2$3d    &    \SLPJ2De{3/2}  &   145549.27   &   -0.465798   &       &   -0.466024   &       \\
16  &   2s$^2$3d    &    \SLPJ2De{5/2}  &   145550.7    &   -0.465785   &   1.4 &   -0.466005   &   2.1 \VS\\

17  &   2p$^3$  &    \SLPJ2Do{5/2}  &   150461.58   &   -0.421034   &       &   -0.412359   &       \\
18  &   2p$^3$  &    \SLPJ2Do{3/2}  &   150466.69   &   -0.420987   &   5.1 &   -0.412388   &   -3.1    \VS\\

19  &   2s$^2$4s    &    \SLPJ2Se{1/2}  &   157234.07   &   -0.359318   &       &   -0.358955   &       \VS\\

20  &   2s$^2$4p    &    \SLPJ2Po{1/2}  &   162517.89   &   -0.311169   &       &   -0.310784   &       \\
21  &   2s$^2$4p    &    \SLPJ2Po{3/2}  &   162524.57   &   -0.311108   &   6.7 &   -0.310723   &   6.6 \VS\\

22  &   2s2p(\SLP3Po)3s     &    \SLPJ4Po{1/2}  &   166967.13   &   -0.270624   &       &   -0.268893   &       \\
23  &   2s2p(\SLP3Po)3s     &    \SLPJ4Po{3/2}  &   166990.73   &   -0.270409   &       &   -0.268645   &       \\
24  &   2s2p(\SLP3Po)3s     &    \SLPJ4Po{5/2}  &   167035.71   &   -0.269999   &   68.6    &   -0.268229   &   72.9    \VS\\

25  &   2s$^2$4d    &    \SLPJ2De{3/2}  &   168123.74   &   -0.260084   &       &   -0.260084   &       \\
26  &   2s$^2$4d    &    \SLPJ2De{5/2}  &   168124.45   &   -0.260078   &   0.7 &   -0.260074   &   1.1 \VS\\

27  &   2p$^3$  &    \SLPJ2Po{1/2}  &   168729.53   &   -0.254564   &       &   -0.245329   &       \\
28  &   2p$^3$  &    \SLPJ2Po{3/2}  &   168748.3    &   -0.254393   &   18.8    &   -0.245083   &   27.1    \VS\\

29  &   2s$^2$4f    &    \SLPJ2Fo{5/2}  &   168978.34   &   -0.252297   &       &   -0.252162   &       \\
30  &   2s$^2$4f    &    \SLPJ2Fo{7/2}  &   168978.34   &   -0.252297   &   0.0 &   -0.252160   &   0.2 \VS\\

31  &   2s$^2$5s    &    \SLPJ2Se{1/2}  &   173347.84   &   -0.212479   &       &   -0.212278   &       \VS\\

32  &   2s$^2$5p    &    \SLPJ2Po{1/2}  &   175287.39   &   -0.194804   &       &   -0.190099   &       \\
33  &   2s$^2$5p    &    \SLPJ2Po{3/2}  &   175294.75   &   -0.194737   &   7.4 &   -0.190072   &   3.0 \VS\\

34  &   2s2p(\SLP3Po)3s     &    \SLPJ2Po{1/2}  &   177774.59   &   -0.172139   &       &   -0.165330   &       \\
35  &   2s2p(\SLP3Po)3s     &    \SLPJ2Po{3/2}  &   177793.54   &   -0.171967   &   19.0    &   -0.165143   &   20.5    \VS\\

36  &   2s$^2$5d    &    \SLPJ2De{3/2}  &   178495.11   &   -0.165573   &       &   -0.165606   &       \\
37  &   2s$^2$5d    &    \SLPJ2De{5/2}  &   178495.71   &   -0.165568   &   0.6 &   -0.165599   &   0.8 \VS\\

38  &   2s$^2$5f    &    \SLPJ2Fo{5/2}  &   178955.94   &   -0.161374   &       &   -0.161296   &       \\
39  &   2s$^2$5f    &    \SLPJ2Fo{7/2}  &   178955.94   &   -0.161374   &   0.0 &   -0.161295   &   0.1 \VS\\

40  &   2s$^2$5g    &    \SLPJ2Ge{7/2}  &   179073.05   &   -0.160307   &       &   -0.160184   &       \\
41  &   2s$^2$5g    &    \SLPJ2Ge{9/2}  &   179073.05   &   -0.160307   &   0.0 &   -0.160184   &   0.0 \VS\\

42  &   2s$^2$6s    &    \SLPJ2Se{1/2}  &   181264.24   &   -0.140339   &       &   -0.140231   &       \VS\\

43  &   2s2p(\SLP3Po)3p     &    \SLPJ4De{1/2}  &   181696.66   &   -0.136399   &       &   -0.136848   &       \\
44  &   2s2p(\SLP3Po)3p     &    \SLPJ4De{3/2}  &   181711.03   &   -0.136268   &       &   -0.136699   &       \\
45  &   2s2p(\SLP3Po)3p     &    \SLPJ4De{5/2}  &   181736.05   &   -0.136040   &       &   -0.136454   &       \\
46  &   2s2p(\SLP3Po)3p     &    \SLPJ4De{7/2}  &   181772.41   &   -0.135709   &   75.8    &   -0.136115   &   80.4    \VS\\

47  &   2s2p(\SLP3Po)3p     &    \SLPJ2Pe{1/2}  &   182023.86   &   -0.133417   &       &   -0.133095   &       \\
48  &   2s2p(\SLP3Po)3p     &    \SLPJ2Pe{3/2}  &   182043.41   &   -0.133239   &   19.6    &   -0.132896   &   21.8    \VS\\

49  &   2s$^2$6p    &    \SLPJ2Po{1/2}  &   182993.23   &   -0.124584   &       &   -0.124376   &       \\
50  &   2s$^2$6p    &    \SLPJ2Po{3/2}  &   182993.66   &   -0.124580   &   0.4 &   -0.124351   &   2.7 \VS\\

51  &   2s$^2$6d    &    \SLPJ2De{3/2}  &   184074.59   &   -0.114730   &       &   -0.114739   &       \\
52  &   2s$^2$6d    &    \SLPJ2De{5/2}  &   184075.28   &   -0.114723   &   0.7 &   -0.114729   &   1.0 \VS\\

53  &   2s$^2$6f    &    \SLPJ2Fo{5/2}  &   184376.06   &   -0.111982   &       &   -0.111924   &       \\
54  &   2s$^2$6f    &    \SLPJ2Fo{7/2}  &   184376.06   &   -0.111982   &   0.0 &   -0.111924   &   0.1 \VS\\

55  &   2s$^2$6g    &    \SLPJ2Ge{7/2}  &   184449.27   &   -0.111315   &       &   -0.111264   &       \\
56  &   2s$^2$6g    &    \SLPJ2Ge{9/2}  &   184449.27   &   -0.111315   &   0.0 &   -0.111264   &   0.0 \VS\\

57  &   2s$^2$6h    &    \SLPJ2Ho{9/2}  &   184466.5    &   -0.111158   &       &   -0.111122   &       \\
58  &   2s$^2$6h    &    \SLPJ2Ho{11/2} &   184466.5    &   -0.111158   &   0.0 &   -0.111122   &   0.0 \VS\\

59  &   2s2p(\SLP3Po)3p     &    \SLPJ4Se{3/2}  &   184690.98   &   -0.109113   &       &   -0.108410   &       \VS\\

60  &   2s$^2$7s    &    \SLPJ2Se{1/2}  &   185732.93   &   -0.099618   &       &   -0.099537   &       \VS\\

61  &   2s2p(\SLP3Po)3p     &    \SLPJ4Pe{1/2}  &   186427.35   &   -0.093290   &       &   -0.092097   &       \\
62  &   2s2p(\SLP3Po)3p     &    \SLPJ4Pe{3/2}  &   186443.69   &   -0.093141   &       &   -0.091950   &       \\
63  &   2s2p(\SLP3Po)3p     &    \SLPJ4Pe{5/2}  &   186466.02   &   -0.092937   &   38.7    &   -0.091717   &   41.8    \VS\\

64  &   2s$^2$7p    &    \SLPJ2Po{1/2}  &   186745.9    &   -0.090387   &       &   -0.090313   &       \\
65  &   2s$^2$7p    &    \SLPJ2Po{3/2}  &   186746.3    &   -0.090383   &   0.4 &   -0.090302   &   1.1 \VS\\

66  &   2s$^2$7d    &    \SLPJ2De{3/2}  &   187353  &   -0.084854   &       &   -0.084804   &       \\
67  &   2s$^2$7d    &    \SLPJ2De{5/2}  &   187353  &   -0.084854   &   0.0 &   -0.084764   &   4.4 \VS\\

68  &   2s$^2$7f    &    \SLPJ2Fo{5/2}  &   187641.6    &   -0.082225   &       &   -0.082171   &       \\
69  &   2s$^2$7f    &    \SLPJ2Fo{7/2}  &   187641.6    &   -0.082225   &   0.0 &   -0.082171   &   0.0 \VS\\

70  &   2s$^2$7g    &    \SLPJ2Ge{7/2}  &   187691.4    &   -0.081771   &       &   -0.081744   &       \\
71  &   2s$^2$7g    &    \SLPJ2Ge{9/2}  &   187691.4    &   -0.081771   &   0.0 &   -0.081744   &   0.0 \VS\\

72  &   2s$^2$7h    &    \SLPJ2Ho{9/2}  &   187701  &   -0.081683   &       &   -0.081645   &       \\
73  &   2s$^2$7h    &    \SLPJ2Ho{11/2} &   187701  &   -0.081683   &   0.0 &   -0.081645   &   0.0 \VS\\

74  &   2s$^2$8s    &    \SLPJ2Se{1/2}  &   --- &   --- &       &   -0.074324   &       \VS\\

75  &   2s2p(\SLP3Po)3p     &    \SLPJ2De{3/2}  &   188581.25   &   -0.073662   &       &   -0.071901   &       \\
76  &   2s2p(\SLP3Po)3p     &    \SLPJ2De{5/2}  &   188615.07   &   -0.073354   &   33.8    &   -0.071582   &   34.9    \VS\\

77  &   2s$^2$8p    &    \SLPJ2Po{1/2}  &   --- &   --- &       &   -0.068349   &       \\
78  &   2s$^2$8p    &    \SLPJ2Po{3/2}  &   --- &   --- &       &   -0.068343   &   0.7 \VS\\

79  &   2s$^2$8f    &    \SLPJ2Fo{5/2}  &   --- &   --- &       &   -0.062873   &       \\
80  &   2s$^2$8f    &    \SLPJ2Fo{7/2}  &   --- &   --- &       &   -0.062872   &   0.0 \VS\\

81  &   2s$^2$8g    &    \SLPJ2Ge{7/2}  &   189794.2    &   -0.062609   &       &   -0.062580   &       \\
82  &   2s$^2$8g    &    \SLPJ2Ge{9/2}  &   189794.2    &   -0.062609   &   0.0 &   -0.062580   &   0.0 \VS\\

83  &   2s$^2$8h    &    \SLPJ2Ho{11/2} &   --- &   --- &       &   -0.062511   &       \\
84  &   2s$^2$8h    &    \SLPJ2Ho{9/2}  &   --- &   --- &       &   -0.062511   &   0.0 \VS\\

85  &   2s$^2$8d    &   \SLPJ2De{3/2}   &   --- &   --- &       &   -0.062612   &       \\
86  &   2s$^2$8d    &   \SLPJ2De{5/2}   &   --- &   --- &       &   -0.062564   &   5.3 \VS\\

87  &   2s$^2$9s    &   \SLPJ2Se{1/2}   &   --- &   --- &       &   -0.057638   &       \VS\\

88  &   2s$^2$9p    &   \SLPJ2Po{1/2}   &   --- &   --- &       &   -0.053490   &       \\
89  &   2s$^2$9p    &   \SLPJ2Po{3/2}   &   --- &   --- &       &   -0.053486   &   0.4 \VS\\

90  &   2s$^2$9d    &   \SLPJ2De{3/2}   &   --- &   --- &       &   -0.049942   &       \\
91  &   2s$^2$9d    &   \SLPJ2De{5/2}   &   --- &   --- &       &   -0.049935   &   0.8 \VS\\

92  &   2s$^2$9f    &   \SLPJ2Fo{5/2}   &   --- &   --- &       &   -0.049651   &       \\
93  &   2s$^2$9f    &   \SLPJ2Fo{7/2}   &   --- &   --- &       &   -0.049650   &   0.0 \VS\\

94  &   2s$^2$9g    &   \SLPJ2Ge{7/2}   &   --- &   --- &       &   -0.049441   &       \\
95  &   2s$^2$9g    &   \SLPJ2Ge{9/2}   &   --- &   --- &       &   -0.049441   &   0.0 \VS\\

96  &   2s$^2$9h    &   \SLPJ2Ho{9/2}   &   --- &   --- &       &   -0.049392   &       \\
97  &   2s$^2$9h    &   \SLPJ2Ho{11/2}  &   --- &   --- &       &   -0.049392   &   0.0 \VS\\

98  &   2s$^2$10s   &   \SLPJ2Se{1/2}   &   --- &   --- &       &   -0.046050   &       \VS\\

99  &   2s$^2$10p   &   \SLPJ2Po{1/2}   &   --- &   --- &       &   -0.042990   &       \\
100 &   2s$^2$10p   &   \SLPJ2Po{3/2}   &   --- &   --- &       &  -0.042987   &   0.3 \VS\\

101 &   2s$^2$10d   &   \SLPJ2De{3/2}   &   --- &   --- &       &   -0.040487   &       \\
102 &   2s$^2$10d   &   \SLPJ2De{5/2}   &   --- &   --- &       &   -0.040484   &   0.3 \VS\\

103 &   2s$^2$10f   &   \SLPJ2Fo{5/2}   &   --- &   --- &       &   -0.040198   &       \\
104 &   2s$^2$10f   &   \SLPJ2Fo{7/2}   &   --- &   --- &       &   -0.040198   &   0.0 \VS\\

105 &   2s$^2$10g   &   \SLPJ2Ge{7/2}   &   --- &   --- &       &   -0.040044   &       \\
106 &   2s$^2$10g   &   \SLPJ2Ge{9/2}   &   --- &   --- &       &   -0.040044   &   0.0 \VS\\

107 &   2s$^2$10h   &   \SLPJ2Ho{9/2}   &   --- &   --- &       &   -0.040007   &       \\
108 &   2s$^2$10h   &   \SLPJ2Ho{11/2}  &   --- &   --- &       &   -0.040007   &   0.0 \VS\\

109 &   2s$^2$11s   &   \SLPJ2Se{1/2}   &   --- &   --- &       &   -0.037746   &       \VS\\

110 &   2s$^2$11p   &   \SLPJ2Po{1/2}   &   --- &   --- &       &   -0.035300   &       \\
111 &   2s$^2$11p   &   \SLPJ2Po{3/2}   &   --- &   --- &       &   -0.035298   &   0.2 \VS\\

112 &   2s$^2$11d   &   \SLPJ2De{3/2}   &   --- &   --- &       &   -0.033449   &       \\
113 &   2s$^2$11d   &   \SLPJ2De{5/2}   &   --- &   --- &       &   -0.033448   &   0.2 \VS\\

114 &   2s$^2$11f   &   \SLPJ2Fo{5/2}   &   --- &   --- &       &   -0.033209   &       \\
115 &   2s$^2$11f   &   \SLPJ2Fo{7/2}   &   --- &   --- &       &   -0.033209   &   0.0 \VS\\

116 &   2s$^2$11g   &   \SLPJ2Ge{7/2}   &   --- &   --- &       &   -0.033092   &       \\
117 &   2s$^2$11g   &   \SLPJ2Ge{9/2}   &   --- &   --- &       &   -0.033092   &   0.0 \VS\\

118 &   2s$^2$11h   &   \SLPJ2Ho{9/2}   &   --- &   --- &       &   -0.033064   &       \\
119 &   2s$^2$11h   &   \SLPJ2Ho{11/2}  &   --- &   --- &       &   -0.033064   &   0.0 \VS\\

120 &   2s$^2$12s   &   \SLPJ2Se{1/2}   &   --- &   --- &       &   -0.031953   &       \VS\\

121 &   2s$^2$12p   &   \SLPJ2Po{1/2}   &   --- &   --- &       &   -0.029501   &       \\
122 &   2s$^2$12p   &   \SLPJ2Po{3/2}   &   --- &   --- &       &   -0.029499   &   0.2 \VS\\

123 &   2s2p(\SLP3Po)3p &   \SLPJ2Se{1/2}   &   --- &   --- &       &   -0.028972   &       \VS\\

124 &   2s$^2$12d   &   \SLPJ2De{3/2}   &   --- &   --- &       &   -0.028090   &       \\
125 &   2s$^2$12d   &   \SLPJ2De{5/2}   &   --- &   --- &       &   -0.028089   &   0.1 \VS\\

126 &   2s$^2$12f   &   \SLPJ2Fo{5/2}   &   --- &   --- &       &   -0.027895   &       \\
127 &   2s$^2$12f   &   \SLPJ2Fo{7/2}   &   --- &   --- &       &   -0.027895   &   0.0 \VS\\

128 &   2s$^2$12g   &   \SLPJ2Ge{7/2}   &   --- &   --- &       &   -0.027804   &       \\
129 &   2s$^2$12g   &   \SLPJ2Ge{9/2}   &   --- &   --- &       &   -0.027804   &   0.0 \VS\\

130 &   2s$^2$12h   &   \SLPJ2Ho{9/2}   &   --- &   --- &       &   -0.027783   &       \\
131 &   2s$^2$12h   &   \SLPJ2Ho{11/2}  &   --- &   --- &       &   -0.027783   &   0.0 \VS\\

132 &   2s$^2$13s   &   \SLPJ2Se{1/2}   &   --- &   --- &       &   -0.025758   &       \VS\\

133 &   2s$^2$13p   &   \SLPJ2Po{1/2}   &   --- &   --- &       &   -0.025021   &       \\
134 &   2s$^2$13p   &   \SLPJ2Po{3/2}   &   --- &   --- &       &   -0.025020   &   0.1 \VS\\

135 &   2s$^2$13d   &   \SLPJ2De{3/2}   &   --- &   --- &       &   -0.023920   &       \\
136 &   2s$^2$13d   &   \SLPJ2De{5/2}   &   --- &   --- &       &   -0.023919   &   0.1 \VS\\

137 &   2s$^2$13f   &   \SLPJ2Fo{5/2}   &   --- &   --- &       &   -0.023762   &       \\
138 &   2s$^2$13f   &   \SLPJ2Fo{7/2}   &   --- &   --- &       &   -0.023762   &   0.0 \VS\\

139 &   2s$^2$13g   &   \SLPJ2Ge{7/2}   &   --- &   --- &       &   -0.023690   &       \\
140 &   2s$^2$13g   &   \SLPJ2Ge{9/2}   &   --- &   --- &       &   -0.023690   &   0.0 \VS\\

141 &   2s$^2$13h   &   \SLPJ2Ho{9/2}   &   --- &   --- &       &   -0.023673   &       \\
142 &   2s$^2$13h   &   \SLPJ2Ho{11/2}  &   --- &   --- &       &   -0.023673   &   0.0 \VS\\

143 &   2s2p(\SLP3Po)3d     &    \SLPJ4Fo{3/2}  &   195752.58   &   -0.008312   &       &       &       \\
144 &   2s2p(\SLP3Po)3d     &    \SLPJ4Fo{5/2}  &   195765.85   &   -0.008191   &       &       &       \\
145 &   2s2p(\SLP3Po)3d     &    \SLPJ4Fo{7/2}  &   195785.74   &   -0.008010   &       &       &       \\
146 &   2s2p(\SLP3Po)3d     &    \SLPJ4Fo{9/2}  &   195813.66   &   -0.007755   &   61.1    &       &       \VS\\

147 &   2s2p(\SLP3Po)3d     &    \SLPJ4Do{1/2}  &   196557.87   &   -0.000974   &       &       &       \\
148 &   2s2p(\SLP3Po)3d     &    \SLPJ4Do{3/2}  &   196563.41   &   -0.000923   &       &       &       \\
149 &   2s2p(\SLP3Po)3d     &    \SLPJ4Do{5/2}  &   196571.82   &   -0.000846   &       &       &       \\
150 &   2s2p(\SLP3Po)3d     &    \SLPJ4Do{7/2}  &   196581.96   &   -0.000754   &   24.1    &       &       \VS\\

\end{longtable}
\begin{list}{}{}
\item[$^{a}$] Index.
\item[$^{b}$] Configuration. The 1s$^{2}$ core is suppressed from all configurations.
\item[$^{c}$] NIST Experimental Energy in Wavenumbers (cm$^{-1}$) relative to the ground state.
\item[$^{d}$] NIST Experimental Energy in Rydberg relative to the C$^{2+}$ \SLPJ1Se0\ limit.
\item[$^{e}$] Fine Structure Splitting from experimental values in cm$^{-1}$. %
\item[$^{f}$] Theoretical Energy in Rydberg from \Rm-matrix calculations relative to the C$^{2+}$ \SLPJ1Se0\ limit. %
\item[$^{g}$] Fine Structure Splitting from \Rm-matrix in cm$^{-1}$. The minus sign indicates that the theoretical levels are in reverse order compared to the experimental. %
\end{list}

}

\newpage

\phantomsection \addcontentsline{toc}{section}{\protect \numberline{} Resonances} %

{\tiny

\begin{longtable}{@{\extracolsep\fill}llllllllllll@{}}
\caption{The available experimental data from NIST for the resonance states of C$^+$ above the
C$^{2+}$ \SLPJ1Se0\ threshold alongside the theoretical results as obtained by \Km-matrix and \QB\
methods.
\label{RTableKQ}}\\
\hline\hline %
{In.$^a$} & {Config.$^b$} & {Level} & {NEEW$^c$} & {NEER$^d$} & {FSS$^e$} & {TERK$^f$} & {FSSK$^g$} & {FWHMK$^h$} & {TERQ$^i$} & {FSSQ$^j$} & {FWHMQ$^k$} \\ %
\hline
\endfirsthead
\caption[]{continued.}\\
\hline\hline %
{In.} & {Config.} & {Level} & {NEEW} & {NEER} & {FSS} & {TERK} & {FSSK} & {FWHMK} & {TERQ} & {FSSQ} & {FWHMQ} \\ %
\hline
\endhead
\hline
\endfoot

1   &   2s2p(\SLP3Po)3d     &    \SLPJ2Do{3/2}  &   198425.43   &   0.016045    &       &   0.017012    &       &   6.00E-10    &   0.017012    &       &   6.00E-10    \\
2   &   2s2p(\SLP3Po)3d     &    \SLPJ2Do{5/2}  &   198436.31   &   0.016144    &   10.9    &   0.017124    &   12.3    &   2.61E-09    &   0.017124    &   12.3    &  2.61E-09    \VS\\

3   &   2s2p(\SLP3Po)3d     &    \SLPJ4Po{5/2}  &   198844  &   0.019859    &       &   0.020655    &       &   2.17E-11    &    ---    &       &    ---    \\
4   &   2s2p(\SLP3Po)3d     &    \SLPJ4Po{3/2}  &   198865.25   &   0.020053    &       &   0.020849    &       &   1.26E-09    &   0.020849    &       &   1.26E-09    \\
5   &   2s2p(\SLP3Po)3d     &    \SLPJ4Po{1/2}  &   198879.01   &   0.020178    &   35.0    &   0.020968    &   34.3    &   5.10E-10    &   0.020968    &       &   5.10E-10    \VS\\

6   &   2s2p(\SLP3Po)3d     &    \SLPJ2Fo{5/2}  &   199941.41   &   0.029860    &       &   0.031693    &       &   5.61E-05    &   0.031689    &       &   5.61E-05    \\
7   &   2s2p(\SLP3Po)3d     &    \SLPJ2Fo{7/2}  &   199983.24   &   0.030241    &   41.8    &   0.032109    &   45.6    &   5.79E-05    &   0.032106    &   45.8    &   5.79E-05    \VS\\

8   &   2s2p(\SLP3Po)3d     &    \SLPJ2Po{3/2}  &   202179.85   &   0.050258    &       &   0.053787    &       &   1.06E-05    &   0.053787    &       &   1.06E-05    \\
9   &   2s2p(\SLP3Po)3d     &    \SLPJ2Po{1/2}  &   202204.52   &   0.050483    &   24.7    &   0.054019    &   25.5    &   1.12E-05    &   0.054019    &   25.5    &   1.12E-05    \VS\\

10  &   2s2p(\SLP3Po)4s     &    \SLPJ4Po{1/2}  &   209552.36   &   0.117441    &       &   0.118682    &       &   2.55E-07    &   0.118682    &       &   2.55E-07    \\
11  &   2s2p(\SLP3Po)4s     &    \SLPJ4Po{3/2}  &   209576.46   &   0.117661    &       &   0.118938    &       &   6.25E-07    &   0.118938    &       &   6.25E-07    \\
12  &   2s2p(\SLP3Po)4s     &    \SLPJ4Po{5/2}  &   209622.32   &   0.118079    &   70.0    &   0.119304    &   68.3    &   $<$1.0E-16  &    ---    &       &    ---    \VS\\

13  &   2s2p(\SLP3Po)4s     &    \SLPJ2Po{1/2}  &    ---    &    ---    &       &   0.137883    &       &  6.13E-03    &   0.137866    &       &   6.02E-03    \\
14  &   2s2p(\SLP3Po)4s     &    \SLPJ2Po{3/2}  &    ---    &    ---    &       &   0.138385    &   55.0    &   6.15E-03    &   0.138365    &   54.8    &   6.03E-03    \VS\\

15  &   2s2p(\SLP3Po)4p     &    \SLPJ2Pe{1/2}  &   214404.33   &   0.161655    &       &   0.162743    &       &   2.12E-08    &   0.162743    &       &   2.12E-08    \\
16  &   2s2p(\SLP3Po)4p     &    \SLPJ2Pe{3/2}  &   214429.95   &   0.161889    &   25.6    &   0.162996    &   27.7    &   9.89E-08    &   0.162996    &   27.8    &   9.89E-08    \VS\\

17  &   2s2p(\SLP3Po)4p     &    \SLPJ4De{1/2}  &   214759.91   &   0.164896    &       &   0.165629    &       &   5.30E-11    &   0.165629    &       &   5.31E-11    \\
18  &   2s2p(\SLP3Po)4p     &    \SLPJ4De{3/2}  &   214772.84   &   0.165014    &       &   0.165760    &       &   5.12E-08    &   0.165760    &       &   5.12E-08    \\
19  &   2s2p(\SLP3Po)4p     &    \SLPJ4De{5/2}  &   214795.27   &   0.165218    &       &   0.165984    &       &   6.95E-08    &   0.165984    &       &   6.95E-08    \\
20  &   2s2p(\SLP3Po)4p     &    \SLPJ4De{7/2}  &   214829.77   &   0.165532    &   69.9    &    ---    &       &    ---    &    ---    &       &   ---    \VS\\

21  &   2s2p(\SLP3Po)4p     &    \SLPJ4Se{3/2}  &   215767.77   &   0.174080    &       &  0.174625    &       &   6.59E-12    &    ---    &       &    ---    \VS\\

22  &   2s2p(\SLP3Po)4p     &    \SLPJ4Pe{1/2}  &   216362.84   &   0.179503    &       &   0.180780    &       &   7.26E-08    &   0.180780    &       &   7.26E-08    \\
23  &   2s2p(\SLP3Po)4p     &    \SLPJ4Pe{3/2}  &   216379.59   &   0.179655    &       &   0.180931    &       &   1.03E-07    &   0.180931    &       &   1.03E-07    \\
24  &   2s2p(\SLP3Po)4p     &    \SLPJ4Pe{5/2}  &   216400.57   &   0.179846    &   37.7    &   0.181145    &   40.0    &   6.05E-07    &   0.181145    &   40.0    &   6.05E-07    \VS\\

25  &   2s2p(\SLP3Po)4p     &    \SLPJ2De{3/2}  &    ---    &    ---    &       &   0.185420    &       &   1.19E-03    &   0.185422    &       &   1.18E-03    \\
26  &   2s2p(\SLP3Po)4p     &    \SLPJ2De{5/2}  &   216927  &   0.184644    &       &   0.185859    &   48.2    &   1.18E-03    &   0.185859    &   48.0    &   1.18E-03    \VS\\

27  &   2s2p(\SLP3Po)4p     &    \SLPJ2Se{1/2}  &    ---    &    ---    &       &   0.199765    &       &   4.28E-04    &   0.199765    &       &   4.28E-04    \VS\\

28  &   2s2p(\SLP1Po)3s     &    \SLPJ2Po{1/2}  &    ---    &    ---    &       &   0.205892    &       &   3.03E-03    &   0.205889    &       &   3.02E-03    \\
29  &   2s2p(\SLP1Po)3s     &    \SLPJ2Po{3/2}  &    ---    &    ---    &       &   0.205912    &   2.2 &   3.01E-03    &   0.205909    &   2.2 &   3.01E-03    \VS\\

30  &   2s2p(\SLP3Po)4d     &    \SLPJ4Fo{3/2}  &   219556.54   &   0.208606    &       &   0.208824    &       &   5.08E-11    &   0.208824    &       &   5.08E-11    \\
31  &   2s2p(\SLP3Po)4d     &    \SLPJ4Fo{5/2}  &   219570.15   &   0.208730    &       &   0.208968    &       &   4.23E-09    &   0.208968    &       &   4.23E-09    \\
32  &   2s2p(\SLP3Po)4d     &    \SLPJ4Fo{7/2}  &   219590.76   &   0.208918    &       &   0.209174    &       &   5.96E-09    &   0.209174    &       &   5.96E-09    \\
33  &   2s2p(\SLP3Po)4d     &    \SLPJ4Fo{9/2}  &   219619.88   &   0.209183    &   63.3    &    ---    &       &    ---    &    ---    &       &    ---    \VS\\

34  &   2s2p(\SLP3Po)4d     &    \SLPJ4Do{1/2}  &   220125.51   &   0.213791    &       &   0.214804    &       &   2.31E-09    &   0.214804    &       &   2.31E-09    \\
35  &   2s2p(\SLP3Po)4d     &    \SLPJ4Do{3/2}  &   220130.86   &   0.213839    &       &   0.214847    &       &   1.06E-09    &   0.214847    &       &   1.06E-09    \\
36  &   2s2p(\SLP3Po)4d     &    \SLPJ4Do{5/2}  &   220139.41   &   0.213917    &       &   0.214923    &       &   1.33E-09    &   0.214923    &       &   1.33E-09    \\
37  &   2s2p(\SLP3Po)4d     &    \SLPJ4Do{7/2}  &   220150.49   &   0.214018    &   25.0    &   0.215042    &   26.1    &   6.19E-09    &   0.215042    &   26.1    &   6.19E-09    \VS\\

38  &   2s2p(\SLP3Po)4d     &    \SLPJ2Do{3/2}  &   220601.53   &   0.218128    &       &   0.219325    &       &   1.99E-08    &   0.219325    &       &   1.99E-08    \\
39  &   2s2p(\SLP3Po)4d     &    \SLPJ2Do{5/2}  &   220614.51   &   0.218247    &   13.0    &   0.219441    &   12.7    &   5.68E-09    &   0.219441    &   12.7    &   5.68E-09    \VS\\

40  &   2s2p(\SLP3Po)4d     &    \SLPJ4Po{5/2}  &   220811.69   &   0.220044    &       &   0.220495    &       &   2.43E-10    &   0.220495    &       &   2.43E-10    \\
41  &   2s2p(\SLP3Po)4d     &    \SLPJ4Po{3/2}  &   220832.15   &   0.220230    &       &   0.220680    &       &   5.32E-10    &   0.220680    &       &   5.32E-10    \\
42  &   2s2p(\SLP3Po)4d     &    \SLPJ4Po{1/2}  &   220845.07   &   0.220348    &   33.4    &   0.220796    &   33.0    &   3.65E-10    &   0.220796    &   33.0    &   3.65E-10    \VS\\

43  &   2s2p(\SLP3Po)4f     &    \SLPJ2Fe{5/2}  &   221088.88   &   0.222570    &       &   0.223542    &       &   9.13E-09    &   0.223542    &       &   9.13E-09    \\
44  &   2s2p(\SLP3Po)4f     &    \SLPJ2Fe{7/2}  &   221097.92   &   0.222652    &   9.0 &   0.223626    &   9.2 &   1.98E-11    &    ---    &       &    ---    \VS\\

45  &   2s2p(\SLP3Po)4f     &    \SLPJ4Fe{3/2}  &   221093.95   &   0.222616    &       &   0.223626    &       &   5.96E-09    &   0.223626    &       &   5.96E-09    \\
46  &   2s2p(\SLP3Po)4f     &    \SLPJ4Fe{5/2}  &   221099.11   &   0.222663    &       &   0.223658    &       &   5.90E-09    &   0.223658    &       &   5.90E-09    \\
47  &   2s2p(\SLP3Po)4f     &    \SLPJ4Fe{7/2}  &   221105.73   &   0.222723    &       &   0.223725    &       &   5.43E-10    &   0.223725    &       &   5.43E-10    \\
48  &   2s2p(\SLP3Po)4f     &    \SLPJ4Fe{9/2}  &   221109.78   &   0.222760    &   15.8    &   0.223779    &   16.8    &   1.26E-09    &   0.223779    &   16.8    &   1.26E-09    \VS\\

49  &   2s2p(\SLP3Po)4d     &    \SLPJ2Fo{5/2}  &   221460.88   &   0.225959    &       &   0.227144    &       &   1.45E-05    &   0.227144    &       &   1.45E-05    \\
50  &   2s2p(\SLP3Po)4d     &    \SLPJ2Fo{7/2}  &   221503.33   &   0.226346    &   42.4    &   0.227556    &   45.1    &   1.51E-05    &   0.227556    &   45.1    &   1.51E-05    \VS\\

51  &   2s2p(\SLP3Po)4f     &    \SLPJ4Ge{5/2}  &   221544.81   &   0.226724    &       &   0.227648    &       &   5.45E-10    &   0.227648    &       &   5.46E-10    \\
52  &   2s2p(\SLP3Po)4f     &    \SLPJ4Ge{7/2}  &   221553.99   &   0.226808    &       &   0.227757    &       &   7.22E-09    &   0.227757    &       &   7.22E-09    \\
53  &   2s2p(\SLP3Po)4f     &    \SLPJ4Ge{9/2}  &   221575.61   &   0.227005    &       &   0.227966    &       &   6.53E-09    &   0.227966    &       &   6.54E-09    \\
54  &   2s2p(\SLP3Po)4f     &    \SLPJ4Ge{11/2} &   221604.9    &   0.227272    &   60.1    &    ---    &       &    ---    &    ---    &       &    ---    \VS\\

55  &   2s2p(\SLP3Po)4f     &    \SLPJ2Ge{7/2}  &   221587.12   &   0.227110    &       &   0.228164    &       &   2.58E-08    &   0.228164    &       &   2.58E-08    \\
56  &   2s2p(\SLP3Po)4f     &    \SLPJ2Ge{9/2}  &   221625.72   &   0.227462    &   38.6    &   0.228541    &   41.4    &   4.16E-08    &   0.228541    &   41.4    &   4.16E-08    \VS\\

57  &   2s2p(\SLP3Po)4f     &    \SLPJ4De{7/2}  &   221698.48   &   0.228125    &       &   0.229105    &       &   2.97E-12    &    ---    &       &   ---     \\
58  &   2s2p(\SLP3Po)4f     &    \SLPJ4De{5/2}  &   221707.71   &   0.228209    &       &   0.229229    &       &   2.39E-06    &   0.229229    &       &   2.39E-06    \\
59  &   2s2p(\SLP3Po)4f     &    \SLPJ4De{3/2}  &   221729.69   &   0.228409    &       &   0.229418    &       &   7.35E-07    &   0.229418    &       &   7.35E-07    \\
60  &   2s2p(\SLP3Po)4f     &    \SLPJ4De{1/2}  &   221742.79   &   0.228528    &   44.3    &   0.229529    &   46.5    &   $<$1.0E-16  &    ---    &       &    ---    \VS\\

61  &   2s2p(\SLP3Po)4f     &    \SLPJ2De{5/2}  &   221729.92   &   0.228411    &       &   0.229459    &       &   4.77E-06    &   0.229459    &       &   4.77E-06    \\
62  &   2s2p(\SLP3Po)4f     &    \SLPJ2De{3/2}  &   221752.26   &   0.228615    &   22.3    &   0.229695    &   25.9    &   6.58E-06    &   0.229696    &   25.9    &   6.58E-06    \VS\\

63  &   2s2p(\SLP3Po)4d     &    \SLPJ2Po{3/2}  &   222258.8    &   0.233231    &       &   0.234626    &       &   2.42E-05    &   0.234626    &       &   2.42E-05    \\
64  &   2s2p(\SLP3Po)4d     &    \SLPJ2Po{1/2}  &   222285.7    &   0.233476    &   26.9    &   0.234870    &   26.8    &   2.46E-05    &   0.234870    &   26.8    &   2.46E-05    \VS\\

\end{longtable}
\begin{list}{}{}
\item[$^{a}$] Index.
\item[$^{b}$] Configuration. The 1s$^{2}$ core is suppressed from all configurations.
\item[$^{c}$] NIST Experimental Energy in Wavenumbers (cm$^{-1}$) relative to the ground state.
\item[$^{d}$] NIST Experimental Energy in Rydberg relative to the C$^{2+}$ \SLPJ1Se0\ limit.
\item[$^{e}$] Fine Structure Splitting from experimental values in cm$^{-1}$. %
\item[$^{f}$] Theoretical Energy in Rydberg from \Km-matrix calculations relative to the C$^{2+}$ \SLPJ1Se0\ limit. %
\item[$^{g}$] Fine Structure Splitting from \Km-matrix in cm$^{-1}$. %
\item[$^{h}$] Full Width at Half Maximum from \Km-matrix in Rydberg. %
\item[$^{i}$] Theoretical Energy in Rydberg from \QB\ calculations relative to the C$^{2+}$ \SLPJ1Se0\ limit. %
\item[$^{j}$] Fine Structure Splitting from \QB\ in cm$^{-1}$. %
\item[$^{k}$] Full Width at Half Maximum from \QB\ in Rydberg. %
\end{list}

}


\clearpage

\phantomsection \addcontentsline{toc}{section}{\protect \numberline{} FB $\OS$-values} %

{\tiny

\begin{landscape}

\begin{longtable}{llllllllllllllll}
\caption{Free-bound $\OS$-values for bound symmetry $J^{\Par} = 1/2^{\rm{o}}$ obtained by
integrating \pcs s from \Rm-matrix calculations. The superscript denotes the power of 10
by which the number is to be multiplied. \label{fValues261o}}\\
\hline\hline %
In.&                \cC{1}&             \cC{12}&            \cC{20}&            \cC{22}&            \cC{27}&            \cC{32}&            \cC{34}&            \cC{49}&            \cC{64}&            \cC{77}&            \cC{88}&            \cC{99}&            \cC{110}&           \cC{121}&           \cC{133}\\
\hline %
\endfirsthead %
\caption[]{continued.}\\ %
\hline\hline %
In.&                \cC{1}&             \cC{12}&            \cC{20}&            \cC{22}&            \cC{27}&            \cC{32}&            \cC{34}&            \cC{49}&            \cC{64}&            \cC{77}&            \cC{88}&            \cC{99}&            \cC{110}&           \cC{121}&           \cC{133}\\
\hline %
\endhead %
\hline %
\endfoot %
\vspace{-0.2cm} \\
15&                 6.12\EE{-3}&        1.57\EE{-3}&        5.60\EE{-4}&        4.58\EE{-5}&        1.92\EE{-2}&        1.01\EE{-3}&        2.89\EE{-2}&        3.93\EE{-3}&        1.55\EE{-3}&        8.99\EE{-4}&        5.98\EE{-4}&        4.26\EE{-4}&        3.17\EE{-4}&        2.44\EE{-4}&        1.92\EE{-4}\\
16&                 3.18\EE{-3}&        8.30\EE{-4}&        2.69\EE{-4}&        1.21\EE{-5}&        9.72\EE{-3}&        5.02\EE{-4}&        1.47\EE{-2}&        2.02\EE{-3}&        7.92\EE{-4}&        4.63\EE{-4}&        3.07\EE{-4}&        2.20\EE{-4}&        1.64\EE{-4}&        1.26\EE{-4}&        9.93\EE{-5}\\
17&                 1.70\EE{-5}&        4.69\EE{-6}&        1.59\EE{-6}&        1.34\EE{-2}&        5.91\EE{-5}&        2.92\EE{-6}&        8.14\EE{-5}&        1.09\EE{-5}&        4.27\EE{-6}&        2.47\EE{-6}&        1.64\EE{-6}&        1.16\EE{-6}&        8.63\EE{-7}&        6.60\EE{-7}&        5.15\EE{-7}\\
18&                 8.53\EE{-6}&        3.43\EE{-6}&        4.94\EE{-7}&        1.35\EE{-2}&        1.19\EE{-5}&        7.05\EE{-7}&        2.68\EE{-5}&        5.00\EE{-6}&        1.95\EE{-6}&        1.19\EE{-6}&        8.00\EE{-7}&        5.89\EE{-7}&        4.64\EE{-7}&        3.50\EE{-7}&        2.90\EE{-7}\\
21&                 4.43\EE{-7}&        1.39\EE{-7}&        5.34\EE{-8}&        1.78\EE{-3}&        1.67\EE{-6}&        7.21\EE{-8}&        2.04\EE{-6}&        2.71\EE{-7}&        1.04\EE{-7}&        5.92\EE{-8}&        3.88\EE{-8}&        2.71\EE{-8}&        1.99\EE{-8}&        1.51\EE{-8}&        1.18\EE{-8}\\
22&                 4.48\EE{-7}&        3.87\EE{-7}&        4.64\EE{-9}&        2.22\EE{-4}&        6.57\EE{-13}&       7.26\EE{-8}&        8.82\EE{-7}&        1.54\EE{-7}&        7.90\EE{-8}&        5.31\EE{-8}&        3.85\EE{-8}&        2.95\EE{-8}&        2.41\EE{-8}&        1.91\EE{-8}&        1.53\EE{-8}\\
23&                 2.44\EE{-6}&        1.65\EE{-6}&        6.16\EE{-7}&        1.01\EE{-3}&        1.65\EE{-7}&        5.61\EE{-8}&        7.62\EE{-7}&        5.76\EE{-7}&        3.75\EE{-7}&        2.74\EE{-7}&        2.08\EE{-7}&        1.61\EE{-7}&        1.27\EE{-7}&        1.02\EE{-7}&        8.32\EE{-8}\\
25&                 1.81\EE{-2}&        1.13\EE{-2}&        3.63\EE{-3}&        3.44\EE{-7}&        2.22\EE{-3}&        1.01\EE{-3}&        1.66\EE{-2}&        7.00\EE{-3}&        4.20\EE{-3}&        2.96\EE{-3}&        2.20\EE{-3}&        1.69\EE{-3}&        1.32\EE{-3}&        1.05\EE{-3}&        8.50\EE{-4}\\
27&                 1.81\EE{-3}&        2.76\EE{-3}&        7.42\EE{-4}&        3.83\EE{-8}&        1.55\EE{-3}&        4.26\EE{-4}&        2.16\EE{-3}&        4.12\EE{-4}&        2.69\EE{-4}&        1.98\EE{-4}&        1.52\EE{-4}&        1.20\EE{-4}&        9.55\EE{-5}&        7.73\EE{-5}&        6.32\EE{-5}\\
45&                 1.31\EE{-9}&        3.51\EE{-8}&        1.80\EE{-8}&        7.18\EE{-7}&        3.61\EE{-6}&        6.21\EE{-6}&        1.84\EE{-5}&        6.64\EE{-7}&        1.40\EE{-7}&        6.59\EE{-8}&        4.25\EE{-8}&        2.98\EE{-8}&        2.26\EE{-8}&        1.82\EE{-8}&        1.53\EE{-8}\\
59&                 8.42\EE{-7}&        3.47\EE{-6}&        1.84\EE{-6}&        6.24\EE{-4}&        4.20\EE{-4}&        6.82\EE{-4}&        1.99\EE{-3}&        6.74\EE{-5}&        1.46\EE{-5}&        6.42\EE{-6}&        3.64\EE{-6}&        2.34\EE{-6}&        1.61\EE{-6}&        1.17\EE{-6}&        8.82\EE{-7}\\
60&                 1.93\EE{-10}&       1.70\EE{-10}&       2.09\EE{-11}&       5.12\EE{-4}&        1.07\EE{-9}&        1.51\EE{-9}&        3.28\EE{-9}&        2.78\EE{-10}&       7.52\EE{-13}&       2.96\EE{-13}&       2.36\EE{-13}&       1.88\EE{-13}&       5.71\EE{-13}&       4.68\EE{-13}&       3.86\EE{-13}\\
62&                 8.30\EE{-6}&        1.94\EE{-5}&        1.09\EE{-5}&        6.60\EE{-5}&        3.76\EE{-3}&        6.16\EE{-3}&        1.74\EE{-2}&        5.94\EE{-4}&        1.29\EE{-4}&        5.62\EE{-5}&        3.17\EE{-5}&        2.03\EE{-5}&        1.39\EE{-5}&        1.00\EE{-5}&        7.51\EE{-6}\\

\end{longtable}


\clearpage

\begin{longtable}{llllllllllllllllll}
\caption{Free-bound $\OS$-values for bound symmetry $J^{\Par} = 3/2^{\rm{o}}$ obtained by
integrating \pcs s from \Rm-matrix calculations. The superscript denotes the power of 10
by which the number is to be multiplied. \label{fValues263o}}\\
\hline\hline %
In.&                \cC{2}&             \cC{13}&            \cC{14}&            \cC{18}&            \cC{21}&            \cC{23}&            \cC{28}&            \cC{33}&            \cC{35}&            \cC{50}&            \cC{65}&            \cC{78}&            \cC{89}&            \cC{100}&           \cC{111}&           \cC{122}&           \cC{134}\\
\hline %
\endfirsthead %
\caption[]{continued.}\\ %
\hline\hline %
In.&                \cC{2}&             \cC{13}&            \cC{14}&            \cC{18}&            \cC{21}&            \cC{23}&            \cC{28}&            \cC{33}&            \cC{35}&            \cC{50}&            \cC{65}&            \cC{78}&            \cC{89}&            \cC{100}&           \cC{111}&           \cC{122}&           \cC{134}\\
\hline %
\endhead %
\hline %
\endfoot %
\vspace{-0.2cm} \\
15&                 1.49\EE{-3}&        3.73\EE{-4}&        3.75\EE{-10}&       2.32\EE{-3}&        1.45\EE{-4}&        6.09\EE{-6}&        4.82\EE{-3}&        2.45\EE{-4}&        7.17\EE{-3}&        9.78\EE{-4}&        3.82\EE{-4}&        2.21\EE{-4}&        1.47\EE{-4}&        1.04\EE{-4}&        7.77\EE{-5}&        5.97\EE{-5}&        4.69\EE{-5}\\
16&                 7.53\EE{-3}&        1.91\EE{-3}&        1.58\EE{-9}&        4.40\EE{-4}&        7.03\EE{-4}&        2.70\EE{-5}&        2.39\EE{-2}&        1.23\EE{-3}&        3.60\EE{-2}&        4.91\EE{-3}&        1.92\EE{-3}&        1.11\EE{-3}&        7.36\EE{-4}&        5.25\EE{-4}&        3.91\EE{-4}&        2.99\EE{-4}&        2.35\EE{-4}\\
17&                 4.15\EE{-6}&        1.12\EE{-6}&        1.96\EE{-10}&       6.73\EE{-6}&        4.15\EE{-7}&        1.33\EE{-3}&        1.48\EE{-5}&        7.06\EE{-7}&        2.02\EE{-5}&        2.73\EE{-6}&        1.06\EE{-6}&        6.09\EE{-7}&        4.02\EE{-7}&        2.85\EE{-7}&        2.11\EE{-7}&        1.61\EE{-7}&        1.26\EE{-7}\\
18&                 9.38\EE{-6}&        2.33\EE{-6}&        1.26\EE{-9}&        2.57\EE{-7}&        9.49\EE{-7}&        8.54\EE{-3}&        3.96\EE{-5}&        1.87\EE{-6}&        5.00\EE{-5}&        6.55\EE{-6}&        2.40\EE{-6}&        1.38\EE{-6}&        9.09\EE{-7}&        6.43\EE{-7}&        4.53\EE{-7}&        3.45\EE{-7}&        2.68\EE{-7}\\
19&                 8.52\EE{-7}&        6.35\EE{-7}&        2.88\EE{-9}&        1.31\EE{-8}&        6.29\EE{-8}&        1.70\EE{-2}&        3.61\EE{-7}&        1.17\EE{-7}&        4.99\EE{-7}&        4.47\EE{-7}&        2.97\EE{-7}&        2.18\EE{-7}&        1.67\EE{-7}&        1.30\EE{-7}&        1.03\EE{-7}&        8.24\EE{-8}&        6.72\EE{-8}\\
21&                 1.07\EE{-6}&        3.31\EE{-7}&        9.71\EE{-9}&        7.28\EE{-8}&        1.41\EE{-7}&        1.73\EE{-3}&        4.13\EE{-6}&        1.77\EE{-7}&        5.05\EE{-6}&        6.70\EE{-7}&        2.56\EE{-7}&        1.45\EE{-7}&        9.47\EE{-8}&        6.65\EE{-8}&        4.88\EE{-8}&        3.70\EE{-8}&        2.88\EE{-8}\\
22&                 8.46\EE{-8}&        1.94\EE{-7}&        3.34\EE{-6}&        2.59\EE{-8}&        8.10\EE{-8}&        6.20\EE{-4}&        5.42\EE{-7}&        5.30\EE{-8}&        2.18\EE{-8}&        3.66\EE{-8}&        3.16\EE{-8}&        2.63\EE{-8}&        2.16\EE{-8}&        1.76\EE{-8}&        1.44\EE{-8}&        1.18\EE{-8}&        8.36\EE{-9}\\
23&                 8.05\EE{-7}&        3.16\EE{-7}&        6.69\EE{-6}&        3.12\EE{-7}&        7.82\EE{-8}&        1.75\EE{-4}&        8.71\EE{-7}&        5.77\EE{-8}&        1.69\EE{-6}&        3.10\EE{-7}&        1.41\EE{-7}&        8.84\EE{-8}&        6.13\EE{-8}&        4.48\EE{-8}&        3.38\EE{-8}&        2.61\EE{-8}&        2.06\EE{-8}\\
24&                 1.21\EE{-5}&        7.19\EE{-6}&        1.02\EE{-5}&        1.33\EE{-7}&        1.97\EE{-6}&        5.70\EE{-4}&        1.39\EE{-6}&        9.26\EE{-7}&        9.35\EE{-6}&        4.66\EE{-6}&        2.71\EE{-6}&        1.93\EE{-6}&        1.44\EE{-6}&        1.11\EE{-6}&        8.68\EE{-7}&        6.92\EE{-7}&        5.60\EE{-7}\\
25&                 2.17\EE{-3}&        1.34\EE{-3}&        4.07\EE{-10}&       2.32\EE{-3}&        3.60\EE{-4}&        1.03\EE{-7}&        2.82\EE{-4}&        1.55\EE{-4}&        1.81\EE{-3}&        7.25\EE{-4}&        4.35\EE{-4}&        3.06\EE{-4}&        2.27\EE{-4}&        1.74\EE{-4}&        1.36\EE{-4}&        1.08\EE{-4}&        8.72\EE{-5}\\
26&                 1.65\EE{-2}&        1.03\EE{-2}&        7.75\EE{-9}&        2.52\EE{-4}&        2.85\EE{-3}&        1.34\EE{-6}&        2.04\EE{-3}&        6.47\EE{-4}&        1.55\EE{-2}&        6.42\EE{-3}&        3.82\EE{-3}&        2.70\EE{-3}&        2.01\EE{-3}&        1.40\EE{-3}&        1.10\EE{-3}&        8.74\EE{-4}&        7.06\EE{-4}\\
27&                 1.83\EE{-3}&        2.69\EE{-3}&        5.75\EE{-9}&        3.73\EE{-8}&        6.92\EE{-4}&        9.90\EE{-8}&        1.46\EE{-3}&        4.33\EE{-4}&        2.26\EE{-3}&        4.40\EE{-4}&        2.77\EE{-4}&        2.29\EE{-4}&        1.76\EE{-4}&        1.39\EE{-4}&        1.11\EE{-4}&        8.97\EE{-5}&        7.34\EE{-5}\\
43&                 8.82\EE{-10}&       3.29\EE{-8}&        2.11\EE{-11}&       1.95\EE{-2}&        4.70\EE{-10}&       3.85\EE{-7}&        4.71\EE{-6}&        9.11\EE{-6}&        2.73\EE{-5}&        9.58\EE{-7}&        2.10\EE{-7}&        9.16\EE{-8}&        5.05\EE{-8}&        3.29\EE{-8}&        2.22\EE{-8}&        1.56\EE{-8}&        1.13\EE{-8}\\
45&                 1.40\EE{-10}&       3.21\EE{-9}&        1.02\EE{-13}&       1.56\EE{-6}&        1.28\EE{-9}&        4.30\EE{-7}&        3.43\EE{-7}&        5.82\EE{-7}&        1.77\EE{-6}&        6.67\EE{-8}&        1.43\EE{-8}&        7.03\EE{-9}&        4.33\EE{-9}&        3.03\EE{-9}&        2.30\EE{-9}&        1.85\EE{-9}&        1.55\EE{-9}\\
46&                 3.92\EE{-10}&       1.91\EE{-8}&        3.10\EE{-12}&       8.73\EE{-4}&        9.29\EE{-9}&        1.59\EE{-6}&        3.18\EE{-6}&        5.65\EE{-6}&        1.69\EE{-5}&        5.77\EE{-7}&        1.32\EE{-7}&        6.04\EE{-8}&        3.58\EE{-8}&        2.41\EE{-8}&        1.77\EE{-8}&        1.37\EE{-8}&        1.11\EE{-8}\\
51&                 2.59\EE{-10}&       2.29\EE{-9}&        4.10\EE{-13}&       2.84\EE{-5}&        1.02\EE{-9}&        2.51\EE{-8}&        2.90\EE{-7}&        4.87\EE{-7}&        1.41\EE{-6}&        4.77\EE{-8}&        1.03\EE{-8}&        4.54\EE{-9}&        2.56\EE{-9}&        1.64\EE{-9}&        1.13\EE{-9}&        8.19\EE{-10}&       6.15\EE{-10}\\
58&                 2.89\EE{-6}&        9.08\EE{-6}&        8.44\EE{-10}&       3.22\EE{-5}&        3.81\EE{-6}&        5.82\EE{-4}&        1.25\EE{-3}&        2.06\EE{-3}&        5.93\EE{-3}&        1.97\EE{-4}&        4.20\EE{-5}&        1.81\EE{-5}&        1.05\EE{-5}&        6.70\EE{-6}&        4.61\EE{-6}&        3.32\EE{-6}&        2.49\EE{-6}\\
59&                 7.73\EE{-9}&        3.06\EE{-7}&        2.55\EE{-11}&       1.70\EE{-4}&        1.65\EE{-7}&        3.98\EE{-4}&        4.12\EE{-5}&        6.90\EE{-5}&        1.98\EE{-4}&        6.91\EE{-6}&        1.45\EE{-6}&        6.29\EE{-7}&        3.54\EE{-7}&        2.39\EE{-7}&        1.64\EE{-7}&        1.19\EE{-7}&        8.89\EE{-8}\\
60&                 1.58\EE{-11}&       1.07\EE{-12}&       1.03\EE{-10}&       3.51\EE{-11}&       9.86\EE{-12}&       5.14\EE{-5}&        2.41\EE{-10}&       3.99\EE{-10}&       8.16\EE{-10}&       1.11\EE{-10}&       6.77\EE{-13}&       5.32\EE{-13}&       4.18\EE{-13}&       3.31\EE{-13}&       3.03\EE{-13}&       2.46\EE{-13}&       2.00\EE{-13}\\
61&                 6.45\EE{-6}&        1.47\EE{-5}&        5.74\EE{-9}&        2.85\EE{-5}&        7.05\EE{-6}&        2.76\EE{-4}&        2.50\EE{-3}&        4.03\EE{-3}&        1.17\EE{-2}&        4.03\EE{-4}&        8.49\EE{-5}&        3.67\EE{-5}&        2.07\EE{-5}&        1.31\EE{-5}&        8.95\EE{-6}&        6.41\EE{-6}&        4.77\EE{-6}\\
62&                 6.54\EE{-7}&        1.93\EE{-6}&        8.56\EE{-10}&       1.53\EE{-3}&        1.60\EE{-6}&        4.23\EE{-5}&        3.67\EE{-4}&        5.87\EE{-4}&        1.70\EE{-3}&        6.05\EE{-5}&        1.31\EE{-5}&        5.72\EE{-6}&        3.24\EE{-6}&        2.07\EE{-6}&        1.42\EE{-6}&        1.02\EE{-6}&        7.67\EE{-7}\\

\end{longtable}


\clearpage

\begin{longtable}{lllllllllllll}
\caption{Free-bound $\OS$-values for bound symmetry $J^{\Par} = 5/2^{\rm{o}}$ obtained by
integrating \pcs s from \Rm-matrix calculations. The superscript denotes the power of 10
by which the number is to be multiplied. \label{fValues265o}}\\
\hline\hline %
In.&                \cC{17}&            \cC{24}&            \cC{29}&            \cC{38}&            \cC{53}&            \cC{68}&            \cC{79}&            \cC{92}&            \cC{103}&           \cC{114}&           \cC{126}&           \cC{137}\\
\hline %
\endfirsthead %
\caption[]{continued.}\\ %
\hline\hline %
In.&                \cC{17}&            \cC{24}&            \cC{29}&            \cC{38}&            \cC{53}&            \cC{68}&            \cC{79}&            \cC{92}&            \cC{103}&           \cC{114}&           \cC{126}&           \cC{137}\\
\hline %
\endhead %
\hline %
\endfoot %
\vspace{-0.2cm} \\
16&                 2.78\EE{-3}&        2.11\EE{-7}&        1.34\EE{-8}&        8.88\EE{-9}&        8.79\EE{-9}&        8.27\EE{-9}&        7.47\EE{-9}&        6.57\EE{-9}&        5.69\EE{-9}&        4.90\EE{-9}&        4.15\EE{-9}&        3.61\EE{-9}\\
18&                 4.47\EE{-6}&        5.27\EE{-4}&        3.69\EE{-9}&        3.39\EE{-9}&        3.30\EE{-9}&        3.25\EE{-9}&        3.14\EE{-9}&        3.18\EE{-9}&        3.12\EE{-9}&        3.09\EE{-9}&        2.98\EE{-9}&        3.21\EE{-9}\\
19&                 1.24\EE{-7}&        4.78\EE{-3}&        4.64\EE{-10}&       4.43\EE{-10}&       4.37\EE{-10}&       4.30\EE{-10}&       4.17\EE{-10}&       4.00\EE{-10}&       3.82\EE{-10}&       3.48\EE{-10}&       3.66\EE{-10}&       3.99\EE{-10}\\
20&                 2.32\EE{-9}&        1.20\EE{-2}&        1.72\EE{-17}&       1.38\EE{-16}&       3.10\EE{-15}&       0&                  0&                  0&                  0&                  0&                  0&                  0\\
21&                 4.56\EE{-7}&        1.61\EE{-3}&        3.64\EE{-13}&       3.65\EE{-13}&       3.21\EE{-13}&       1.59\EE{-13}&       1.15\EE{-13}&       4.49\EE{-15}&       7.60\EE{-14}&       7.09\EE{-14}&       7.56\EE{-14}&       1.44\EE{-13}\\
23&                 2.24\EE{-8}&        5.24\EE{-4}&        9.47\EE{-9}&        8.94\EE{-9}&        9.08\EE{-9}&        8.40\EE{-9}&        6.36\EE{-9}&        5.54\EE{-9}&        4.76\EE{-9}&        4.07\EE{-9}&        3.93\EE{-9}&        3.43\EE{-9}\\
24&                 1.24\EE{-6}&        1.07\EE{-3}&        3.20\EE{-9}&        2.91\EE{-9}&        2.67\EE{-9}&        2.43\EE{-9}&        2.17\EE{-9}&        1.92\EE{-9}&        1.69\EE{-9}&        1.48\EE{-9}&        1.30\EE{-9}&        1.14\EE{-9}\\
25&                 1.57\EE{-4}&        1.52\EE{-10}&       5.59\EE{-5}&        4.91\EE{-5}&        3.89\EE{-5}&        3.48\EE{-5}&        3.05\EE{-5}&        2.63\EE{-5}&        2.26\EE{-5}&        1.93\EE{-5}&        1.65\EE{-5}&        1.25\EE{-5}\\
26&                 2.29\EE{-3}&        3.79\EE{-8}&        3.50\EE{-6}&        3.16\EE{-6}&        2.83\EE{-6}&        2.51\EE{-6}&        2.20\EE{-6}&        1.90\EE{-6}&        1.63\EE{-6}&        1.39\EE{-6}&        1.19\EE{-6}&        9.00\EE{-7}\\
43&                 8.17\EE{-4}&        1.17\EE{-7}&        3.26\EE{-7}&        5.10\EE{-7}&        6.46\EE{-7}&        7.25\EE{-7}&        7.59\EE{-7}&        7.58\EE{-7}&        7.35\EE{-7}&        6.97\EE{-7}&        6.51\EE{-7}&        6.02\EE{-7}\\
44&                 1.46\EE{-2}&        2.83\EE{-6}&        9.13\EE{-8}&        7.43\EE{-8}&        9.39\EE{-8}&        1.06\EE{-7}&        1.12\EE{-7}&        1.14\EE{-7}&        1.12\EE{-7}&        1.08\EE{-7}&        1.03\EE{-7}&        9.68\EE{-8}\\
45&                 1.09\EE{-7}&        2.81\EE{-8}&        1.63\EE{-8}&        2.04\EE{-10}&       2.08\EE{-10}&       2.01\EE{-10}&       1.94\EE{-10}&       1.87\EE{-10}&       1.83\EE{-10}&       1.98\EE{-10}&       2.20\EE{-10}&       2.44\EE{-10}\\
46&                 2.49\EE{-5}&        4.66\EE{-7}&        3.70\EE{-9}&        2.17\EE{-8}&        2.74\EE{-8}&        3.06\EE{-8}&        3.17\EE{-8}&        3.13\EE{-8}&        2.99\EE{-8}&        2.77\EE{-8}&        2.50\EE{-8}&        2.22\EE{-8}\\
47&                 4.70\EE{-3}&        9.64\EE{-7}&        4.28\EE{-8}&        9.19\EE{-9}&        1.03\EE{-8}&        1.06\EE{-8}&        1.05\EE{-8}&        9.61\EE{-9}&        8.45\EE{-9}&        7.13\EE{-9}&        5.78\EE{-9}&        4.46\EE{-9}\\
51&                 5.49\EE{-7}&        7.39\EE{-9}&        1.91\EE{-8}&        2.24\EE{-10}&       2.40\EE{-10}&       2.20\EE{-10}&       1.69\EE{-10}&       1.04\EE{-10}&       4.31\EE{-11}&       4.57\EE{-12}&       1.03\EE{-11}&       9.47\EE{-11}\\
52&                 3.19\EE{-7}&        1.35\EE{-7}&        2.87\EE{-7}&        7.10\EE{-7}&        9.12\EE{-7}&        1.04\EE{-6}&        1.11\EE{-6}&        1.13\EE{-6}&        1.12\EE{-6}&        1.09\EE{-6}&        1.04\EE{-6}&        9.99\EE{-7}\\
55&                 5.58\EE{-5}&        6.96\EE{-8}&        3.70\EE{-6}&        5.17\EE{-6}&        6.57\EE{-6}&        7.40\EE{-6}&        7.76\EE{-6}&        7.76\EE{-6}&        7.51\EE{-6}&        7.11\EE{-6}&        6.62\EE{-6}&        6.08\EE{-6}\\
57&                 1.23\EE{-5}&        9.69\EE{-4}&        1.91\EE{-9}&        3.11\EE{-10}&       4.07\EE{-10}&       4.68\EE{-10}&       5.02\EE{-10}&       5.15\EE{-10}&       5.15\EE{-10}&       5.06\EE{-10}&       4.92\EE{-10}&       4.75\EE{-10}\\
58&                 6.08\EE{-4}&        1.62\EE{-4}&        1.33\EE{-8}&        5.77\EE{-9}&        5.26\EE{-9}&        4.56\EE{-9}&        3.83\EE{-9}&        3.16\EE{-9}&        2.57\EE{-9}&        2.10\EE{-9}&        1.70\EE{-9}&        1.41\EE{-9}\\
59&                 1.27\EE{-5}&        2.48\EE{-5}&        8.56\EE{-9}&        2.70\EE{-8}&        2.53\EE{-8}&        2.30\EE{-8}&        2.04\EE{-8}&        1.77\EE{-8}&        1.52\EE{-8}&        1.31\EE{-8}&        1.11\EE{-8}&        9.56\EE{-9}\\
61&                 1.26\EE{-3}&        8.08\EE{-5}&        1.74\EE{-9}&        4.53\EE{-9}&        3.89\EE{-9}&        3.29\EE{-9}&        2.76\EE{-9}&        2.31\EE{-9}&        1.95\EE{-9}&        1.65\EE{-9}&        1.41\EE{-9}&        1.24\EE{-9}\\
62&                 1.12\EE{-4}&        2.75\EE{-6}&        2.04\EE{-7}&        2.38\EE{-7}&        2.25\EE{-7}&        2.05\EE{-7}&        1.81\EE{-7}&        1.57\EE{-7}&        1.36\EE{-7}&        1.17\EE{-7}&        9.93\EE{-8}&        8.59\EE{-8}\\

\end{longtable}

\end{landscape}

}


\clearpage

\phantomsection \addcontentsline{toc}{section}{\protect \numberline{} Sample of \SSo} %

\begin{table} [!h]

\vspace{-1.5cm}

\caption{Sample of the \SSo\ line list where several lines have been observed astronomically. The
first column is for experimental/theoretical energy identification for the upper and lower states
respectively where 1 refers to experimental while 0 refers to theoretical. The other columns are
for the atomic designation of the upper and lower states respectively, followed by the air
wavelength in angstrom and effective dielectronic recombination rate coefficients in
cm$^3$.s$^{-1}$ for the given logarithmic temperatures. The superscript denotes the power of 10 by
which the number is to be multiplied. The 1s$^{2}$ core is suppressed from all configurations.
\label{ListSample}}
 \vspace{-0.7cm}
\begin{center}
{\tiny
\begin{tabular}{llllllllll}
\hline
    {\bf } &     {\bf } &     {\bf } &     {\bf } &                                           \multicolumn{ 6}{c}{{\bf log10(T)}} \\

  {\bf ET} & {\bf Upper} & {\bf Lower} & {\bf $\WL_{air}$} &  {\bf 2.0} &  {\bf 2.5} &  {\bf 3.0} &  {\bf 3.5} &  {\bf 4.0} &  {\bf 4.5} \\
\hline
        11 & 2s2p(\SLP3Po)4f \SLPJ4De{1/2} & 2s2p(\SLP3Po)3d \SLPJ2Po{1/2} &    5116.73 &  5.53\EE{-190} &   2.59\EE{-81} &   1.85\EE{-47} &   2.89\EE{-37} &   1.48\EE{-34} &   3.27\EE{-34} \\

        11 & 2s2p(\SLP3Po)4f \SLPJ4De{5/2} & 2s2p(\SLP3Po)3d \SLPJ2Po{3/2} &    5119.46 &  5.14\EE{-168} &   9.10\EE{-62} &   1.11\EE{-28} &   9.88\EE{-19} &   4.25\EE{-16} &   8.88\EE{-16} \\

        11 & 2s2p(\SLP3Po)3p \SLPJ2Pe{3/2} & 2s$^2$4p \SLPJ2Po{1/2} &    5120.08 &   2.50\EE{-26} &   1.64\EE{-19} &   7.84\EE{-18} &   3.26\EE{-17} &   3.79\EE{-17} &   1.76\EE{-17} \\

        11 & 2s2p(\SLP3Po)4f \SLPJ4De{3/2} & 2s2p(\SLP3Po)3d \SLPJ2Po{1/2} &    5120.17 &  6.88\EE{-169} &   1.51\EE{-62} &   1.97\EE{-29} &   1.80\EE{-19} &   7.78\EE{-17} &   1.63\EE{-16} \\

        11 & 2s2p(\SLP3Po)3p \SLPJ2Pe{3/2} & 2s$^2$4p \SLPJ2Po{3/2} &    5121.83 &   1.20\EE{-25} &   7.87\EE{-19} &   3.75\EE{-17} &   1.56\EE{-16} &   1.81\EE{-16} &   8.40\EE{-17} \\

        11 & 2s$^2$7f \SLPJ2Fo{5/2} & 2s$^2$4d \SLPJ2De{3/2} &    5122.09 &   6.05\EE{-29} &   7.95\EE{-22} &   2.03\EE{-18} &   9.32\EE{-18} &   6.66\EE{-18} &   3.55\EE{-18} \\

        11 & 2s$^2$7f \SLPJ2Fo{7/2} & 2s$^2$4d \SLPJ2De{5/2} &    5122.27 &   2.11\EE{-28} &   1.95\EE{-21} &   3.05\EE{-18} &   1.44\EE{-17} &   1.03\EE{-17} &   5.43\EE{-18} \\

        11 & 2s$^2$7f \SLPJ2Fo{5/2} & 2s$^2$4d \SLPJ2De{5/2} &    5122.27 &   4.32\EE{-30} &   5.68\EE{-23} &   1.45\EE{-19} &   6.66\EE{-19} &   4.76\EE{-19} &   2.54\EE{-19} \\

        11 & 2s2p(\SLP3Po)3p \SLPJ2Pe{1/2} & 2s$^2$4p \SLPJ2Po{1/2} &    5125.21 &   1.46\EE{-26} &   8.64\EE{-20} &   4.82\EE{-18} &   5.15\EE{-17} &   7.05\EE{-17} &   3.50\EE{-17} \\

        11 & 2s2p(\SLP3Po)3p \SLPJ2Pe{1/2} & 2s$^2$4p \SLPJ2Po{3/2} &    5126.96 &   7.08\EE{-27} &   4.20\EE{-20} &   2.34\EE{-18} &   2.50\EE{-17} &   3.42\EE{-17} &   1.70\EE{-17} \\

        11 & 2s2p(\SLP3Po)3p \SLPJ4Pe{3/2} & 2s2p(\SLP3Po)3s \SLPJ4Po{1/2} &    5132.95 &   4.96\EE{-28} &   9.80\EE{-20} &   1.65\EE{-17} &   2.49\EE{-16} &   2.26\EE{-15} &   1.48\EE{-15} \\

        11 & 2s2p(\SLP3Po)3p \SLPJ4Pe{5/2} & 2s2p(\SLP3Po)3s \SLPJ4Po{3/2} &    5133.28 &   1.69\EE{-27} &   1.09\EE{-19} &   1.71\EE{-17} &   1.66\EE{-16} &   1.50\EE{-15} &   1.08\EE{-15} \\

        11 & 2s2p(\SLP3Po)3p \SLPJ4Pe{1/2} & 2s2p(\SLP3Po)3s \SLPJ4Po{1/2} &    5137.26 &   1.15\EE{-28} &   5.07\EE{-20} &   8.50\EE{-18} &   9.92\EE{-17} &   8.62\EE{-16} &   5.55\EE{-16} \\

        11 & 2s2p(\SLP3Po)3p \SLPJ4Pe{3/2} & 2s2p(\SLP3Po)3s \SLPJ4Po{3/2} &    5139.17 &   1.57\EE{-28} &   3.10\EE{-20} &   5.23\EE{-18} &   7.89\EE{-17} &   7.15\EE{-16} &   4.68\EE{-16} \\

        01 & 2s$^2$11p \SLPJ2Po{3/2} & 2s$^2$5s \SLPJ2Se{1/2} &    5141.70 &   9.29\EE{-33} &   1.36\EE{-25} &   3.26\EE{-21} &   1.25\EE{-19} &   6.35\EE{-19} &   6.92\EE{-19} \\

        01 & 2s$^2$11p \SLPJ2Po{1/2} & 2s$^2$5s \SLPJ2Se{1/2} &    5141.76 &   2.10\EE{-33} &   4.79\EE{-26} &   1.60\EE{-21} &   6.26\EE{-20} &   3.16\EE{-19} &   3.48\EE{-19} \\

        11 & 2s2p(\SLP3Po)3p \SLPJ4Pe{1/2} & 2s2p(\SLP3Po)3s \SLPJ4Po{3/2} &    5143.49 &   5.77\EE{-28} &   2.55\EE{-19} &   4.27\EE{-17} &   4.99\EE{-16} &   4.34\EE{-15} &   2.79\EE{-15} \\

        11 & 2s2p(\SLP3Po)3p \SLPJ4Pe{5/2} & 2s2p(\SLP3Po)3s \SLPJ4Po{5/2} &    5145.16 &   3.97\EE{-27} &   2.56\EE{-19} &   4.01\EE{-17} &   3.90\EE{-16} &   3.54\EE{-15} &   2.53\EE{-15} \\

        11 & 2s2p(\SLP3Po)3p \SLPJ4Pe{3/2} & 2s2p(\SLP3Po)3s \SLPJ4Po{5/2} &    5151.08 &   5.59\EE{-28} &   1.11\EE{-19} &   1.86\EE{-17} &   2.81\EE{-16} &   2.55\EE{-15} &   1.67\EE{-15} \\

        11 & 2s2p(\SLP3Po)4f \SLPJ4Ge{5/2} & 2s2p(\SLP3Po)3d \SLPJ2Po{3/2} &    5162.53 &  1.92\EE{-171} &   6.83\EE{-66} &   5.01\EE{-33} &   3.80\EE{-23} &   1.56\EE{-20} &   3.20\EE{-20} \\

        11 & 2s2p(\SLP3Po)3d \SLPJ2Do{5/2} & 2s$^2$5g \SLPJ2Ge{7/2} &    5162.98 &   7.68\EE{-33} &   5.07\EE{-26} &   2.23\EE{-24} &   2.27\EE{-24} &   7.55\EE{-25} &   2.38\EE{-25} \\

        00 & 2s2p(\SLP3Po)4s \SLPJ2Po{3/2} & 2s$^2$11s \SLPJ2Se{1/2} &    5172.37 &  8.25\EE{-108} &   1.12\EE{-43} &   6.57\EE{-24} &   3.60\EE{-18} &   7.20\EE{-17} &   5.71\EE{-17} \\

        00 & 2s2p(\SLP3Po)4s \SLPJ2Po{1/2} & 2s$^2$11s \SLPJ2Se{1/2} &    5187.14 &  9.10\EE{-108} &   7.19\EE{-44} &   3.55\EE{-24} &   1.84\EE{-18} &   3.63\EE{-17} &   2.86\EE{-17} \\

        10 & 2s2p(\SLP3Po)4s \SLPJ4Po{3/2} & 2s$^2$9s \SLPJ2Se{1/2} &    5196.92 &   7.54\EE{-98} &   1.97\EE{-43} &   9.74\EE{-27} &   5.69\EE{-22} &   5.63\EE{-21} &   3.57\EE{-21} \\

        10 & 2s2p(\SLP3Po)4s \SLPJ4Po{1/2} & 2s$^2$9s \SLPJ2Se{1/2} &    5203.44 &   1.74\EE{-98} &   3.58\EE{-44} &   1.65\EE{-27} &   9.41\EE{-23} &   9.23\EE{-22} &   5.84\EE{-22} \\

        11 & 2s2p(\SLP3Po)3p \SLPJ4De{5/2} & 2s$^2$4p \SLPJ2Po{3/2} &    5203.77 &   7.63\EE{-33} &   2.44\EE{-24} &   4.07\EE{-22} &   7.12\EE{-21} &   1.35\EE{-19} &   1.71\EE{-19} \\

        11 & 2s2p(\SLP3Po)4p \SLPJ4Se{3/2} & 2s2p(\SLP3Po)3d \SLPJ4Do{1/2} &    5204.20 &  2.74\EE{-137} &   2.03\EE{-56} &   2.33\EE{-31} &   6.55\EE{-24} &   5.86\EE{-22} &   8.17\EE{-22} \\

        11 & 2s2p(\SLP3Po)4p \SLPJ4Se{3/2} & 2s2p(\SLP3Po)3d \SLPJ4Do{3/2} &    5205.70 &  1.73\EE{-136} &   1.28\EE{-55} &   1.48\EE{-30} &   4.15\EE{-23} &   3.71\EE{-21} &   5.18\EE{-21} \\

        11 & 2s2p(\SLP3Po)4p \SLPJ4Se{3/2} & 2s2p(\SLP3Po)3d \SLPJ4Do{5/2} &    5207.98 &  3.34\EE{-136} &   2.47\EE{-55} &   2.84\EE{-30} &   8.00\EE{-23} &   7.15\EE{-21} &   9.98\EE{-21} \\

        11 & 2s2p(\SLP3Po)3p \SLPJ4De{3/2} & 2s$^2$4p \SLPJ2Po{1/2} &    5208.74 &   1.87\EE{-31} &   3.80\EE{-24} &   4.99\EE{-22} &   1.05\EE{-20} &   1.62\EE{-19} &   1.83\EE{-19} \\

        11 & 2s2p(\SLP3Po)3p \SLPJ4De{3/2} & 2s$^2$4p \SLPJ2Po{3/2} &    5210.56 &   2.50\EE{-30} &   5.10\EE{-23} &   6.70\EE{-21} &   1.41\EE{-19} &   2.17\EE{-18} &   2.46\EE{-18} \\

        11 & 2s2p(\SLP3Po)3p \SLPJ4De{1/2} & 2s$^2$4p \SLPJ2Po{1/2} &    5212.65 &   1.15\EE{-30} &   2.43\EE{-23} &   3.42\EE{-21} &   1.01\EE{-19} &   1.06\EE{-18} &   8.41\EE{-19} \\

        11 & 2s2p(\SLP3Po)3p \SLPJ4De{1/2} & 2s$^2$4p \SLPJ2Po{3/2} &    5214.46 &   5.52\EE{-31} &   1.17\EE{-23} &   1.64\EE{-21} &   4.85\EE{-20} &   5.09\EE{-19} &   4.03\EE{-19} \\

        11 & 2s2p(\SLP3Po)4p \SLPJ4De{7/2} & 2s2p(\SLP3Po)3d \SLPJ4Fo{5/2} &    5244.05 &  1.17\EE{-158} &   1.89\EE{-61} &   3.48\EE{-31} &   4.16\EE{-22} &   9.58\EE{-20} &   1.65\EE{-19} \\

        11 & 2s2p(\SLP3Po)4p \SLPJ4De{7/2} & 2s2p(\SLP3Po)3d \SLPJ4Fo{7/2} &    5249.53 &  2.53\EE{-157} &   4.08\EE{-60} &   7.50\EE{-30} &   8.98\EE{-21} &   2.07\EE{-18} &   3.55\EE{-18} \\

        11 & 2s2p(\SLP3Po)4p \SLPJ4De{5/2} & 2s2p(\SLP3Po)3d \SLPJ4Fo{3/2} &    5249.90 &  1.29\EE{-126} &   6.68\EE{-50} &   3.73\EE{-26} &   3.70\EE{-19} &   1.86\EE{-17} &   1.97\EE{-17} \\

        11 & 2s2p(\SLP3Po)4p \SLPJ4De{5/2} & 2s2p(\SLP3Po)3d \SLPJ4Fo{5/2} &    5253.56 &  2.49\EE{-125} &   1.29\EE{-48} &   7.20\EE{-25} &   7.15\EE{-18} &   3.59\EE{-16} &   3.80\EE{-16} \\

        11 & 2s2p(\SLP3Po)4p \SLPJ4De{3/2} & 2s2p(\SLP3Po)3d \SLPJ4Fo{3/2} &    5256.09 &  2.63\EE{-125} &   1.09\EE{-48} &   5.69\EE{-25} &   5.52\EE{-18} &   2.75\EE{-16} &   2.91\EE{-16} \\

        11 & 2s2p(\SLP3Po)4p \SLPJ4De{7/2} & 2s2p(\SLP3Po)3d \SLPJ4Fo{9/2} &    5257.24 &  2.45\EE{-156} &   3.96\EE{-59} &   7.27\EE{-29} &   8.71\EE{-20} &   2.01\EE{-17} &   3.44\EE{-17} \\

        11 & 2s2p(\SLP3Po)4p \SLPJ4De{5/2} & 2s2p(\SLP3Po)3d \SLPJ4Fo{7/2} &    5259.06 &  1.26\EE{-124} &   6.53\EE{-48} &   3.64\EE{-24} &   3.61\EE{-17} &   1.81\EE{-15} &   1.92\EE{-15} \\

        11 & 2s2p(\SLP3Po)4p \SLPJ4De{1/2} & 2s2p(\SLP3Po)3d \SLPJ4Fo{3/2} &    5259.66 &  1.46\EE{-126} &   5.35\EE{-50} &   2.67\EE{-26} &   2.65\EE{-19} &   1.51\EE{-17} &   1.78\EE{-17} \\
\hline
\end{tabular}
}
\end{center}
\end{table}

\chapter{Input Data for DVR3D Code} \label{AppInDataM}
In this appendix we present a sample of the DVR3D input data files that have been used to generate
the \htdp\ line list. This is for the purpose of thoroughness and to enable the interested
researcher to reproduce and check our results. The contents of the missing files can be easily
deduced by comparing the given files. It should be remarked that the structure of the DVR3D data
files is fully explained in the write-up of this code as given by Tennyson \etal\
\cite{TennysonKBHPRZ2004}.

\section{Input Files for Stage DVR3DRJZ} \label{DVR3DRJZ}

For $J=0$ even

{\tiny
\begin{spacing}{1}
\begin{verbatim}
 &PRT ZR2R1=.FALSE.,
      zembed=.false.,zmors2=.false.,ztran=.true., /
    3
   44    0 2000  36 1000 2000    2    1   20    0
 H2D+: J=0

        2.013814000         1.007537200         1.007537200
        2.013553200         1.007276470         1.007276470
       66000.000         45000.000000
         1.7100              0.10000             0.01080
         0.0000              0.00000             0.00750
 4073.99136362
\end{verbatim}
\end{spacing}
}

For $J=0$ odd

{\tiny
\begin{spacing}{1}
\begin{verbatim}
 &PRT ZR2R1=.FALSE.,
      zembed=.false.,zmors2=.false.,ztran=.true., /
    3
   44    0 2000  36 1000 2000    2    1   20    1
 H2D+: J=0

        2.013814000         1.007537200         1.007537200
        2.013553200         1.007276470         1.007276470
       66000.000         45000.000000
         1.7100              0.10000             0.01080
         0.0000              0.00000             0.00750
 4073.99136362
\end{verbatim}
\end{spacing}
}

For $J=10$ even

{\tiny
\begin{spacing}{1}
\begin{verbatim}
 &PRT ZR2R1=.FALSE.,
      zembed=.false.,zmors2=.false.,ztran=.true., /
    3
   44   10 2000  36 1000 2000    2    1   20    0
 H2D+: J=10

        2.013814000         1.007537200         1.007537200
        2.013553200         1.007276470         1.007276470
       66000.000         45000.000000
         1.7100              0.10000             0.01080
         0.0000              0.00000             0.00750
 4073.99136362
\end{verbatim}
\end{spacing}
}

For $J=10$ odd

{\tiny
\begin{spacing}{1}
\begin{verbatim}
 &PRT ZR2R1=.FALSE.,
      zembed=.false.,zmors2=.false.,ztran=.true., /
    3
   44   10 2000  36 1000 2000    2    1   20    1
 H2D+: J=10

        2.013814000         1.007537200         1.007537200
        2.013553200         1.007276470         1.007276470
       66000.000         45000.000000
         1.7100              0.10000             0.01080
         0.0000              0.00000             0.00750
 4073.99136362
\end{verbatim}
\end{spacing}
}

For $J=20$ even

{\tiny
\begin{spacing}{1}
\begin{verbatim}
 &PRT ZR2R1=.FALSE.,
      zembed=.false.,zmors2=.false.,ztran=.true., /
    3
   44   20 2000  36 1000 2000    2    1   20    0
 H2D+: J=20

        2.013814000         1.007537200         1.007537200
        2.013553200         1.007276470         1.007276470
       66000.000         45000.000000
         1.7100              0.10000             0.01080
         0.0000              0.00000             0.00750
 4073.99136362
\end{verbatim}
\end{spacing}
}

For $J=20$ odd

{\tiny
\begin{spacing}{1}
\begin{verbatim}
 &PRT ZR2R1=.FALSE.,
      zembed=.false.,zmors2=.false.,ztran=.true., /
    3
   44   20 2000  36 1000 2000    2    1   20    1
 H2D+: J=20

        2.013814000         1.007537200         1.007537200
        2.013553200         1.007276470         1.007276470
       66000.000         45000.000000
         1.7100              0.10000             0.01080
         0.0000              0.00000             0.00750
 4073.99136362
\end{verbatim}
\end{spacing}
}

\section{Input Files for Stage ROTLEV3} \label{ROTLEV3}

For $J=9$

{\tiny
\begin{spacing}{1}
\begin{verbatim}
 &PRT toler = 1.0d-4, ztran=.true.,zdcore=.true.,zpfun=.true. /
2000 585  2    18000
  Rotation for H2D+ J=9
 4073.99136362
\end{verbatim}
\end{spacing}
}

For $J=13$

{\tiny
\begin{spacing}{1}
\begin{verbatim}
 &PRT toler = 1.0d-4, ztran=.true.,zdcore=.true.,zpfun=.true. /
2000 550  2    25200
  Rotation for H2D+ J=13
 4073.99136362
\end{verbatim}
\end{spacing}
}

For $J=17$

{\tiny
\begin{spacing}{1}
\begin{verbatim}
 &PRT toler = 1.0d-4, ztran=.true.,zdcore=.true.,zpfun=.true. /
2000 407  2    32400
  Rotation for H2D+ J=17
 4073.99136362
\end{verbatim}
\end{spacing}
}

\section{Input Files for Stage DIPOLE} \label{DIPOLE}

{\tiny
\begin{spacing}{1}
\begin{verbatim}
&PRT   zprint=.true.,zstart=.true., /
  Dipole Transitions for H2D+
  50
 4073.99136362
\end{verbatim}
\end{spacing}
}

\section{Input Files for Stage SPECTRA} \label{SPECTRA}

For $T=300$~K

{\tiny
\begin{spacing}{1}
\begin{verbatim}
 &PRT zsort=.true., zpfun=.false., GZ=4073.99136362  /
Spectra for dipole moment of h2d+, T=300k
     1.0     3.0
 300.0      0.00001     1.0     18000.0    1.0     76.485
 &SPE zemit=.false.,zplot=.true.,zfreq=.true.,zene=.true.,zprof=.false.,prthr=0.0d0, /
\end{verbatim}
\end{spacing}
}

\section{Input Files for Spectra-BT2} \label{SpectraBT2}

These files are used to generate temperature-dependent synthetic spectra data used in plotting
Figure~\ref{SynthSpec}.

For $T=100$~K

{\tiny
\begin{spacing}{1}
\begin{verbatim}
&PRT zsort=.true., zout=.false., zpfun=.false., GZ=0.0,smin=1.0d-28, emin=-1.0d10, emax=1.0d5,
wsmax=10000.0d0, wsmin=1000.0d0,zband=.false., / Spectra_H2D+_at_100K
       1.0       3.0
     100.0    1.0d-4    1000.0   10000.0     0.500    15.119
 &SPE  zplot=.true.,zlist=.true.,z90=.false.,zassign=.true.,zemit=.false.,emax1=1.0d6,emax2=1.0d6,
 zprof=.true.,zdop=.false.,npoints=90000,zfreq=.true.,jmax=20 /
\end{verbatim}
\end{spacing}
}

For $T=500$~K

{\tiny
\begin{spacing}{1}
\begin{verbatim}
&PRT zsort=.true., zout=.false., zpfun=.false., GZ=0.0,smin=1.0d-28, emin=-1.0d10, emax=1.0d5,
wsmax=10000.0d0, wsmin=1000.0d0,zband=.false., / Spectra_H2D+_at_500K
       1.0       3.0
     500.0    1.0d-4    1000.0   10000.0     0.500    164.53
 &SPE  zplot=.true.,zlist=.true.,z90=.false.,zassign=.true.,zemit=.false.,emax1=1.0d6,emax2=1.0d6,
 zprof=.true.,zdop=.false.,npoints=90000,zfreq=.true.,jmax=20 /
\end{verbatim}
\end{spacing}
}

For $T=1000$~K

{\tiny
\begin{spacing}{1}
\begin{verbatim}
&PRT zsort=.true., zout=.false., zpfun=.false., GZ=0.0,smin=1.0d-28, emin=-1.0d10, emax=1.0d5,
wsmax=10000.0d0, wsmin=1000.0d0,zband=.false., / Spectra_H2D+_at_1000K
       1.0       3.0
    1000.0    1.0d-4    1000.0   10000.0     0.500    516.31
 &SPE  zplot=.true.,zlist=.true.,z90=.false.,zassign=.true.,zemit=.false.,emax1=1.0d6,emax2=1.0d6,
 zprof=.true.,zdop=.false.,npoints=90000,zfreq=.true.,jmax=20 /
\end{verbatim}
\end{spacing}
}

For $T=2000$~K

{\tiny
\begin{spacing}{1}
\begin{verbatim}
&PRT zsort=.true., zout=.false., zpfun=.false., GZ=0.0,smin=1.0d-28, emin=-1.0d10, emax=1.0d5,
wsmax=10000.0d0, wsmin=1000.0d0,zband=.false., / Spectra_H2D+_at_2000K
       1.0       3.0
    2000.0    1.0d-4    1000.0   10000.0     0.500   2487.98
 &SPE  zplot=.true.,zlist=.true.,z90=.false.,zassign=.true.,zemit=.false.,emax1=1.0d6,emax2=1.0d6,
 zprof=.true.,zdop=.false.,npoints=90000,zfreq=.true.,jmax=20 /
\end{verbatim}
\end{spacing}
}

\onehalfspace

\phantomsection \addcontentsline{toc}{chapter}{\protect \numberline{} Index} %
\printindex

\end{document}